\documentclass[openright,twoside,a4paper,english,12pt]{book}

\renewcommand{\baselinestretch}{1.5}

\usepackage{microtype}
\usepackage[T1]{fontenc}
\usepackage[utf8]{inputenc}
\usepackage[english]{babel}

\usepackage[a4paper,inner=4cm,outer=2.5cm,twoside]{geometry}

\usepackage{layout}
\usepackage{lineno}
\usepackage{setspace}

\usepackage{enumitem}

\usepackage{graphicx}
\usepackage{subcaption}
\usepackage{listings,multicol}

\usepackage[toc,acronym,nonumberlist]{glossaries}
\usepackage[toc,page]{appendix}

\usepackage[textfont={small},labelfont={small,bf}]{caption}

\usepackage[nottoc]{tocbibind}

\usepackage{comment}
\usepackage{todonotes}

\usepackage{xcolor}

\usepackage[ruled]{algorithm2e}

\usepackage{pdfpages}
\usepackage{atbegshi}

\usepackage{csquotes}
\usepackage[
    natbib=true,
    backend=biber,
    bibstyle=phys,
    biblabel=brackets,
    url=false,
    giveninits=true,
    uniquename=false,
    uniquelist=false,
    minbibnames=6,
    maxbibnames=6,
    maxcitenames=2,
    sorting=none,
]{biblatex}

\DeclareSourcemap{
  \maps[datatype=bibtex]{
    \map{
      \step[fieldsource=author, match=\regexp{(.*?)\s+and\s+}]
      \step[fieldset=sortname, fieldvalue={$1}]
    }
    \map{ 
        \step[fieldsource=doi,
              match=\regexp{\{\\\_\}|\{\_\}|\\\_},
              replace=\regexp{\_}]
        }
    \map{ 
        \step[fieldsource=doi,
              match=\regexp{\{\$\\sim\$\}|\{\~\}|\$\\sim\$},
              replace=\regexp{\~}]
    }
  }
}

\renewbibmacro*{doi+eprint+url}{%
    \printfield{doi}%
    \newunit\newblock%
    \iftoggle{bbx:eprint}{%
        \usebibmacro{eprint}%
    }{}%
    \newunit\newblock%
    \iffieldundef{doi}{%
        \usebibmacro{url+urldate}}%
        {}%
    }

\DeclareBibliographyDriver{misc}{%
    \printnames{author}%
    \newunit%
    \printfield{title}%
    \newunit%
    \printfield{url}%
    \printfield{booktitle}
    \newunit%
    \printfield{issn}%
    \newunit%
    \printfield{doi}%
    \finentry
}

\AtEveryBibitem{\clearfield{issn}}

\DeclareDelimFormat[cbx@textcite]{nameyeardelim}{\addspace}

\makeglossaries

\newcommand{\glossterm}[2]{
    \newglossaryentry{#1}{
        name=#1,
        description={#2}
    }
}

\glossterm{HPC}{High-Performance Computing}
\glossterm{FLOPS}{Floating-Point Operations per Second}
\glossterm{AI}{Artificial Intelligence}
\glossterm{I/O}{Input-Output}
\glossterm{PCIe}{Peripheral Component Interconnect Express}
\glossterm{SSD}{Solid-State Drive}
\glossterm{HDD}{Hard Disk Drive}
\glossterm{DIMM}{Dual Inline Memory Module}
\glossterm{DNE}{Distributed Name Space}
\glossterm{DRAM}{Dynamic Random-Access Memory}
\glossterm{SCM}{Storage Class Memory}
\glossterm{S3}{Amazon Simple Storage Protocol}
\glossterm{API}{Application Programming Interface}
\glossterm{NWP}{Numerical Weather Prediction}
\glossterm{ECMWF}{European Centre for Medium-Range Weather Forecasts}
\glossterm{NVMe}{Non-Volatile Memory Express}
\glossterm{NVM}{Non-Volatile Memory}
\glossterm{PMEM}{Persistent Memory}
\glossterm{DCPMM}{Data Center Persistent Memory Module}
\glossterm{POSIX}{Portable Operating System Interface}
\glossterm{NFS}{Network File System}
\glossterm{RDMA}{Remote Direct Memory Access}
\glossterm{MDS}{Metadata Server}
\glossterm{OSS}{Object Storage Server}
\glossterm{OST}{Object Storage Target}
\glossterm{DAOS}{Distributed Asynchronous Object Store}
\glossterm{OFI}{OpenFabrics Interface}
\glossterm{OID}{Object Identifier}
\glossterm{FUSE}{File System in User Space}
\glossterm{MVCC}{Multi-Version Concurrency Control}
\glossterm{ACL}{Access Control List}
\glossterm{OSD}{Object Storage Daemon}
\glossterm{RADOS}{Reliable Autonomous Distributed Object Store}
\glossterm{PG}{Placement Group}
\glossterm{ACID}{Atomicity, Consistency, Isolation, and Durability}
\glossterm{URI}{Uniform Resource Identifier}
\glossterm{TOC}{Table Of Contents}
\glossterm{PGEN}{Product Generation}
\glossterm{RPC}{Remote Procedure Call}
\glossterm{REST}{Representative State Transfer}
\glossterm{HTTP}{Hypertext Transfer Protocol}
\glossterm{AWS}{Amazon Web Services}
\glossterm{SDK}{Software Development Kit}
\glossterm{GCP}{Google Cloud Platform}
\glossterm{VM}{Virtual Machine}
\glossterm{MTU}{Maximum Transmission Unit}
\glossterm{RAID}{Redundant Array of Independent Disks}
\glossterm{IOPS}{I/O Operations Per Second}
\glossterm{HDF5}{Hierarchical Data Format 5}
\glossterm{CI/CD}{Continuous Integration and Continuous Development}
\glossterm{SMT}{Simultaneous Multi-Threading}
\glossterm{WAL}{Write-Ahead Log}

\glsaddall

\begin{document}

    \includepdf{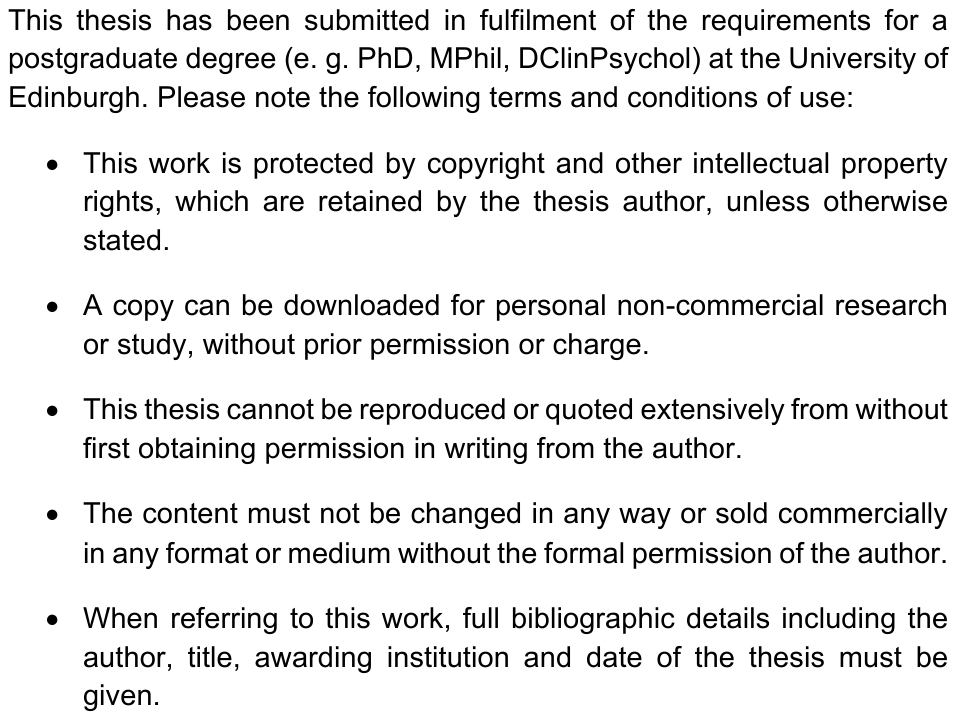}
    \AtBeginShipoutNext{\AtBeginShipoutDiscard}

    \frontmatter
    \begin{titlepage}
  \vspace*{\fill}
    \begin{center}
        \Huge{\textsc{\textbf{Exploring Novel Data Storage Approaches for Large-Scale Numerical Weather Prediction}}}\\[2cm]
        \Large{\textsc{Nicolau Manubens Gil}}\\[2.5cm]
        \includegraphics[width=0.3\linewidth]{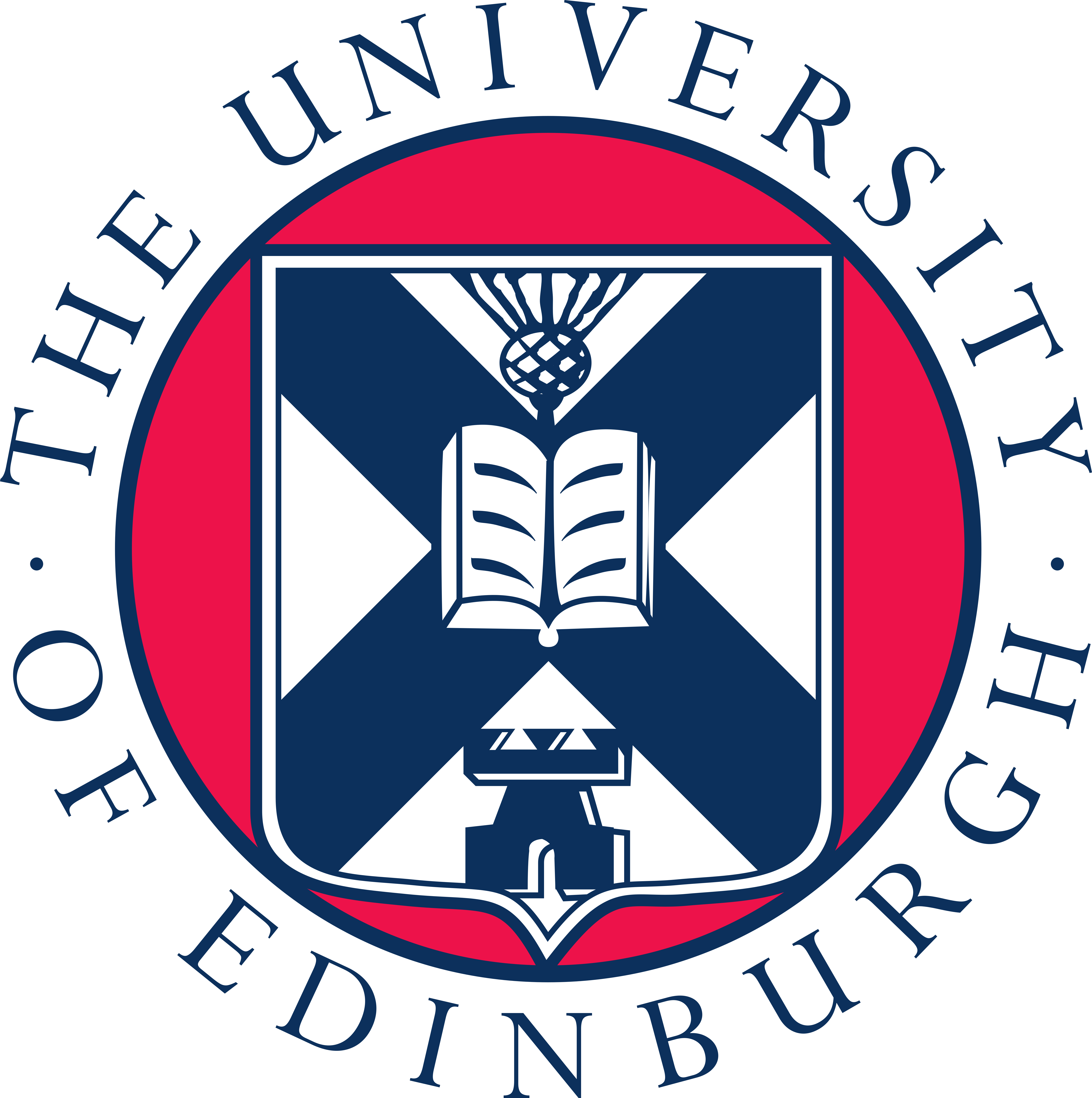}\\[2cm]
        \large{Submitted for the degree of Doctor of Philosophy}\\
        \large{\textsc{The University of Edinburgh\\2025}}\\[0cm]
    \end{center}
  \vspace*{\fill}
\end{titlepage}

    \newpage
\noindent {\Huge \bf Abstract}\\ \\

Driven by scientific and industry ambition, HPC and AI applications such as operational Numerical Weather Prediction (NWP) require processing and storing ever-increasing data volumes as fast as possible.
Whilst POSIX distributed file systems and NVMe SSDs are currently a common HPC storage configuration providing I/O to applications,~new storage solutions have proliferated or gained traction over the last decade with potential to address performance limitations POSIX file systems manifest at scale for certain I/O workloads.

This work has primarily aimed to assess the suitability and performance of two object storage systems ---namely DAOS and Ceph--- for the ECMWF's operational~NWP as well as for HPC and AI applications in general.
New software-level adapters have been developed which enable the ECMWF's NWP to leverage these systems, and extensive I/O benchmarking has been conducted on a few computer systems, comparing the performance delivered by the evaluated object stores to that of equivalent Lustre file system deployments on the same hardware.
Challenges of porting to object storage and its benefits with respect to the traditional POSIX I/O approach have been discussed and, where possible, domain-agnostic performance analysis has been conducted, leading to insight also of relevance to I/O practitioners and the broader HPC community.

DAOS and Ceph have both demonstrated excellent performance, but DAOS stood out relative to Ceph and Lustre, providing superior scalability and flexibility for applications to perform I/O at scale as desired. This sets a promising outlook for DAOS and object storage, which might see greater adoption at HPC centres in the years to come, although not necessarily implying a shift away from POSIX-like I/O.

\newpage
\noindent{\Huge \bf Lay summary}\\ \\

All consumer-level computers and cellphones nowadays keep data such as images and text in what are commonly known as files and directories, and these are in turn kept in hard disk drives (HDDs) and solid-state drives (SSDs) in computers, or in internal flash memory in cellphones.
Files and directories were invented in the early years of computer systems and are both a necessary concept for computer systems to function, and a practical way for us humans to interact with computers.

Larger computers, for example in big research institutions, also use files and directories, but because the data volumes involved are so large, these are kept across many dedicated computers with several HDDs or SSDs each.
The problems that need to be solved in such large computers have progressively become more ambitious and difficult, and scientists and programmers are struggling to move the data fast enough in and out between the large computer system and the dedicated storage computers.
This is the case for example at the European Centre for Medium-Range Weather Forecasts (ECMWF), where very large simulations of the Earth system are run to predict the weather.

Different approaches to this issue are being taken in all fields of research and industry, such as producing and storing less data where possible, compressing them, and improving the programs or applications running on the large computer as well as those managing the the dedicated storage computers to make the most of the available resources.
Some experts say that this slow data movement issue is in part due to using files and directories, which were designed for small computers in the early days, and are now not suitable anymore.
Some have worked on a new alternative way of encapsulating and storing data in so called \textit{objects}, leaving behind the files and directories and all the associated standards and legacy programs, which they deemed a limiting factor.
One well-known example of this was the early adoption of object-based storage by the Amazon company, in 2006, to manage and provide access to huge amounts of data distributed across their data centers.

Both file systems and object storage systems are very complex, with many moving parts, and it is therefore not easy to estimate the benefits of using one over the other, least when object storage requires carefully adapting the existing programs or applications to the mechanisms provided by this new paradigm.

This work has adapted the ECMWF's operational weather forecasting software to function on two open-source object storage solutions ---namely DAOS and Ceph---, and evaluated their suitability and performance benefits, both for the ECMWF's use case and for other applications similarly entailing large data movement.
While both were found to be suitable and able to provide advantages with respect to file systems, DAOS stood out reaching higher performance levels.
This work shed some light on the performance aspect of object storage, and opened new doors for the ECMWF for future computer and storage system acquisitions.

\newpage
\noindent{\Huge \bf Declaration}\\ \\

I declare that this thesis has been composed solely by myself, and that it contains only my work except where otherwise specified, or where the work is explicitly indicated below to have formed part of a jointly-authored publication. This work has not been submitted for any other degree or professional qualification.\newline

\noindent

\begin{flushright}
\hspace*{\fill}Nicolau Manubens Gil\newline
July 2025
\end{flushright}

\newpage
\noindent{\Huge \bf Acknowledgements}\\ \\

This work has been conducted as part of a collaboration between the ECMWF and the EPCC at The University of Edinburgh. I am very grateful to both institutions for this unique opportunity.

Special thank you to Adrian Jackson for his constant guidance and patience during all these years, his invaluable insight, and his very substantial technical and paper contribu\-tions, as well as seeking systems where to conduct the analysis. It has been a privilege and a pleasure to have such an involved supervisor.

Special thank you to Tiago Quintino and Simon Smart; Tiago, for proposing this research idea and making this project possible, and for his invaluable guidance, insight, and contributions during these years; Simon, for his guidance, brilliant insight, and substantial contributions to the published papers and technical developments --- particularly, the implementation of the FDB DAOS backends.

Thank you to Emanuele Danovaro for his insight, contributions, and dedication to the software infrastructure surrounding the FDB backends, without which this research would not have been possible.

Thank you to all the above and the rest of my colleagues at the ECMWF, from whom I have had the privilege of learning so much, and have all contributed to an ideal atmosphere for technical discussion and professional growth.

Thank you to the DAOS development team, especially Johann Lombardi and Mohamad Chaarwi, for their invaluable help and inputs provided for the development and evaluation of the FDB DAOS backends.

Thank you to Google Cloud and Dean Hildebrand for providing resources to conduct large part of the analysis, and for their help and insight.

\clearpage
Thank you to David Henty and Michele Guidolin for thoroughly examining this thesis and for their very valuable feedback, and thank you to Michael Bareford for his follow-up and feedback.

Last but most subjectively important, thank you to my parents, siblings, and close friends for their support and patience during these challenging years. Special thank you to my mother Isabel Gil González, my father Domingo Manubens Bertran, and my brother Domingo Manubens Gil.

\newpage

\begin{spacing}{1}
    \tableofcontents
    \listoffigures
    \listoftables
    \printglossary
\end{spacing}

    \mainmatter

    \chapter{Introduction}

Driven by scientific and industry ambition, computer systems have steadily and rapidly evolved since their inception nearly eight decades ago, with frontline High-Performance Computing (HPC) systems currently providing on the order of one exaFLOPS (10\textsuperscript{18} floating-point operations per second) of processing power and tens of TB/s (10\textsuperscript{12} bytes per second) of storage bandwidth\cite{TimelineComputers,TimelineStorage,WernerMeuerTOP500Lists,Bez2024IO500Lists}.
Traditionally, HPC practitioners have striven to adopt new hardware technology and adapt software and applications to efficiently leverage the increasingly capable resources, facing new challenges along the way as the technology landscape evolved.

As we enter the Exascale era, one of the most prominent challenges, besides handling the unprecedented high degree of complexity and heterogeneity of HPC systems, is that of handling the extremely large and ever-increasing data volumes that the greater processing power now makes possible to process and produce. Although this challenge was also present to an extent in previous epochs, it is more marked this time due to a bigger leap in processing power unmatched by storage capability, and due to certain data access patterns becoming more commonly used, mainly by Artificial Intelligence (AI) applications, which can add significant strain on current storage systems\cite{Acquaviva2025ETP4HPCStorage}.

In consequence, HPC systems ---both large and small--- now need to be more carefully designed to achieve the right balance between processing and storage capability, maximizing overall system performance while minimizing total cost of ownership.
This requires exploring best ways of combining the storage hardware options available in the market, as well as ensuring the software layer of the storage system ---typically a parallel file system nowadays--- is aware and makes efficient use of the hardware topology.
Simultaneously, an additional effort is required to ensure HPC and AI applications make the most of the available storage resources by revisiting and if necessary optimizing their Input-Output (I/O) logic, or even shifting to new I/O approaches both at the application and storage software levels.

Regarding storage hardware, despite the current gap between processing and storage capability, there certainly have been advances in the recent years impacting the design of modern-day HPC systems.
Most remarkably, there were upgrades in the Peripheral Component Interconnect Express (PCIe) standard and in network technologies such as Ethernet and Infiniband, enabling large sets of storage devices such as Solid-State Drives (SSDs) distributed over a high-performance network to provide higher aggregate I/O throughput than ever before.
Currently, SSDs are usually attached to a set of nodes in the HPC system devoted to high-performance storage and exposed as a single view to compute nodes, but are also often used in other configurations including compute-node-local storage, fast caching tiers interfacing slower storage devices such as Hard Disk Drives (HDDs), or pools of disaggregated fast storage devices which can be mounted and exploited individually by any one node in the network.
HDDs, on the other hand, were relegated to providing affordable capacity in high-performance storage systems with fast SSD-based cache tiers, as well as providing fast caching themselves in slower tape-based data archives\cite{Luttgau2018SurveyComputing}.

Also noteworthy, a new non-volatile memory technology, namely 3D Xpoint, was developed and marketed in PCIe and Dual Inline Memory Module (DIMM) form factors, offering a new latency and capacity trade-off and thus filling the gap in the storage hierarchy between Dynamic Random-Access Memory (DRAM) and SSDs\cite{Weiland2019AnApplications}. Storage devices within this echelon of the hierarchy are also commonly referred to as Storage Class Memory (SCM).
3D Xpoint, however, was unfortunately discontinued in 2022 due to a lack of demand.

Regarding storage software, parallel file systems such as Lustre and IBM Storage Scale remained the most widespread software-level approach for high-performance storage in HPC systems. These file systems also evolved in recent years with new capability and performance enhancements, and were made able to exploit fast storage hardware tiers as those became available.
Despite their popularity, parallel file systems can manifest performance limitations at scale for certain I/O workloads due to fundamental design aspects (see Chapter \ref{chap:chap2}), calling for I/O optimization at the application level ---which is often challenging--- or even consideration of alternative or complementary storage software approaches to ensure efficient usage of the hardware resources.

In that respect, of particular relevance to this work, the past two decades have seen the object storage paradigm flourish, giving way to technologies such as the S3 (Amaon Simple Storage Service) protocol and the Ceph and DAOS object stores, which all gained traction both in Cloud and HPC environments.
Object stores break away from the parallel file system standards, implementations, and approaches, and offer new Application Programming Interfaces (APIs) and semantics revolving around objects, which hold blobs of data of any size and are lightweight in metadata.
They also generally provide key-value or dictionary objects such that entries can be inserted ---each associating a key to a value--- and queried with strong consistency guarantees.
To varying extents, on a case-by-case basis, object stores are designed in ways which potentially enable them to overcome the performance limitations of parallel file systems at scale.

All these hardware and software developments facilitated storage tiering and boosted the storage capability per node and per cost unit, ultimately providing means to combat the imbalance relative to processing power. Nevertheless, the challenge still persists and requires constant review of the I/O logic and exploration of new storage approaches at HPC centres.

One exemplar HPC application this work has particularly focused on is large-scale Numerical Weather Prediction (NWP) as conducted operationally at the European Centre for Medium-Range Weather Forecasts (ECMWF).
The application produces large amounts of time-critical forecast data which need to be stored into the distributed file system that is currently part of their HPC system, and simultaneously read back and processed during an operational run to then be disseminated to stakeholders within strict deadlines.
The large scale of the simulation and the high degree of write-read contention require the various components of the application to be implemented following certain I/O programming best practices to ensure high performance and data consistency.

Over the past decades, the ECMWF has developed and used operationally a middleware I/O library, namely the FDB, which abstracts away the I/O programming complexity and reduces the overall software complexity of the application. The FDB has been iteratively and aggressively optimised to make as efficient use as possible of the storage resources available.
Notwithstanding these efforts, at present, a significantly large and costly storage system is still necessary to handle the operational I/O workload with enough margin to avoid performance issues that can manifest under excessive contention.

Looking forward, the ECMWF plans to increase the resolution of the operational simulations in the coming years\cite{Bauer2015ThePrediction}, likely giving continuity to the so far exponential growth in data volumes and further exacerbating the I/O performance issue.
Because the ECMWF newly procures its HPC system every four years, there might be opportunities in the future to increase I/O capability and re-balance the system, potentially mitigating this.

In light of these facts, exploration and evaluation of new storage hardware and alternative or complementary storage software approaches have become paramount duties for the success of the ECMWF's mission.

Latest generations of storage hardware should be evaluated and adopted, even though new generations of processing hardware may potentially offset the benefits of the former.
Regarding storage software, given parallel file systems can manifest performance limitations at scale, simply procuring one of larger size than the current might be less cost-effective than adopting other more efficient storage software approaches, if there were any.

This dissertation work has primarily aimed to evaluate the suitability and performance of the Ceph and DAOS object stores, both specifically for the ECMWF's operational NWP as well as for HPC applications in general.
This work has also aimed at evaluating, on one hand, the performance of 3D Xpoint DIMM devices and, on the other hand, the options and challenges of porting the ECMWF's NWP to the S3 storage protocol for additional compatibility with other storage solutions.

In the course of this work, new software-level adapters have been developed which enable the FDB, and by extension the ECMWF's NWP workflows, to operate on the evaluated storage systems and protocols. Extensive benchmarking has been conducted on two HPC systems where Ceph and DAOS were deployed, using both generic and NWP I/O benchmarks, and the observed performance has been compared to that of equivalent Lustre parallel file system deployments on the same hardware. The challenges and advantages of porting to object storage have been evaluated along the way.

\section{Contributions}

The contributions of this work consist of:

\begin{enumerate}[leftmargin=*,label=\Alph*]

    \item{\textbf{Documentation of the ECMWF's storage software stack}:}
        The ECMWF's operational systems, applications, and access patterns have been further documented, primarily focusing on the FDB library and its techniques to make efficient use of parallel file systems at scale. Whilst already covered in existing literature\cite{Smart2017AData,Smart2019AClimate,Weiland2019AnApplications}, this thesis has more thoroughly documented these matters with additional information collated from discussion with a number of the ECMWF's staff members, review of internal documentation, and analysis of the FDB's source code and applications. This will provide useful information and guidance for future I/O-related developments at the centre.

        Part of this work has been presented as a poster at a technical meeting\cite{Manubens2024ECMWFsData}.
        
    \item{\textbf{Object storage adapters for the ECMWF's NWP workflows}:}
        New adapters, or backends, have been developed which enable the FDB, and by extension the ECMWF's NWP workflows, to operate on the DAOS and Ceph object stores, as well as on storage systems conforming to the S3 storage protocol. This has opened new doors for the ECMWF to consider and exploit these storage solutions if made available in future systems, projects, or collaborations. The different options and challenges encountered in the course of these developments have been discussed, providing insight of relevance to other HPC practitioners needing to port applications to these storage systems and protocols.

        Best practices for porting HPC applications to object stores have been documented and presented at a technical user meeting\cite{Jackson2023ProfilingDAOS} and as part of a conference tutorial\cite{Jackson2024High-PerformanceEra}. The development of the FDB DAOS backend has been documented and published in a conference paper\cite{Manubens2024ReducingDAOS}, and presented in a scientific workshop\cite{Lombardi2021AcceleratingDAOS} and a technical user meeting\cite{Manubens2021AssessmentFDB}. The development of the other FDB backends has been documented in this dissertation, and the source code of all backends has been made publicly available in the ECMWF's FDB GitHub repository\cite{FDB}.

    \item{\textbf{I/O performance assessment methodology}:}
        A methodology has been defined for the performance assessment of I/O systems and applications, which provides guidance on how to approach exploration and optimization of the extensive parameter domain in this type of assessments. It is sufficiently general that it can be used to guide assessment of other I/O systems and developments.
        
    \item{\textbf{Suitability and performance assessment of Ceph and DAOS for the ECMWF's operational NWP}:}
        A comprehensive analysis of the Ceph and DAOS object stores including functionality, consistency, and performance aspects has been conducted, aiming to determine the suitability of these stores for operational NWP at the ECMWF. The types of storage hardware tested included Non-Volatile Memory Express (NVMe) SSDs and 3D Xpoint DIMMs. This analysis has resulted in relevant insight for the ECMWF to consider for future HPC system procurements or other acquisitions.
        
        Part of this work has been published in a conference paper\cite{Manubens2023DAOSPrediction} and a workshop paper\cite{Manubens2024ExploringPerformance}, and the rest has been documented in this dissertation.

    \item{\textbf{Fair performance comparison of DAOS, Ceph, and Lustre}:}
        The DAOS, Ceph, and Lustre storage systems have been deployed on the same hardware and exercised both with generic and NWP I/O benchmarks, and their performance and scalability have been measured and compared. This had not been documented previously in such a rigorous apples-to-apples comparison, and will provide insight to the HPC community and I/O practitioners.

        Part of this work has been published in the workshop paper mentioned in the second contribution and in another workshop paper\cite{Manubens2022PerformanceApproaches}, and presented at a technical user meeting\cite{Jackson2024PerformanceSharding}. The rest has been documented in this dissertation.
    
\end{enumerate}

\section{Outline}

Chapter 2 provides background information on the storage systems evaluated in this work, and compares them in terms of their features and limitations. It also provides a detailed explanation of the ECMWF's operational systems, workflows, and storage software stack, particularly focusing on the FDB library and its backend for efficient operation on parallel file systems.

Chapter 3 describes the newly developed FDB backends that enable operation on the selected object stores and storage protocols, including discussion on the options and challenges encountered along the development process.

Chapter 4 presents an I/O performance assessment of the DAOS and Ceph object stores, including performance and scalability results obtained from running various benchmarks against these object stores deployed on two different HPC systems. The performance of the object stores is compared to that of equivalent Lustre file system deployments on the same hardware. The analysis includes a performance evaluation of the data redundancy features in DAOS and Ceph, and an analysis of the performance impact of using different I/O interfaces provided by DAOS.

Chapter 5 concludes with a summary of the most relevant findings and higher-level conclusions, and suggests avenues for future work.

The conference and workshop papers mentioned in 1.1 Contributions have been inserted in this dissertation as Appendix A\cite{Manubens2022PerformanceApproaches}, Appendix B\cite{Manubens2023DAOSPrediction}, Appendix C\cite{Manubens2024ReducingDAOS}, and Appendix D\cite{Manubens2024ExploringPerformance}, as these document a substantial part of the accomplished work and contributions. The most relevant material and results in these papers have been included and discussed again in the body of this dissertation, and the dissertation further expands those with additional results and discussion. The rest of the results in the papers have been referenced where relevant with a pointer to the corresponding paper and section.
    \chapter{Background}
\label{chap:chap2}

This chapter introduces concepts relevant to the understanding of the contributions of this dissertation, ranging from file systems, through high-performance object stores, to the ECMWF's current storage libraries used for operational NWP. This also includes a comparison of features and limitations across file systems and object stores, as well as a review of related work.

Part of this chapter therefore addresses the first contribution of the dissertation, \textit{"Documentation of ECMWF's storage software stack"}.

Beyond the concepts introduced in this chapter, this dissertation also assumes prior knowledge of high-performance computer and storage systems. An excellent overview of some of the most relevant introductory concepts is given in Section 2 of Sarah Neuwirth's PhD. dissertation\cite{Neuwirth2019AcceleratingEnvironments}.

\section{State of The Art}

There have been several research works on how to exploit and utilise novel storage technology such as SCM and object storage on the path to adapting I/O to Exascale.

Some have focused on the benefits of object storage as opposed to commonly used file systems and their limitations\cite{Liu2018EvaluationSystems,Aghayev2019FileEvolution,Gadban2021AnalyzingWorkloads}.
Others have focused on adapting I/O middleware to exploit object storage, with some reports of satisfactory performance\cite{Lofstead2016DAOSSystem,Soumagne2022AcceleratingDAOS,SarpangalaVenkatesh2023EnhancingContext,Jackson2023DAOSInterfaces}.
There have also been reports of successful outcomes using object storage via file system\cite{Hennecke2023UnderstandingScalability} and block device interfaces\cite{Nelson2024Ceph:TiB/s}.

Some have gone beyond file system and I/O middleware APIs and developed domain-specific object stores building on top of general-purpose object stores and low-level non-volatile memory (NVM) storage libraries, and verified the benefits of that approach\cite{RajaChandrasekar2017AnSupercomputers}.
This includes the ECMWF's previous work on the FDB\cite{Smart2017AData}, which originally exploited POSIX file systems, and was later adapted to make native use of SCM\cite{Smart2019AClimate} and the Ceph RADOS and Cortx Motr object stores.
Part of the conclusions of the ECMWF's work were that a PMEM (Persistent Memory) backend was difficult to maintain, and RADOS did not at the time provide sufficient performance. Of the remaining storage options, Motr was discontinued in 2023.

The Intel's Optane Data Center Persistent Memory Module (DCPMM) ---the only ever relevant SCM option in the market--- was unexpectedly discontinued in 2022 due to a lack of demand.
Object stores, nevertheless, have continued to gain traction, and this is also true for DAOS despite the fact that it originally required SCM to function --- it shifted to supporting NVMe-only systems shortly after Optane was discontinued\cite{Hennecke2023DAOSResults}.
A few object stores became increasingly present in IO-500 performance rankings\cite{Bez2024IO500Lists}, with a few large institutions at the top after adopting object storage often as a backend for the file system APIs commonly used by their applications\cite{Latham2025InitialAurora}.

The work in this dissertation describes newly developed FDB backends which enable native FDB operation on DAOS and Ceph, and discusses the implementation and performance differences between running realistic NWP I/O workflows with the new object storage backends vis-à-vis the traditional POSIX backend on Lustre.
This research significantly advances the understanding of the place of object storage in the storage landscape for HPC systems and discusses some of the features, configurations, and challenges this approach presents both for the ECMWF's NWP and for HPC applications in general.

\section{POSIX file systems}

In the early years of computer systems, standard subroutines were developed to facilitate data I/O from and to persistent storage, and these I/O developments were integrated and continued as part of the UNIX family of operating systems ---first released in the 70s--- and the C programming language ---first released in 1972---, crystallizing eventually as the standard file system API, largely as we know it today.
This API revolves around the file and directory concepts, which enable encapsulation and organisation of data.

The file system API was included as part of the ISO POSIX.1 (Portable Operating System Interface) standard, first published in 1990\cite[Chapter~2]{Stevens2013AdvancedEnvironment}. POSIX was broadly adopted by most computer vendors, and the file system API thus became tightly integrated with most operating systems and programming languages, paving the way for the development of vast amounts of applications relying on this API in the following decades. To this day it is still the most widely used I/O API for persistent storage.

POSIX defined a standard interface for file and directory operations such as \verb!open!, \verb!close!, \verb!write!, \verb!read!, \verb!stat!, and so forth, but left flexibility for file systems to implement these as desired.
Since the publication of POSIX, file systems have evolved significantly to support growing storage capacities and provide higher performance, resulting in the creation of a range of file systems designed for specific operating systems, workloads, and hardware configurations.

Early file systems relied on relatively simple mechanisms. File data was stored in fixed-size \textit{blocks} ---the smallest unit of allocation; typically 4 KiB or smaller---, and an \textit{inode} data structure was maintained for every file, containing file metadata and pointers to the corresponding data blocks.
The blocks corresponding to a file could be contiguous or scattered in storage media depending on fragmentation and allocation strategy.

A region at the beginning of the storage medium ---the superblock--- was usually reserved for metadata describing the overall file system layout and contained a table of inodes and directory entries mapping file names to inode structures.
When a file was accessed, the operating system typically cached its inode and block mappings in memory to accelerate subsequent operations. 
Blocks were fetched from storage as atomic units and mapped and cached into system memory via pages ---typically 4 KiB in size---, although some systems also supported hugepages ---larger memory pages from a few MiBs up to a few GiBs in size--- to reduce the overheads associated to the management of many small pages for large file access.

As file systems matured, new mechanisms were introduced to improve reliability and performance. One major development was the introduction of log-structured file systems. These systems treated storage as a circular log, appending all updates sequentially. This minimized disk seeks and allowed fast writes, especially on spinning disks. However, log-structured systems often required complex garbage collection logic to reclaim space and suffered from write overheads as the system was further populated with data.

Journaled file systems later emerged as a solution to the problem of data corruption due to unexpected shutdowns. In these systems a journal ---or log--- records metadata updates before they are applied to the file system. This allows the system to recover to a consistent state after a crash by replaying the journal. The journal is typically held in a dedicated area on disk and is managed by the file system. Some implementations ---e.g. ext3--- only journal metadata, while others ---e.g. NTFS and XFS--- can optionally journal both metadata and data.
Among local file systems, ext4 is one of the most widely used in Linux. It is a journaled file system and provides features for efficient block allocation and reduced fragmentation.

More recent innovations include copy-on-write (COW) file systems like Btrfs and ZFS, which enhance reliability by never overwriting existing data. Instead, updates are written to new blocks, and only after success are block pointers updated.
Also, these systems use advanced techniques for more efficient block indexing.

In parallel to local file systems, the need for shared access in distributed computing led to the development of networked file systems.
One of the earliest and most influential was the Network File System (NFS), developed by Sun Microsystems. NFS enables clients to mount remote directories over a network as if they were local. It relies on caching and periodic refreshes to maintain consistency across clients and servers, with the drawbacks that this can lead to consistency issues and cause significant overheads in highly concurrent workloads. Its successor, NFSv4, provides solutions to some of these issues, although it is still in the process of being fully implemented and adopted.

For HPC and data-intensive workloads, specialized distributed file systems such as Lustre and GPFS emerged. These file systems combine storage resources distributed across multiple networked nodes, leveraging local file system instances on each node to manage the local resources, and expose these as a single view while preserving the POSIX semantics and consistency guarantees of a single-node local file system.
This enables distributed file systems to scale to larger capacities while providing higher throughput by concurrently serving data and I/O operations from multiple storage nodes.

Typically, distributed file systems are deployed as separate metadata and data servers. When a client requires accessing a file, it first establishes a connection with a metadata server, which handles file namespaces and access control, and is then redirected to the relevant data servers where file data is to be written or read from.

To avoid the communication overheads when repeatedly accessing the same file, or when accessing a recently written file, most distributed file systems by default cache written and read data pages in client memory.
In turn, this implies that written data is by default not persisted immediately on storage media, and applications need to issue an explicit system call or close the open file to ensure data persistence.

Due to this caching approach, maintaining strict consistency across clients and servers and fully complying with the POSIX semantics is one of the most challenging aspects of the design of distributed file systems.
POSIX mandates that all write and read operations must be consistent. That is, a read operation initiated right after a write operation on the same file extent must see the data being written by the write operation, and a write operation initiated right after a read operation must not modify the original data before it is fully returned to the reader.
Distributed file systems commonly accomplish this by implementing a distributed locking mechanism such that every client process opening a file for write or read must request a lock from a lock server ---involving a network round-trip--- for the target file extent before writing or reading that extent from storage, and in case of conflicting I/O the last racing process blocks until it obtains the lock it requested. Any issued set of conflicting write and read operations is thus guaranteed to be consistent and free of interruptions or failures due to I/O conflicts, although this can result in large lock communication overheads on the client nodes for highly contentious workloads\cite{Paul2020UnderstandingStatistics,Schmuck2002GPFS:Clusters,GeorgeUnderstandingInternals}.

To mitigate this, some recent distributed file systems such as GlusterFS, GFS, or VAST ---when accessed via its file system interface---, have given up some of the POSIX consistency guarantees, often deemed as excessive for write-once-read-many workloads, in favor of performance.

\subsection{Lustre}

Lustre\cite{Braam2019TheArchitecture} is a fully POSIX-compliant distributed file system widely used nowadays in HPC environments which can exploit HDDs and SSDs as well as Remote Direct Memory Access (RDMA) networks.
Lustre has been used in this work as a reference for performance comparison vis-à-vis other storage approaches due to its popularity and the fact that it is currently being used and has been used operationally at the ECMWF for more than a decade.

Lustre is deployed as a set of Metadata Servers (MDSs) and Object Storage Servers (OSSs) on the available storage nodes, with either an MDS or an OSS typically deployed on every storage node.
An MDS handles metadata aspects such as file namespaces and access control, and can often benefit from being deployed on a faster storage hardware tier if available.
An OSS, on the other hand, is usually deployed on a node with large-capacity storage hardware, and serves bulk file data and handles negotiation of locks on local files with the clients.
Both the MDSs and the OSSs leverage underlying local file systems, such as ldiskfs or ZFS, to manage the storage hardware.

While Lustre deployments in most cases encompass multiple OSSs on multiple ---often several--- storage nodes, a single MDS on a single node can be sufficient for small to medium-scale instances. Nevertheless, multiple MDSs can be deployed on multiple nodes for larger instances if required, and the metadata workload can be distributed over the available MDSs by utilizing the Lustre's Distributed Name Space (DNE) feature. DNE supports two different modes of operation; DNE1, which allows system administrators to delegate metadata handling for a selected file system directory ---or sub-tree--- to a specific MDS other than the first; and DNE2, which allows the user to enable balanced distribution of the metadata workload for a given directory over a configurable number of MDSs.

Clients interact with both the MDSs and OSSs to write and read files in parallel. Lustre does not implement algorithmic placement, and this implies that for every metadata operation performed, such as a file \verb!open!, \verb!stat!, \verb!unlink!, or \verb!mkdir!, a client needs to interact with an MDS. Once a file is opened, \verb!write! and \verb!read! operations are performed directly against the corresponding OSSs.

Every OSS manages a set of Object Storage Targets (OSTs) deployed on the same node, with every OST serving data for a separate local file system deployed on a subset of the storage resources in the node.
Files in Lustre can be sharded ---or \textit{striped}---, such that the file is split in parts ---or \textit{stripes}--- and every stripe is placed in a separate OST. The size of the stripes to split the file into and the number of OSTs to distribute the resulting stripes over can be configured at a per-directory or file level.
Striping enables parallel and therefore faster access to a same file, exploiting the aggregate bandwidth of the storage devices and network interfaces of multiple storage nodes.

Lustre does not currently provide software-level fault tolerance mechanisms such as erasure-coding. For that purpose, either specialised hardware can be employed, or \verb!mdadm! used at the local file system level to create RAID setups.

\clearpage

\section{DAOS}

The Distributed Asynchronous Object Store (DAOS)\cite{Liang2020DAOS:Memory} is an open-source high-performance object store designed for massively distributed NVM, including SCM ---which resides in the memory domain of a compute node--- and NVMe SSDs. DAOS was originally developed by Intel but was recently transferred to the DAOS foundation, with HPE ---one of the founding members--- actively maintaining and developing it.

DAOS provides low-level key-value storage functionality on top of which other higher-level APIs, also provided as part of DAOS, are built. DAOS's features include transactional non-blocking I/O, fine-grained I/O operations with zero-copy I/O to SCM, end-to-end data integrity, and advanced data protection. The OpenFabrics Interfaces (OFI) library is used for low-latency communications over a wide range of network backends.

DAOS is deployable as a set of I/O processes or \textit{engines} ---generally one per physical socket in a server node---, each managing access to NVM devices associated with the socket. Graphical examples of how engines, storage devices, and network connections can be arranged in a DAOS system can be found at \cite{DAOSArchitecture}.

An engine partitions the storage it manages into \textit{targets} to optimize concurrency, each target being managed and exported by a dedicated group of threads. DAOS allows reserving space distributed across targets in \textit{pools}, a form of virtual storage. A pool can serve multiple transactional object stores called \textit{containers}, each with their own address space and transaction history.

The low-level key-value DAOS API is provided in the \texttt{libdaos} library, and it exposes what is commonly known as a map or dictionary data structure, with the feature that every value indexed in it is associated to two keys: the distribution key (\textit{dkey}) and the attribute key (\textit{akey}). All entries indexed under a same dkey are collocated in the same target, and the akeys identify the different entries under a same dkey. Listing dkeys and akeys is supported. This is an advanced API and not commonly used by third-party libraries using DAOS.

\begin{figure}[htbp]
\centerline{\includegraphics[width=250pt,trim={0 0pt 0 0pt},clip]{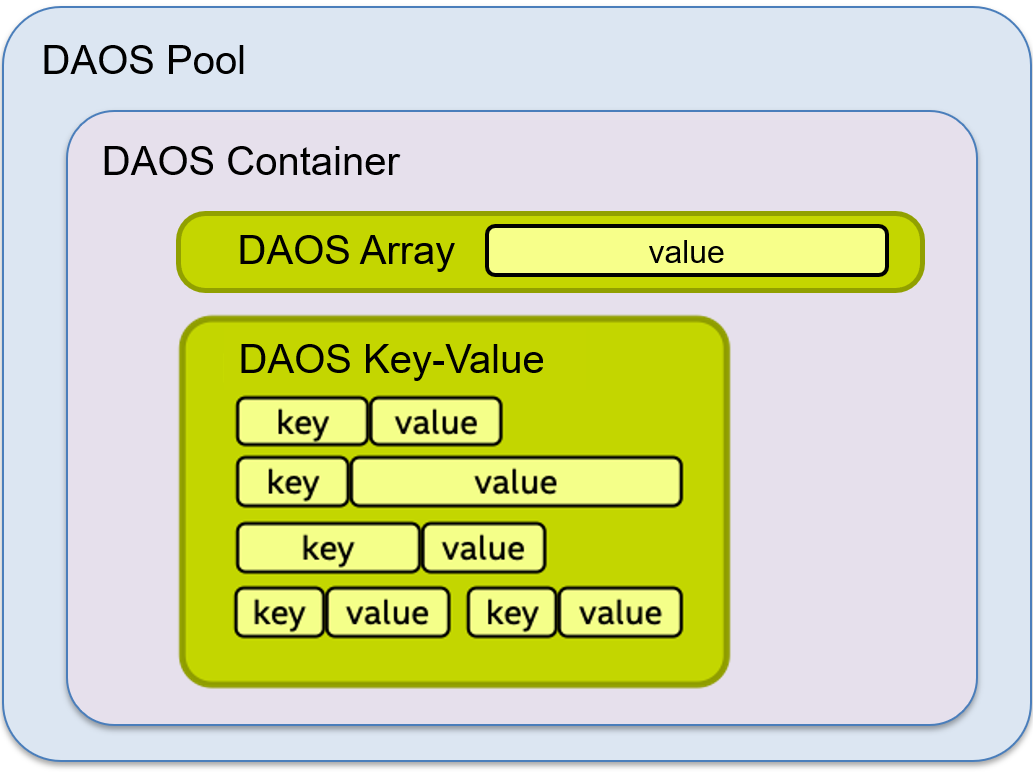}}
\caption{Illustration of the high-level APIs provided by libdaos.}
\label{fig:libdaos_apis}
\end{figure}

High-level key-value and array APIs are also provided as part of \texttt{libdaos}, both building on top of the low-level API. These are illustrated in Fig. \ref{fig:libdaos_apis}.

The high-level key-value API exposes a single-key dictionary structure, where limited-length character strings (the \textit{keys}) can be mapped onto byte strings of any length (the \textit{values}). Entries can be added or queried with the transactional \texttt{daos\_kv\_put} and \texttt{daos\_kv\_get} API calls. Querying the size of an entry and listing keys is also supported.

The array API is intended for bulk-storage of large one-dimensio\-nal data arrays. An in-memory buffer can be stored into DAOS or populated with data retrieved from DAOS with the transactional \texttt{daos\_array\_write} and \texttt{daos\_array\_read} API calls. These operations support targeting one or multiple byte ranges with arbitrary offset and length.

Upon creation, key-value and array objects are assigned a 128-bit unique object identifier (OID), of which 96 bits are user-managed. These objects can be configured for replication, erasure-coding, and striping across pool targets by specifying their \textit{object class}. If configured with striping, for instance, they are transparently stored by parts in different low-level dkeys and thus distributed across targets, enabling concurrent access analogous to Lustre file striping.

DAOS also provides a \texttt{libdfs}\cite{DAOSSystem} library which implements POSIX directories, files and symbolic links on top of the described APIs, such that an application including this library can perform common file system calls ---with slightly different function names--- which are transparently mapped to DAOS. \texttt{libdfs} is, however, not fully POSIX-compliant and does not support features such as the atomicity of \texttt{mkdir}, appending to a file after opening it with the \texttt{O\_APPEND} flag, and advisory locks. A File System in User Space (FUSE) daemon ---namely DFUSE--- and interception library are also distributed as part of DAOS for use by existing applications using the POSIX file system API without modification.

Other interfaces provided by DAOS include S3 and NVMe-oF.

An important feature of DAOS, if compared to POSIX distributed file systems, is that contention between writer and reader processes is resolved server-side with Multi-version Concurrency Control (MVCC) rather than via distributed locking on the clients.
When a write operation is issued, it is immediately persisted by the server in a new region or object in storage, with no read-modify-write operations. The new object is then atomically indexed in a persistent index kept in low-latency SCM or in a write-ahead log on NVMe, and the write operation returns successfully. Any subsequent read operation for that object triggers visitation of the index and returns the associated data from the corresponding storage regions across the servers. This way, writes always occur in new regions without modifying data potentially being read, and reads always find the latest fully written version of the requested object in the corresponding index entry. This mechanism not only ensures strong consistency guarantees but also avoids the use of locks and the associated overheads.

Another relevant feature of DAOS is that it implements algorithmic placement ---that is, algorithmically determining which server node an object should be written to or read from before any interaction with the storage system occurs---, and this means metadata operations are performed against all server nodes in DAOS, rather than on dedicated metadata servers which can potentially bottleneck and hinder performance at scale.

In terms of authentication and authorisation, the DAOS pools and containers can be configured via Access Control Lists (ACLs) to allow read-only or write-read access to different users. The client processes attach their effective UNIX user and group in every I/O operation sent to the server, and these are used to determine access permissions according to the ACLs.

\section{Ceph}

Ceph\cite{Weil2006Ceph:System,WelcomeCeph} is an open-source distributed storage system designed to operate on commodity hardware, with a focus on resilience and data safety to support failures of such hardware. Although data safety was prioritised over high performance, Ceph was designed to scale up to large amounts of storage resources.

It became popular in Cloud environments mainly due to the fact that, among other purposes, it can be used as a backend to provide virtual machine images, in a way that the resources used for compute and storage can be detached and managed independently. Also, the fact that it can be accessed via the S3 object storage protocol contributed to its popularity in Cloud.
 
Ceph is software-defined. It can be deployed on any commodity hardware and is not dependent on specific hardware.
The types of storage device it can exploit include HDDs, SATA or SAS SSDs, and NVMe SSDs.
Regarding networking, it can only exploit TCP/IP networks out of the box, although there are ongoing developments to support native operation on low-latency networks with DPDK and RDMA\cite{CephReference}.

Ceph was architected as a set of services or daemons that can be deployed on different nodes and scaled independently, and these services provide the core storage functionality. Additional daemons can be deployed on additional nodes or on top of the existing ones to provide additional functionality.
A minimal Ceph cluster has at least one Object Storage Daemon, or OSD, one Monitor daemon, and one Manager daemon.
 
An OSD daemon manages storage devices in a storage node. One or more OSDs may be deployed in a single node, each managing a subset of the devices. OSDs can exploit raw devices directly without the need of an intermediate local file system.
The OSD daemon stores object data and maintains a metadata index in the devices it manages. Each OSD requires between 2 and 4 GiB of DRAM to function, and can be configured to build the index on faster storage hardware if available.

A Monitor daemon keeps an up-to-date map of the state of the cluster ---also known as \textit{OSD map}---, which describes which OSDs are available in the system and which are down, and how they are organised in nodes, racks, and so forth. The Monitor serves the OSD map to clients when required.
For high availability, multiple Monitors should be deployed in a Ceph system, from 3 to 7. The different Monitor replicas reach agreement on a consistent view of the OSD map with the Paxos algorithm.
The monitor also acts as manager for the other daemons in the system, and acts as the central authority for authentication.
A Ceph monitor node requires 32 GiB of DRAM for small clusters, and up to 128 GiB for very large ones.
 
The Manager daemon aggregates and exposes system metrics. Two instances should be deployed for high availability, and each requires approximately 32 GiB of DRAM.
 
A system with these three daemons is also referred to as a Reliable Autonomous Distributed Object Store (RADOS), or also as a Ceph Storage Cluster.
 
\begin{figure}[htbp]
\centerline{\includegraphics[width=250pt,trim={0 0pt 0 0pt},clip]{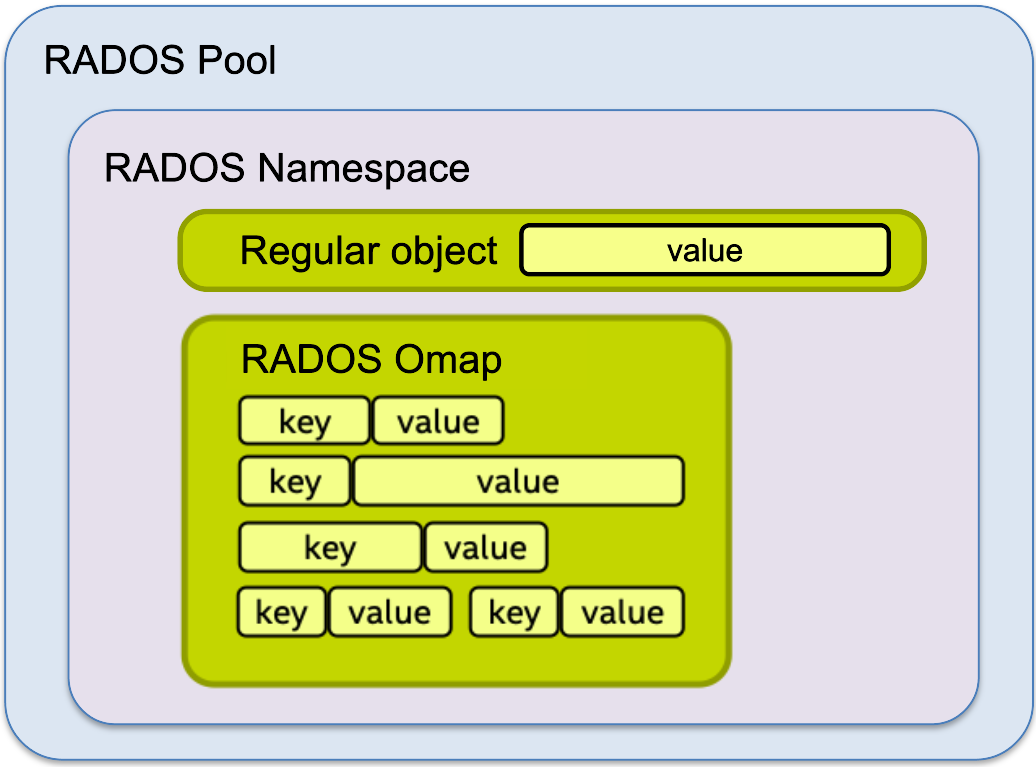}}
\caption{Illustration of the high-level APIs provided by librados.}
\label{fig:librados_apis}
\end{figure}

RADOS exposes an object storage API, and client applications can interact with it via the \verb!librados! library\cite{IntroductionLibrados}.
Every object in RADOS is identified with a name, and can have attributes. There are two types of objects: regular objects, which are similar to files or DAOS arrays in the sense that they store a string of bytes, and \verb!Omaps!, which provide key-value functionality akin to the DAOS key-values.
These are illustrated in Fig. \ref{fig:librados_apis}.

Objects are stored within \textit{pools}, which are a form of partitioning of the storage often used to contain data for different applications. For example, one pool could contain virtual machine images for Cloud infrastructure, another could contain media or data files, and so forth.
Pools are identified by name, and by default span all OSDs in the system, although they can be assigned to specific subsets of OSDs on creation.

Pools can be configured to replicate or erasure-code the objects they contain, resulting in replicas or chunks of the objects distributed across multiple OSDs.
In contrast to DAOS, RADOS objects cannot be replicated or erasure-coded individually as it is a per-pool setting. Also, RADOS objects cannot be sharded other than by enabling erasure-coding, and Omaps cannot be erasure-coded.
 
Objects in a RADOS pool can be placed in \textit{namespaces}, such that the names of the objects in a given namespace do not collide with the names of those in another namespace, akin to DAOS containers.

Like DAOS, RADOS implements algorithmic placement and achieves strong consistency with a mechanism similar to MVCC.
When an object write is performed, the client first retrieves an up-to-date copy of the OSD map if it does not have one already, and then calculates a list of OSDs the object should be written to based on the object name and the OSD map. The first OSD in the list is selected as primary.
The client then transfers the object data directly to the calculated primary OSD without prior roundtrips to a centralised set of metadata servers. Once the OSD has persisted the data in storage media, it sends copies to any other OSDs that should keep replicas or erasure-coding chunks, and only after all replicas or parts are persisted the index on the primary OSD is updated and the write operation is acknowledged.
A reader requesting that same object does the same calculations to determine the primary OSD, which is then queried to retrieve the object data, finding always the latest version made available in the index.
Since RADOS does not implement client-side caching, there are no concerns regarding consistency of any such caches.

To achieve efficient recovery in case of hardware failures, RADOS does not determine the placement of every object individually. Instead, objects in a pool are internally grouped in \textit{placement groups} (PGs), and all objects assigned to a same PG end up being placed by the algorithm in the same set of OSDs. Why this grouping is necessary is further explained in \cite{Weil2019CephCeph}.

By default, RADOS automatically adjusts how many PGs are used ---always aiming to have 100 PGs per OSD---, although the PG count can also be set manually on pool creation. The overall RADOS performance can be very sensitive to this parameter.

A drawback of the PG concept is that, because OSDs usually manage storage devices of a few TiB of capacity, PGs usually end up being a few GiB in size, thus the objects in them cannot be very large. In fact, RADOS is designed to handle objects of up to a few MiB in size, and it imposes an object size limit of 128 MiB by default. This limit can be adjusted when deploying a RADOS cluster, but using larger sizes is discouraged and results in low performance.
Because of this, if a large data element such as a video file of a few GiB needs to be stored in RADOS, the application must first split it in parts, and assign a different name to each part.
 
Ceph provides a few higher-level interfaces or layers which build on top of RADOS, and this includes the Rados Gateway, or RGW, providing S3 object storage functionality; the Rados Block Device, or RBD, which tightly integrates RADOS with the Linux kernel to provide virtual machine images and block devices; and the Ceph File System, or CephFS, which provides POSIX distributed file systems.
 
Both the S3 and POSIX layers solve the limitation of small objects in RADOS in a transparent way, allowing applications to use S3 objects and POSIX files as large as desired and breaking them down into small RADOS objects under the hood. However, when using any of these, the key-value functionality that RADOS Omaps provide if using \verb!librados! directly is lost.

\clearpage

\section{Comparison of object storage and file system features}

\begin{figure}[htbp]
\centerline{\includegraphics[width=280pt,trim={0 0pt 0 0pt},clip]{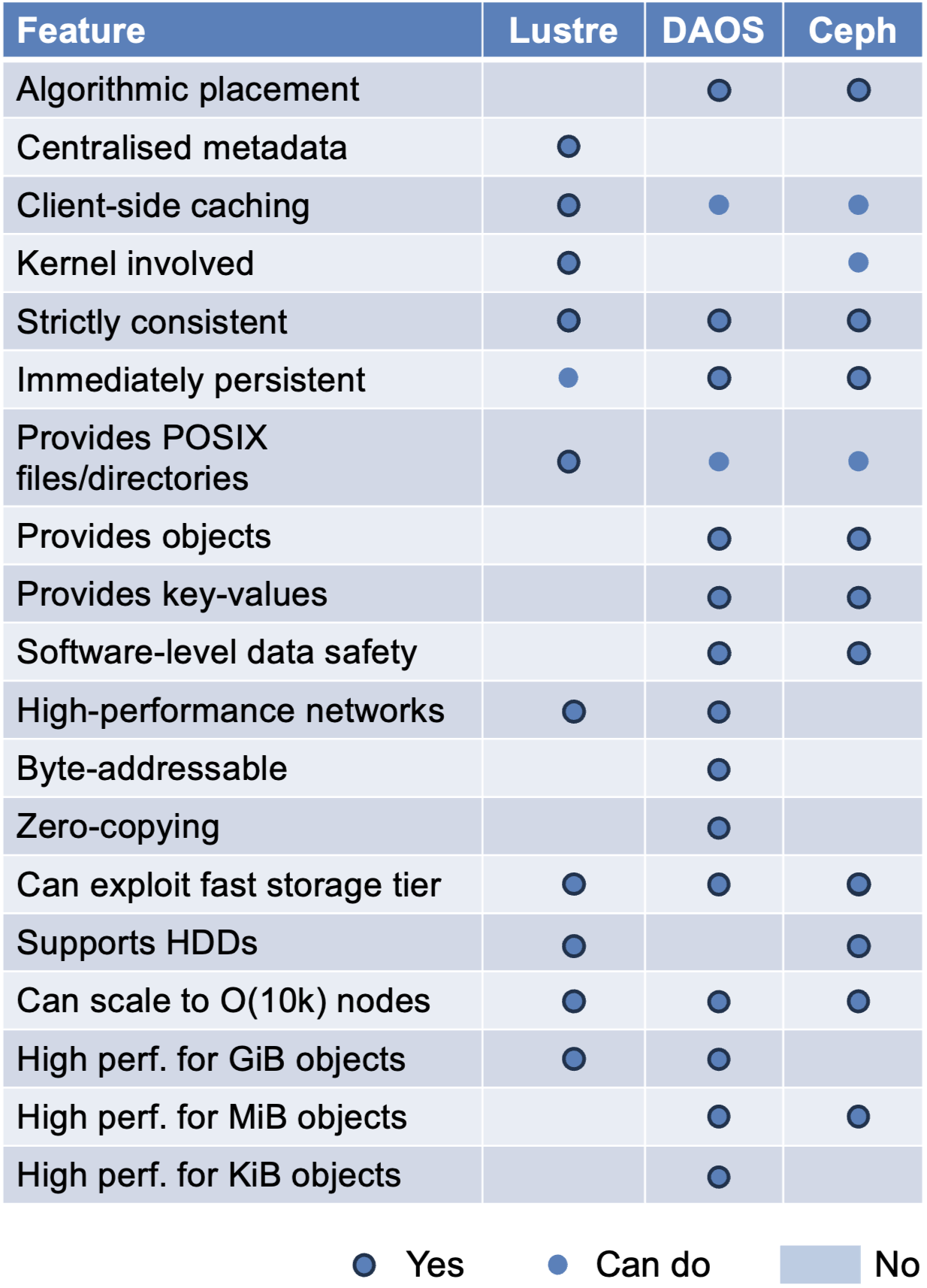}}
\caption{Comparison of features provided by Lustre, DAOS, and Ceph.}
\label{fig:feature_comparison}
\end{figure}

Fig. \ref{fig:feature_comparison} summarises and compares the most relevant features of Lustre ---the exemplar distributed file system chosen for this analysis---, DAOS, and Ceph --- the two open-source high-performance object stores considered.
From top to bottom:

\begin{itemize}[leftmargin=*]

    \item Lustre delegates metadata serving to a set of designated nodes separate from the bulk storage nodes, whereas both DAOS and Ceph implement algorithmic placement such that clients communicate directly with the relevant storage nodes which also serve the associated metadata. Object stores therefore have potential to more efficiently handle large-scale metadata-intensive workloads.

    \item Lustre implements client-side caching and it is enabled by default, whereas the object stores do not implement it in their core, although their corresponding file system interfaces do implement caching.
    This means Lustre can provide a performance advantage for small-scale applications that do not overflow the caches, which object stores could never provide if accessed natively.

    \item Lustre is tightly ingrained in the operating system, and this implies many context switches between user space and the kernel are required both on the client and server sides while performing I/O, both for file operations and network communications. DAOS, instead, was carefully implemented to operate fully user-space to avoid these overheads. If using TCP communications over the fabric, however, DAOS will involve the kernel. Ceph is also largely designed to run user-space, but it does not currently support RDMA networks, and therefore always involves the kernel to a certain degree for TCP communications.

\end{itemize}

\begin{itemize}[leftmargin=*]

    \item In terms of consistency, all storage systems provide strong guarantees. Lustre achieves this via distributed locking, and DAOS and Ceph do so with MVCC.

    \item Lustre does not by default guarantee immediate persistence of writes, although there are flags and system calls available for applications to ensure persistence. DAOS and Ceph, instead, persist all I/O operations before returning unless using their file system interfaces.

    \item While Lustre provides file and directory APIs, and DAOS and Ceph provide object and key-value APIs, the latter storage systems also provide file and directory APIs for POSIX I/O access with varying degrees of compliance.

    \item DAOS and Ceph provide software-level redundancy and erasure-coding, wheres Lustre does not.

    \item Lustre and DAOS can exploit high-performance fabrics natievly out of the box, whereas Ceph can not.

    \item Lustre builds on top of the block device infrastructure and therefore fetches data from storage in entire blocks even if requesting only one ore a few bytes. DAOS, instead, is designed to provide byte-level access to data. Ceph, at least if accessing erasure-coded objects, currently fetches the full object extent even if a partial range is accessed.

    \item DAOS is the only storage system carefully implemented to avoid unnecessary copies between storage media, intermediary library buffers, and kernel and network buffers. Data is transferred directly from storage media to client memory.

    \item The three storage systems can exploit low-latency storage hardware, if available, for superior overall performance.

    \item DAOS cannot exploit HDDs ---it is exclusively designed for NVM--- whereas Lustre and Ceph can.

    \item The three storage systems are designed to scale to tens of thousands of storage nodes.

    \item Regarding performance at different I/O sizes, DAOS is the only storage system that provides high performance for small, medium, and large object sizes, as demonstrated in Chapter \ref{chap:chap4}. Lustre does not perform well when spreading data across many small or medium files, and Ceph does not perform well when using a few large objects nor large amounts of very small objects.

\end{itemize}

\section{NWP at the ECMWF}

The ECMWF is an intergovernmental organization aiming, among other duties, to provide its member and cooperating states with global weather forecasts of time periods ranging from a few days to a few weeks into the future\cite{AboutForecasts}.
The states ---specifically their national meteorological and hydrological services--- then use these global forecasts as boundary conditions for their own regional NWP models which are run periodically to produce localized forecasts and take action accordingly.

As part of the ECMWF's operational procedure, their NWP models are run in an in-house HPC system composed of thousands of compute nodes and a Lustre parallel file system providing 1.5 PiB of SSD-based storage\cite{ECMWF2025SupercomputerFacility}. During an operational run, which lasts approximately one hour ---but cannot last longer than that as per committed deadlines---, the models presently produce approximately 95 TiB of forecast data, which are stored into the parallel file system. Those data are read back from storage, while the models continue to run, and post-processed to generate derived products, of much smaller size, to be delivered to member states together with the forecast data\cite{Weiland2019AnApplications}.

This application presents two challenges. On one hand, the large problem size requires partitioning the model domain into sub-domains which are processed in parallel across the compute nodes. In turn, every node employs several parallel processes for the execution of its corresponding model sub-domain to fully exploit the computing resources, and all these processes need to store their output into the parallel file system as it is produced, concurrently. This high degree of I/O parallelism needs to be supported by the file system while maintaining high performance, and requires I/O to be performed in particular ways at the application level to minimize strain on the file system. In practice, at the ECMWF, this is addressed in part by funneling I/O through a set of  designated I/O nodes to reduce the degree of parallelism before I/O hits the storage system ---a technique also known as I/O forwarding---, but also by adhering to certain best practices for high-performance I/O on parallel file systems\cite{LustrePractices} such as employing a separate file per writer process and configuring large files to be distributed across multiple storage nodes ---also known as sharding or striping--- to unlock their aggregate bandwidth. These approaches ensure writing of model output is performed as fast as the file system allows, causing the least possible impact on the progress of model processes, which sit in the critical path of an operational run.

On the other hand, post-processing of model output cannot be performed on the fly in the compute nodes where model sub-domains are run, and this is because post-processing requires visibility of the full model domain. Instead, model output is forwarded to I/O servers as it is produced and written into the storage system, and separate post-processing jobs read data back for the full domain from storage while model processes continue to produce and write output. This results in a highly contentious write-read access pattern where every post-processing job reads data across a large number of files being continually written into by model processes.
Again, the storage system needs to be able to handle such contention and large number of file access operations without deteriorating performance, and I/O needs to be performed ---and is performed--- in particular ways at the application level to avoid data and metadata corruption and minimize performance impact.

The ECMWF developed an I/O library over the past few decades, namely the FDB\cite{Smart2019AClimate}, which exposes a scientifically meaningful API for applications to easily write and read forecast data, abstracting away the underlying file system and the associated I/O programming complexity.
A significant effort was made over the years to optimize the FDB's internal I/O logic and mechanisms, maximizing I/O performance and reducing strain on the parallel file system during operational runs whilst ensuring data consistency under contention.

\section{The FDB}
\label{sec:fdb}

The FDB\cite{Smart2017AData,Smart2019AClimate} is a domain-specific object store for meteorological data, which sits between various data producing and consuming components in wider NWP workflows.
In practice, it is a software library which is compiled into these components, and provides them with functionality to easily and efficiently store and retrieve meteorological data elements or objects into and from persistent storage systems.
The library is provided along with a set of management command-line tools, and configurations and data governance rules allowing adjustment of the patterns in which the storage is accessed.
The FDB exists to abstract away the specific behavioural details of various underlying storage systems, and to provide a standardised API for high-performance data use.

\begin{figure}[htbp]
\centerline{\includegraphics[width=400pt,trim={0 20pt 0 20pt},clip]{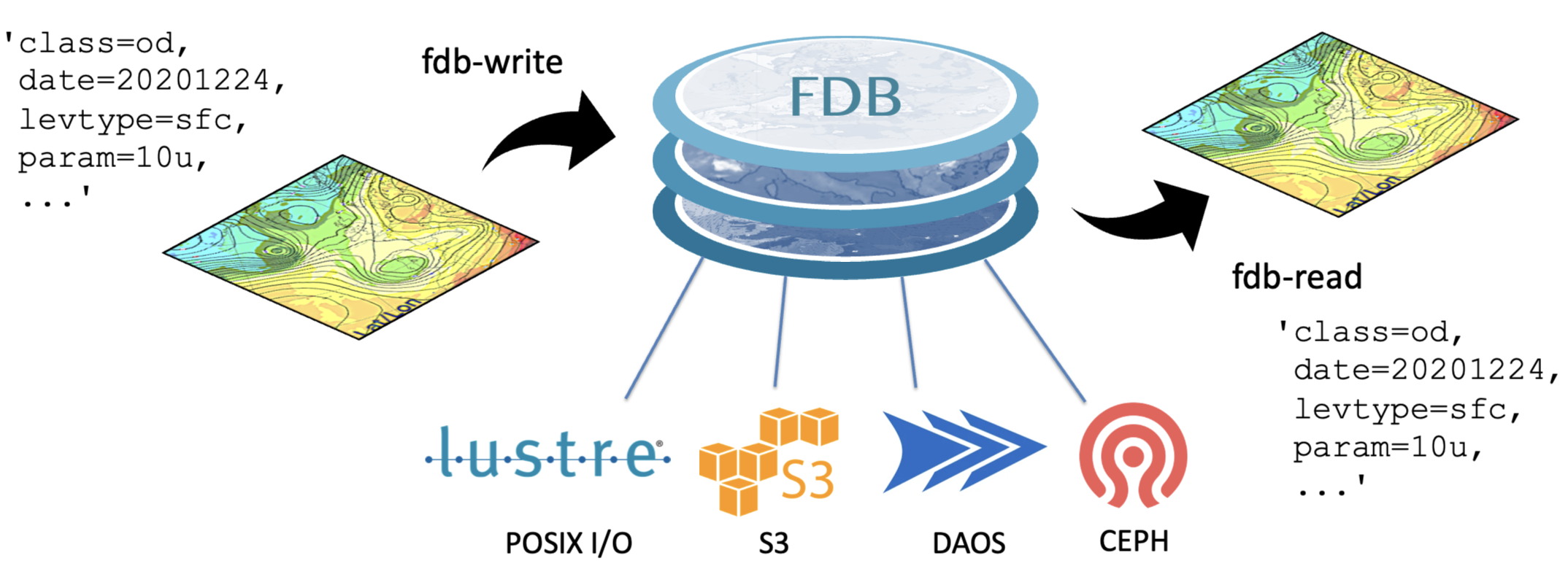}}
\caption{Illustration of the FDB, a domain-specific object storage library abstracting away underlying storage systems, providing functionality to store and retrieve meteorological data objects.}
\label{fig:fdb_intro}
\end{figure}

The FDB, under the hood, creates and maintains an index of stored objects in persistent storage. The need for this index stems from the fact that it is usually not possible to map small meteorological data objects one-to-one to the APIs or storage units (e.g. files) provided by storage systems whilst maintaining high performance, and therefore objects need to be collocated in shared storage units. Discovery and querying of stored objects ---a necessary feature for ECMWF's operational workflows--- would be too expensive to perform without an index.

The FDB API is metadata-driven, that is, all API actions are invoked using scientifically-meaningful metadata describing the data to be acted upon. All data objects ---usually weather fields--- are identified by a globally unique metadata identifier, formed as a set of key-value pairs conforming to a user-defined schema. An example identifier for a weather field from an operational ensemble forecast is illustrated in Listing\ \ref{listing:mars-request}.

\vspace{10pt}
\begin{multicols}{3}[\captionof{lstlisting}{An example metadata identifier of a weather field in the FDB.}]
\begin{lstlisting}[label=listing:mars-request]
class = od,
date = 20231201,
levtype = sfc,
levelist = 1,
expver = 0001,
time = 1200,
step = 1,
param = v
stream = oper,
type = ef,
number = 13,
\end{lstlisting}
\end{multicols}

The schema defines not only the valid object identifier keys and values, but also how the FDB will internally split the identifiers provided by the user application into three sub-identifiers which control how the FDB storage logic lays out data within the storage system, and therefore control the storage access patterns:
\begin{enumerate}
    \item \textbf{Dataset key}: describes the dataset an object belongs to. For instance, a forecast produced today starting at midday. For example, if the schema is configured to recognise \texttt{class}, \texttt{stream}, \texttt{expver}, \texttt{date} and \texttt{time} as dataset dimensions, the following dataset key would result for Listing\ \ref{listing:mars-request}:
    \begin{multicols}{2}
    \begin{lstlisting}
    class = od,
    stream = oper,
    time = 1200
    expver = 0001,
    date = 20231201,
    \end{lstlisting}
    \end{multicols}
    \item \textbf{Collocation key}: the object should be collocated in storage with other objects sharing the same collocation key. E.g.:
    \begin{multicols}{2}
    \begin{lstlisting}
    type = ef,
    levtype = sfc
    \end{lstlisting}
    \end{multicols}
    \item \textbf{Element key}: identifies the object within a collocated dataset. E.g.:
    \begin{multicols}{2}
    \begin{lstlisting}
    step = 1,
    levelist = 1,
    number = 1,
    param = v
    \end{lstlisting}
    \end{multicols}
\end{enumerate}

Aside from a number of administrative functions, there are four primary functions in the FDB API, namely \verb!archive()!, \verb!flush()!, \verb!retrieve()!, and \verb!list()!. Their signatures, with some simplification, are shown in Listing \ref{listing:fdb-api}.

\begin{lstlisting}[caption={Simplified signatures of the primary FDB API methods.},language=C++,basicstyle=\linespread{2}\small,label=listing:fdb-api]
   void archive(const Identifier& id, DataHandle& data);
   void flush();
   DataHandle* retrieve(const IdentifierList& query);
   IdentifierIterator list(const PartialIdentifier& query);
\end{lstlisting}

\noindent
Applications can use the \verb!archive()! method to request archival ---that is, writing and indexing--- of a meteorological data object by providing its metadata identifier together with a data handling object which manages a pointer to the object data in memory.
Usually, data producers have multiple objects available for archival simultaneously, and in that case a more efficient variant of the \verb!archive()! method can be used which accepts as only input a single data handling object managing the full set of data objects and identifiers.
Producer applications are expected to \verb!archive()! data for one or a few dataset keys, on the order of ten collocation keys, and up to millions of element keys.

\verb!flush()! must be called to ensure any previously \verb!archive()!d objects are persisted into storage and made visible to consumers. This method should typically be called as infrequently as possible, as persisting small amounts of data separately and frequently may cause large overheads. For example, in the ECMWF's operational workflow, \verb!flush()! is called by I/O servers only at the end of a simulation step, to ensure the step data is made visible before signaling the workflow manager to trigger execution of the post-processing task for that step.

Consumers call the \verb!retrieve()! method providing as only input a list of identifiers of the objects to be read back. This only fetches information from the index on the location of the objects in storage, and that information is encapsulated in a data handling object and returned to the application. The actual object data can then be read from storage through this data handle.

The \verb!list()! method enables applications to query which objects are available in storage with identifiers matching a provided partial identifier.

The FDB API has precisely determined semantics. In particular\cite{Smart2019AClimate}:

\begin{enumerate}
    \item Data is either visible, and correctly indexed, or not. The FDB adheres to the ACID\cite{Haerder1983PrinciplesRecovery} (Atomicity, Consistency, Isolation, and Durability) semantics commonly used to define database transactions.
    \item \verb!archive()! blocks until the FDB has taken control of (a copy of) the data. Data is not necessarily visible to consumers or persisted in storage devices at this point, but it is permitted to be.
    \item \verb!flush()! blocks until all data \verb!archive()!ed from the current process is persisted into the underlying storage medium, correctly indexed, and made visible and accessible to any reading process via \verb!retrieve()! and \verb!list()!.
    \item Once data is made visible, it is immutable.
    \item Data can be replaced by \verb!archive()!ing a new piece of data with the same metadata. This second \verb!archive()! shares the semantics of the first, such that the old data is visible until the new data is fully persisted and indexed.
\end{enumerate}

The FDB provides adapters or backends for operation on a range of storage systems. These backends are responsible for ensuring that the described semantics are correctly implemented on top of whatever provisions are made by the underlying storage system.

The FDB internally implements indexing functionality in what is known as a Catalogue backend, and functionality for storage ---bulk write and read--- of meteorological objects in a Store backend. The FDB defines abstract interfaces for these backends, such that specific Catalogue or Store instances can be implemented to operate on top of a given type of storage system. If backends conform to the established interfaces and semantics, the FDB will guarantee its external API semantics. Any pair of conforming Catalogue and Store backends can then be used in conjunction even if they operate on different underlying storage systems.

When an FDB API method is called, the FDB library determines which Store and Catalogue backends to use according to configuration provided by the FDB administrator at deployment time.

\subsection{Store and Catalogue Interfaces}

Both the Store and Catalogue backend interfaces define \verb!archive()!, \verb!flush()! and \texttt{retrieve()} methods. Leaving aside some complexity, the Catalogue interface also defines \verb!list()! and \verb!axis()! methods. A call to the high-level FDB API will result in internal calls to the corresponding methods in the lower-level backend interfaces. Specifically:
\begin{itemize}
    \item an FDB \verb!archive()! call internally calls Store \verb!archive()! and then Catalogue \verb!archive()!.
    \item \verb!flush()! calls Store \verb!flush()! and then Catalogue \verb!flush()!.
    \item \verb!retrieve()! calls Catalogue \verb!axis()! and \verb!retrieve()!, and then Store\\\texttt{retrieve()}.
    \item \verb!list()! calls Catalogue \verb!list()! (neglecting some complexity).
\end{itemize}

\subsubsection{The Store Interface}

\noindent
\textbf{\textit{archive()}}

A Store backend must implement an \verb!archive()! method accepting, as arguments, a pointer to in-memory data, a dataset key, and a collocation key. The data is taken control of and optionally persisted into storage before the method returns. An object location descriptor ---equivalent to a Uniform Resource Identifier (URI)--- is returned describing where the data is to be persisted. This location should be collocated with other fields sharing the same collocation key. The storage location must be unique, avoiding collisions with concurrent processes --- \verb!archive()! will potentially be called repeatedly with the same dataset and collocation key, and previously archived objects must not be overwritten or modified.

\noindent
\textbf{\textit{flush()}}

The \verb!flush()! method blocks until all data which has been \texttt{ar\-chive()}ed has been persisted to permanent storage and made accessible to external reading processes.

\noindent
\textbf{\textit{retrieve()}}

Given an object location descriptor, the Store \verb!retrieve()! method builds and returns a \verb|DataHandle| ---a backend-specific instance of an abstract reader object---, allowing the calling process to read the object from storage without knowledge of the backend implementation.

When multiple objects are requested via the top-level FDB \verb!retrieve()! method, that method may merge the \verb!DataHandle!s returned by the corresponding Store \verb!retrieve()! calls, such that as few I/O operations as possible are issued to the storage system when reading data from the merged \verb!DataHandle!. This can help avoiding unnecessary overheads if the requested objects are collocated within a same storage data structure (e.g. a same file or object) or placed in consecutive regions in storage. For this, the \verb!DataHandle!s returned by Store \verb!retrieve()! must support merging.

\subsubsection{The Catalogue Interface}

A Catalogue backend has more complexity than a Store backend, not only because it addresses a more complex problem ---that of maintaining a consistent index under contention--- but also because it is involved in more operations, including archival, retrieval, removal, and listing. The Catalogue interface comprises more than ten methods, but only the ones required for field archival, retrieval, and listing are covered here, with some simplification.

\noindent
\textbf{\textit{archive()}}

Catalogue backends require an \verb!archive()! method such that, given a dataset key, a collocation key, an element key, and an object location descriptor (from an earlier Store \verb!archive()! call), an entry is inserted in an indexing structure, either in local memory or shared storage, mapping the element key to the provided location descriptor. Once the entry is inserted, the the method returns with no return value. This may be a local in-memory operation, and is not guaranteed to be persistent or visible to reading processes on return. The backend may make use of the dataset and collocation keys, and knowledge of the process performing the write operation, to index related information together, or separately, as makes most sense for performance on the targeted storage system.

\noindent
\textbf{\textit{flush()}}

\verb!flush()! must also be implemented, which blocks until all indexing information has been persisted and made visible to any external \verb|retrieve()|ing processes.

One of the requirements of Catalogue archival is that all indexing information \verb!archive()!d and \verb!flush()!ed by any parallel writers should be accessible to \verb!retrieve()!ing and \verb!list()!ing processes in a reasonably efficient manner. This may require helper data structures to be present, such as summaries of the indexed data which can be retrieved from a superficial layer of the index. If \verb!archive()!ing processes persist indexing information in separate data structures or storage units (e.g. files), higher level shared indexing structures may be required such that all indexing information can be discovered and reached from a single entry point.

The index must always be consistent from the perspective of an external reading process, even under read and write contention. If multiple \verb!archive()! calls with same dataset, collocation, and element keys occur, the older data should be replaced by newer in a transactional manner from the perspective of any reading processes.

\noindent
\textbf{\textit{axis()}}

Although not captured by Listing \ref{listing:fdb-api} due to oversimplification, the identifiers provided to the top-level FDB \verb!retrieve()! method can contain expressions to request retrieval of multiple objects matching one such expression. When called, the FDB \verb!retrieve()! method first expands those expressions into fully specified object identifiers, using both the schema definition and the output from Catalogue \verb!axis()! calls, and forwards those identifiers to the Catalogue \verb!retrieve()! method.

The \verb!axis()! method accepts as inputs a dataset key, a collocation key, and the name of one element key dimension, and returns a vector with all indexed values for that dimension for the given dataset and collocation keys. As suggested earlier, this information should ideally be made available and retrieved from a superficial summary rather than requiring a deep scan of the entire indexing structure.

\noindent
\textbf{\textit{retrieve()}}

The Catalogue \verb!retrieve()! method must return the object location descriptor of an object given its dataset, collocation, and element keys. Due to the potential for the use of the FDB as a cache in a larger data infrastructure, failing to find a field is not an error and results only in no data being returned. The returned location descriptor, if any, can then be passed as an argument to the Store \verb!retrieve()! method to obtain a \verb!DataHandle! through which the actual data can be read.

\noindent
\textbf{\textit{list()}}
\noindent

The Catalogue \verb!list()! method must return a list of all \verb!archive()!d object identifiers and location descriptors for a given dataset matching a partial identifier provided as input. The partial identifier represents a span of the domain of possible object identifiers, and the information for all indexed objects within this span must be returned.

When the FDB \verb!list()! method is called, the FDB scans a registry of possible storage systems and locations within them where FDB datasets may be found. This registry is maintained by the FDB administrator and must be previously made available to applications using the FDB API. The Catalogue \verb!list()! method is then called internally for every dataset found within the registered storage systems and locations which matches the dataset key part of the partial identifier provided as input, and the results of all calls are merged and returned.

\subsection{POSIX I/O backends}
\label{sec:posix_backends}

The logic for the FDB to operate on file systems originally constituted the core of the library. As the ECMWF's HPC systems evolved over the years, the FDB was adapted to operate on the new file systems made available. The POSIX standard was first released in 1990, and in the late 90s and early 2000s POSIX distributed file system software such as the GPFS (today known as IBM Storage Scale) and Lustre gained traction and were adopted by HPC centres including the ECMWF.
When the FDB Catalogue and Store abstractions were put in place in 2020, the logic of the FDB to operate on POSIX file systems was encapsulated as the POSIX I/O Catalogue and Store backends.

These backends aim to make as efficient use as possible of the file and directory APIs provided by POSIX distributed file systems to implement the required meteorological object indexing and bulk storage functionality. 
To that end, the backends were designed following two guiding principles.
The first lay in adhering to I/O programming best practices for high performance on file systems, and this included avoiding use of many small files, reducing the number of I/O operations, and avoiding file contention where possible. The second principle lay in favoring write performance at the expense of read performance where there was the choice, to avoid throttling model progress as much as possible.

The POSIX backends store data and indexing information in a directory per dataset key. That is, all \verb!archive()!d objects identified with a same dataset key are placed, together with their indexing information, under a same directory.
For example, in the ECMWF's operational case, all weather fields for an operational run use the same dataset key and are therefore placed in a same directory along with indexing information.

For every parallel process performing FDB \verb!archive()! operations for data objects for a same dataset and collocation key pair, four files are created within the dataset directory. One contains the data of the objects \verb!archive()!d, and the other three contain indexing information for those objects.
To avoid immediately persisting many small objects and indexing information very frequently, which would slow down the writing, these objects are only made persistent in storage ---in the per-process files--- when \verb!flush()! is called.

Also, a high-level shared indexing file is created in the dataset directory which contains an index of all per-process indexing files, binding them together and making them easily and efficiently discoverable.

Fig. \ref{fig:fdb_posix_flush_1step} illustrates the directory and files created for an example case where a single process calls \verb!archive()! four times for four different objects ---all sharing a same dataset key--- and then calls \verb!flush()! once. The same collocation key is used for the first two \verb!archive()! calls, and another collocation key is used for the other two calls.

How these files are populated in a consistent and efficient way and used for retrieval and listing is explained in the following.

\subsubsection{The FDB POSIX I/O Store}

\noindent
\textbf{\textit{archive()}}

When the POSIX Store \verb!archive()! method is called, 
the dataset directory is created, if it does not exist, using a stringified representation of the input dataset key as directory name. This is done with a \verb!mkdir! system call which guarantees atomicity of the operation even if multiple processes call \verb!archive()! for an object with a new dataset key concurrently.

Also, the first time
\verb!archive()! is called by a process for a new dataset and collocation key pair, the backend creates a unique file within the dataset directory where all data objects \verb!archive()!d by that process for that key pair will be written. To ensure the file is unique, it is named as a function of the input collocation key, current system time, host name, process ID, and a process-local counter. The data file is then opened in append mode (using the \verb!O_APPEND! flag of the \verb!open! system call) and kept open for the lifetime of the \verb!archive()!ing process.

Every time \verb!archive()! is called, the object data provided as input is written at the end of the data file. However, this is not done directly. To avoid the overheads of performing large amounts of small write operations if a long sequence of small data objects are \verb!archive()!d by a process ---as the I/O servers do in the ECMWF's operational runs---, the data is buffered in process memory, and only written in large bursts as the buffer becomes full. To that end, streamed I/O as provided by the \verb!stdio! library is employed (i.e. using the \verb!fopen!, \verb!setvbuf!, and \verb!fwrite! functions). Furthermore, if the operating system is configured in write-back mode, the data may be temporarily held in the kernel's page cache before it is sent to the storage system and persisted. The \verb!archive()!d data is therefore not guaranteed to be persisted immediately as it may be held in process or kernel memory.

A location descriptor for the object is finally returned, composed of the path name of the containing data file, the file offset the object data was written at (but not necessarily persisted), and the object data length. Because the file is exclusive to the process, the file region allocated for a new object is guaranteed to be only allocated for that object.

If the underlying POSIX file system is a Lustre distributed file system, the POSIX I/O backends by default configure the data files with a sharding factor of 8 and a shard size of 8 MiB, such that the data is written to or read from multiple Lustre OSTs simultaneously, resulting in network and server saturation, and therefore best performance, in \verb!archive()! or \verb!retrieve()! workloads at scale.

\begin{figure}[htbp]
\centerline{\includegraphics[width=415pt,trim={0 0 0 0},clip]{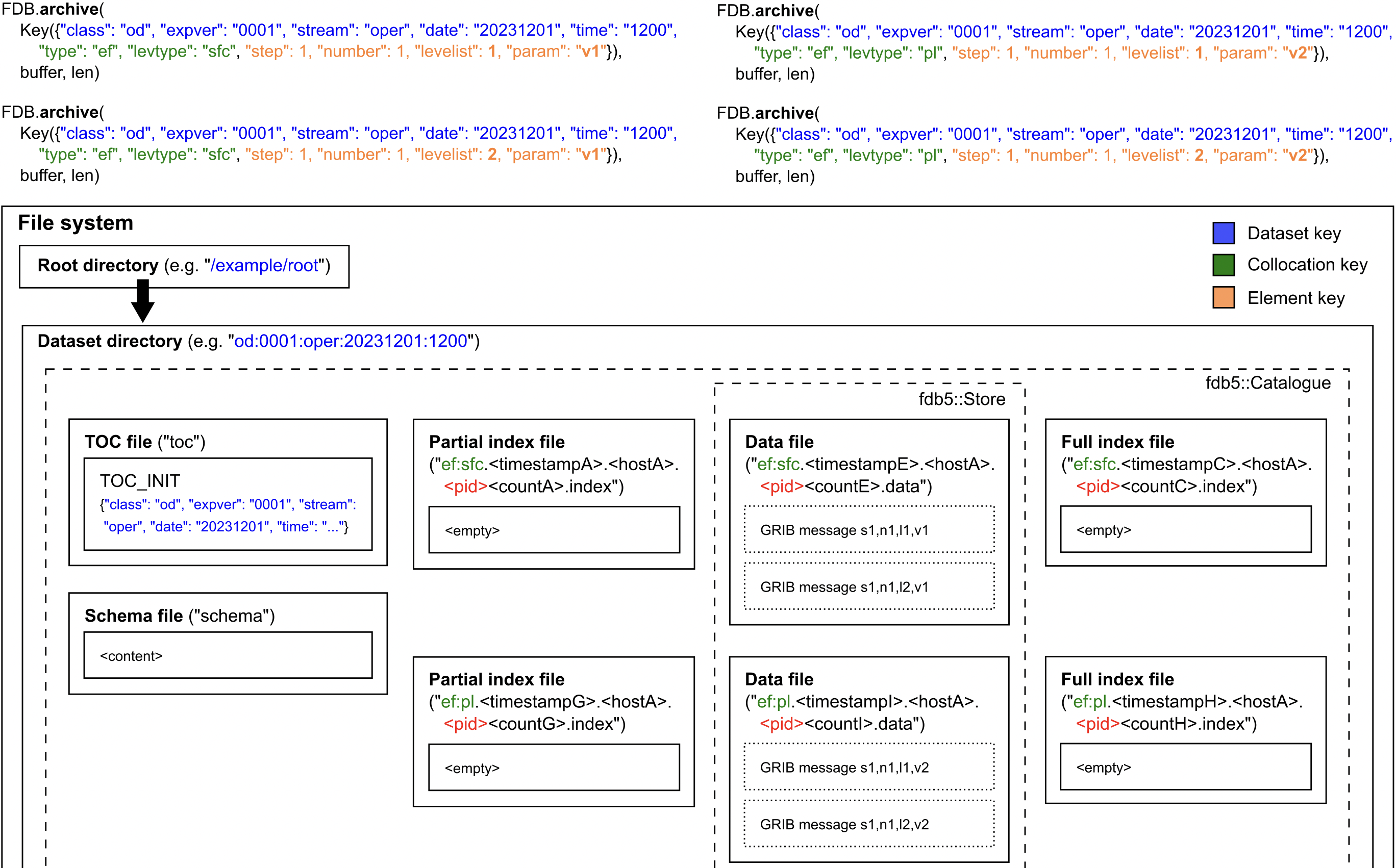}}
\caption{Snapshot of the content of a dataset directory after a process has called FDB archive() four times for four meteorological objects for two different collocation keys. The archive() calls are shown at the top, using different colours for the dataset, collocation, and element keys. flush() has not been called yet, thus the index files are empty as the indexing structures are still held in memory at this point. The object data may or may not have been persisted into storage media at this point.}
\label{fig:fdb_posix_archive_before_flush}
\end{figure}

Fig. \ref{fig:fdb_posix_archive_before_flush} shows an example of the state of a dataset directory after a process has called FDB \verb!archive()! a few times for two different collocation keys, before \verb!flush()! has been called. The data files may or may not yet contain the object data, and this is illustrated with dotted edges for the objects.

\noindent
\textbf{\textit{flush()}}

\verb!flush()! ensures all object data \verb!archive()!d by the calling process is made persistent with \verb!fflush! and \verb!fdatasync! function calls for each data file written by the process --- one per dataset and collocation key.

\noindent
\textbf{\textit{retrieve()}}

When \verb!retrieve()! is called, the input field location descriptor is used to build ---without performing any I/O operation--- and return a \verb|DataHandle| object which enables the application to transparently read data from file range specified in the descriptor. The \verb!DataHandle! internally uses the \verb!open!, \verb!seek!, and \verb!read! system calls for reading.

This handle supports merging with other handle instances returned by the POSIX Store \verb!retrieve()! method. When FDB \verb!retrieve()! is called for multiple objects which are located in sparse ranges within a same data file, a merged handle is returned which internally opens the file only once even if the application reads all data ranges. Also, if multiple requested objects are collocated in adjacent ranges in the file, a single \verb!read! operation is performed.

\subsubsection{The FDB POSIX I/O Catalogue}

\noindent
\textbf{\textit{archive()}}

When the Catalogue \verb!archive()! method is called, the dataset directory is created if it does not exist, in the same way as described for Store \verb!archive()!. The top-level shared index file ---also referred to as the \textit{TOC} (Table Of Contents) file--- is also created within the directory if it does not exist, and some header data is written in it. A copy of the schema ---previously made available to applications by the FDB administrator--- is written in a file within the dataset directory. Because multiple concurrent processes may race to perform the first \verb!archive()! operation for a given dataset key, mechanisms were put in place to ensure consistency of these dataset initialisation operations under contention.

Also, the first time \verb!archive()! is called by a process for a new dataset and collocation key pair, a pair of unique files are created within the dataset directory where the indexing information for all data objects \verb!archive()!d by that process for that key pair will be written. The two index files are ensured to be unique with the same mechanisms as in Store \verb!archive()!, and both are opened and kept open for the lifetime of the \verb!archive()!ing process.

One of the index files is used to store small indexes of the objects \verb!archive()!d between the last two \verb!flush()! calls ---that is, one indexing structure per \verb!flush()! call---, and the other index file is used to store, at the end of the process lifetime, a single index of all objects \verb!archive()!d by the process. Although both files eventually contain equivalent indexing information, they have different characteristics. The indexing structures in the first file are made available sooner, on \verb!flush()!, but are be slower to use for retrieval and listing as many more files need to be accessed ---this is explained below--- and multiple indexing structures need to be visited. The indexing structure in the second file is faster and requires fewer file accesses, but it is only made available at the very end of the producer application lifetime.

Every time FDB \verb!archive()! is called, the Catalogue \verb!archive()! method associates the element key of the archived object with the location descriptor returned by Store \verb!archive()! within a pair of B*-Tree indexes built in memory. One of the indexes holds the indexing information for a given dataset and collocation key pair between two \verb!flush()! calls. On \verb!flush()!, this index is written to the first index file, and the instance in memory is reset. The other index holds the full indexing information for the same key pair, and is eventually written to the second index file at the end of the process lifetime.

It is likely that many of the location descriptors inserted in these B*-Trees will point to regions within a same data file, as producer applications usually \verb!archive()! many objects for a given dataset and collocation key pair. These descriptors will all be identical except for the offset and length of the objects they point to. To avoid repeatedly inserting that common part of the location descriptors and unnecessarily enlarging the indexes, a helper data structure is built in memory, referred to as the \textit{URI store}, such that every file URI is mapped to an integer, and that integer is inserted in the index in place of the URI.

Also on every \verb!archive()! call, the values of the element key are recorded into another set of helper data structure in memory, referred to as the \textit{axes}, such that a summary is kept of all values inserted in an index for every element key dimension. These axes enable the \verb!retrieve()! method to quickly skip indexes that do not contain the targeted objects.

As the producer processes continue to \verb!archive()! more objects, these in-memory structures will grow. Processes issuing many consecutive \verb!archive()! calls ---as is done in operational runs--- should not result in any I/O performance issues caused by this indexing mechanism as all the work is done in memory.

\begin{figure}[htbp]
\centerline{\includegraphics[width=415pt,trim={0 0 0 0},clip]{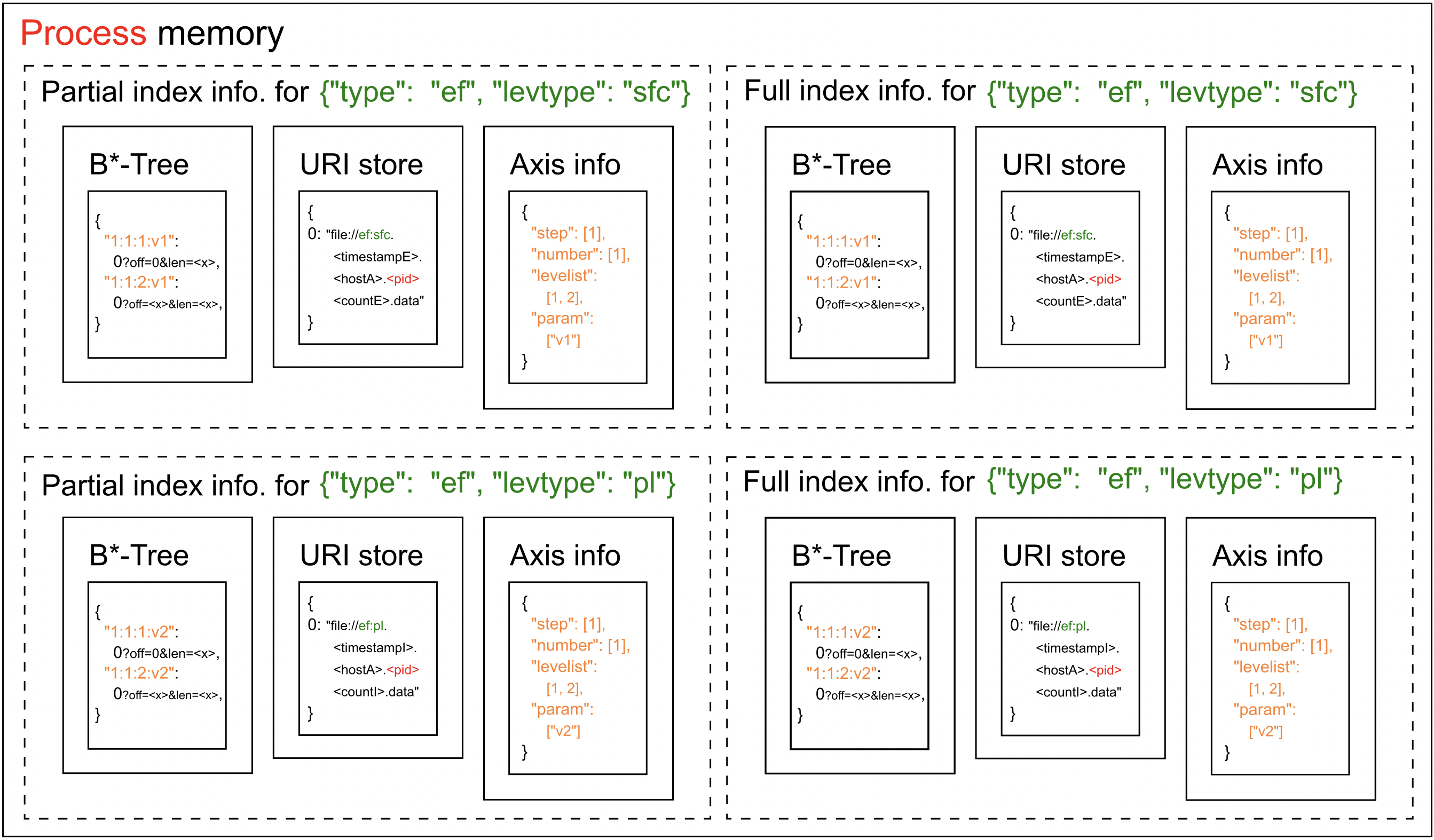}}
\caption{Content held in memory by a process after calling FDB archive() four times for four meteorological objects for two different collocation keys, before flush() is called.}
\label{fig:fdb_posix_memory_before_flush}
\end{figure}

Fig. \ref{fig:fdb_posix_archive_before_flush} shows an example of the state of a dataset after \verb!archive()! has been called but before \verb!flush()! is called. Most indexing files are empty, and sub-TOC files have not been created yet. Fig. \ref{fig:fdb_posix_memory_before_flush} illustrates the data structures held in process memory after these \verb!archive()! calls.

\noindent
\textbf{\textit{flush()}}

The POSIX Catalogue \verb!flush()! method writes all partial B*-Tree indexes held by a process, potentially for multiple dataset and collocation key pairs, at the end of the corresponding partial index files. For every index, the associated URI store and axes are written at the end of a separate per-process file ---referred to as \textit{sub-TOC} file--- which is first created if it does not exist. The collocation key associated to the index and a descriptor of the location of the index within the index file (composed of a file URI, offset, and length) are also written in the sub-TOC file.

On creation of a sub-TOC file ---which occurs only on the first \verb!flush()! call for a given process and dataset key---, a pointer to that file is appended to the shared TOC file, making all sub-TOC and index files easily discoverable by \verb!list()!ing and \verb!retrieve()!ing processes. Given the TOC file was opened with \verb!O_APPEND!, this is simply done with a \verb!write! system call. 
Multiple concurrent process may race to append their pointers to the TOC file on their first \verb!flush()! call, but POSIX guarantees atomicity and therefore consistency as long as the contending \verb!write! operations are for data buffers smaller than the configured system block size, as is the case here.

The sub-TOC files exist separately from the index files to make high-level indexing information efficiently accessible without requiring scanning of the index data.
Because every process writes exclusively to its own index and sub-TOC files, there is no contention or room for inconsistency when writing data at the end of these files.

After the indexing information has been persisted and the pointers appended to the TOC file, the partial in-memory indexes are reset, the full indexes are kept intact, and \verb!flush()! returns.

If more objects are \verb!archive()!d after a \verb!flush()! call, a subsequent Catalogue \verb!flush()! call will only result in indexing information being appended to the corresponding index and sub-TOC files, as long as no objects are previously archived for new dataset and collocation key pairs.

\begin{figure}[htbp]
\centerline{\includegraphics[width=415pt,trim={0 0 0 0},clip]{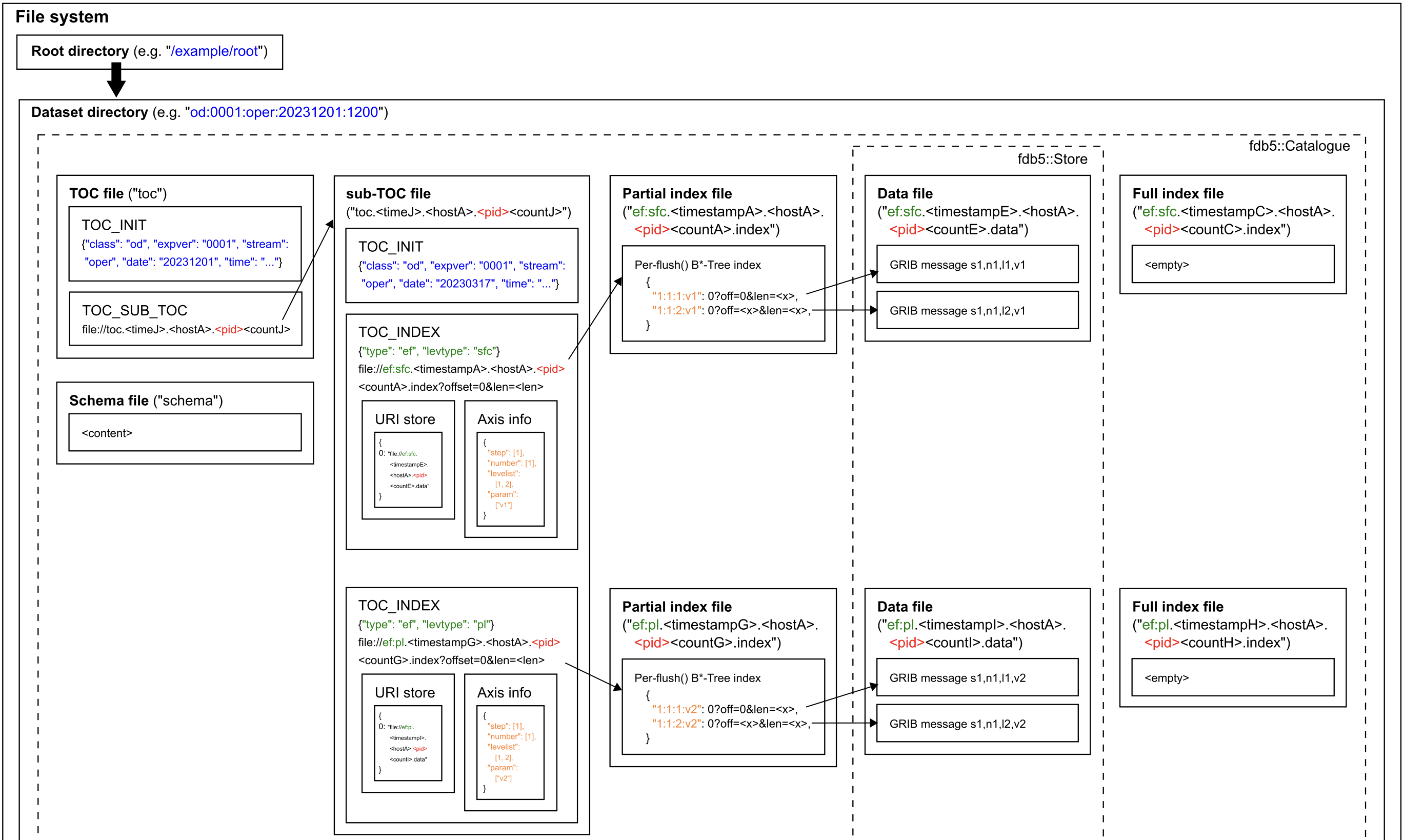}}
\caption{Snapshot of the content of a dataset directory after a process has called FDB archive() four times for four meteorological objects for two different collocation keys, after flush() has been called. All data objects and indexing information required to access them has been persisted into storage media at this point. close() has not been called yet and therefore the full index files are still empty.}
\label{fig:fdb_posix_flush_1step}
\end{figure}

Fig. \ref{fig:fdb_posix_flush_1step} shows the state of the example dataset from Fig. \ref{fig:fdb_posix_archive_before_flush} after \verb!flush()! is called, with the mentioned indexing information persisted in storage media.

\begin{figure}[htbp]
\centerline{\includegraphics[width=415pt,trim={0 0 0 0},clip]{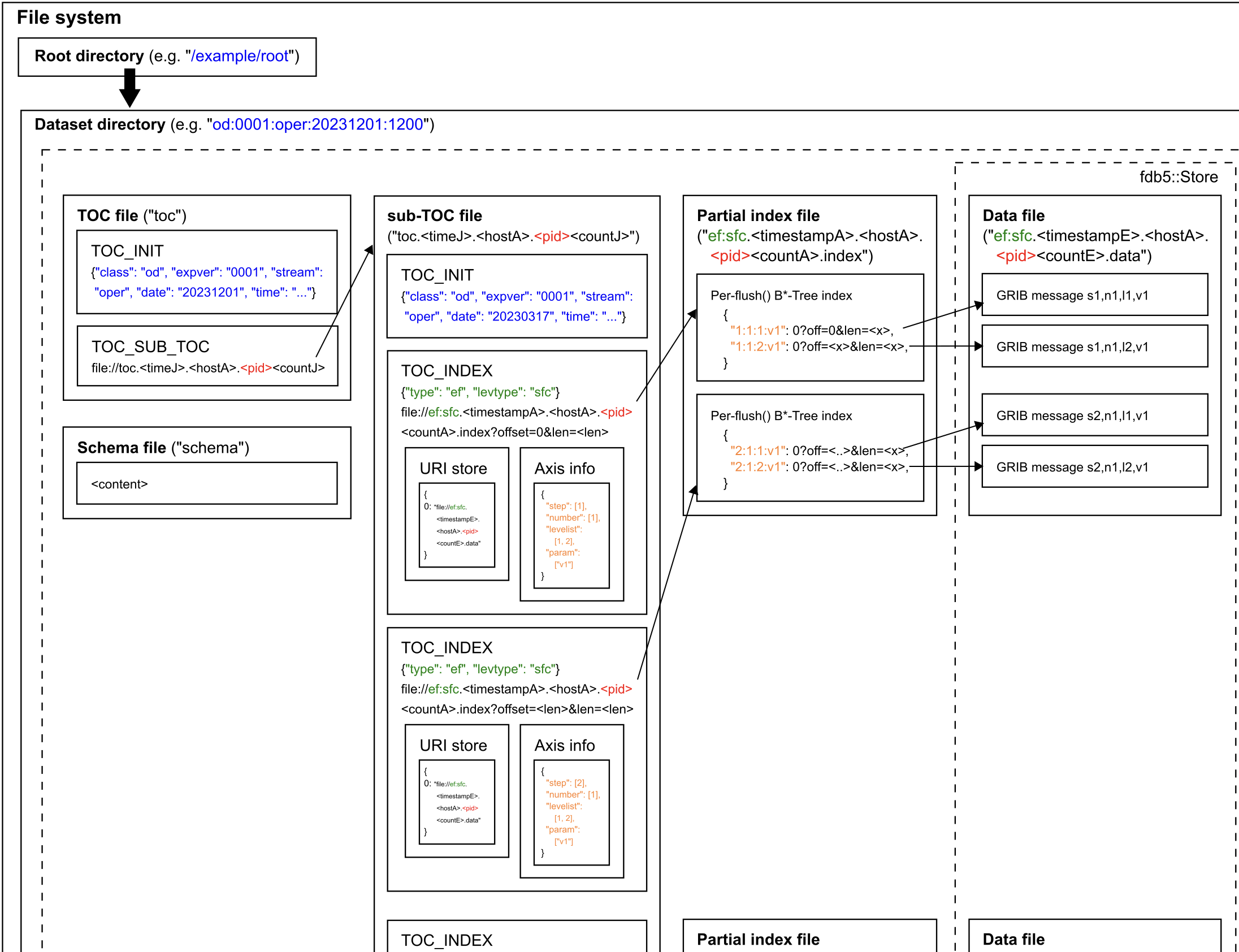}}
\caption{Snapshot of the content of a dataset directory after a process has called FDB archive() a few times, then flush(), then archive() a few more times, and then flush() again. The partial indexing information for all archive() calls before a flush() call is appended to the partial index and sub-TOC files.}
\label{fig:fdb_posix_flush_2step}
\end{figure}

Fig. \ref{fig:fdb_posix_flush_2step} shows part of the dataset state after additional objects are archived and flushed, using a value of 2 instead of 1 for the \verb!step! element key.

\begin{figure}[htbp]
\centerline{\includegraphics[width=415pt,trim={0 0 0 0},clip]{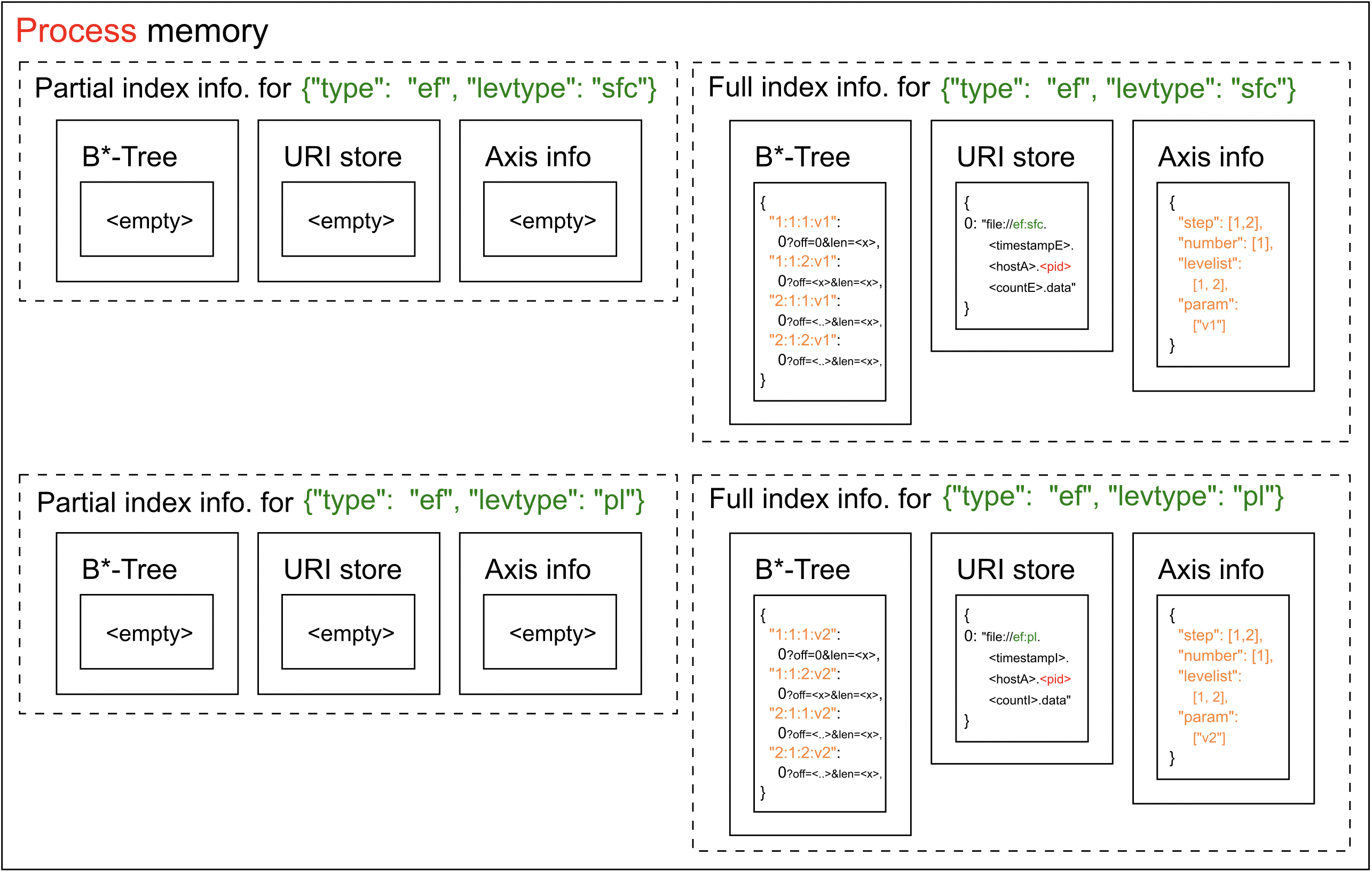}}
\caption{Content held in memory by a process after calling FDB archive() a few times, then flush(), then archive() a few more times, and then flush() again. The partial indexes and information are reset, and the full indexes and information accumulate new entries for all archive() calls during the lifetime of the process.}
\label{fig:fdb_posix_memory_after_flush}
\end{figure}

Fig. \ref{fig:fdb_posix_memory_after_flush} shows the structures held in memory after such second flush. Full indexes have grown, and partial indexes are empty.

\noindent
\textbf{\textit{close()}}

At the end of the lifetime of an \verb!archive()!ing process, the Catalogue \verb!close()! method is called, which persists the full indexes into the corresponding files created on the first \verb!archive()! call. The URI store and axes associated to every full index are encapsulated together with the collocation key and a pointer to the corresponding full index file, and appended as an entry to the shared TOC file using the same mechanisms as in \verb!flush()! to ensure consistency.

An additional \verb!TOC_MASK! entry is appended to the TOC file after every full index entry. This entry signals any \verb!retrieve()!ing or \verb!list()!ing processes to skip the pointers to sub-TOC files made obsolete by the new full index.
Consumer processes are therefore expected to read the TOC file in reverse order to ensure any masking entries are found before scanning unnecessary sub-TOCs.

\begin{figure}[htbp]
\centerline{\includegraphics[width=415pt,trim={0 0 0 0},clip]{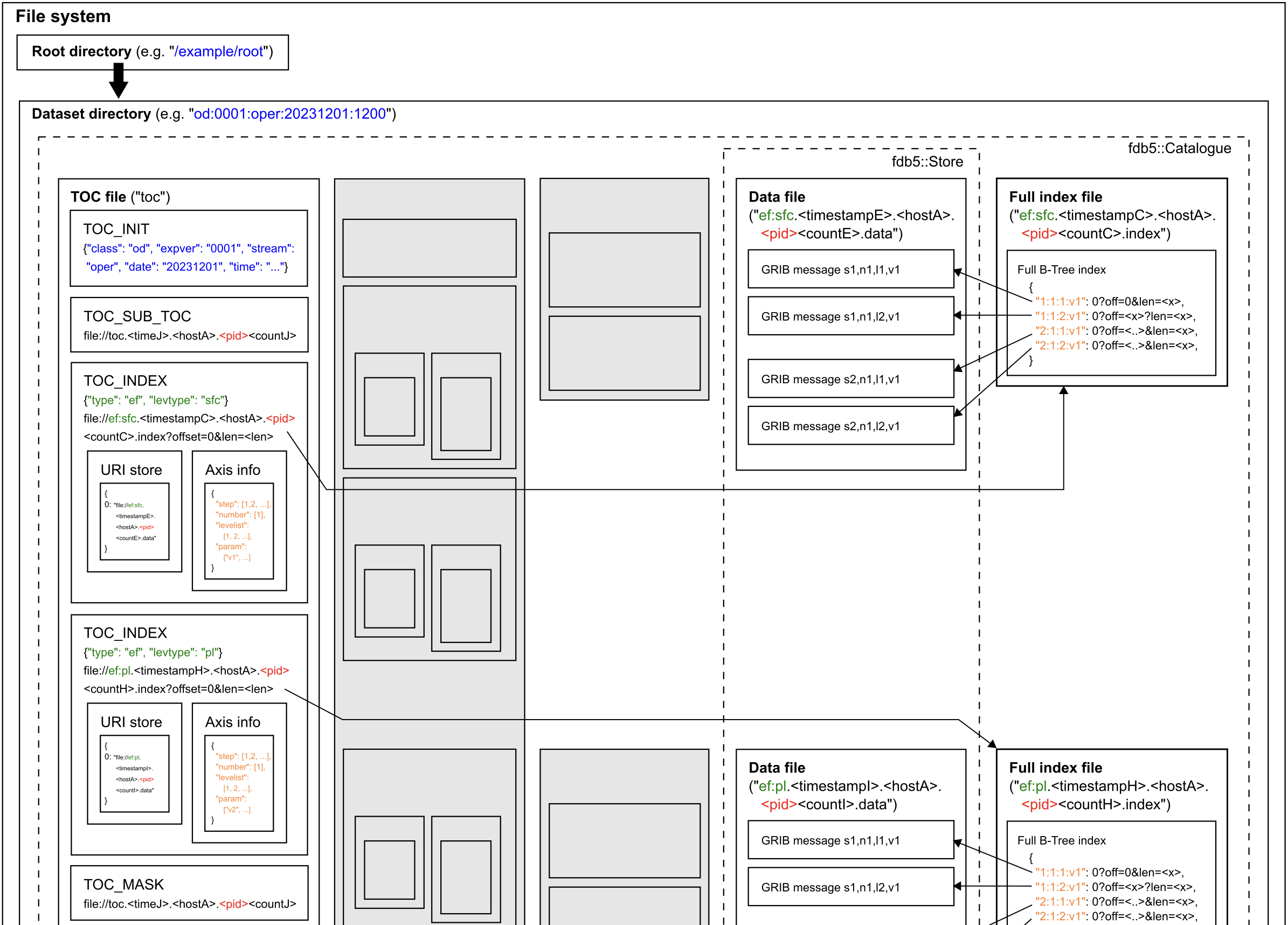}}
\caption{Snapshot of the content of a dataset directory after an archive()ing process has terminated and Catalogue close() has been called. New entries are appended to the TOC file pointing to the full indexes, and the sub-TOCs and partial indexes are masked.}
\label{fig:fdb_posix_close}
\end{figure}

Fig. \ref{fig:fdb_posix_close} shows the state of the dataset after \verb!close()! is called, with new full index entries present in the TOC file, and with the sub-TOC file masked.

\newpage

\noindent
\textbf{\textit{TOC pre-loading}}

The first time a process calls FDB \verb!retrieve()! or \verb!list()! for a given dataset key, the POSIX Catalogue backend reads the dataset TOC file, scans it in reverse order, and reads all sub-TOC files ---except for the masked ones---, rebuilding all URI stores and axes in memory. The TOC and sub-TOC files are loaded entirely with a single \verb!read! system call per file.

This loading and scanning in advance of the full content of all TOC and non-masked sub-TOC files is necessary for \verb!list()! to provide consistent results, and desirable for \verb!retrieve()! as otherwise, if loading the sub-TOC files only when strictly necessary, the loading would potentially happen in many sparse \verb!read! system calls, not benefiting from I/O parallelism.

One caveat of this pre-loading mechanism is that, if a consumer application \verb!retrieve()!s a dataset using multiple parallel processes ---as is the case in operational runs---, all of them will read all TOC and sub-TOC files on the first \verb!retrieve()! call, potentially adding significant strain on the file system if many processes are used.
This can be addressed by using threads instead of processes or, as done in operational runs, by calling \verb!retrieve()! from a single process and node for the full set of identifiers of objects to be retrieved, and distributing the resulting location descriptors or \verb!DataHandle!s to other nodes and processes.

\newpage

\noindent
\textbf{\textit{axis()}}

As explained in the Catalogue interface introduction, the \verb!axis()! method is called by FDB \verb!retrieve()! in a preliminary step to expand any multi-object expressions provided as input.

When \verb!axis()! is called, a dataset key, collocation key, and the name of an element key dimension are provided as inputs. Also, due to TOC pre-loading, all axes of the dataset are available in memory for the \verb!axis()! method to use.
The method visits the axes of all indexes matching the input dataset and collocation keys, extracts the values for the specified dimension from each of them, and merges the obtained values. This results in a single set containing all values indexed across all indexes in the dataset for the specified dimension, and this set is returned.

This operation should be relatively fast, as it is performed entirely in memory and does not require scanning the content of the B*-trees.

\noindent
\textbf{\textit{retrieve()}}

The \verb!retrieve()! method visits all index entries pre-loaded from the TOC and sub-TOC files which have a collocation key matching the one provided as input. For every index entry visited, the axes are queried to determine whether the B*-tree index may contain a match for the element key provided as input. If so, the B*-tree index is then loaded from the corresponding index file, and this is done with multiple \verb!read! system calls. The B*-tree index is then queried to obtain an object location descriptor associated to the element key. If not found, visiting of other index entries is resumed. If found, the descriptor is expanded using the information in the URI store and returned.

A \verb!retrieve()! operation only has visibility of the index entries and axes pre-loaded on first call of a \verb!retrieve()! or \verb!list()! operation for any objects with the same dataset key. Any newly \verb!archive()!d or \verb!flush()!ed objects will not be visible.

Leaving TOC pre-loading aside, a \verb!retrieve()! operation usually requires accessing the file system to load the matching B*-Tree, and this is likely to dominate the latency of this operation. However, the in-memory visiting of the TOC and sub-TOCs can also impact performance, particularly if the number of indexes ---and therefore number of parallel \verb!archive()!ing processes--- is large, or if the \verb!archive()!ing processes have not yet finalized and persisted the full indexes. For isolated \verb!retrieve()! operations for one or a few object identifiers, TOC pre-loading will likely dominate the latency if full indexes are not yet available.

\noindent
\textbf{\textit{list()}}

The \verb!list()! method loads the B*-tree indexes, for a given dataset, which have a collocation key matching the partial identifier provided as input. Multiple \verb!read! system calls are used to load each index.
For every index, all of its entries are visited, and those matching the partial identifier are added to a list which is returned at the end of the visiting.

Like \verb!retrieve()!, a \verb!list()! operation only has visibility of the pre-loaded indexing information of the dataset, and will suffer from the same TOC pre-loading performance impact if full indexes are not available. Aside from TOC pre-loading, the latency of a list operation will largely vary depending on the number of indexes matched and loaded.

\subsubsection{Operational NWP I/O pattern}
\label{subsubsec:operational_nwp_pattern}

Currently, operational runs at the ECMWF consist of an ensemble of 52 NWP model members. For each member, 5 HPC nodes are allocated to forward I/O from model processes to HPC storage, adding up to a total of 260 I/O nodes for all members. 2600 I/O server processes run across these nodes, each performing long sequences of FDB \verb!archive()! calls.
During a model time step, every I/O server process receives and \verb!archive()!s 65 weather fields of an average size of 4.3 MiB, and calls FDB \verb!flush()! at the end of the archival of the data for that step. This is repeated for 144 time steps.
A full operational model run produces approximately 24 million fields, which are distributed as approximately 9300 fields per I/O server process.
The same dataset and collocation keys are used for the identifiers of most of the \verb!archive()!d fields.

\begin{figure}[htbp]
\centerline{\includegraphics[width=415pt,trim={0 0 0 0},clip]{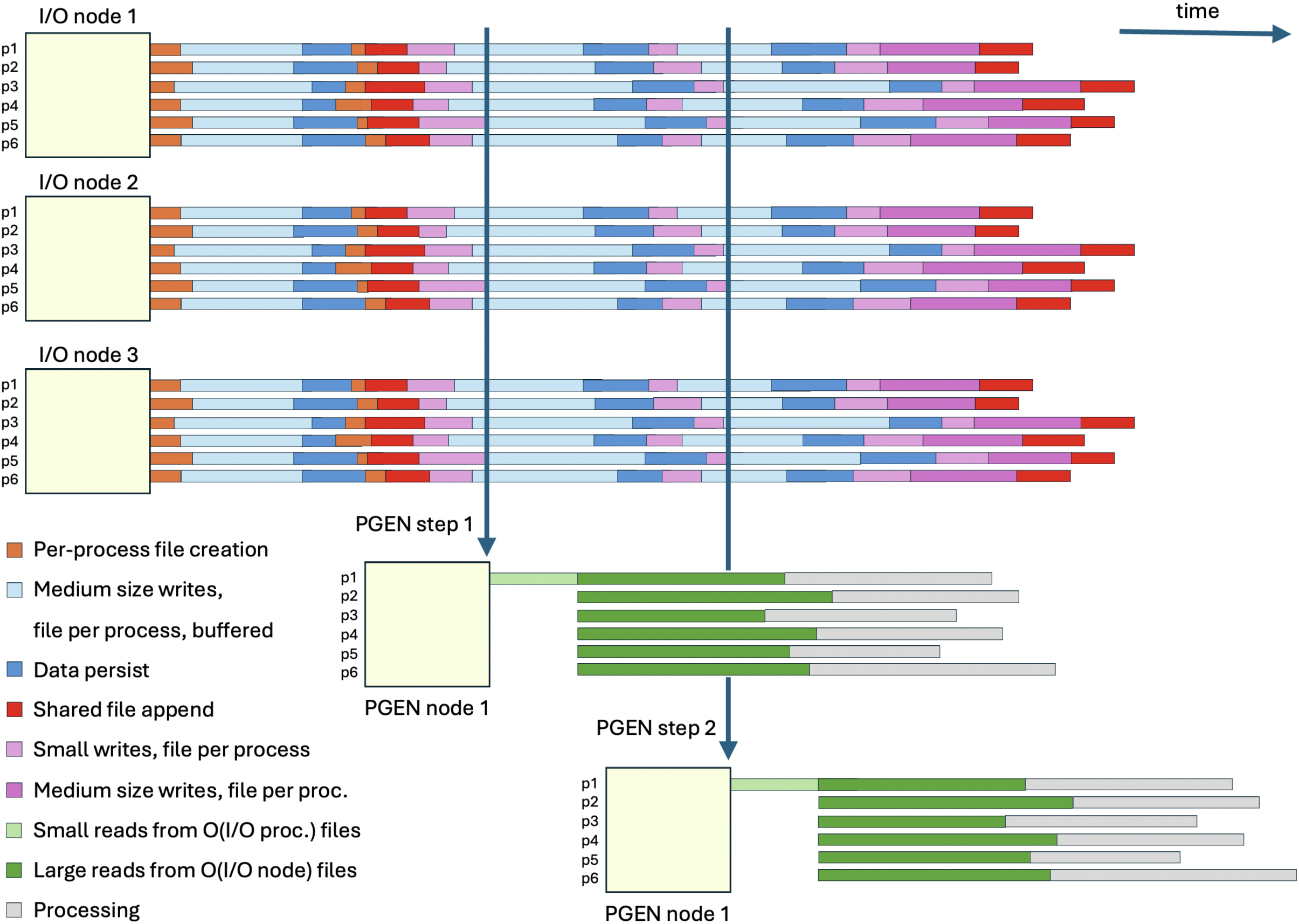}}
\caption{Not based on real profiling. Illustration of a simplification of the ECMWF's operational NWP storage access pattern. Three I/O server nodes archive weather fields produced by the NWP model, using multiple parallel processes per node (p1-p6), for three simulation time steps. At the end of every simulation step the archived fields are flushed, and a post-processing job (PGEN) is launched for that step. The third PGEN job has been omitted. Every PGEN job runs on one node and lists and retrieves data using multiple processes per node (p1-p6). The proportions of the durations for the different types of operation in real operational runs might significantly differ from the ones displayed in this illustration. Real operational runs currently use 260 I/O nodes and 4 to 8 nodes per PGEN job, and archive data for 144 time steps.}
\label{fig:operational_nwp_pattern}
\end{figure}

On first \verb!archive()! call, the I/O server processes would normally contend to initialise the dataset directory, including directory and TOC file creation, TOC file initialisation, and schema writing. This is avoided, however, by manually initializing the directory in advance.
Therefore, the first operation performed by \verb!archive()!ing processes is the creation of their per-process data and index files. This means several file create operations are issued simultaneously, putting metadata operation strain on the file system, although this is a one-off event. This initialisation phase is illustrated with orange stripes at the start of the parallel processes, on the left end in Fig. \ref{fig:operational_nwp_pattern}.

Writing of field data into per-process data files then begins. Object data is first written into memory buffers and kernel page caches, which start becoming full after a few \verb!archive()! calls, and their content is then written to storage in large transfers in the background.
Because objects are first written in memory, the internal Store \verb!archive()! calls return location descriptors quickly, without waiting for the file system to persist the data, and these descriptors are passed to the Catalogue \verb!archive()! method, which performs in-memory indexing of the location descriptors.
This phase lasts as long as required for all processes to archive and index their 65 assigned fields, and combines writing object data to buffers and in-memory indexing work, while, in the background, some of the data is being transferred from buffers to storage. This phase is illustrated with light blue stripes in Fig. \ref{fig:operational_nwp_pattern}.

As soon as processes have archived their fields for a step, they call FDB \verb!flush()!, where the data buffers are flushed and persisted into storage. This adds bulk I/O pressure on the system, and processes wait for this to complete. Once object data is persisted, partial indexes are then written into the per-process partial index files, and processes also wait for this operation to be fully persisted. This blocking write phase is illustrated with dark blue stripes in Fig. \ref{fig:operational_nwp_pattern}.

Still as part of \verb!flush()!, processes then create a per-process sub-TOC file, adding metadata pressure on the system ---illustrated with orange stripes---, then contend to append an entry to the shared TOC file ---illustrated with red stripes---, and finally write the sub-TOC index entries, including the axes and URI store, at the end of their sub-TOC file, and wait for the write to be persisted --- illustrated with purple stripes.

Processes continue to \verb!archive()! and \verb!flush()! fields for subsequent simulation steps. For these, only the light blue, dark blue, and purple phases are performed.

At the end of the model run, Catalogue \verb!close()! is called and all processes write their full indexes into the per-process full index files, waiting for the writing to be fully persisted ---illustrated with dark purple stripes---. Finally, they contend to append the corresponding index and mask entries to the shared TOC file --- illustrated with red stripes.

At the end of most model steps, as soon as the straggler I/O process has \verb!flush()!ed its data for the step, the workflow manager is signaled to execute a post-processing job ---also referred to as Product Generation (PGEN) job--- for that step.

Post-processing jobs run each on 4 to 8 HPC nodes, using 8 processes per node.
First, the entire indexing information for the step is loaded from a single node and process by calling FDB \verb!list()!, providing a partial request which matches all weather fields \verb!archive()!d for that step. This reads the full content of the TOC and the several sub-TOC files, as well as the last partial index in all index files, adding I/O pressure to the system as many files ---two per \verb!archive()!ing process--- need to be accessed and many I/O operations are performed.
This can create write-read contention with any \verb!flush()!ing processes appending entries to the sub-TOC files or indexes to the partial index files.
The listing process then calls FDB \verb!retrieve()! for all field identifiers, which efficiently extracts the full set of location descriptors from the indexes made available in memory by \verb!list()!.
This phase, dominated by the retrieval of indexing information, is illustrated with light green stripes in Fig. \ref{fig:operational_nwp_pattern}.

The location descriptors are then distributed across all job nodes and processes such that each process is assigned descriptors of fields for a subset of parameters across all ensemble members, as the full ensemble data is required for the generation of derived products. The processes then build \verb!DataHandle!s for their assigned descriptors, merge the handles into one, and read the data from storage through that handle. 
This results in many read operations being issued. Although the size of the data transfers is large ---thanks to handle merging---, these operations hit all data files written by I/O servers, adding significant I/O pressure on the file system.
Additionally, write-read contention may occur between these and any \verb!flush()!ing processes appending data to the data files. This phase, characterised by several large reads, is illustrated with dark green stripes in Fig. \ref{fig:operational_nwp_pattern}.

The processes then apply computations as required to generate the derived products. This is illustrated with grey stripes.

\subsubsection{fdb-hammer I/O pattern}
\label{subsubsec:fdbh_pattern}

\verb!fdb-hammer! is an FDB performance benchmarking tool provided as part of the FDB Git repository\cite{FDB}, which can be built alongside the other FDB command-line tools.

\verb|fdb-hammer| takes a single weather field as input, supplied as argument on the command line, and uses its encoded metadata as a template to generate a sequence of field identifiers which are then archived ---alongside a copy of the field data---, retrieved or listed. By default, \verb|fdb-hammer| runs independently as a single process, without synchronisation with any other parallel processes. However, multiple instances of \verb|fdb-hammer| can be executed in parallel mimicking the ECMWF's operational I/O behaviour, where model I/O server processes \verb!archive()! independent streams of data, and post-processing processes \verb!retrieve()! subsets of the archived data.
\verb|fdb-hammer| creates an ``I/O-pessimised'' benchmark, that is, a worst possible case for I/O as all relevant computation that would occur in the operational workflows has been removed.

For the performance analysis in this dissertation, \verb|fbd-hammer| has been run in two phases: a write phase, and a read phase. In each phase, multiple parallel processes are launched across a set of nodes with access to the storage system to benchmark against.

During the write phase, each parallel process issues a sequence of FDB \verb!archive()! calls.
A sequence of weather field identifiers corresponding to a range of NWP model parameters and levels ---adjusted via the \verb!--nparams! and \verb!--nlevels! options--- is iterated, and every identifier is passed as input to an FDB \verb!archive()! call along with a fixed weather field used as data object for archival.

The \verb!number! key of the identifiers, which identifies the ensemble member a field belongs to, is set to a fixed value depending on the node the writer process runs on. This is to mimic operational behavior, where an I/O server node \verb!archive()!s several fields for a single member. Although in operations the data for a given member is managed by multiple I/O server nodes, fdb-hammer is designed to have each writer node \verb!archive()! fields for a single member.

Initially, every parallel writer process sets the \verb!step! key of the identifiers to a value of 1. When the full range of identifiers for the different parameters and levels has been iterated, \verb!flush()! is called. The \verb!step! value is then increased by 1, and another sequence of \verb!archive()! calls is issued for all parameters and levels, followed by a \verb!flush()!. This is repeated for as many steps as configured via the \verb!--nsteps! option, and the write phase ends with a Catalogue \verb!close()! call after the \verb!flush()! call for the last step.

The \verb!param!, \verb!levelist!, \verb!number!, and \verb!step! keys are therefore the only field identifier keys adjusted by \verb!fdb-hammer! and, when run on POSIX file systems ---using the default FDB schema described in the introduction of Section \ref{sec:fdb}---, this results in the benchmark \verb!archive()!ing fields with unique element keys which all share the same dataset and collocation key.

\begin{figure}[htbp]
\centerline{\includegraphics[width=415pt,trim={0 0 0 0},clip]{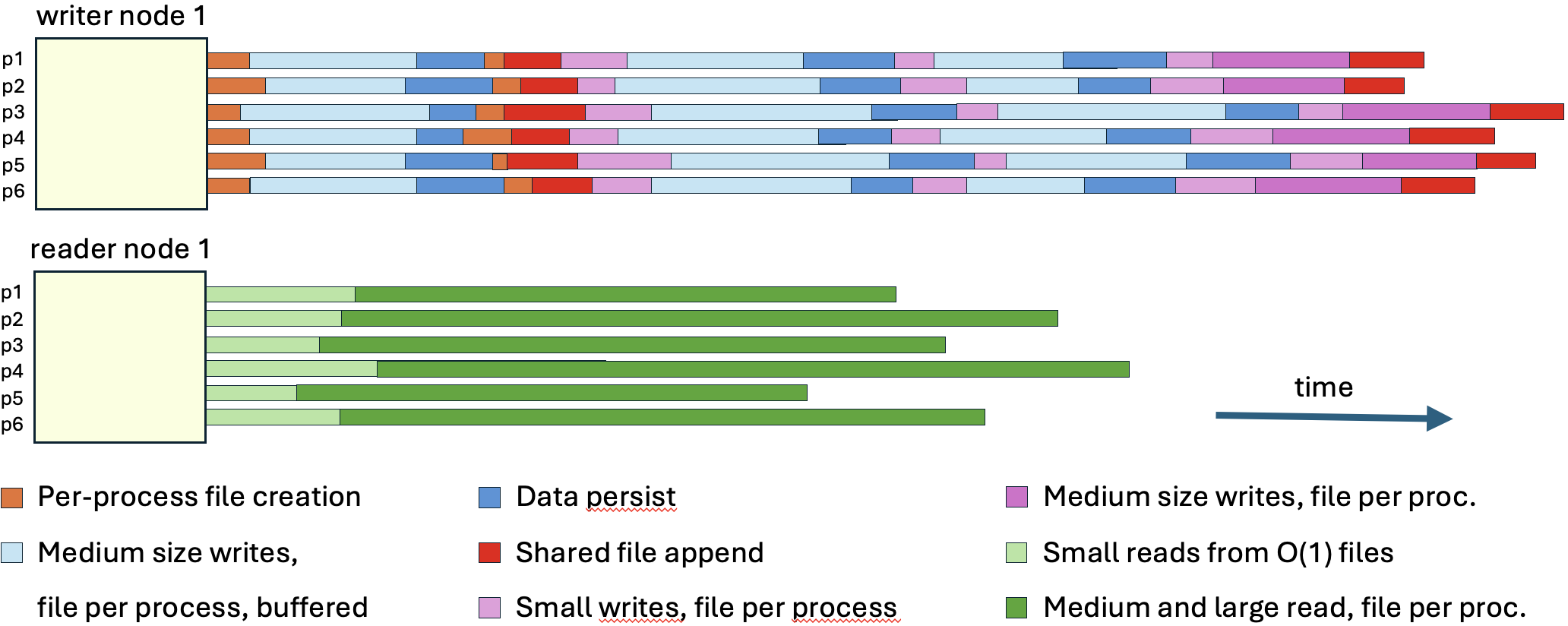}}
\caption{Not based on real profiling. Illustration of a simplification of the storage access pattern of the fdb-hammer benchmark. One writer node archive()s weather fields for three time steps using multiple parallel processes per node (p1-p6). At the end of every step the archived fields are flushed. One reader node simultaneously retrieves pre-archived data using an equivalent set of parallel processes (p1-p6) performing equivalent sequences of retrieve() operations. Each reader process loads the content of the TOC file, then loads one full index from one index file, and reads all fields from a single data file --- these index and data files have been written by the corresponding writer process. The proportions of the durations for the different types of operation in real fdb-hammer runs might significantly differ from the ones displayed in this illustration. Real fdb-hammer runs performed in this analysis employed up to 24 writer and 24 reader nodes, with up to 48 processes per node, and were run for 100 time steps.}
\label{fig:fdbh_pattern}
\end{figure}

Overall, the write phase issues a set of FDB API calls and produces a storage access pattern very similar to that of operational runs of similar size. Fig. \ref{fig:fdbh_pattern} illustrates the storage access pattern of a simple \verb!fdb-hammer! run using one writer node and one reader node. The write pattern is very similar to that of operational runs, shown in Fig. \ref{fig:operational_nwp_pattern}.

Despite the write phase being very similar to operational runs, it is worth noting that all \verb!fdb-hammer! runs performed in Chapter \ref{chap:chap4} were configured to produce a different number of fields than would be produced operationally. As shown in table \ref{tab:ope_fdbh_size_comparison}, \verb!fdb-hammer! runs were configured to simulate not only fewer members and steps, but also fewer parameters and levels per step. This resulted in less data being written per step --- that is, in \verb!fdb-hammer! runs having proportionally shorter light and dark blue stripes compared to operational runs.

\begin{table}[htbp]
\caption{Comparison of the dimension of ECMWF's operational NWP runs and fdb-hammer runs performed in Chapter 4.}
\begin{center}
\begin{tabular}{|c|c|c|}
\hline
\textbf{}&\textbf{operational run}&\textbf{fdb-hammer} \\
\hline
members & 52 & 1 to 24  \\
\hline
steps & 144 & 100  \\
\hline
levels & 150 & 10  \\
\hline
parameters & 20 & 10  \\
\hline
\end{tabular}
\label{tab:ope_fdbh_size_comparison}
\end{center}
\end{table}

The read phase is run on as many nodes and using as many processes as the write phase. The processes in the read phase issue sequences of \verb!retrieve()! calls equivalent to those in the write phase.

This phase can be run either independently once the write phase has ended, or concurrently with the write phase on a separate but equally sized set of nodes, although the latter option requires previously running the write phase once in isolation to populate the storage system for the readers to always find data in case they are faster than the writers. Both modes have been employed for the performance analysis in this dissertation.

The reader processes by default check that all previously written and indexed fields are found and can be de-indexed and their data retrieved. If this consistency check is not fulfiled, the benchmark execution fails with an error. Additionally, the readers can optionally be configured to verify that the read data matches exactly the data that was originally written. This requires readers to load and process the read data in memory, which can impact benchmark performance, and it is therefore advisable to execute the benchmark separately, with this option enabled, to fully test storage system consistency.

The \verb!retrieve()! calls issued during the \verb|fdb-hammer| read phase load a set of field location descriptors equivalent to the set loaded by all \verb!list()!ing PGEN processes in an operational run of the same size. However, the timing and the storage access pattern generated are different in a few ways.

The first difference is that all \verb!fdb-hammer! reader nodes and processes start issuing \verb!retrieve()! calls and fetching data at the very beginning of the read phase.
On one hand, this implies that \verb!fdb-hammer! readers
mostly hit data files and full indexes from a previously completed write phase. Therefore, there is no write-read contention on sub-TOC or data files. This is in contrast to operational readers, which are started in a staggered way, one subset of processes at a time after the data for each step has been \verb!flush()!ed and before the full indexes have been made available, and this results in write-read contention on sub-TOC and data files.
On the other hand, this means that all the index accesses occur at the beginning of the \verb!fdb-hammer! read phase, making the first part of a run more challenging in terms of metadata operations than the rest of the run. In operational runs, instead, indexes are accessed in multiple bursts, one at the beginning of every PGEN job, thus having the index access workload spread more evenly along the duration of the run.

Also, although TOC pre-loading is performed by every \verb!fdb-hammer! reader process on the first \verb!retrieve()! operation, it involves very few or no partial indexes, resulting in a small number of \verb!read! system calls overall for TOC pre-loading. In operational runs, instead, many sub-TOC files are \verb!read! during TOC pre-loading, and although only a single process in every PGEN job pre-loads, the total number of \verb!read! system calls performed for pre-loading is larger than in \verb!fdb-hammer! runs.

Another difference is that, because every \verb!fdb-hammer! reader process issues a sequence of \verb!retrieve()! calls for the entire set of fields \verb!archive()!d by its corresponding writer process ---for a single member and the full sequence of \mbox{steps---,} every reader process only loads the full index made available by the corresponding writer process from a previously completed write phase. This means as many \verb!read! system calls are issued overall for index loading as there are reader process.
In operational runs, instead, every PGEN job has a single \verb!list()!ing process load the last partial index from every index file for every writer process, thus issuing as many \verb!read! system calls as the number of writer processes times the number of PGEN jobs. This is a larger number of \verb!read! operations for index loading than in \verb!fdb-hammer!, by approximately two orders of magnitude.

Every \verb!retrieve()! call performed by an \verb!fdb-hammer! reader processes returns a location descriptor wrapped as a \verb!DataHandle!, which is then merged into one with subsequent \verb!DataHandle!s. The resulting merged handle is then used to read the entire content of the data file written by the corresponding writer process. This results in a few large \verb!read! system calls per reader process.
In operational runs, instead, the processes in a PGEN job read data of fields for a single step for all ensemble members, resulting in many \verb!read! system calls hitting all data files at every step. The amount of data read is the same as in \verb!fdb-hammer! runs of equivalent size, but data is read using a much larger number of \verb!read! system calls, by approximately one order of magnitude.

The read phase is illustrated at the bottom of Fig. \ref{fig:fdbh_pattern}. All processes first load the TOC file ---represented with light green stripes---, and then load the full content of a single index file and a single data file --- illustrated with dark green stripes.

In summary, whereas the write access pattern of \verb!fdb-hammer! very closely mimics that of the operational runs, the read phase has a few differences, the main ones being that all reading starts simultaneously at the beginning of the run, and that the transposed access of operational runs is not reproduced. This results in \verb!fdb-hammer! producing a more condensed I/O workload at the beginning of the run and an easier I/O workload overall, as write-read contention is partly not captured, fewer small \verb!read! operations are performed for index access, and fewer and larger \verb!read! operations are performed for bulk data reading.

Despite these differences, \verb!fdb-hammer! is still challenging, as it performs only I/O ---excluding all processing---, and is representative of operational runs, as it captures well the I/O in terms of FDB API calls and reproduces the operational contention in terms of volume of data being stored and retrieved simultaneously from the file system. Also, \verb!fdb-hammer! has other advantages such as being a much simpler tool to configure and run than other more sophisticated benchmarks like Kronos or real operational workflows, and not requiring large amounts of nodes to meaningfully reproduce the desired contentious patterns.

This analysis has informed further developments at the ECMWF to make \verb!fdb-hammer! more representative of operational runs.

    \chapter{Object Storage Backends}
\label{chap:chap3}

This chapter describes newly developed FDB backends for operation on the DAOS and Ceph object stores, as well as backends to support other storage solutions implementing the S3 storage protocol. The options these storage approaches provided and the challenges encountered during the development are discussed.

This addresses the second contribution of the dissertation --- the development of \textit{"Object storage adapters for the ECMWF's NWP workflows"}.

\section{DAOS backends}
\label{sec:daos_backends}

Before the DAOS Catalogue and Store backends for the FDB were developed, a preliminary assessment was conducted to determine whether DAOS provided sufficient functionality for the implementation of both backends, as well as the required consistency guarantees and performance.
This assessment, which resulted in positive answers to all these questions, is presented in Appendix B.
As part of the assessment, a proof-of-concept library named Field I/O was developed which provides a pair of functions to write and read weather fields from DAOS, including creation and use of an internal index.
During this development, best practices for high performance on DAOS were identified\cite{Manubens2023DAOSPrediction,Jackson2023ProfilingDAOS} and implemented in the Field I/O library.

The implementation of the DAOS backends was heavily informed by the preliminary assessment and the design of the Field I/O functions.
Essentially, the DAOS backends store object data and indexing information in a container per dataset key.
Every \verb!archive()!d meteorological object is written into a new DAOS array ---with its own object identifier--- within the corresponding dataset container, and a network of key-value objects is created and populated in the container which in conjunction form an index of all objects \verb!archive()!d, enabling efficient discovery and access on \verb!retrieve()! and \verb!list()!.
The transactionality of the \texttt{daos\_kv\_put} and \texttt{daos\_kv\_get} operations on the DAOS key-value objects is critical to ensuring consistency of the index under \verb!archive()! and \verb!retrieve()! contention, and such contention is resolved on the DAOS servers rather than by negotiating locks between clients and servers as is the case in the POSIX I/O backends.

The consistent behavior of DAOS and the FDB backends was verified by executing the \verb!fdb-benchmark! at scale, with the consistency check option enabled, on both computer systems employed for the performance analysis in Chapter \ref{chap:chap4}.

The end-to-end implementation of the FDB DAOS backends spanned a period of one to two years. This was in part due to the need to familiarise with the DAOS libraries and systems, and to assess any overall consistency or performance limitations ---covered in detail in the preliminary assessment---; but also due to the need to fully analyse and map the FDB's internals onto the object storage paradigm for the first time; the need for developing auxiliary unit testing libraries ---discussed in Appendix A---; the multiple required iterations of performance testing at scale and optimisation ---discussed in this section and in Chapter \ref{chap:chap4}---; and the fact that the developments were tested initially on a system with novel storage and network hardware, which posed a number of challenges. 

\begin{figure}[htbp]
\centerline{\includegraphics[width=415pt,trim={0 0 0 0},clip]{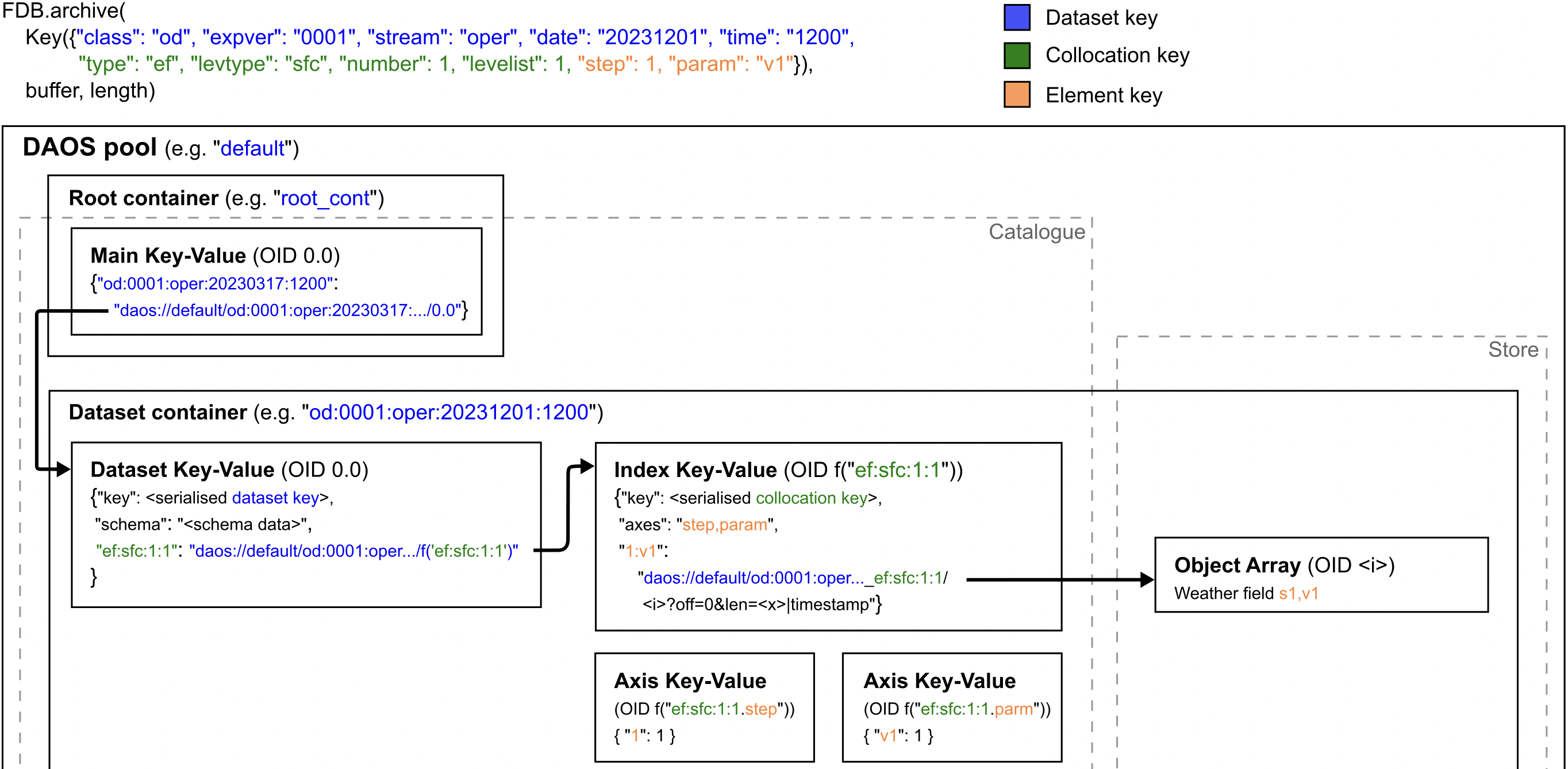}}
\caption{Diagram of DAOS entities resulting from an FDB archive() call for one meteorological object. The archive() call is shown at the top, using different colours for the dataset, collocation, and element keys. The object data and indexing information are persisted immediately. A flush() call would result in no additional changes to these entities.}
\label{fig:fdb_daos_backend}
\end{figure}

A graphical representation of the different DAOS entities involved in an FDB \verb!archive()! call is shown in Fig. \ref{fig:fdb_daos_backend}.

At the deepest layer of the key-value hierarchy, an \textit{index key-value} is created for every collocation key, holding an index of all objects \verb!archive()!d for that key within the dataset.
A few ancillary key-values, referred to as \textit{axis key-values}, hold summaries of the entries inserted into each index key-value for increased efficiency of the \verb!retrieve()! calls.
One layer above, a \textit{dataset key-value} is created within every dataset container, which holds an index of collocation keys for which objects were \verb!archive()!d, associating every collocation key to a pointer to the corresponding index key-values.
Finally, a top-level index of the available datasets, also implemented with a DAOS key-value, is kept in a separate \textit{root container}.
By default, all containers created by the DAOS backends are placed within a same DAOS pool configured by the FDB administrator at deployment time.

By default, all key-values and arrays created by the DAOS backends are configured with no sharding, that is, with an object class of \verb!OC_S1!, which results in them being stored each in a single DAOS storage target. In the preliminary assessment, this configuration was found to result in best performance for \verb!archive()!al and \verb!retrieve()!al at scale, as shown in Appendix B - Figure 8.
This choice assumes \verb!archive()! and \verb!retrieve()! are mainly used operationally and performed in parallel on multiple nodes and several processes per node, with every process writing or reading several small objects, on the order of 1 MiB. In such a scenario, because the several DAOS arrays are evenly distributed across DAOS targets, all targets are hit by all client nodes at all times, resulting in network and server saturation, and therefore good performance, despite the arrays not being sharded.

Sharding arrays might nevertheless be desirable for other use cases, for example if using a single process or working with larger objects. To that end, the FDB user can override the object class used for indexing key-values or object arrays, separately, choosing values such as \verb!OC_S2! to shard over two targets or \verb!OC_SX! to shard across all targets.
Depending on the type of storage hardware managed by the DAOS servers, and the power failure resilience mechanisms of the system, it might also be desirable to replicate or erasure-code the key-values or arrays over multiple storage targets, and this can be attained with object classes such as \verb!OC_RP_2G1! or \verb!OC_EC_2P1G1!. The performance of these DAOS data redundancy options is briefly analysed in 4.5 DAOS and Ceph data redundancy.

Regarding the overall design of the DAOS backends, as discussed in the preliminary assessment, a slightly different design was first considered such that the objects for every dataset and collocation key pair were placed in a separate container, as this mapped more closely to the design of the POSIX I/O backends.
However, the use of so many containers was found to significantly deteriorate performance, and the design was revisited to make use of a single container per dataset, plus one container for the top-level index of datasets, which addressed the performance issue.
This final design assumes applications will \verb!archive()! meteorological objects for only one or a few dataset keys, therefore resulting in only a few containers and thus avoiding deteriorating performance. This assumption is valid for an operational NWP run at the ECMWF, where objects are \verb!archive()!d for a single dataset key.

A potential alternative design to completely eliminate this issue would consist in using a single container for all datasets, also including the top-level dataset index.
However, having each dataset in a separate container is desirable as it simplifies maintenance tasks such as removing a specific dataset, which can be done with a simple DAOS container destroy operation, that is, without requiring knowledge and manipulation of the internal structures of the FDB backends.
Due to this, together with the fact that some performance tests revealed that using on the order of a few tens of containers did not significantly deteriorate performance, the current design was considered to be the best compromise.

Given every \verb!archive()!d FDB object is assigned a unique meteorological metadata identifier on \verb!archive()!, and given every object is stored in a separate DAOS array, the indexing component of the DAOS backends ---that is, the network of key-values--- could seem unnecessary, as every array could simply be created with an OID resulting from applying a hash function to the metadata identifier, and any consumer process requiring to \verb!retrieve()! a particular object could calculate the same hash to determine the corresponding OID, and read the object content directly without the need of querying an index to determine the OID.
However, without an index, it would be impossible to implement efficient listing of objects matching a partial identifier, or functionality to request retrieval of large sets of objects based on compact expressions with special features such as wildcards.
If this type of functionality is required, as is the case in the ECMWF's operational NWP, an index must be built on \verb!archive()!al. 
Nevertheless, even if an index is implemented, as is the case in the current DAOS backend design, the arrays could still be assigned OIDs based on hashes of the identifiers, and this would allow further optimisation in \verb!retrieve()! and \verb!list()!. This optimisation was not pursued and remains as potential future work.

\subsubsection{Differences w.r.t. the POSIX I/O backends}

One major difference with respect to the POSIX I/O backends is that no per-process structures are created in the DAOS backends, neither for storage of object data nor for indexing.
Regarding storage of object data, as demonstrated in the preliminary assessment, DAOS supports creating a DAOS array for every object ---that is, several objects per process--- without deteriorating performance, whereas using a file per object instead of a file per process in POSIX file systems would have resulted in an excessive strain on the metadata servers.
Regarding indexing, DAOS provides key-value objects which enable easy implementation of efficient, persistent, and strictly consistent indexes shared across processes, removing the need for building in-memory per-process indexes or implementing custom mechanisms to persist these and ensure their consistency under contention, as was done in the POSIX I/O backends.
These DAOS features resulted in notably simpler logic and thus better readability and maintainability of the code base with respect to the POSIX I/O backends.

The preliminary assessment showcased the implementation of an index based on DAOS key-values, and demonstrated this approach does not deteriorate performance as long as excessive multi-process contention on same key-value objects is avoided --- the performance impact of such contention is shown in Appendix B - Figures 6 and 7.

The root and dataset key-values are not sensitive to contention as, although concurrent producer and consumer processes interact with them, this interaction generally happens only shortly at the beginning of the processes' lifetimes. Index key-values are the most sensitive to contention.
Because all objects \verb!archive()!d by all concurrent writers for a same collocation key are indexed in the same index key-value, avoiding contention on these key-values requires avoiding \verb!archive()!ing objects for a same collocation key from many parallel processes simultaneously. This can be attained, if necessary, either by modifying the application accordingly, or by adjusting the FDB schema to recognize as part of the collocation key those object identifier components that vary the most across parallel processes.
As an example, in the ECMWF's operational case, the default FDB schema used in conjunction with the POSIX I/O backends defines that a collocation key is formed by the \verb!type! and \verb!levtype! components of an identifier, as illustrated in Listing \ref{listing:mars-request}. As explained in \ref{subsubsec:operational_nwp_pattern}, many parallel processes use the same collocation key in the operational case, and this is not an issue when using the POSIX I/O backends as a separate set of index and data files are created for every process, however this would be problematic if using the DAOS backends. To avoid contention on indexing key-values, a modified schema is used in conjunction with the DAOS backends which recognises the \verb!number! and \verb!levelist! components as part of the collocation key. This results in every parallel processes \verb!archive()!ing objects for a different collocation key, thus never contending with other processes. The use of this modified schema is illustrated in Fig. \ref{fig:fdb_daos_backend}.

One last noteworthy difference with respect to the POSIX I/O backends is that, because DAOS does not implement client-side caching and provides high performance despite performing many small I/O operations, all object writes and key-value updates are always persisted immediately on the server side, without the need for buffered I/O.
Objects are available for consumers to \verb!retrieve()! or \verb!list()! on return of \verb!archive()!, and the \verb!flush()! implementation does not perform any operation internally.
Although this does not immediately benefit operational runs at the ECMWF, it opens doors to explore further optimisation of the operational workflows, and may be beneficial for other FDB use cases.
Nevertheless, this also suggests that the current implementation of the DAOS backends may be excessively conservative, and there might be room for further performance gains if implementing a buffered and persistent-on-\verb!flush()! approach.

Similarly, because DAOS performs well for small I/O operations, only the strictly necessary information is extracted from the indexing key-values for \verb!retrieve()! and \verb!list()!, rather than loading and caching large chunks of indexing information as is done in the POSIX I/O backends. Due to this, the DAOS backends have different performance behaviour for these two operations, providing lower latency for retrieval or listing of one or a few meteorological objects, but higher latency when listing many of them.

\subsection{The DAOS Store}

\textit{\textbf{archive()}}

Data is stored in the DAOS backends in containers identified by a stringified representation of the dataset key. When the DAOS Store \verb|archive()| method is called, the corresponding dataset container is created if it does not exist within the DAOS pool. This is done with the \texttt{daos\_cont\_create\_with\_label} function of the \verb!libdaos! API, which guarantees atomicity of the operation even if multiple processes call \verb!archive()! for objects with a same new dataset key concurrently. Once opened for use, the pool and dataset container handles are cached for the process lifetime.

Every \verb|archive()|d object is written into a new DAOS array object.
A new unique OID is first obtained from DAOS with the \texttt{daos\_cont\_\-alloc\_oids} function to avoid collisions with any concurrent processes. This requires a round trip to the DAOS server and, as such, a large set of OIDs are pre-allocated in a single call and cached by a process before object creation.
The array is then created and opened with the \verb!daos_array_open_with_attr! function, which in fact does not perform any Remote Procedure Call (RPC). This was found to improve write performance at scale in contrast to using \verb!daos_array_create! which does perform an RPC.
Once the array is created, the input object data is written in it with \verb!daos_array_write! and the array is closed.
A unique location descriptor to this array is finally returned, composed of the pool name, the dataset container name, the OID of the array containing the written object, the offset the object data was written at within the array ---always zero given the array-per-object design---, and the object data length.

Note that the collocation key provided as input to this method has not been used to determine data placement. In a previous version of the backend this key was used to create separate containers that collocated data, but the additional containers resulted in significant performance overheads and they were removed. As such, all the object data associated with a single dataset key is collocated in the same container. The collocation key is nonetheless still used in the Catalogue backend for structuring the indexing information.

Another choice that had to be made for the implementation of this method was whether to store multiple FDB objects into a single DAOS array, as this would reduce the number of I/O operations with respect to the initial design and potentially result in better backend performance.
This could be attained by either collocating all objects for a same collocation key in a same array, which would require any concurrently \verb!archive()!ing processes to atomically append to such array, or writing any objects \verb!archived()! by a same process at the end of a per-process array, either individually and immediately on \verb!archive()!, or in batches on \verb!flush()!.
However, DAOS does not currently implement atomic array appends, and the first option was therefore discarded. The approach of writing immediately at the end of a per-process array would result in the same number of I/O operations being issued as in the initial design with an array per FDB object, and it was thereby discarded as well. And writing in batches on \verb!flush()! would require \verb!archive()!ing processes to accumulate all data in memory until \verb!flush()! is called and, particularly if using multiple processes within a node, these would easily run out of memory. Given these limitations and the good performance of the array-per-object approach observed in the preliminary assessment, these alternatives were not further pursued.

\textit{\textbf{flush()}}

By contrast to the POSIX I/O backend implementation, as the DAOS API immediately persists objects and makes them available, the DAOS Store makes objects available to external readers immediately on \verb!archive()!, and there is no further action to be taken in the \verb|flush()| method.

In the future, the \verb|flush()| method may be useful if, for example, the non-blocking DAOS APIs are used in \verb!archive()! to write object data, in which case the \verb!flush()! method would block until those operations were complete.

\textit{\textbf{retrieve()}}

When \verb!retrieve()! is called, the input object location descriptor is used to build ---without performing any I/O operation--- and return a \verb|DataHandle| object which enables the application to transparently read data from the corresponding DAOS array. The \verb!DataHandle! internally uses the \verb!daos_array_read! function when the application requests data content.

Note that \verb!daos_array_read! does not fail if the requested length to read is larger than the object and does not report back the actual number of bytes read. Due to this, \verb!daos_array_get_size! must be called first to discover the array size.
However, in this case, because the object length is encoded in the location descriptor on \verb!archive()!, no call needs to be made to obtain the array size. Removing unnecessary \verb!daos_array_get_size! calls was found to have a noticeable positive impact on the performance of the backends at scale.

By contrast to the POSIX I/O backend, the handles returned by the DAOS Store \verb!retrieve()! method do not support merging because there would be no benefit in this case. Every FDB object is stored in a separate DAOS array and it is therefore not possible to merge \verb!daos_array_read! calls or avoid \verb!daos_array_open_with_attr! calls.

\subsection{The DAOS Catalogue}

\textit{\textbf{archive()}}

Fig. \ref{fig:fdb_daos_backend} can be used as visual guidance to follow the steps described here.
When the Catalogue \verb|archive()| method is called, a connection to the DAOS pool is opened, and the \textit{root container} ---whose name is configured by the FDB administrator at deployment time--- is created if it does not exist, and opened.
The container creation is performed with the \verb!daos_cont_create_with_label! function, which guarantees it is created only once even if multiple process race on their first \verb!archive()! call. 
As there is a significant overhead in opening pools and containers, once any have been opened they are cached for the lifetime of the process.

A key-value object with OID \verb!0.0! is then opened in the root container with the \verb!daos_kv_open! function. This issues no RPCs and does not fail if the key-value does not exist yet. In fact, objects are generally always considered to exist in DAOS, although they might have no content associated yet.
This key-value will hold an index of dataset keys for which objects have been \verb!archive()!d, associating each dataset key with a pointer to the corresponding dataset key-value.

If the dataset key provided as input is not found in the root key-value ---queried using the \verb!daos_kv_get! function---, a new dataset container is created in the same way as described for the Store backend.
A dataset key-value is then created in the container, again at OID \verb|0.0|, where the dataset key and the schema are inserted as new entries.
An entry is then inserted in the root key-value which maps the dataset key to (the URI of) the dataset key-value just created.
These insertions are performed with the \verb!daos_kv_put! function.
Any processes racing on their first \verb!archive()! call for a same dataset key may insert the same key-value entry in the root key-value, but this does not break the consistency of the Catalogue.

The collocation key is handled in the same manner. If the supplied collocation key is not found in the dataset key-value, a new index key-value is created with an OID generated as a function of the collocation key, and an entry is added to the dataset key-value mapping the collocation key to (the URI of) the new index key-value. Otherwise, the found index key-value is used.

Finally, an entry is added to the index key-value mapping the element key to the supplied object location description, and the axis key-values are built if not yet present, which will describe the span of values indexed in this index key-value.
To avoid collision between axes for different collocation keys, the axis key-value OIDs are generated as a function of the collocation key and the name of the element key dimension they store values for.
For every dimension of the \verb!archive()!d element key, an entry is inserted into the axis key-value for that dimension, where the keyword is the value of the element key for that dimension, and a placeholder value of \verb!1! is used.
This way, racing processes \verb!archive()!ing objects for a same element key dimension will consistently register their dimension values as keywords of the corresponding axis key-value, and repeatedly inserted values will only be present once, resulting in a concise summary of the values indexed.
This information will be available for consumers to use by calling the \verb!daos_kv_list! function on the relevant axis key-value.

Every \verb!archive()!ing process keeps an in-memory history of values inserted into the axis key-values to avoid repeatedly inserting the same value in the same axis key-value in the likely case that multiple objects with similar identifiers are \verb!archive()!d.

Note that the URIs inserted into the dataset and index key-values share identical roots, and the size of these entries could be significantly reduced by implementing a mechanism similar to the URI store used in the POSIX I/O Catalogue. This would reduce the latency of the \verb!daos_kv_put! function calls and free up valuable space in the fast storage tier of the DAOS servers.

\begin{figure}[htbp]
\centerline{\includegraphics[width=415pt,trim={0 0 0 0},clip]{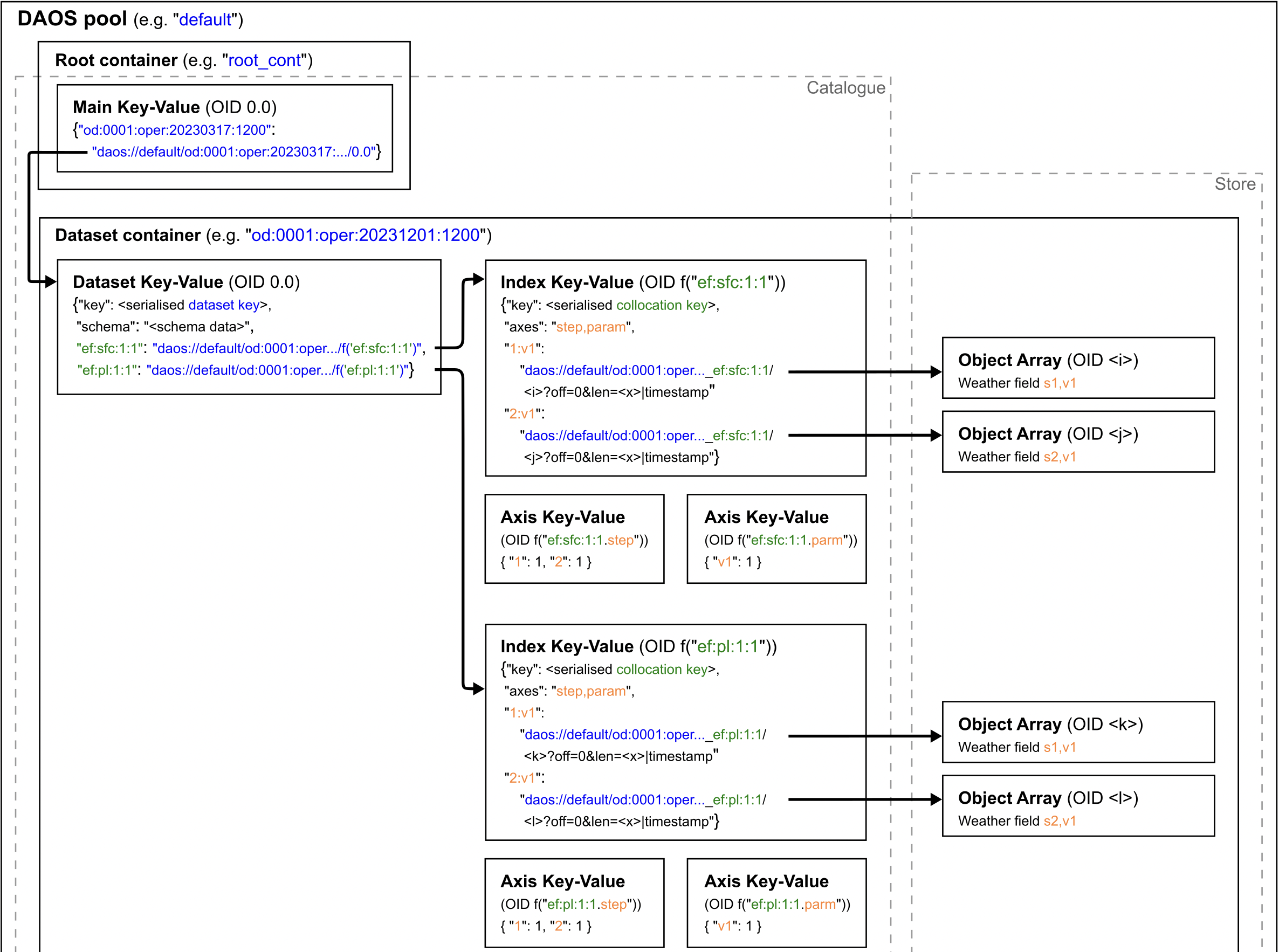}}
\caption{Diagram of DAOS entities resulting from four FDB archive() calls for four meteorological objects spanning two different collocation keys (one having levtype=sfc and the other levtype=pl) and two different element keys (one having step=1 and the other step=2).}
\label{fig:fdb_daos_backend_multiple}
\end{figure}

As producer processes continue to \verb!archive()! more objects, new arrays are written and new entries inserted into the index and axis key-values. If objects are \verb!archive()!d for new collocation keys, new sets of index and axis key-values are created, and new entries added to the dataset key-value. An example of the state of a dataset container after four different objects have been \verb!archive()!d for two different collocation keys is shown in Fig. \ref{fig:fdb_daos_backend_multiple}.

\textit{\textbf{flush()}}

After \verb!archive()! has returned, the indexing information has already been persisted and the indexed objects have been made visible to reader processes. There is no further action to be taken in the \verb!flush()! method.

\textit{\textbf{close()}}

By contrast to the POSIX I/O Catalogue, where \verb!close()! appended full indexes to the TOC file and masked out previously inserted partial indexes, the DAOS Catalogue takes no action on \verb!close()!. Therefore, consumer processes will use the same indexing structures for accessing objects regardless of producers being still active or not, resulting in more uniform \verb!retrieve()! and \verb!list()! performance.

\textit{\textbf{Connection caching and axis pre-loading}}

The first time a process calls FDB \verb!retrieve()! or \verb!list()! for a given dataset key, the DAOS pool and the root container are opened. The root key-value is then queried to obtain the URI(s) of the corresponding dataset key-value(s), and the dataset container is opened. Once any pools or containers are opened, they are kept open for the lifetime of the process.

The first time a process calls FDB \verb!retrieve()! for a given dataset and collocation key pair, the DAOS Catalogue backend queries the dataset key-value to obtain the location of the corresponding index key-value, and then retrieves the values associated to the \verb!key! and \verb!axes! keywords from it. The \verb!axes! entry contains a list of the names of all axes available for that index. The OIDs for all axis key-values are calculated using these names, and a \verb!daos_kv_list! function call is then issued to retrieve the values from each axis key-value. The obtained axis data is stored in memory for use in the Catalogue \verb!axis()! and \verb!retrieve()! methods.

\textit{\textbf{axis()}}

The Catalogue \verb!axis()! method is called internally at the beginning of every FDB \verb!retrieve()! call --- more detail on this is given in the Catalogue interface description. A dataset key, collocation key, and the name of an element key dimension are supplied as inputs. This method extracts the list of values from the pre-loaded axis corresponding to the element key dimension and collocation key provided as inputs, and returns this list.

\textit{\textbf{retrieve()}}

This method queries the pre-loaded axes corresponding to the supplied dataset and collocation keys to determine whether the index key-value for these keys may contain a match for the element key provided as input. If so, a \verb!daos_kv_get! operation is issued on the index key-value to obtain the object location descriptor associated to the element key and, if found, the descriptor is returned.

Although \verb!retrieve()! has immediate visibility of any \verb!archive()!d objects, if a new object is \verb!archive()!d for a given dataset and collocation key pair after a process has \verb!retrieve()!d any object for that key pair, a subsequent \verb!retrieve()! call by that process for the newly \verb!archive()!d object may determine the object does not exist based on the pre-loaded axis information on first \verb!retrieve()!.

By contrast to the POSIX I/O Catalogue, the DAOS Catalogue only loads the strictly necessary bits of information from the indexing structures in storage as needed.
The POSIX backend pre-loads all TOC and sub-TOC files ---including all axes and URI stores--- on first \verb!retrieve()! call for a dataset key, and loads the entire index when \verb!retrieve()! is called for a given collocation key.
The DAOS backend, instead, only pre-loads the axes on first \verb!retrieve()! call for a given collocation key, but performs a few \verb!daos_kv_get! I/O operations on the dataset and index key-values on each \verb!retrieve()! call.
This means \verb!retrieve()!al of location descriptors for small sets of few FDB objects spread across different dataset and collocation keys should be more efficient on DAOS than on POSIX file systems, but less so for requests of large spans of objects for one or a few dataset and collocation key pairs.

\textit{\textbf{list()}}

This method loads all entries from the dataset key-value corresponding to the dataset key provided as input. To that end, a \verb!daos_kv_list! function call is first issued, followed by a \verb!daos_kv_get! call for each entry in the key-value, resulting in a list of index key-value URIs.
Ideally, this loading should be performed in a single I/O operation, but DAOS currently does not provide the option of loading all key-value keys and values in a single operation.

The collocation key present in each index key-value (under the \verb!key! keyword) is then retrieved with a \verb!daos_kv_get! call, and if it matches the partial identifier provided as input to this method, all keywords of the corresponding index key-value are loaded with a \verb!daos_kv_list! call.
All found keywords are turned into element keys, and for every one matching the input partial identifier, the associated object location descriptor is retrieved from the index key-value with a \verb!daos_kv_get! call. All found location descriptors are finally returned by \verb!list()!.

The DAOS \verb!list()! method fetches indexing information from storage slightly more selectively than the POSIX I/O backends do, but performs many more I/O operations.
This means \verb!list()!ing on DAOS might be more efficient for requests of one or a few FDB objects, but less efficient for requests of large sets of objects.

This method has immediate visibility of any \verb!archive()!d objects without exception.

\subsection{Operational NWP I/O pattern on DAOS}

If a DAOS system and the new DAOS backends were used for an operational run at the ECMWF, the type of I/O operations issued and the I/O patterns generated would significantly differ from those in operational runs using the POSIX I/O backends.

Consider the operational workflow described in \ref{subsubsec:operational_nwp_pattern}.

\begin{figure}[htbp]
\centerline{\includegraphics[width=415pt,trim={0 0 0 0},clip]{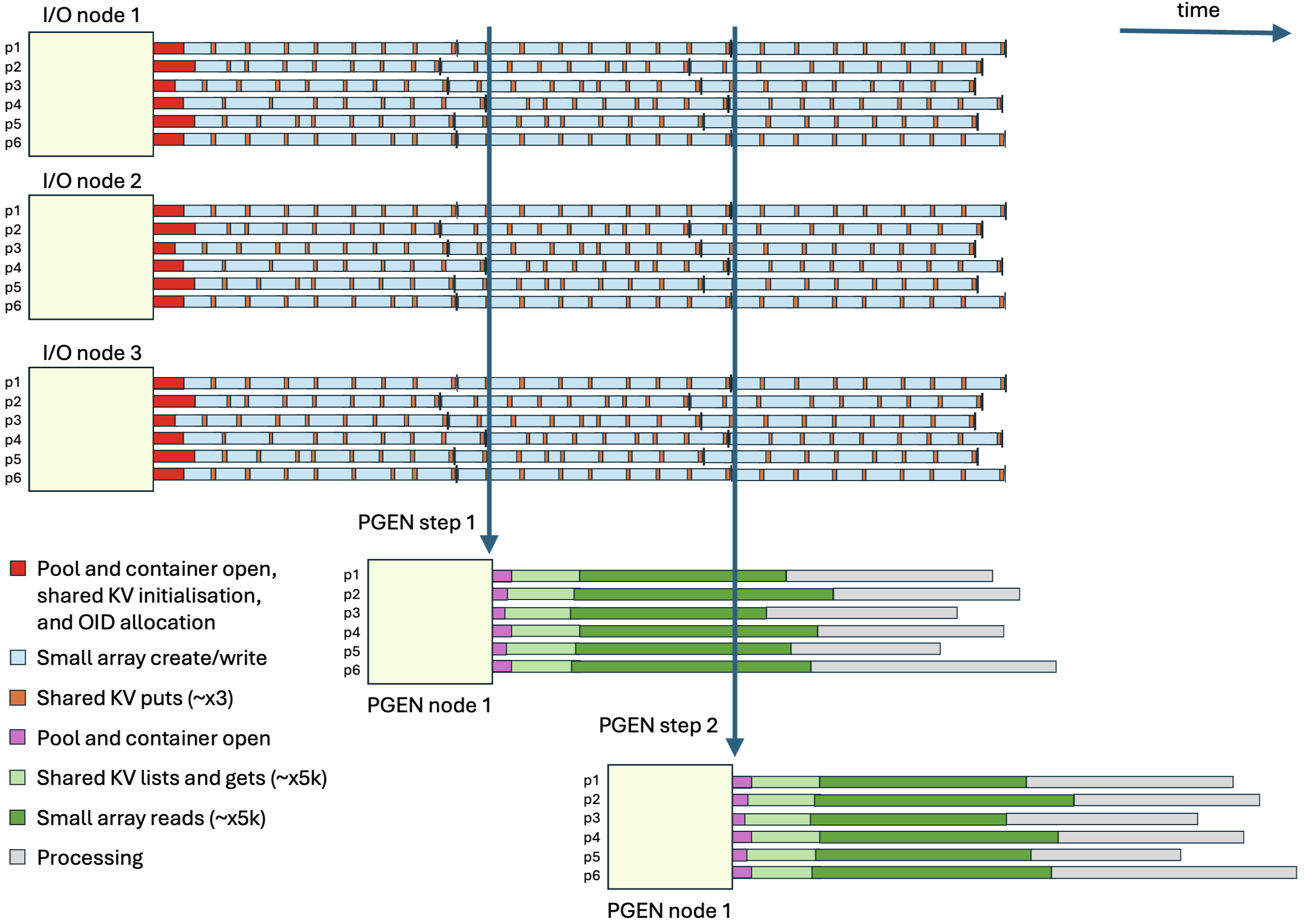}}
\caption{Not based on real profiling. Illustration of a simplification of the ECMWF's operational NWP storage access pattern using the FDB DAOS backends. The PGEN job for the third simulation step has been omitted. The proportions of the durations for the different types of operation in real operational runs might significantly differ from the ones displayed in this illustration.}
\label{fig:daos_operational_nwp_pattern}
\end{figure}

At the end of the first simulation step, on first \verb!archive()! call, the I/O server processes would contend to connect to the DAOS pool, open the root container ---likely already created in previous runs---, create and open the dataset container, populate the dataset key-value, and insert the corresponding entry in the root key-value. The straggler processes would directly find the dataset key-value indexed in the root key-value. The processes would then allocate a wide range of OIDs each before writing begins. This initialisation phase, likely dominated by the more expensive pool and container connection operations, is represented with red stripes in the diagram in Fig. \ref{fig:daos_operational_nwp_pattern}.

Writing of field data into DAOS arrays ---one per field--- would then begin as part of Store \verb!archive()!. For every field to be archived, a writer process would pick the next available OID, and call \verb!daos_array_write!
which, on return, would make the field data available for consumers to access via the relevant OID. Store \verb!archive()! would then return a location descriptor, pointing to the new array, which would be passed on to Catalogue \verb!archive()!.

As part of Catalogue \verb!archive()!, the process would attempt getting a match for the collocation key from the dataset key-value. On first \verb!archive()! call no match would be found, and the process would initialise an index key-value and insert corresponding entries in the dataset key-value. In subsequent \verb!archive()! calls a match would be found and this initialisation would be skipped. In both cases, the process would then insert the new field location descriptor into the index key-value with a \verb!daos_kv_put! call, and insert the relevant axis elements into the axis key-values with one \verb!daos_kv_put! call per dimension in the element key.
Due to the FDB being configured to recognize the \verb!number! and \verb!levelist! components of the object identifiers as part of the collocation key, processes would never contend to insert entries into the same index and axis key-values.

Every process would repeat these archival operations for each of their 65 assigned step fields, for 144 simulation steps. At the end of every step \verb!flush()! would be called, although it does not perform any operation in the DAOS backend implementation.

This writing plus indexing phase, with long sequences of \verb!daos_array_write! calls followed by a few \verb!daos_kv_put! calls, is represented in Fig. \ref{fig:daos_operational_nwp_pattern} with blue stripes for the object writing and orange stripes for the indexing. \verb!flush()! calls are represented with small black segments.

At the end of every simulation step, once the straggler \verb!archive()!ing process \verb!flush()!ed, the workflow manager would be signaled to trigger execution of a PGEN job.

The entire indexing information for the step could first be loaded by a single node and process via \verb!list()!, as done with the POSIX I/O backends. This would entail opening the pool and the relevant dataset container, then listing all keys of the dataset key-value and getting the content for each (there would be on the order of 8000 keys), then getting the collocation key from each index key-value and listing its keys, and finally getting the content for each key matching the current step (there would be on the order of 20 matching keys per index key-value). This would amount to a total of approximately 8000 \verb!daos_kv_list! and 175000 \verb!daos_kv_get! operations.

However, in this case it would be preferable to have each reader process \verb!retrieve()! directly its corresponding subset of weather fields to be processed, which would result in a similar number of \verb!daos_kv_list! and \verb!daos_kv_get! operations being performed in parallel, and would not incur large per-process initialisation overheads --- distributing the retrieval or listing workload over multiple processes had to be avoided with the POSIX I/O backends as all processes would perform TOC pre-loading, resulting in significant overheads.

Every \verb!daos_kv_list! and \verb!daos_kv_get! operation performed on an index key-value by a reader process would contend with up to one writer process performing a \verb!daos_kv_put! operation on the same key-value.

This parallel \verb!retrieve()! phase, likely dominated by key-value operations, is represented in Fig. \ref{fig:daos_operational_nwp_pattern} with purple stripes for pool and container connection, and light green stripes for \verb!daos_kv_list! and \verb!daos_kv_get! operations.

Every PGEN process would obtain a subset of field location descriptors from the previous phase, and would then read the data from these locations issuing a sequence of on the order of 5000 \verb!daos_array_read! operations --- that is, approximately as many as the total number of fields \verb!archive()!d per step (2600 times 65) divided by the number of processes per PGEN job (4 times 8). The reading of these arrays would not cause contention with writer processes as any newly \verb!archive()!d objects would be written into separate new arrays. This bulk read phase is represented with dark green stripes in Fig. \ref{fig:daos_operational_nwp_pattern}.

\subsection{fdb-hammer I/O pattern on DAOS}

The \verb!fdb-hammer! benchmark can be executed against DAOS storage systems by configuring the FDB to use the DAOS backends. The various benchmark processes \verb!archive()! or \verb!retrieve()! sequences of weather fields exactly as described in Section \ref{sec:posix_backends}, with the only notable difference being that, because the FDB schema used for operation on DAOS systems recognises the \verb!number! and \verb!levelist! identifier components as part of the collocation key, the benchmark writes or reads fields not only for different element keys, but also for different collocation keys.

The write phase of \verb!fdb-hammer! on DAOS produces a storage access pattern very similar to that of an operational run on DAOS, with the only significant differences being the configured dimension sizes ---as summarised in Table \ref{tab:ope_fdbh_size_comparison}---, the timing of the operations, and their distribution across processes.

Regarding timing of operations, all reader nodes and processes immediately start fetching their assigned subset of fields at the beginning of the run. Every process first issues a sequence of \verb!retrieve()! calls for all the relevant fields, to obtain their location descriptors. Overall, this results in a large number of pool and container connection operations, as well as index key-value list and get operations, all taking place at the beginning of the read phase. Therefore, during this part of the read phase, \verb!fdb-hammer! creates more contention on the index key-values than there would be in an operational run, but no contention is created during the rest of the \verb!fdb-hammer! run.
For the rest of the read phase, reader processes build \verb!DataHandle!s from the obtained location descriptors and read data from the corresponding arrays --- one per field. There is no write-read contention on access to these arrays, just as in an operational run.

Regarding distribution, every process reads exactly the subset of fields its corresponding writer process \verb!archive()!d for the full sequence of simulation steps and for a single member, rather than reading fields for a single step across multiple members.
In consequence, a smaller number of key-value operations per reader process are performed than in operational runs, but the total set of index key-value operations performed across all processes does not change.

\begin{figure}[htbp]
\centerline{\includegraphics[width=415pt,trim={0 0 0 0},clip]{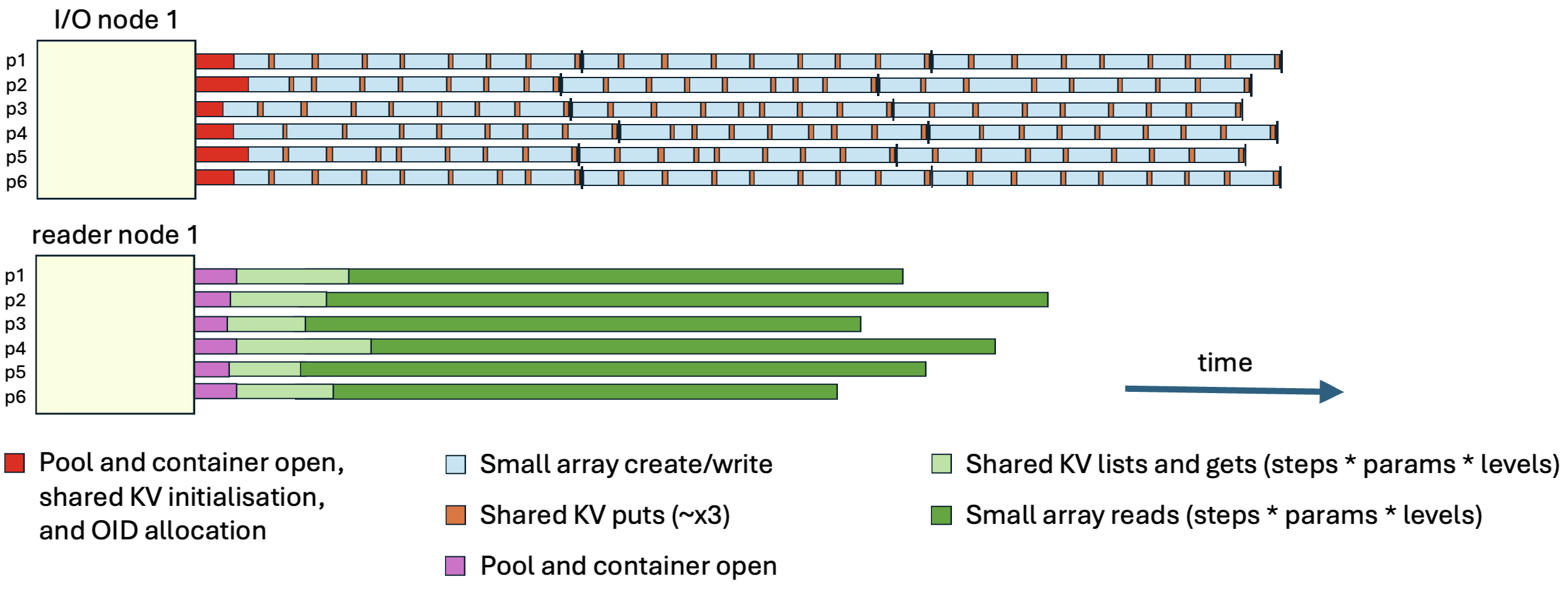}}
\caption{Not based on real profiling. Illustration of a simplification of the storage access pattern of the fdb-hammer benchmark on DAOS. The proportions of the durations for the different types of operation in real fdb-hammer runs on DAOS might significantly differ from the ones displayed in this illustration. Real fdb-hammer runs performed in this analysis employed up to 24 writer and 24 reader nodes, with up to 48 processes per node, and were run for 100 time steps.}
\label{fig:fdbh_pattern_daos}
\end{figure}

An illustration of an example \verb!fdb-hammer! run on DAOS, employing a single node for each phase, is shown in Fig. \ref{fig:fdbh_pattern_daos}.

All in all, \verb!fdb-hammer! on DAOS is more representative of operational runs on DAOS than \verb!fdb-hammer! on POSIX is representative of operational runs on a POSIX file system, and this is mainly due to the fact that the writer processes in \verb!fdb-hammer! on DAOS do access and reuse the indexing data structures created on the preliminary write phase, thus reproducing contention with readers. When using the POSIX backends, instead, \verb!fdb-hammer! writer processes mostly create new indexing structures which the readers do not access, thus partly not capturing the contention of operational runs.

\section{Ceph backends}

Ceph's RADOS provides functionality, APIs, and semantics similar to those of DAOS. Most relevant, RADOS provides regular objects where byte strings can be stored ---similar to DAOS's arrays---, and Omap objects for key-value functionality, very similar to DAOS's key-values.

In 2019, the ECMWF attempted implementing FDB RADOS backends due to Ceph's increasing popularity in cloud environments and its adoption in part of the ECMWF's cloud infrastructure. A RADOS Store backend was implemented which used large RADOS regular objects, each containing multiple weather fields, and as new weather fields were \verb!archive()!d and the objects reached the RADOS object size limit ---usually 128 MiB--- these were expanded with additional objects which were linked to the previous ones using object attributes. The backend implementation at that time resulted in lower I/O performance than expected, potentially due to the overheads caused by attribute manipulation, and was in consequence discontinued, with the implementation of a RADOS Catalogue never being pursued.

A few years later, as part of this thesis, new RADOS Store and Catalogue backends were implemented heavily based on the design of the DAOS backends, requiring only replacing DAOS API calls by RADOS API calls for large part of the development thanks to the similarities between DAOS and RADOS. In turn, this resulted in an end-to-end implementation and evaluation time of only 3 to 6 months.
Nevertheless, there were significant differences between the two storage systems and APIs, which raised a few questions regarding the overall design of the new backends, and required further evaluation, decision making, and modifications in the backend code base with respect to the DAOS backends.

The first question was related to data organization. Encapsulating all data and indexing information for a given dataset key within a separate logical division is desirable as it makes some administrative tasks easier, such as removing entire datasets without knowledge of the backend internals. This was achieved with containers in the DAOS backends, but RADOS does not provide containers. Instead, RADOS provides the option of creating namespaces within pools, akin to DAOS containers. Although RADOS namespaces seemed an ideal resource for encapsulation, pools were also an option worth exploring as it is more difficult to accidentally remove a RADOS pool than a namespace. Also, the data redundancy of RADOS objects can only be configured at the pool level, and having a pool per dataset would allow more flexibility in that regard. Conversely, one strong advantage of RADOS namespaces is that these are lightweight and should in principle have a smaller performance impact than pools.

The next question concerned data collocation and object size. Using a RADOS object per meteorological object, just as with the DAOS backends, seemed a more natural and easier way of implementing the Store backend than storing multiple meteorological objects in a single RADOS object. However, the RADOS Store backend attempt a few years earlier resulted in insufficient performance despite using large RADOS objects where multiple meteorological objects were collocated, suggesting that using a RADOS object per meteorological object could potentially result in even worse performance as a larger number of small operations would be performed over the network between RADOS servers and clients.

In turn, if choosing the approach involving large RADOS objects, there were two options. The first ---and the one followed in the existing RADOS Store implementation--- consisted in having each process write all meteorological objects \verb!archive()!d for a same dataset and collocation key into a single RADOS object, but spanning additional objects as needed if exceeding the default maximum object size of 128 MiB. The second option consisted in enlarging the maximum permitted object size on the RADOS servers ---at deployment time--- to a value large enough to fit all meteorological objects to be \verb!archive()!d for a same dataset and collocation key by a single process, thus avoiding spanning multiple medium-sized objects. Although altering this setting is not possible in third-party-managed Ceph systems, it might be an option worth of consideration if using a Ceph system solely devoted to the application at hand.

The last question was whether to immediately persist \verb!archive()!d objects and indexing information ---as done in the DAOS backends---, or deferring ensuring persistence until \verb!flush()! is called ---as done in the POSIX I/O backends---, potentially resulting in better backend performance. RADOS provides options to implement both approaches. Regular objects can be written either immediately on \verb!archive()! with \verb!rados_write!, or asynchronously with \verb!rados_aio_write! and have their persistence ensured later on \verb!flush()! with \verb!rados_aio_wait_for_complete!. Omap entries could also be inserted asynchronously, but this feature is not necessary as, if object data is written asynchronously, index Omap insertions must be performed in blocking mode on \verb!flush()! to ensure a consistent view is provided to consumers at all times. If object data is written immediately, Omap insertions ---much cheaper than the data writing--- should preferably be performed immediately on \verb!archive()!.

\begin{figure}[htbp]
\centerline{\includegraphics[width=415pt,trim={0 0 0 0},clip]{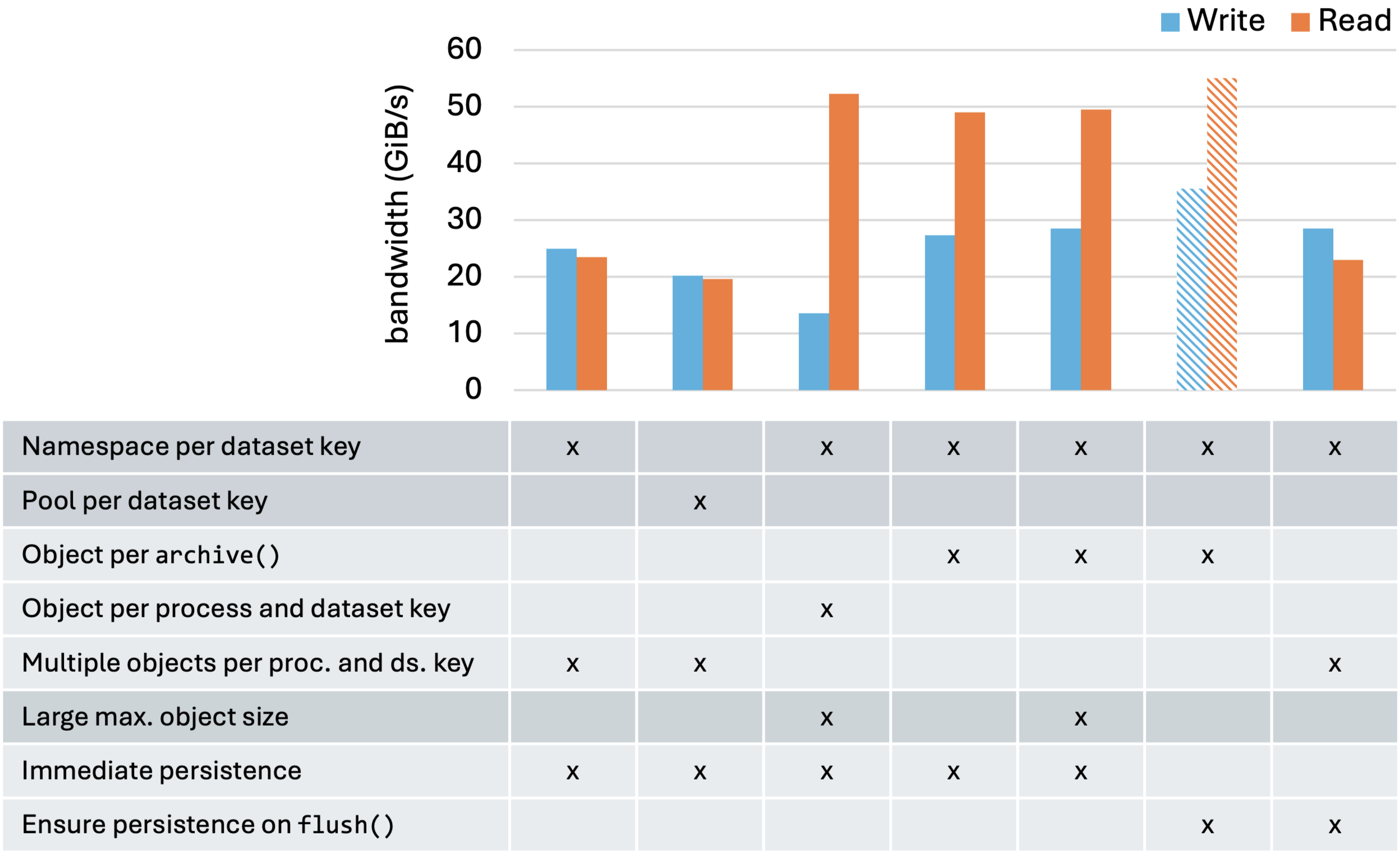}}
\caption{Performance of the FDB Ceph/RADOS backends with different options enabled. Measured with fdb-hammer runs on 32 client nodes, using 16 processes per node, against a Ceph deployment on 16 OSD nodes with 6 TiB of NVMe SSDs each. RADOS pools were configured with 512 placement groups and no replication or erasure-coding. Every process archive()d or retrieve()d 10000 weather fields of 1 MiB each. The mean of the bandwidths obtained for 3 repetitions are shown for each configuration. Using a RADOS object per archive()d FDB object resulted in best performance. Using a RADOS object per archive() call and ensuring persistence on flush(), shown with patterned columns, did not fulfill the consistency requirements.}
\label{fig:ceph_backend_options_and_performance}
\end{figure}

Given no in-depth preliminary assessment was conducted for RADOS ---unlike for DAOS--- and there was no insight available as to which of these options would result in better performance, all of of them were implemented in the new RADOS backends, with the possibility of enabling or disabling them via compilation flags. All options were then tested at scale with the \verb!fdb-hammer! benchmark, with the consistency check option enabled, to determine their performance level and verify their consistent behavior. The results are shown in Fig. \ref{fig:ceph_backend_options_and_performance}.

The first test set, shown in the left-most column, had the Ceph backends configured to use a single pool with multiple namespaces in it --- one per dataset key. All objects \verb!archive()!d by a same process for a same dataset and collocation key spanned multiple RADOS objects of default size (128 MiB), and all Omap insertions and object writes were persisted immediately.

The second test set was identical except a pool per dataset key was used instead of namespaces. Specifically, two pools were created by these tests instead of just one.
The achieved bandwidths were slightly lower than with the previous configuration likely due to the fact that RADOS is sensitive performance-wise to the total number of placement groups (PGs) in the system. Having an additional pool meant there were double as many PGs, which could have caused the performance downgrade.
This performance issue could potentially be mitigated by enabling PG auto-scaling or manually adjusting the PG count per pool to an appropriate value as a function of the number of pools in the system. Although the pool-per-dataset configuration was deemed valid and suitable, all following tests were configured to use a single pool and a namespace per dataset key as this avoided the complexity associated with PG counts and resulted in slightly better performance.

The third set of tests were configured to use a single large object per process and collocation key. Although this resulted in significantly better read performance, the write performance halved, implying this option would be inconvenient for applications such as the ECMWF's operational runs where write performance is critical. 

The fourth set used a RADOS object per \verb!archive()!d FDB object. This resulted in write performance as high as with the approach using multiple objects per process and collocation key, and read performance nearly as high as with a single large object per process and collocation key. The object-per-archive-call approach thus provided the best performance balance among the tested configurations. The high read bandwidths obtained with this configuration ---using several objects per process and collocation key---, in combination with the high read bandwidths in the third set of tests ---using a single object per process and collocation key--- suggested there might be inefficiencies potentially worth addressing, impacting read performance, in the option employing multiple objects per process and collocation key --- tested in the first, second, and seventh sets. Nevertheless, even if removing such inefficiencies, the read performance would unlikely exceed the bandwidths observed in the third test set ---which used as few objects as possible---, and the configuration with an object per \verb!archive()! call would therefore remain as one of the best performing options.

The fifth set was identical to the fourth except a larger maximum object size (1024 instead of 128 MiB) was set on the RADOS servers. The performance did not vary significantly with respect to the previous test set, suggesting that enlarging the maximum object size should enable applications requiring storing larger objects without deteriorating performance for smaller objects.

The sixth set used an object per \verb!archive()! and the default maximum object size, just as the fourth set, but was configured to use the asynchronous family of functions in \verb!librados! and ensured persistence of Omap and object operations on \verb!flush()!. This configuration performed better than the rest, but unfortunately the objects \verb!archive()!d by writers were not always accessible to readers if attempting \verb!retrieve()!al shortly after \verb!flush()!. This did not fulfill the consistency guarantees expected from the RADOS asynchronous API, and this configuration was therefore discarded as a viable option. Nevertheless, further analysis would be required to confirm the asynchronous RADOS APIs were not misused in the backend implementation.

The seventh and last set of tests used multiple objects per process and collocation key, just as in the first set, but this time using the asynchronous APIs and ensuring persistence on \verb!flush()!. This configuration did provide the expected consistency guarantees, and resulted in slightly higher write performance than equivalent tests using the blocking APIs.

Among all tested configurations, the one using a namespace per dataset key, an object per \verb!archive()! call, and blocking I/O with immediate persistence, was the one that performed best. These options were thereby fixed as defaults for the Ceph backends, and used for all \verb!fdb-hammer! runs on Ceph in Chapter \ref{chap:chap4}.

\begin{figure}[htbp]
\centerline{\includegraphics[width=415pt,trim={0 0 0 0},clip]{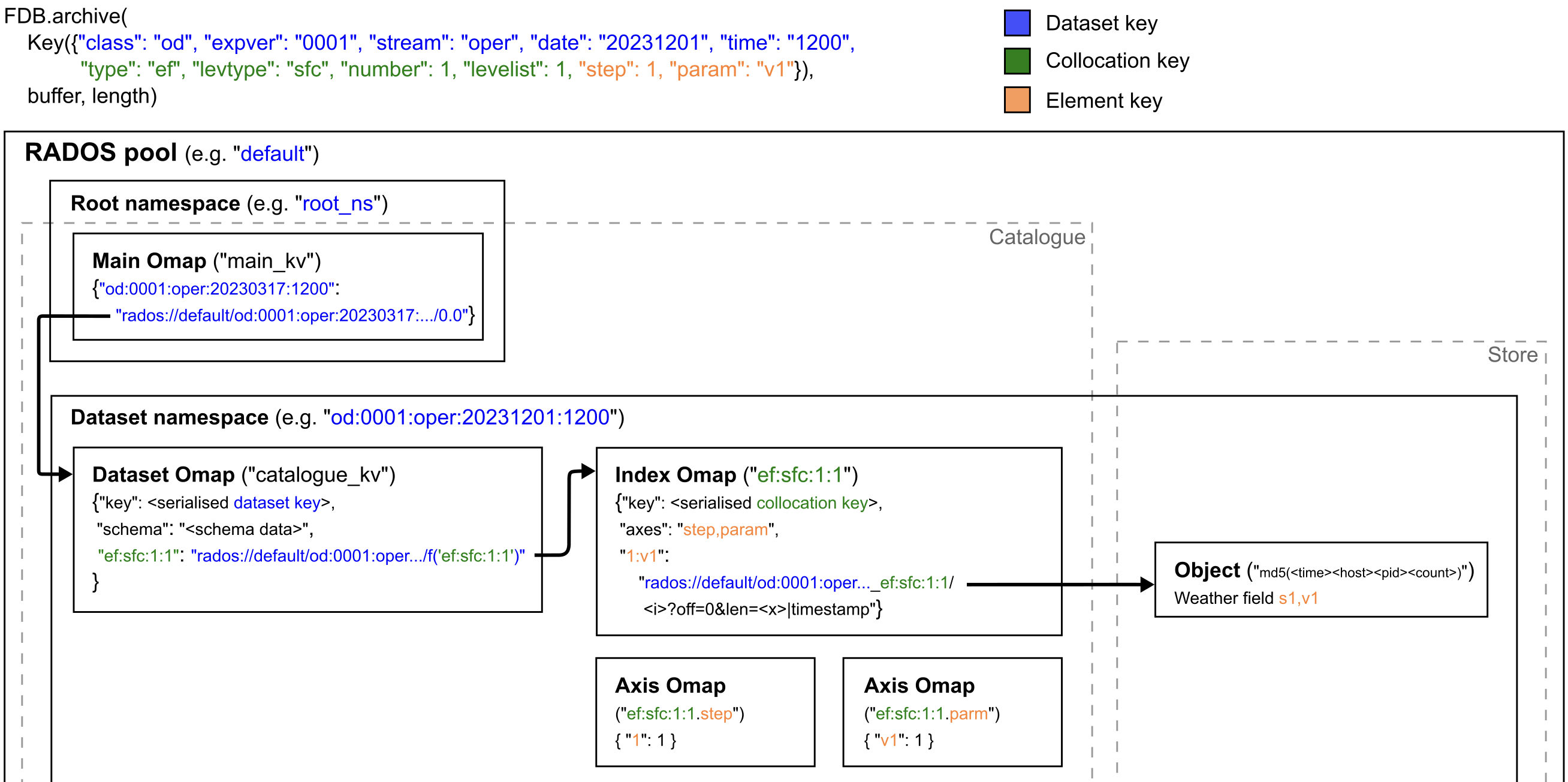}}
\caption{Diagram of Ceph/RADOS entities resulting from an FDB archive() call for one meteorological object. The archive() call is shown at the top, using different colours for the dataset, collocation, and element keys. The object data and indexing information are persisted immediately. A flush() call would result in no additional changes to these entities.}
\label{fig:fdb_ceph_backend}
\end{figure}

A diagram of RADOS entities resulting from an FDB \verb!archive()! call using the Ceph backends with the selected configuration is shown in Fig. \ref{fig:fdb_ceph_backend}.

Due to the design of the Ceph backends ---with the chosen configuration--- being nearly identical to that of the DAOS backends, the overall object layout and logic of both backends are very similar, and most of the explanations and discussion in Section \ref{sec:daos_backends} concerning the DAOS backends are directly applicable to the Ceph backends.

All in all, more design options were explored and tested for the Ceph backends than for DAOS, and this was justified partly because the previous work on the DAOS backends paved the way for a more in-depth exploration on Ceph, but also because past backend attempts on Ceph resulted in insufficient performance, making such deeper exploration necessary to ensure a well-performing design was found for the Ceph backends.
Nevertheless, the main findings ---i.e. the fact that using an object per \verb!archive()! performs very well, and the fact that using asynchronous I/O and ensuring persistence on \verb!flush()! may perform slightly better than persisting immediately--- should also be applicable to the DAOS backends, meaning that the current object per \verb!archive()! design of the DAOS backends probably performs equally or better than a hypothetical alternative design using large per-process objects, and that DAOS asynchronous I/O might be worth exploring in the future to achieve sightly higher performance with the DAOS backends.

Eventually, the design of the backends for both DAOS and Ceph have used the same design approaches, making them a suitable tool to fairly compare the performance of both storage systems.

\subsection{The Ceph Store and Catalogue}

The logic of the developed Ceph Store and Catalogue backends is very similar to that of the DAOS backends, described in Section \ref{sec:daos_backends}, with only the following differences:

\begin{itemize}[leftmargin=*]

    \item Namespaces are used instead of containers. Namespaces are more lightweight and do not require creation or opening.

    \item RADOS regular objects are used instead of DAOS arrays. The \texttt{rados\_write\-\_full} and \verb!rados_read! calls are used instead of \verb!daos_array_write! and \texttt{daos\-\_array\-\_read}. Arrays are richer in the ways they provide to access the data in them, but these features were not used in the DAOS backends and are not necessary either for the Ceph backends.

    \item Omaps are used instead of key-values. The \verb!rados_write_op_omap_set2! and \verb!rados_read_op_omap_get_vals_by_keys2! calls are used instead of \texttt{daos\_kv-\-\_put} and \verb!daos_kv_get!. The functionality they provide is very similar. Omaps require creation with \verb!rados_op_write_create! before use.

    \item Omaps are richer than DAOS key-values in the ways they provide to retrieve, query, and filter entries. Most relevant, \texttt{rados\_read\_op\_omap\_get\_vals\_by\-\_keys2} allows retrieving the full set of keys and values for an Omap in a single RPC, which was not possible in DAOS. This resulted in a more efficient FDB \verb!list()! implementation in the Ceph backends.

    \item RADOS regular objects and Omaps are identified with character strings instead of 128-bit OIDs.

    \item DAOS containers provided functionality for the DAOS backends to allocate a set of unused OIDs. In the Ceph backends, instead, allocating a new object name is done by creating unique character strings based on the technique used for per-process file creation in the POSIX I/O backends. An MD5 checksum of the unique character string is then calculated and used as object name to avoid objects being placed repeatedly in the same OSD if the character strings share a common root.

\end{itemize}

\subsection{Operational and fdb-hammer I/O patterns on Ceph}

Both the operational NWP and \verb!fdb-hammer! I/O patterns using the Ceph backends are very similar to those described for the DAOS backends, illustrated in Fig. \ref{fig:daos_operational_nwp_pattern} and Fig. \ref{fig:fdbh_pattern_daos}, with only the following differences:

\begin{itemize}[leftmargin=*]

    \item There are no container connection nor OID allocation overheads at the beginning of operational nor \verb!fdb-hammer! writers or readers --- the red and purple stripes in both Fig. \ref{fig:daos_operational_nwp_pattern} and Fig. \ref{fig:fdbh_pattern_daos} would be shorter if using the Ceph backends.

    \item Since \verb!list()!ing is more efficient with the Ceph backends, the operational workflows on Ceph might benefit from the approach used for operational runs with the POSIX I/O backends, illustrated in Fig. \ref{fig:operational_nwp_pattern}, where a single process per PGEN job first lists the full set of FDB objects to be read.

\end{itemize}

\verb!fdb-hammer! on Ceph is more representative of operational runs on Ceph than \verb!fdb-hammer! on POSIX is representative of operational runs on a POSIX file system.

\section{S3 Store}

S3 is a very popular object storage protocol which originated as a Representational State Transfer (REST) Hypertext Transfer Protocol (HTTP) API for Amazon to expose their cloud object storage services.
Although it is mainly used in Cloud rather than HPC setups, and there are challenges associated to using it for high-performance\cite{Gadban2021AnalyzingWorkloads}, many high-performance storage systems such as DAOS, Ceph, VAST, Weka, and MinIO, provide S3 interfaces.

Due to the relevance of this protocol, the development of S3 backends for the FDB was explored with the aim of broadening the range of storage systems supported by the FDB rather than assessing S3 as a potential HPC storage candidate for the more highly demanding operational workloads at the ECMWF.

S3-compatible storage systems serve a REST HTTP API which conforms to the S3 API\cite{AmazonReference},
exposing endpoints for applications to request GET, POST, PUT and DELETE operations to be performed on buckets ---storage partitions with their own namespace--- or objects --- independent data blobs belonging to a bucket. Both buckets and objects are identified with a name.

Credentials for authorisation are typically negotiated first through dedicated endpoints (see the CreateSession S3 operation) and these credentials can then be provided in further S3 API requests via HTTP headers.

Buckets and objects can be created, listed and deleted, among others. Of most relevance here, are the object write and read operations.

An object can be created with a PUT request specifying, in the headers, the name of an existing target bucket and the desired object name ---also referred to as \textit{key}--- for the object to be created, and attaching the desired object data in the body. According to the S3 semantics, either the PUT operation will succeed and result in the object being fully written, or the operation will fail. If multiple concurrent PUT operations are submitted for a same bucket and object, the last racing operation prevails, replacing the object data for the previous operations. An object cannot be modified or updated. An object can be locked such that further PUT operations on it fail, or configured for versioning such that old versions of the object are preserved and identified with a version number.

While it is not possible to expand or append data to an object, multipart uploads are supported. A multipart upload must first be initiated with a POST operation specifying the desired key for the full object upon completion. This returns an upload ID which must then be provided in subsequent part upload PUT operations, each returning a part ID. Finally, a multipart upload completion operation must be POSTed, containing the upload and part IDs. Upon success, the full assembled object is made available to readers.

An existing object can be read with a GET request specifying, in the headers, the name of the containing bucket and the object key. The current version of the object is fully returned by default. Partial byte ranges of the object can be requested via the HTTP Range header.

S3 defines many other operations, each with its semantics, and with multiple options and variants. Implementing a fully compliant HTTP client can be a challenging task. Many S3 client libraries exist to make this easier, with the Amazon Web Services (AWS) S3 Software Development Kit (SDK) being the one used by the ECMWF to implement higher-level S3 functionality which was then leveraged for the development of the S3 FDB backends.

Implementation of an S3 Catalogue backend was considered but eventually discarded due to a lack of features in S3 to do so easily. The previously described POSIX I/O and object storage Catalogue backends relied on the atomic append and the key-value features to implement a consistent and expandable index of archived objects.
S3, however, did not provide these features.
It should theoretically be possible to implement an S3 Catalogue backend very similar to the POSIX I/O one by simply replacing index and sub-TOC files by S3 objects, and using S3 object listing features to provide an inventory of available sub-TOC and index objects in place of the TOC file.
However, this would have likely resulted in an excessive and non-well-performing use of such listing features, and was therefore not pursued.

Conversely, it was possible to implement an S3 Store backend such that each FDB meteorological object is stored in a separate S3 object, and all objects for a same dataset key are stored in a same bucket.

Every time \verb!archive()! is invoked, it generates a unique ID based on the current time, hostname, and ID of the process invoking it. That ID is used as key for a new S3 object which is created and populated with the meteorological object data with an S3 PutObject operation. \verb!archive()! blocks until the S3 operation returns successfully, and therefore until the data is made available to reader processes.

\verb!flush()! performs no operation as the data is already ensured to be persistent upon \verb!archive()!.

\verb!retrieve()! creates and returns a \verb!DataHandle! object which wraps the S3 object containing the requested field. \verb!DataHandle! provides methods for reading the data which issue GetObject operations under the hood.

Two major design options arose while developing the S3 Store backend. The first was whether to use a single bucket to contain all fields or objects for all dataset keys, as opposed to using a separate bucket for each. The single-bucket approach is more convenient for Cloud environments where the number of buckets a user can have is usually limited, or where a single bucket with a name chosen by the system administration is provided. The bucket-per-dataset approach seems cleaner as the data structures and data are kept separately for each dataset key, which not only makes it easier and safer wipe a specific dataset, but also makes the backend code simpler. The bucket-per-dataset approach was chosen, although code for the single-bucket approach was also drafted and should be easy to fully implement if necessary.

The second design option was whether to store every FDB object in a separate object, as opposed to having a common object for all FDB objects sharing dataset and collocation keys, where newly archived objects would be "appended" with a part upload operation and the resulting multipart object would be aggregated and made visible upon \verb!flush()!. The object-per-field approach was chosen due to ease of implementation, although the multipart upload approach would likely result in better performance as it would substantially reduce the number of S3 objects --- and this is known to have a significant overhead in S3 stores. The code infrastructure required for the multipart upload approach was also drafted and should equally be easy to fully implement if needed.

The developed S3 Store backend was verified to work consistently against local deployments of Ceph ---via the Object Gateway interface--- and MinIO.

Although this was not pursued due to lack of time, the FDB S3 Store backend should be tested against high-performance DAOS and Ceph deployments exposing an S3 API, and the performance results compared to equivalent tests operating against the storage systems natively. Some performance loss would be expected for tests using the S3 backend due to the inherent overheads of the HTTP protocol, and also due to HTTP servers and clients not necessarily being designed for optimal performance in HPC environments.

    \chapter{Performance Assessment}
\label{chap:chap4}

So far, this work has looked mainly into the functionality and consistency of the DAOS and Ceph object stores. However, performance and scalability are also key factors to determine the overall suitability of these object stores for the ECMWF's operational NWP and for HPC and AI applications in general.

This chapter presents performance and scalability measurements obtained by running certain I/O-intensive applications, also referred to as benchmarks, against both object stores and the Lustre distributed file system deployed on the same hardware, and discusses the observed differences.
A range of I/O benchmarks were employed, some of which perform generic I/O workloads, and some others perform NWP-specific I/O workloads. Among the latter, some make use of the FDB library and the backends described in Chapters \ref{chap:chap2} and \ref{chap:chap3}.

The benchmarks were run in two different systems with different storage hardware. One of them ---the NEXTGenIO system--- was equipped with Intel's Optane DCPMM, and the other ---the Google Cloud Platform (GCP)--- provided Virtual Machines (VMs) equipped with NVMe SSDs.

This addresses the third, fourth, and fifth contributions of this dissertation: \textit{"I/O performance assessment methodology"}, \textit{"Suitability and performance assessment of Ceph and DAOS for the ECMWF's operational NWP"}, and \textit{"Fair performance comparison of DAOS, Ceph, and Lustre"}.

\clearpage
\section{Methodology}

The main question this assessment aimed to address is whether object storage systems and approaches are suitable and can provide an advantage, in terms of performance, over traditional POSIX file systems both for the ECMWF's operational NWP as well as for HPC and AI applications in general.

The adopted methodology consisted in conducting I/O benchmarking experiments against DAOS and Ceph object storage systems as well as against Lustre file systems deployed on the same hardware, and comparing the resulting performance measurements.

I/O benchmarking consists in measuring how a given storage system performs under the I/O workload of a given application or benchmark where I/O operations usually predominate.
This has several degrees of freedom, described below, which need to be navigated and adjusted in function of the benchmarking context and aims to ensure that the benchmark configuration and the produced I/O workloads are as representative as possible of a real production setup,
and that the resulting bandwidth measurements are actually and fairly indicative of what performance the storage system can provide under these I/O workloads.

\begin{itemize}[leftmargin=*]

    \item \textbf{The type of I/O operations issued by the benchmark}. These can include metadata operations such as file or object creation, deletion, listing, writing, or reading, among others.
    Write and read operations, in turn, may be larger or smaller in size, and this can have a significant impact on performance.
    Which I/O benchmark is selected largely determines the type of I/O operations that will be issued, although many benchmarks provide parameters to modulate, to varying extents, the composition of the I/O workload.

    \item \textbf{The I/O patterns produced by the benchmark}.
    Some benchmarks issue I/O operations from all client nodes and processes for the entire duration of the run, and some others have different subsets of nodes or processes start issuing I/O operations at different times. 
    Some have all processes issue I/O operations in a synchronised fashion, all at the same instant, and some others have the processes issue sequences of operations independently.
    Some combine I/O operations with client-side sleeping or processing, and some do not.
    Some have all processes perform the same type of I/O operation at any one time (e.g. write operations only), and some combine different types of potentially contending I/O (e.g. writes and reads simultaneously).
    Some access large regions of contiguous data in storage, and some access small data units at random locations.
    All these aspects characterise the I/O patterns produced by a benchmark.
    Again, what I/O patterns are produced is largely determined by the selected benchmark, but can often be modified through parameters.

    \item \textbf{The degree of parallelism of the benchmark}. While some real applications (as opposed to I/O benchmarks) run massively in parallel, some others are constrained to one or a few processes or daemons, and the I/O benchmark should capture this aspect according to the intended use of the storage system.
    If the intended use is for massively parallel applications, the degree of parallelism of the benchmark can repeatedly be altered and the benchmark rerun to discover the optimal configuration and the performance potential of the storage system.
    
    \item \textbf{The size of the storage system used for benchmarking}.
    It is often not possible to access and benchmark against a storage system large enough and therefore representative of a production setup.
    Instead, I/O benchmarking can be conducted against increasingly large partitions of the storage resources available for testing to determine the scalability behaviour of the storage system, and this result extrapolated to estimate the performance of a system of a scale as large as required for production.

    \item \textbf{The size and duration of the I/O workload produced by the benchmark}.
    A benchmark may run for a very short time period if intending to test the performance of applications performing small one-off queries or updates where low latency is desirable, or for a long time period if intending to test the performance of parallel applications writing or reading large data volumes where high throughput is more important than low latency.
    In the latter case, the duration of a run may be modulated by adjusting the I/O size, process count, and number of I/O operations performed per process. This degree of freedom is therefore tightly linked to the previously described ones.
    If the size of the storage system is varied to determine scalability, the I/O workload can be scaled together with the storage system if aiming to test weak scaling ---that is, the ability of the storage system to deliver higher performance as more storage resources are added given a proportionally large I/O workload---, or fixed to a constant size if aiming to test strong scaling ---that is, the ability to deliver higher performance given a fixed I/O workload---, although in the latter case the duration of the I/O workload will usually reduce when run against larger and faster storage system deployments.
    
    \item \textbf{The amount of client resources used for benchmarking}.
    Similarly to the parallelism degree of freedom, some applications are designed to run on a single machine with a single network adapter, whereas some others are designed to run on as many machines and exploiting as many network adapters as necessary to obtain the highest performance the storage system can deliver.
    These resources may be scaled weakly or strongly relative to the storage system, independently of how the I/O workload is scaled.

    \item \textbf{Other client and server-side configuration}.
    I/O performance can be very sensitive to a wide array of other client and server-side settings, both at the operating system and application levels. Determining when and which such settings exactly may be negatively impacting I/O performance, and what might be their optimal values, can be one of the most challenging aspects of I/O benchmarking.
    Some examples of relevant settings are the operating system block size both on the client and server sides (mostly relevant when operating against a parallel file system), the network maximum transmission unit (MTU), whether storage devices in the server machines are interleaved or arranged as a redundant array of independent disks (RAID) configuration, how many processes are deployed on these machines to serve such devices, how parallel processes are pinned to CPU cores both client and server-side, whether client-side caches are utilized or bypassed, and whether the storage system and the benchmark are configured to stripe and/or create redundant copies of the files or objects.
    These are often set to their default, or a guess made as to which might be the best value, and their optimal value is then explored by readjusting and rerunning the benchmark.

    \item \textbf{The events to log and performance metrics to use}.
    At the very least, an I/O benchmark should record timestamps when I/O starts and when it ends so that the total elapsed time can be calculated and used to compare the performance of different benchmark runs against different storage systems or using different configuration. For simple benchmarks issuing only a single metadata operation or one-off query this measurement is referred to as \textit{latency}, and for more complex benchmarks issuing multiple operations, potentially concurrently, this is referred to as \textit{wall-clock time}. If the benchmark runs as multiple non-synchronised processes, all processes must record start and end timestamps for it to be possible to calculate the overall wall-clock time.
    
    The wall-clock time metric has the disadvantage that when the degree of parallelism or the size of the I/O workload are repeatedly altered and the benchmark is rerun, the wall-clock time measurements for the successive runs may not be indicative of how well the storage system and the benchmark performed. For instance, a benchmark run taking 10 seconds to perform a single I/O operation performs worse than a benchmark run taking 12 seconds to complete 10 of the same I/O operations in parallel in terms of work accomplished per unit of time, but the wall-clock time measurement does not capture this.
    In these cases, calculating the rate of I/O operations per time unit ---also referred to as \textit{IOPS}--- or bytes transferred per time unit ---referred to as \textit{bandwidth}--- is preferable.

    Often, mainly due to the high degree of complexity of I/O systems, variance can occur among latency, wall-clock time, IOPS, or bandwidth measurements for subsequent runs of a same I/O benchmark on a same system with identical configuration. It is therefore advisable to perform multiple benchmark runs and calculate the average, maximum, or any other suitable statistic across the set of obtained measurements.

\end{itemize}

\noindent
This methodology defined ways to fix or narrow down a few of these degrees of freedom, and ways to explore and optimise the rest of degrees of freedom, such that the I/O benchmarking experiments would result in meaningful measurements of the performance the different storage systems can provide for operational NWP at the ECMWF as well as for other similar data-intensive HPC applications.

The methodology ---particularly the parameter optimisation and scaling strategies described below--- is sufficiently generic that it can be used for other storage and I/O performance analyses looking into similar applications issuing many I/O operations with a high degree of parallelism.

\subsection{Type of I/O operations and benchmarks}

Regarding the type of I/O operations, this analysis mainly focused on file or object write and read operations of 1 MiB in size. This operation size was selected because, at the time the first part of this performance assessment was conducted, operational runs at the ECMWF wrote and read data elements ---weather fields--- from storage of approximately 1 MiB in size. The operational model resolution was later increased, resulting in fields of an average size of 4.3 MiB, but given both the previous and the new field sizes were within the same order of magnitude, the performance assessment adhered to the previous one for consistency and comparability across results.
A set of I/O benchmarks were selected which produce I/O workloads where these types of operation predominate.

As described earlier, operational NWP at the ECMWF not only performs bulk writing and reading but also requires creating an index of written meteorological data objects. Some of the selected benchmarks also exercise the storage systems with the necessary operations to implement the index --- these operations include small file writes and reads, file appends, and key-value or Omap gets and puts.

The selected benchmarks are described in the following.

\subsubsection{IOR}

IOR\cite{HPCRepository} is a popular open-source I/O benchmark developed by the HPC community, originally intended to measure the I/O performance of parallel file systems, but expanded over time with new I/O backends to support operation on other storage systems like DAOS and Ceph.

IOR runs as a parallel MPI application where the concurrent processes create a file or object each, wait for each other, and commence issuing a sequence of write or read operations against their corresponding file or object. This mode, where processes do not contend for access to a same file, is known as \textit{IOR easy}\cite{IO500}.
IOR can also be configured such that all processes operate against a single shared file or object, and this is known as \textit{IOR hard}.
For this analysis, only the IOR easy mode was used as it is more representative of what a domain-specific object store like the FDB would do, avoiding multi-process contention on the same files or objects.

IOR provides parameters to adjust the number and size of the I/O operations performed by each process as well as their distribution in the file.
The following algorithm shows a simplification of the logic and I/O operations performed by IOR easy, highlighting some of the parameters.

\begin{algorithm}
\caption{IOR easy main loop}
\SetKwInput{KwData}{Inputs}
\KwData{nIterations, nSegments, blockSize, transferSize}
\For{i in 1..nIterations}
{
    barrier\\
    create and open per-process file or object\\
    \For{s in 1..nSegments}
    {
        \For{transfer in 1..(blockSize / transferSize)}
        {
            write/read transferSize consecutive bytes from segment s\\
        }
    }
    close\\
    barrier\\
}
\label{alg:algorithm1}
\end{algorithm}

For all IOR tests in this analysis, a single iteration (\verb!-i 1!), multiple segments (\verb!-s! > 1), and equal block and transfer size (\verb!-b! = \verb!-t!) were used. This means every process performed a single barrier, a single create and open operation, and multiple write or read I/O operations. Therefore, I/O operations across processes were not synchronised.

\subsubsection{HDF5}

The IOR benchmark provides a backend for operation via Hierarchical Data Format 5 (HDF5).
HDF5\cite{TheHDFGroupHierarchicalSoftware} is an I/O middleware library for efficient storage of complex and voluminous datasets used in a range of disciplines, and supports features such as data compression and encryption. HDF5 operates, by default, on POSIX file systems, but has adapters to support operation on other storage systems including DAOS\cite{Soumagne2022AcceleratingDAOS}.

When IOR is run with the HDF5 backend on POSIX, a file is created per writer process where the process metadata, indexing information, and data are stored. If the DAOS adapter is enabled, a DAOS container is created per writer process, and the data from every write operation stored in a separate object in the container and an index created using key-value objects. This results in IOR and the HDF5 backend producing a more complex set of I/O patterns ---both on POSIX and on DAOS--- than those produced by IOR and its POSIX or DAOS backends.

\subsubsection{Field I/O}

Field I/O is a standalone I/O benchmark developed as part of this research to evaluate the performance a DAOS system can provide for operational NWP at the ECMWF without involving the full complexity of the operational I/O stack. It runs as a set of independent processes, each writing and indexing a sequence of weather variables, or fields, into DAOS with a combination of array and key-value operations. If configured in read mode, the processes retrieve the same sequence of fields by querying the key-values and reading the array data. Field I/O processes write each field in a separate array, and store indexing information in a set of key-values --- some of them exclusive to the process, and some of them shared among all processes.

The Field I/O benchmark is described and discussed in more detail in Appendix B - Section IV.

Although Field I/O was designed to run on DAOS only, it can also be run on POSIX file systems via a mock \verb!libdaos! library, also developed as part of this research, which implements the \verb!libdaos! API using files and directories under the hood. More detail on this library is provided in Appendix A - Section IV.

\subsubsection{fdb-hammer}

\verb!fdb-hammer! is an FDB performance benchmarking tool provided as part of the FDB source repository which can be built alongside the other FDB command-line tools.
In the same way as Field I/O, \verb!fdb-hammer! runs as a set of independent processes, each archiving, retrieving, or listing ---depending on supplied arguments--- a sequence of weather fields via the FDB. 

When run with the POSIX I/O FDB backends, \verb!fdb-hammer! writer processes create a few dedicated files each, which are expanded incrementally with indexing information and field data, respectively. Reader processes then open and read these files to retrieve the written weather fields. This is described thoroughly at the end of Section \ref{sec:posix_backends}, including a discussion of the differences between the I/O workload and patterns generated by \verb!fdb-hammer! and those of operational NWP runs at the ECMWF.

When run with the DAOS or Ceph FDB backends, \verb!fdb-hammer! writers use a set of arrays (or regular objects) and key-values (or Omaps) to store and index the weather fields, which are then accessed by readers. This is described thoroughly at the end of Section \ref{sec:daos_backends}.

\subsection{I/O patterns of interest}

Three generic I/O patterns of interest were identified and defined, and the benchmarks configured where relevant to adhere to these.

\begin{itemize}[leftmargin=*]

    \item \textbf{No write+read contention}.
    A writing phase is run first, where the benchmark is run in parallel from a number of parallel processes, and each process writes (and indexes, if relevant) a sequence of 1 MiB data units. Following this, a reading phase is run where the benchmark is executed from as many processes, with each process reading back the same sequence of data units its corresponding writer process wrote. There is no contention between writers and readers as they run separately, optionally on separate sets of nodes.
    If run on the same set of nodes, the readers in a node must read data written from a different node to avoid accessing locally cached data, if any.
    This I/O pattern is suitable to determine the performance potential of the storage system for either writing or reading in isolation and in parallel.

    \item \textbf{Write+read contention}.
    A writing phase is run initially to populate the storage, and then a writer and a reader phase ---as described in \textit{no write+read contention}--- are run simultaneously. In this variant there is contention between writers and readers.
    This I/O pattern is intended to mimic the I/O patterns of the ECMWF's operational NWP, where parallel processes independently write and read sequences of meteorological objects.

    \item \textbf{Contending repeated writes and reads}.
    A writing phase is run initially to populate the storage, and then a writer and a reader phase are run simultaneously where every process repeatedly writes or reads a same data unit, with the same identifier. Every parallel process uses a different identifier.
    This aims to test a slightly more aggressively contentious I/O pattern, although this was seldom used as this is not representative of the ECMWF's operational NWP.

\end{itemize}

These I/O patterns do not fully specify all properties of an I/O pattern mentioned earlier.
Beyond the specified constraints, all selected benchmarks were used with their default settings. This meant all benchmarks issued I/O operations from all parallel processes for the entire duration of the run, without synchronisation, and avoided non-I/O overheads.

Given the non-synchronised nature of the benchmarks, significant delays could occur between the start times of the parallel processes of a phase, as well as between the start times of the write and the read phases in the contentious patterns. Such delays would result in unfair performance measurements, and should therefore be monitored and corrected if they occur.
A methodology to characterise and determine whether delays require correcting was defined in Appendix B - Section V - F.

\subsection{Parameter optimisation}

A procedure was defined to explore and fix the "Size and duration of the I/O workload", "Degree of parallelism", "Amount of client resources", and "Other client and server-side configuration" degrees of freedom.

Firstly, an appropriate run length is determined. The benchmark is run against a fixed-size system with fixed configuration, only varying the number of I/O operations per process, which is increased progressively until the variability in bandwidth measurements becomes small (less than 5\%). This helps balance test reproducibility whilst also keeping run times as short as possible.

After this, the client-to-server node ratio and the number of benchmark processes per client node are varied to find the point of diminishing returns, where the server connections approach saturation and adding additional client nodes or processes has little benefit. Note that each additional process added increases the overall size of the I/O workload as number of I/O operations per process is now fixed.

Once these parameters are fixed, the I/O startup timestamps are then analysed to ensure that all processes in the benchmark are really working in parallel. If this is not true, the startup processes of the benchmark should be modified, the number of I/O operations per process needs to be increased further to minimise the impact of startup time, or explicit I/O startup synchronisation needs to be implemented.

Following this, any other client and server-side configuration parameters can be tested and optimised.

\subsection{Scaling the system and benchmark runs}

Once the previous procedure is completed, the number of storage nodes employed for the storage system can be progressively increased, and the benchmark run at each scale with a corresponding number of client nodes ---according to the previously determined client-to-server-node ratio---, with all I/O patterns of interest. The bandwidths obtained for the sequence of steps will provide insight into the scaling behaviour for the different access patterns.
This approach effectively scales the I/O workload and client resources weakly relative to server size.

This addresses the "Size of the storage system" degree of freedom.

\subsection{Performance metrics}

All selected benchmarks report per-process I/O start and end timestamps.
Bandwidths are calculated as the ratio between the total volume of data written or read ---that is, bytes transferred between the benchmark processes and the storage system--- and the time elapsed between the start time of the first I/O operation performed by the benchmark and the end time of the last operation.
This ratio, measured in bytes per second, is referred to as \textit{global timing bandwidth} in some of the appendices but referred to as \textit{bandwidth} hereunder.

\begin{figure}[htbp]
\centerline{\includegraphics[width=330pt,trim={0pt 0 0 0},clip]{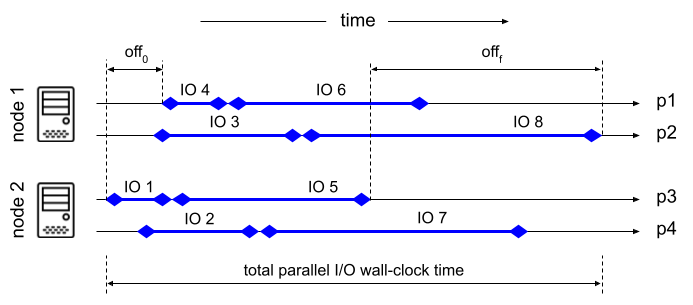}}
\caption{Plausible timeline of I/O operations and delays in a non-synchronised parallel I/O benchmark run. Only the writer phase of a run of the "no write+read contention" access pattern is represented. \copyright 2023 IEEE.}
\label{fig:bandwidth_diagram}
\end{figure}

Fig. \ref{fig:bandwidth_diagram} shows a plausible timeline of I/O operations in a parallel benchmark run with no synchronisation between processes, highlighting how the I/O wall-clock time is measured.

To account for potential variation in system performance, the benchmark is rerun at least three times for any given configuration, and the average of the bandwidths is calculated.

\clearpage
\section{DAOS and Lustre on SCM}
\label{sec:daos_lustre_scm}

This section presents I/O benchmarking experiments conducted against DAOS and Lustre deployments in a HPC system with SCM.

\subsection{The NEXTGenIO system}

NEXTGenIO\cite{NEXTGenIOApplications} is a research HPC system composed of 34 dual-socket nodes with Intel Xeon Cascade Lake processors, with 48 cores and 192 GiB of DRAM per node.
Each socket has six 256 GiB first-generation Intel Optane DCPMMs configured in App Direct interleaved mode, amounting to 3 TiB of SCM per node.
There are no NVMe SSDs.

\begin{figure}[htbp]
\centerline{\includegraphics[width=400pt,trim={8pt 0 0 0},clip]{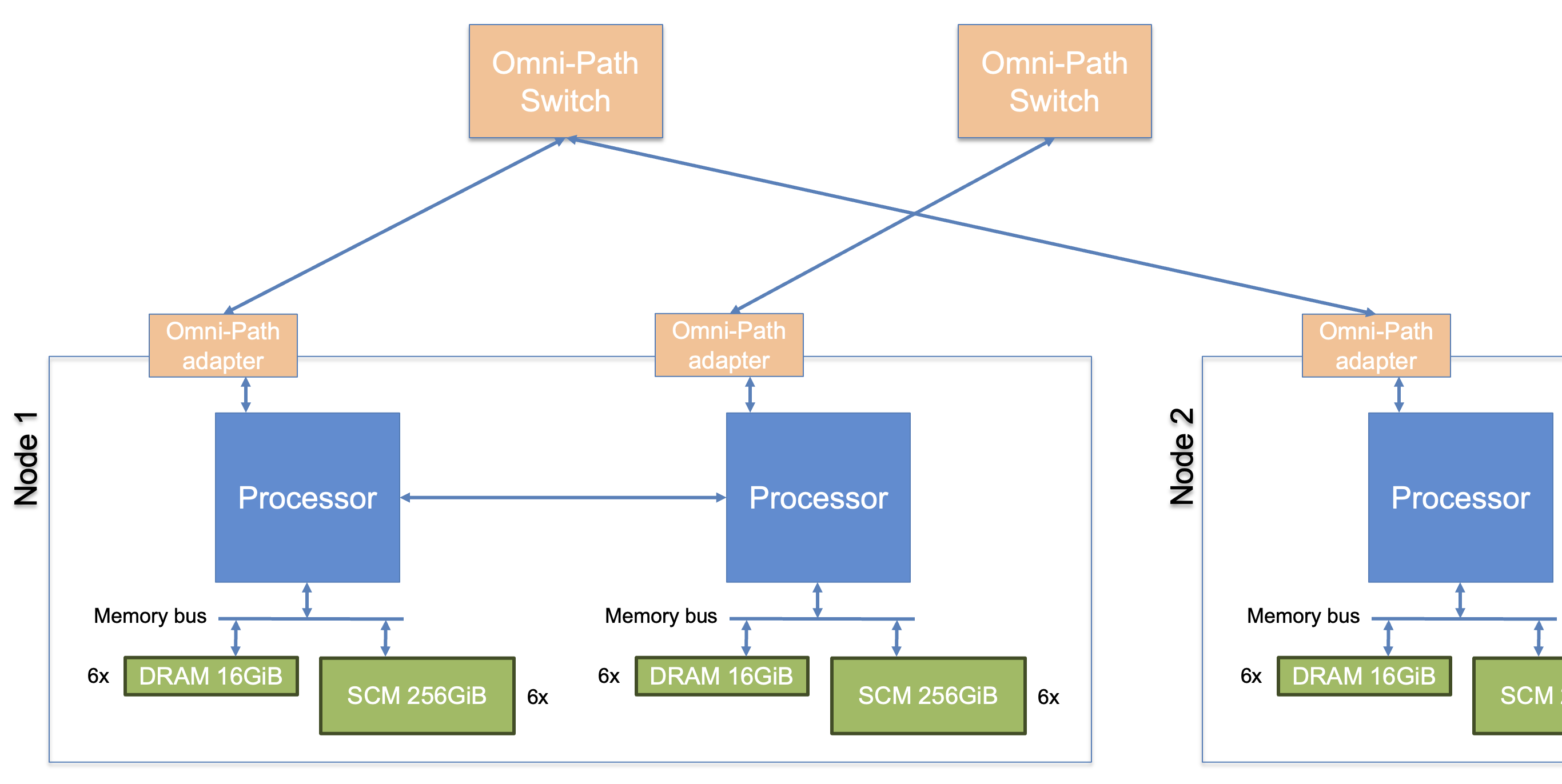}}
\caption{NEXTGenIO architecture.}
\label{fig:ngio_architecture}
\end{figure}

As shown in Fig. \ref{fig:ngio_architecture}, each processor is connected via its own integrated network adapter to a low-latency OmniPath fabric, with each of these adapters providing a maximum bandwidth of 12.5 GiB/s.
The fabric is configured in dual-rail mode. That is, there are two separate high-performance networks --- one connecting the first processor sockets of all nodes via an OmniPath switch, and the other connecting the second processor sockets via another OmniPath switch.

The system nodes use a CentOS7 operating system, and batch jobs can be submitted to the nodes via a Slurm workload manager.

For the DAOS benchmarking, DAOS v2.4 was deployed on different numbers of nodes and configured to execute a single DAOS engine per socket (i.e. two per node), with each engine using an ext4 file system spanning the entire SCM for the corresponding socket.
Each engine was configured to use the full set of cores, fabric interface, and interleaved SCM devices associated to its corresponding socket, and was configured to create 12 DAOS targets to manage these resources.
DAOS did not support using PSM2 --- a low-latency communication protocol implementing RDMA on OmniPath. Instead, it was configured to use the TCP protocol.

For the Lustre benchmarking, Lustre was deployed as well on different numbers of nodes, with one node always devoted exclusively to the metadata service, and the rest having one OST deployed per socket.
Both the OSTs and the MDTs mounted an ext4 file system on the SCM attached to their respective sockets, providing 1.5 TiB of high-performance storage per OST and MDT, with servers and clients connected using the high-performance PSM2 protocol.

\begin{figure}[htbp]
\centerline{\includegraphics[width=250pt,trim={0 0 0 0},clip]{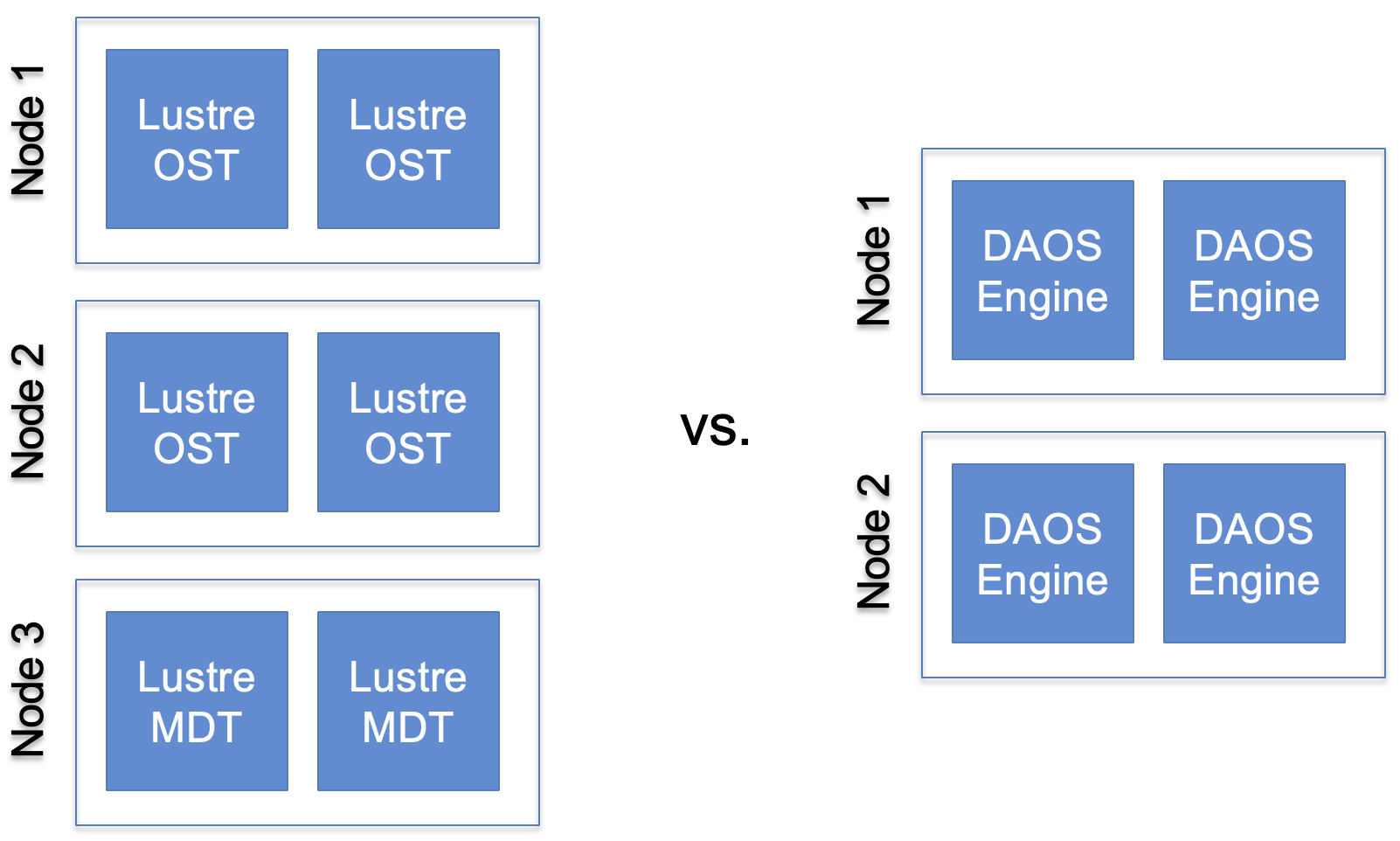}}
\caption{Example Lustre and DAOS configurations for comparison. An additional node was employed for Lustre MDTs.}
\label{fig:lustre_vs_daos}
\end{figure}

Lustre deployments used one more node than corresponding DAOS deployments. For instance, to compare to a Lustre deployment on two OST nodes ---with 4 OSTs overall--- and one MDT node ---with 2 MDTs---, a two-node DAOS deployment ---with 4 engines--- was also provisioned and benchmarked. This example is illustrated in Fig. \ref{fig:lustre_vs_daos}.

DAOS and Lustre deployments were made on an ad-hoc basis, having them deployed on the desired number of nodes as needed and removed once all relevant tests were completed. Part or all of the remaining system nodes were used to run the benchmarks. Up to 20 nodes were employed as clients to execute the benchmark processes, exploiting both sockets and network interfaces in each node. The SCM in the client nodes was not used and did not have any effect on I/O performance.

\subsection{Hardware performance measurements}

Measuring the raw bandwidths the network and storage devices can provide is a crucial first step in any I/O benchmarking experiment, as these bandwidths can later be used to determine whether any software-level storage system managing and providing access to these devices is being able to use them efficiently.

The raw bandwidth of the SCM devices in this system was analyzed thoroughly in \cite{Jackson2023EvaluatingSwansong}. As shown in Table 2, Figure 3, and Figure 4 in the referenced paper, the SCM devices in one NEXTGenIO node can provide approximately up to 10 GiB/s of write bandwidth and 80 GiB/s of read bandwidth, although these bandwidths can be much lower depending on the I/O size and how often data persistence is enforced.

Network bandwidth was determined by measuring data transfer rates between a pair of MPI processes, each pinned to the first socket of a different node to ensure only a single network adapter on each node was utilized.
Having MPI configured to use the PSM2 high-performance communication protocol, and using transfer sizes of 4 MiB or larger, the observed transfer rates were of approximately 12 GiB/s --- very close to the 12.5 GiB/s the OmniPath network adapters should provide according to specifications. This result is shown in the first row of Table \ref{tab:ngio_process_pairs} below.
However, because DAOS did not support PSM2, measuring network performance with the TCP protocol was also relevant. The measured bandwidths with TCP were of approximately 3 GiB/s between a single pair of processes, much lower than with PSM2. Nevertheless, using 8 process pairs simultaneously, with all processes pinned to the first socket of a single node pair, resulted in transfer rates of up to 9.5 GiB/s, much closer to the PSM2 bandwidths.

\begin{table}[htbp]
\caption{Process-to-process transfer rates with PSM2 and TCP.}
\begin{center}
\begin{tabular}{|c|c|c|c|c|}
\hline
\textbf{fabric}&\textbf{process}&\textbf{optimal trans-}&\textbf{bandwidth} \\
\textbf{provider}&\textbf{pairs}&\textbf{fer size (MiB)}& \textbf{(GiB/s)} \\
\hline
PSM2 & 1 & 8 & 12.1  \\
\hline
TCP & 1 & 2 & 3.1  \\
\hline
TCP & 2 & 1 & 4.1  \\
\hline
TCP & 4 & 2 & 6.9  \\
\hline
TCP & 8 & 16 & 9.5  \\
\hline
TCP & 16 & 2 & 9.0 \\
\hline
\end{tabular}
\label{tab:ngio_process_pairs}
\end{center}
\end{table}

This meant that, for a storage system such as DAOS, which did not support PSM2, having multiple processes transferring data both on the server and client sides could largely, albeit not entirely, mitigate the performance downgrade caused by the use of TCP. The performance impact of using TCP versus PSM2 was further analysed in Appendix B - Section VI C.

\begin{figure}[htbp]
\centerline{\includegraphics[width=270pt,trim={0 0 0 0},clip]{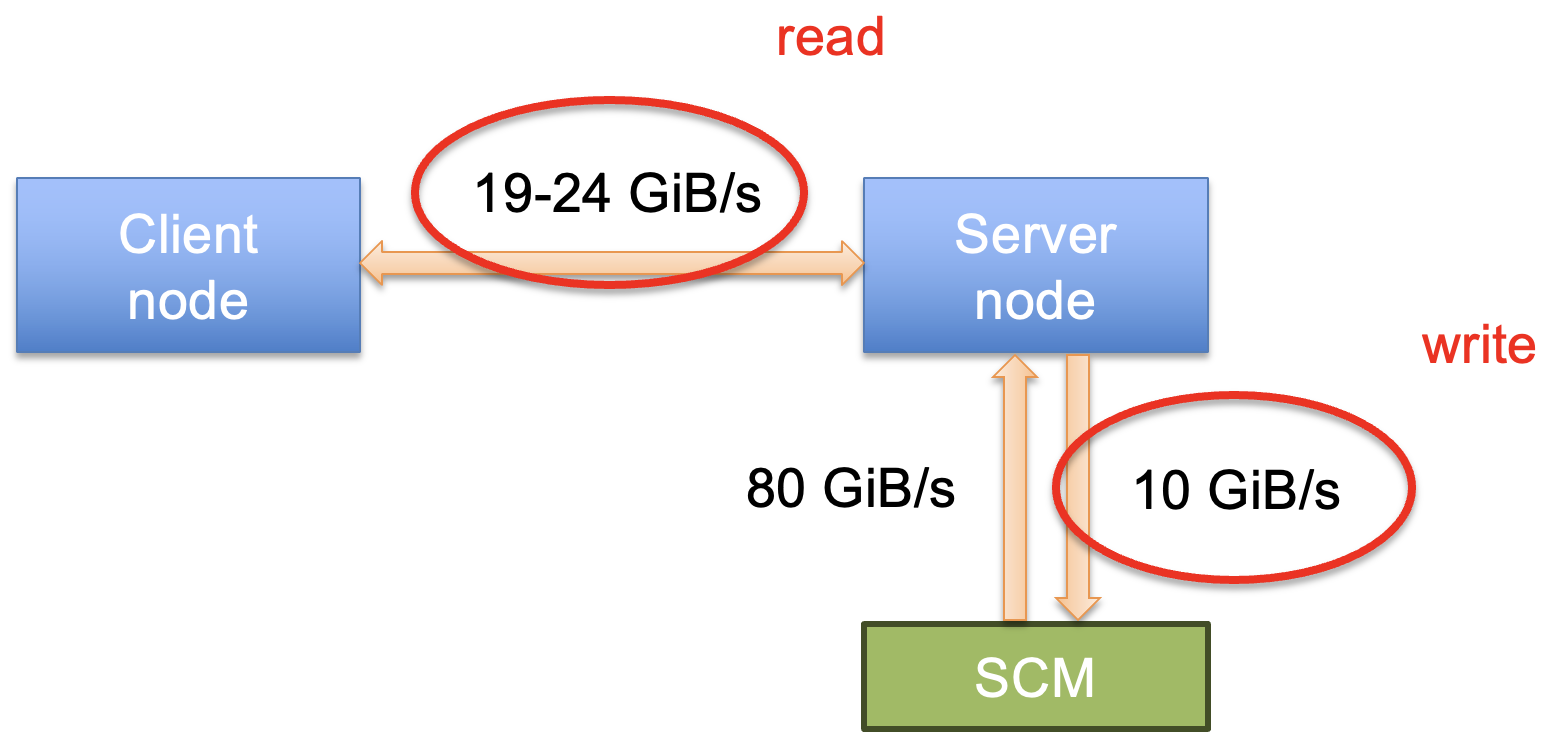}}
\caption{Ideal write and read bandwidths of a NEXTGenIO node used as networked storage server.}
\label{fig:ngio_bottlenecks}
\end{figure}

All in all, assuming optimal transfer size, amount of multiprocessing, and persistence horizon were used by a software-level networked storage system deployed in NEXTGenIO using the TCP protocol, every additional node used as server should provide approximately 10 GiB/s of write bandwidth, limited by SCM, and 19 GiB/s of read bandwidth, limited by the two network interfaces. These bottlenecks are illustrated in Fig. \ref{fig:ngio_bottlenecks}. If using PSM2, the read bandwidth per node should be 5 GiB/s higher. However, the assumption that such storage system would consistently use optimal values for all the parameters is very optimistic --- significant variation is likely to occur for any of these, which would result in significantly lower bandwidths per server node.

\subsection{IOR performance}

Due to its simplicity, IOR was the first benchmark used to measure I/O performance of the DAOS and Lustre deployments in NEXTGenIO, with the expectation that close-to-ideal bandwidths would be reached.
This benchmark produces an I/O access pattern with \textit{no w+r contention}.

The full parameter optimisation and scaling strategy, as defined in the methodology, was followed for the IOR benchmark.
The procedure is explained in the following.

IOR was configured to have every parallel process perform 100 I/O operations, of 1 MiB each, within a dedicated per-process file or object. This required setting \verb!-b! and \verb!-t! to \verb!1m!, \verb!-s! to 100, and \verb!-i! to 1, and specifying \verb!-F!. The 1 MiB I/O size was motivated by the order of magnitude of the size of weather fields currently produced in operational runs at the ECMWF. The amount of 100 operations per process was selected as part of the first step of the parameter optimisation strategy --- i.e. determining an appropriate run length. This value was found to result in little variability in bandwidths across consecutive benchmark runs, both for Lustre and DAOS, whilst resulting in short benchmark run times. Files for IOR runs on Lustre were sharded across all OSTs using the \verb!setstripe! command on the parent directory, but objects for IOR runs against DAOS were not sharded across engines as their object class was set to \verb!OC_S1!.

\begin{figure*}[htbp]
    \centering
    \begin{subfigure}[b]{212pt}
        \includegraphics[width=212pt]{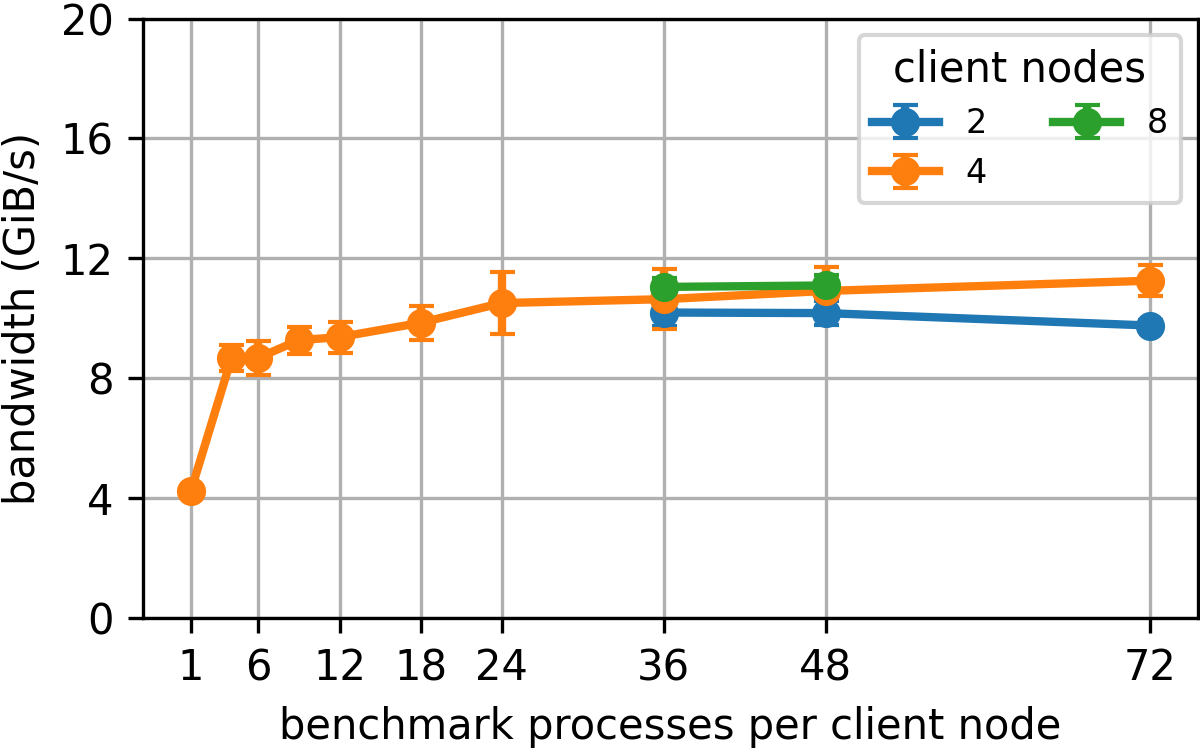}
        \caption{Write}
    \end{subfigure}
    \begin{subfigure}[b]{190pt}
        \includegraphics[width=190pt,trim={30pt 0 0 0},clip]{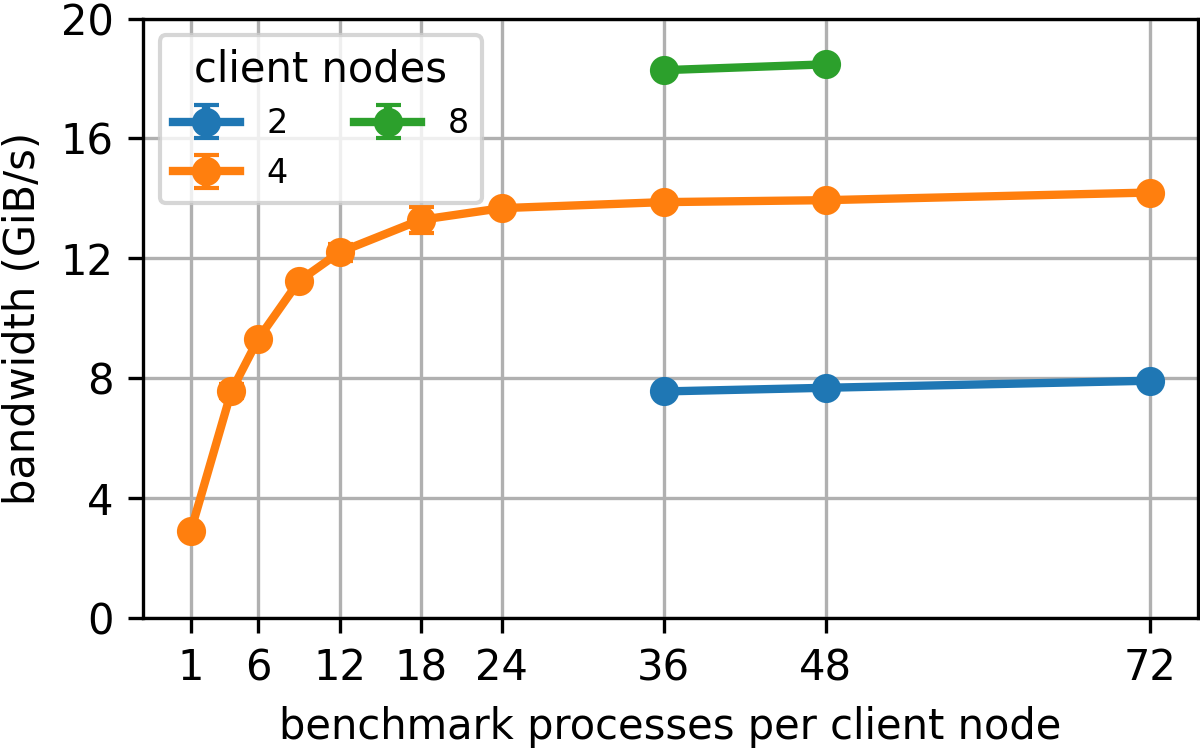}
        \caption{Read}
    \end{subfigure}
    \caption{Bandwidths for IOR runs against a 2+1-node Lustre deployment in NEXTGenIO. Every process accessed a separate file and performed 100 x 1MiB I/O operations. Tests were repeated 5 times.}
    \label{fig:ngio_ior_lustre_2sn_cn_cpcn}
\end{figure*}

The next step consisted in determining the optimal benchmark process count and client-to-server node ratio. For this, Lustre was deployed on 2 NETXGenIO nodes plus one node for the MDTs, and IOR was run on 2, 4, and 8 client nodes using a range of different process counts. The results are shown in Fig. \ref{fig:ngio_ior_lustre_2sn_cn_cpcn}. Every test was repeated 5 times, and the average and standard deviation are shown for each test with a dot and a segment around it.

As shown in the orange curve, 24 processes per client node or more resulted in best bandwidths. Employing 8 client nodes resulted in best bandwidths, reaching approximately 12 GiB/s for write and 19 GiB/s for read. A client-to-server node ratio of 4 was therefore found to be optimal for IOR on Lustre, although a ratio of 2 was also acceptable at the expense of a 25\% of read bandwidth.

\begin{figure*}[htbp]
    \centering
    \begin{subfigure}[b]{212pt}
        \includegraphics[width=212pt]{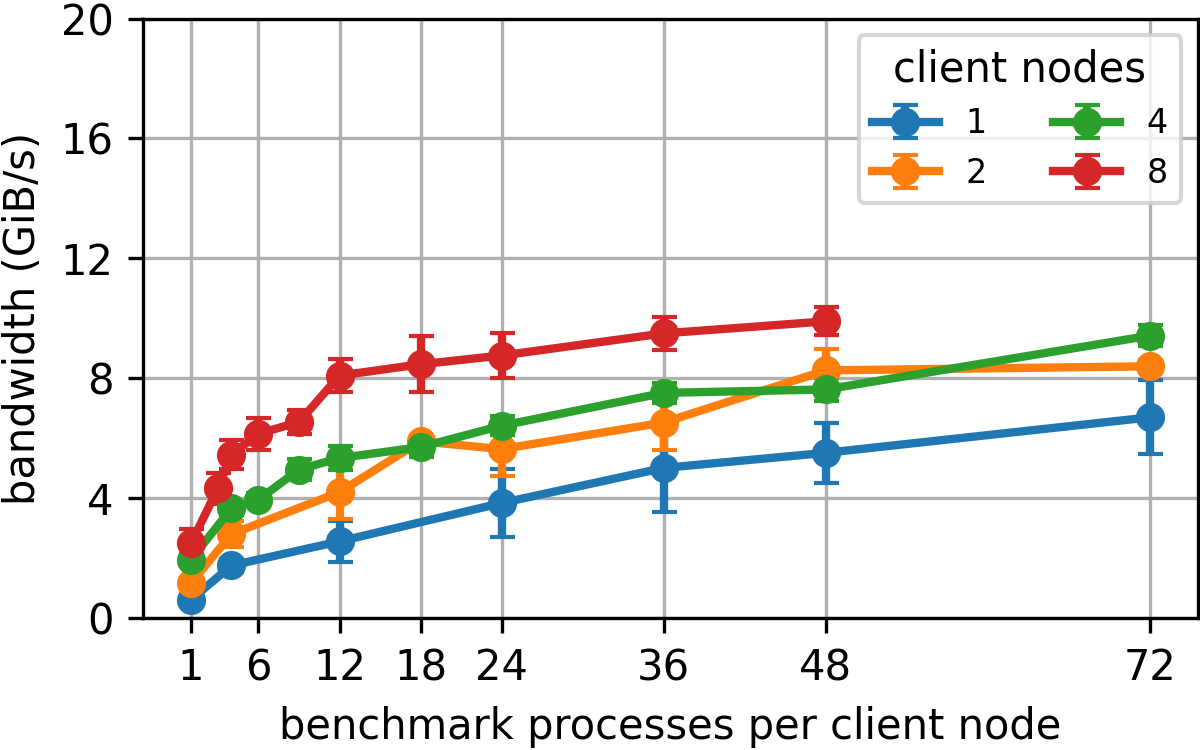}
        \caption{Write}
    \end{subfigure}
    \begin{subfigure}[b]{190pt}
        \includegraphics[width=190pt,trim={30pt 0 0 0},clip]{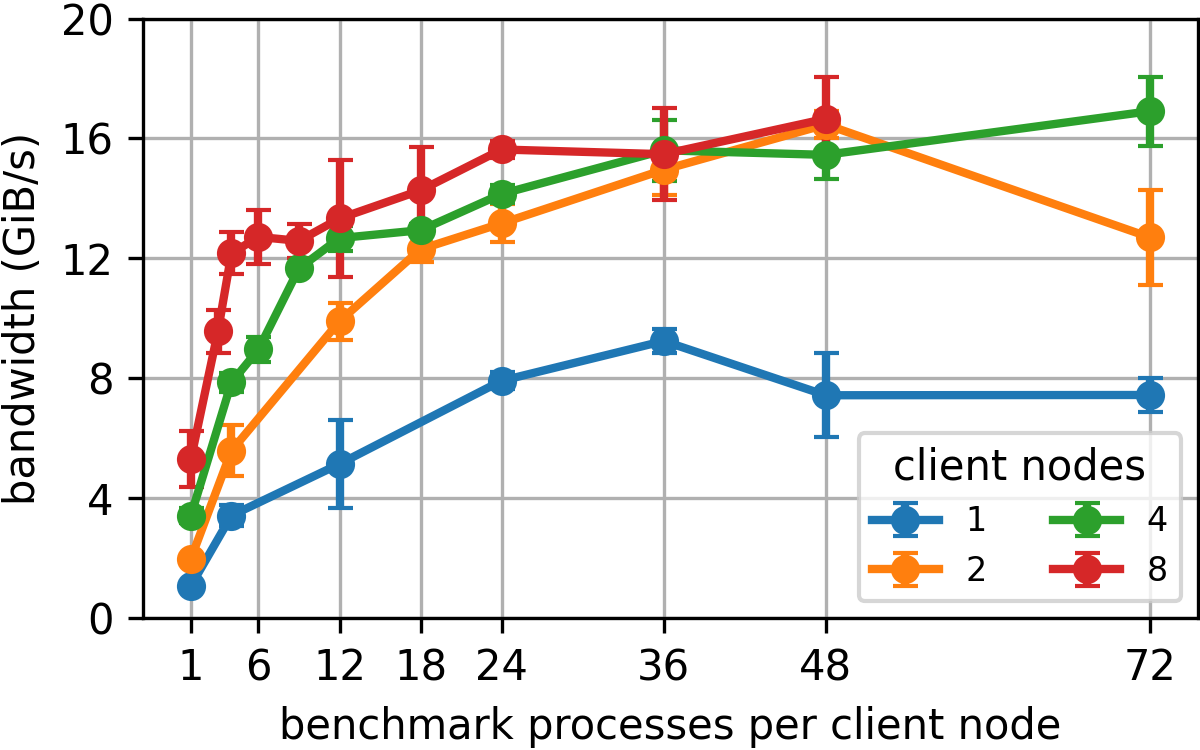}
        \caption{Read}
    \end{subfigure}
    \caption{Bandwidths for IOR runs against a 2-node DAOS deployment in NEXTGenIO. Every process wrote or read 100 x 1MiB objects. Tests were repeated 5 times.}
    \label{fig:ngio_ior_daos_2sn_cn_cpcn}
\end{figure*}

The same client count optimisation steps were followed for IOR runs against a DAOS system deployed on 2 NEXTGenIO nodes. The results are shown in Fig. \ref{fig:ngio_ior_daos_2sn_cn_cpcn}.

These results showed DAOS could reach bandwidths similar to Lustre's, approaching 11 GiB/s for write and 18 GiB/s for read, although with a smaller client-to-server node ratio of 2. Even with a ratio of 1 DAOS could reach bandwidths within the higher range. Large process counts were found to be more beneficial for DAOS than for Lustre, and this was likely due to the fact that DAOS did not use the PSM2 protocol.

The achieved IOR bandwidths, both for Lustre and DAOS, were notably far from the ideal bandwidths two NEXTGenIO nodes should provide --- that is, 20 GiB/s for write and 38 to 48 GiB/s for read.
This gap was likely mostly due to the high sensitivity of the hardware to small variations in transfer size, process count, and persistence horizon, as this seemed to affect both storage systems equally.
In hindsight, fully sharding the per-process objects across storage servers ---i.e. using an \verb!SX! object class instead of \verb!S1!--- for IOR runs on DAOS might have resulted in better utilisation of the network interfaces, and therefore higher bandwidths.

\begin{figure*}[htbp]
    \centering
    \begin{subfigure}[b]{212pt}
        \includegraphics[width=212pt]{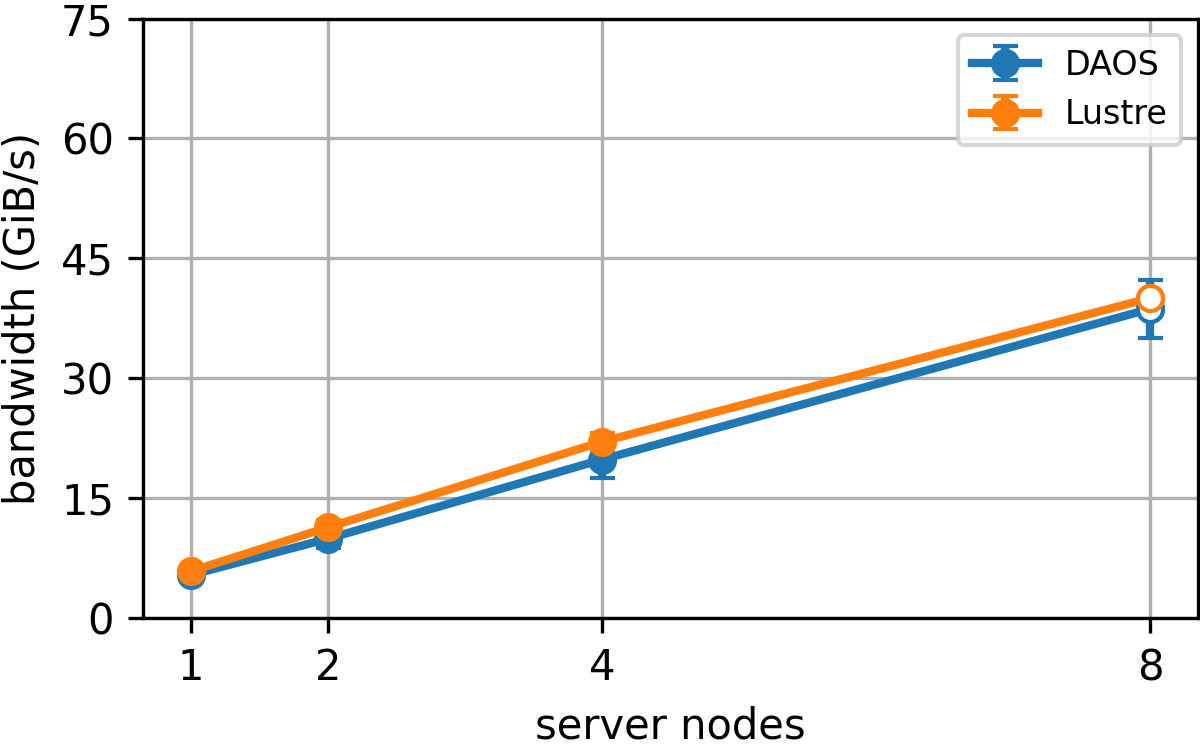}
        \caption{Write}
    \end{subfigure}
    \begin{subfigure}[b]{190pt}
        \includegraphics[width=190pt,trim={30pt 0 0 0},clip]{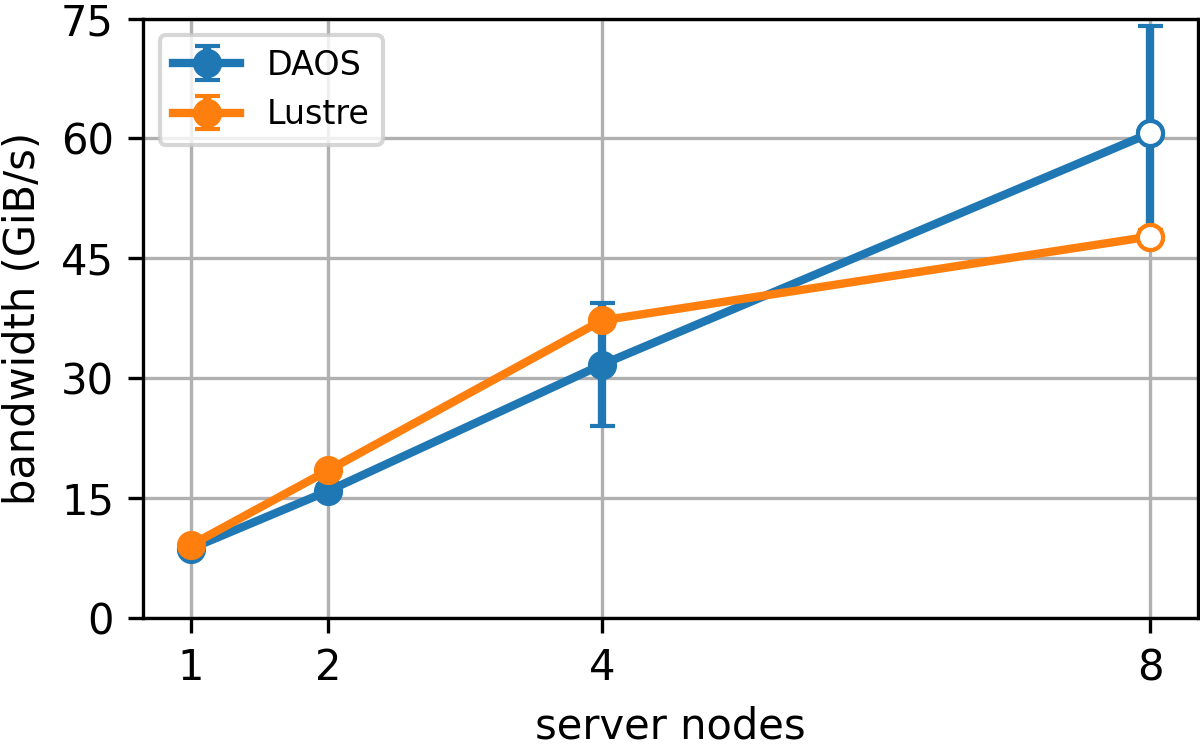}
        \caption{Read}
    \end{subfigure}
    \caption{IOR bandwidth scalability against increasingly large Lustre and DAOS deployments in NEXTGenIO. A ratio of 4-to-1 client-to-server nodes was used except for hollow dots, and 36 to 72 processes were run in each client node. Every process performed 100 x 1MiB I/O operations. Tests were repeated 5 times. \copyright 2022 IEEE.}
    \label{fig:ngio_ior_lustre_daos_scalability}
\end{figure*}

Following the scaling strategy, further tests were conducted having Lustre and DAOS deployed on increasing numbers of server nodes, up to 8 server nodes, and running IOR using the optimal process counts (i.e. 36 to 72) and a client-to-server node ratio of 4 where possible. The results are shown in Fig. \ref{fig:ngio_ior_lustre_daos_scalability}. Hollow dots indicate where it was not possible to test with a client-to-server node ratio of 4 due to the limited size of the system. Instead, a ratio of 2 was used in these cases.

Both Lustre and DAOS showed very good linear scalability both for write and read. For write, their performance was nearly identical, reaching 40 GiB/s in both cases. For read, DAOS showed significant variance in the achieved bandwidths, whereas Lustre consistently reached bandwidths at the top of the bandwidth range achieved by DAOS up to 4 server nodes, likely due to Lustre using RDMA instead of TCP. Beyond 4 server nodes, Lustre showed a scalability decline for read due to the fact that, as observed previously, IOR on Lustre requires a client-to-server node ratio of 8 to fully saturate the storage servers for read-intensive workloads, but it was only possible to test with a ratio of up to 2 for configurations with 8 server nodes.

The fact that both systems showed similar performance behavior was a good sign indicating that both were equally configured to make use of the available storage and network resources. This was verified by monitoring resource usage during some of the test runs.

Given DAOS was able to reach high performance with a client-to-server node ratio of 2 or even 1 in some cases, the scalability curve for IOR on DAOS was further expanded with more tests at larger scales. The results are shown in Appendix B - Section VI B.

Overall, despite the gap relative to hardware bandwidths, these IOR results were encouraging as they showed DAOS could perform and scale well, at a level very similar to Lustre, at least for easy IOR workloads in this system.

One relevant setting that required adjustment before the bandwidths shown here were reached, was the pinning of server and client process across available cores for optimal network use. This was found to have substantial impact in I/O performance, with up to 50\% reduction in performance when DAOS server engines were not optimally pinned across processors in a node, and up to 90\% reduction when benchmark client processes were not optimally pinned. For best performance, each DAOS engine was configured to pin engine processes to a single socket, and to target the corresponding fabric adapter. This was achieved via the \verb!pinned_numa_node! DAOS server configuration item. On the client side, processes were distributed in a balanced way across sockets in each client node with the \verb!mpirun! option \verb!--bind-to socket!.

\subsection{Field I/O performance}
\label{sec:ngio_fieldio}

Field I/O was the next benchmark used. Field I/O was developed aiming to mimic and perform the ECMWF's operational I/O patterns against DAOS systems in a minimalist way, without involving the complexity of the ECMWF's software stack. This would provide insight on the performance potential as well as the options and challenges of porting to DAOS.

Field I/O has each parallel process write or read a sequence of weather fields of 1 MiB into or from DAOS via the \verb!libdaos! API, using a separate DAOS array for every field. The writer processes create a hierarchy of key-value objects where the stored fields are indexed, and these key-values are accessed by reader processes to locate the field array before reading it. For every field in the sequence, a few key-value operations are performed and one array object is written or read. 

\begin{figure*}[htbp]
    \centering
    \begin{subfigure}[b]{214pt}
        \includegraphics[width=214pt]{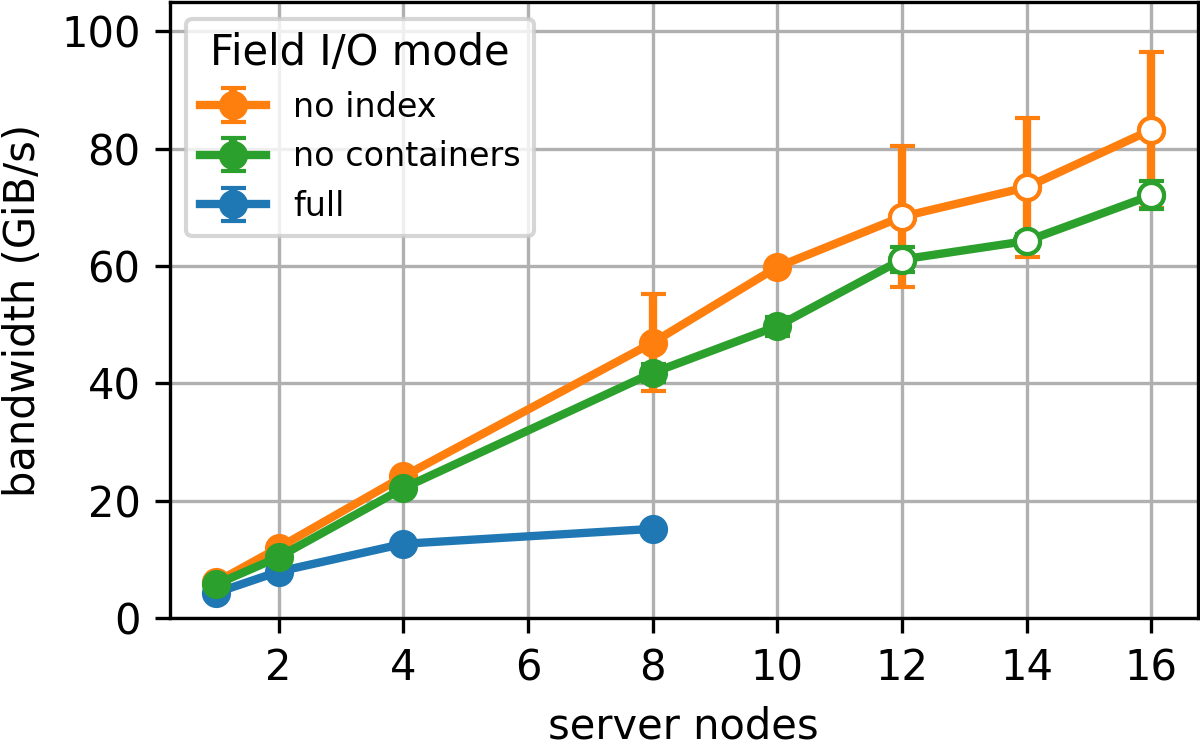}
        \caption{Write, no w+r contention}
    \end{subfigure}
    \begin{subfigure}[b]{188pt}
        \includegraphics[width=188pt,trim={35pt 0 0 0},clip]{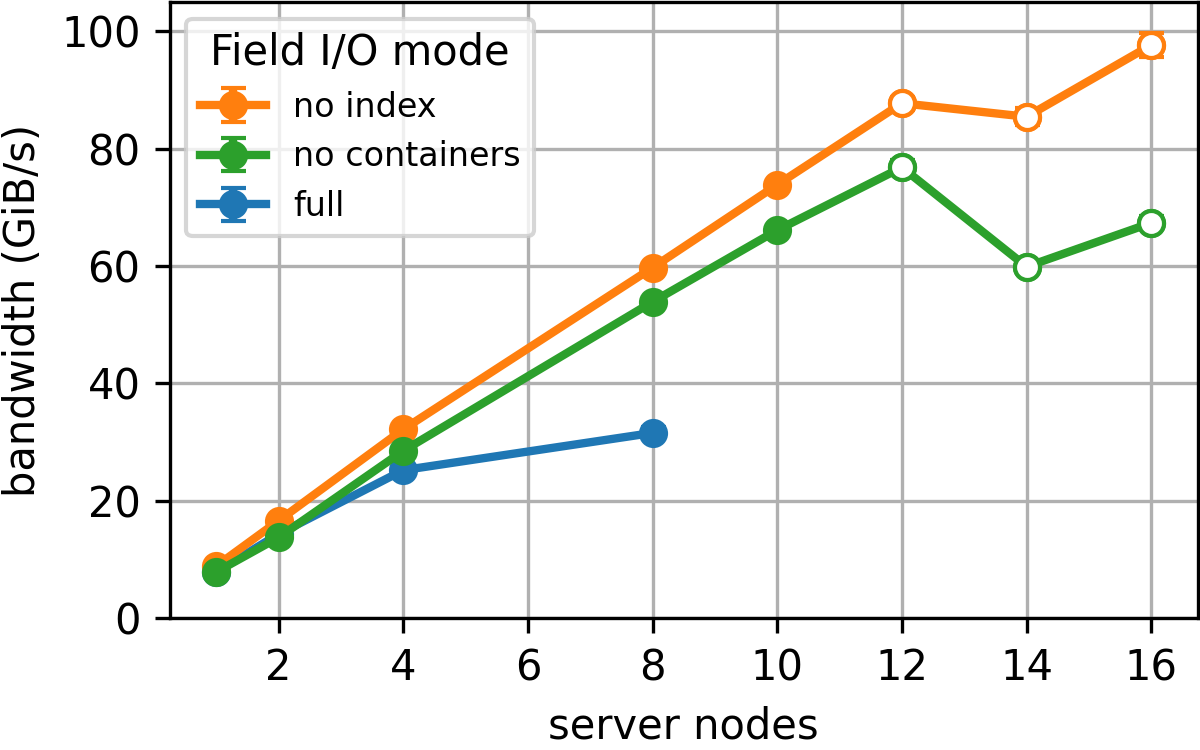}
        \caption{Read, no w+r contention}
    \end{subfigure}
    \caption{Bandwidth scalability of Field I/O runs, with no write+read contention, against increasingly large DAOS deployments in NEXTGenIO. A ratio of 2-to-1 client-to-server nodes was used except for hollow dots, and 36 to 48 processes were run in every client node. Every process wrote and indexed, or de-referenced and read, 2000 x 1MiB weather fields. Tests were repeated 10 times. \copyright 2023 IEEE.}
    \label{fig:ngio_fieldio_daos_scalability_pattern_a}
\end{figure*}

The parameter optimisation and scaling strategy, as defined in the methodology, was followed from scratch for Field I/O. This was documented thoroughly in Appendix B. One of the most relevant results, also shown here in Fig. \ref{fig:ngio_fieldio_daos_scalability_pattern_a}, were the scalability curves obtained for Field I/O runs with \textit{no write+read contention} against DAOS deployments on up to 16 NEXTGenIO nodes. As shown with different curve colours, Field I/O was tested with different configuration options. The \textit{full} mode, in blue, used several indexing key-value objects and had the array and key-value objects distributed across several DAOS containers. The \textit{no containers} mode, in green, was identical except it placed all objects in a single container. The \textit{no index} mode, in orange, skipped the creation and use of indexing key-values. Hollow dots indicate where the optimal client-to-server node ratio of 2 could not be fulfilled due to resource limitations. Every process stored and indexed, or de-referenced and read, a sequence of 2000 fields.

These results showed that DAOS and Field I/O, if configured to use a single container, scaled linearly and reached IOR bandwidths despite using several small objects and performing several key-value operations per process. Also, the results demonstrated that the implementation of an index based on DAOS key-value objects had little impact on overall performance and scalability, whilst using several containers did have a significant impact.

These and other findings materialised as a set of recommendations for the development of high-performance DAOS applications, which were presented in \cite{Jackson2023ProfilingDAOS}.

The work in Appendix B - Section VI D also measured the Field I/O performance and scalability under \textit{contending repeated writes and reads}. This result demonstrated the advantages of implementing domain-specific object stores on top of general-purpose object stores for such type of access pattern. The work in Appendix B, however, did not analyse \textit{write+read contention} patterns, which are more representative of the ECMWF's operational workloads.

\begin{figure*}[htbp]
    \centering
    \begin{subfigure}[b]{214pt}
        \includegraphics[width=214pt]{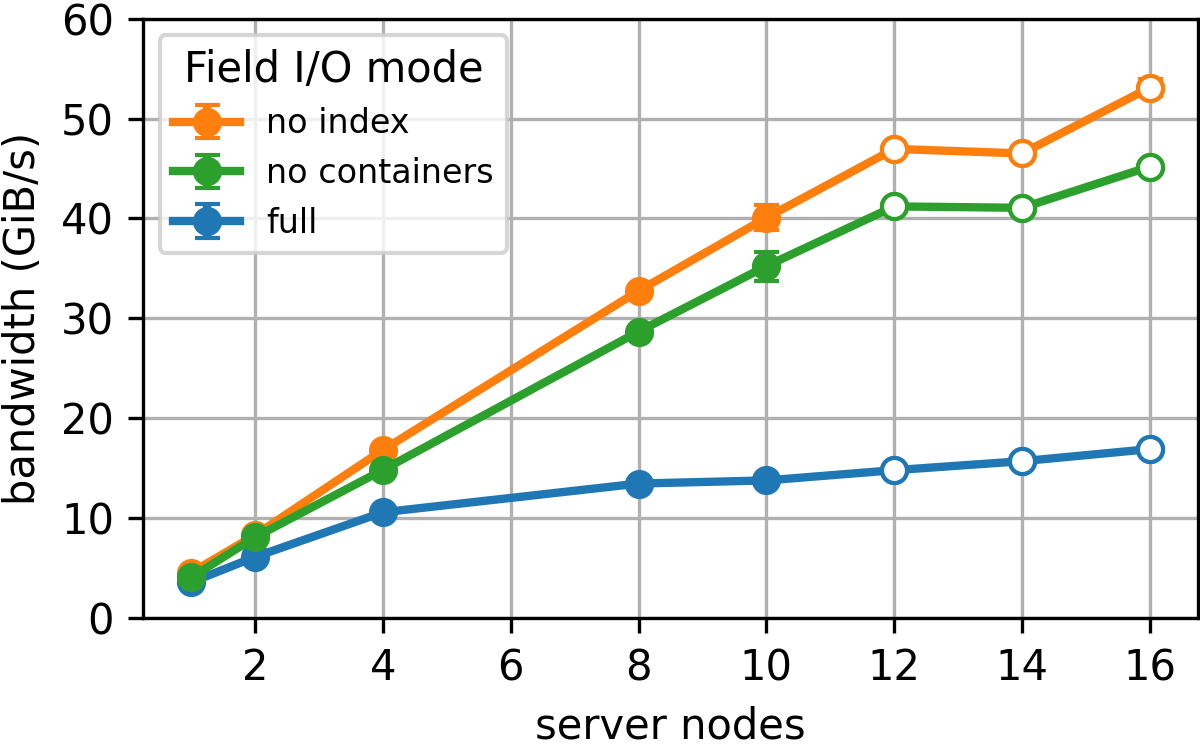}
        \caption{Write, w+r contention}
    \end{subfigure}
    \begin{subfigure}[b]{188pt}
        \includegraphics[width=188pt,trim={35pt 0 0 0},clip]{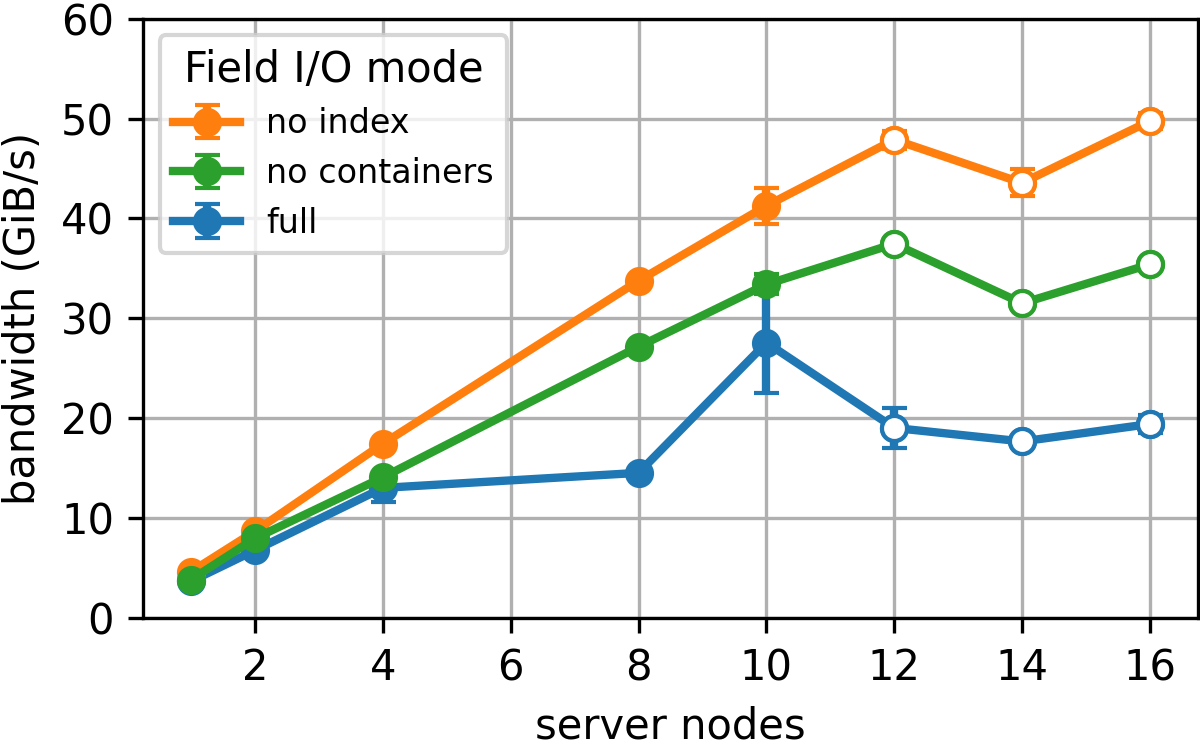}
        \caption{Read, w+r contention}
    \end{subfigure}
    \caption{Bandwidth scalability of Field I/O runs, with write+read contention, against increasingly large DAOS deployments in NEXTGenIO. A ratio of 2-to-1 client-to-server nodes was used except for hollow dots, and 36 to 48 processes were run in each client node. Every process wrote and indexed (or de-referenced and read) 2000 x 1MiB weather fields. Tests were repeated 10 times.}
    \label{fig:ngio_fieldio_daos_scalability_pattern_c}
\end{figure*}

Field I/O was later run with \textit{write+read contention}, and the measured bandwidths are shown here in Fig. \ref{fig:ngio_fieldio_daos_scalability_pattern_c}. These runs were configured to use a half of the allocated client nodes to execute Field I/O writer processes, and the other half to execute reader processes simultaneously. For example, for a run against a 8-node DAOS deployment, 16 client nodes were allocated according to the optimal client-to-server node ratio of 2 ---determined earlier in the analysis---, and 8 of these nodes were used to run writer processes and the other 8 to run reader processes. This means the bandwidths for a given test in graphs a) and b) of Fig. \ref{fig:ngio_fieldio_daos_scalability_pattern_c} can be added to determine the total aggregate bandwidth the system delivered during the execution of that test.

The Field I/O mode using a single container ---i.e. the \textit{no containers} mode---, scaled linearly under operational-like contention both for write and read, providing aggregate (write plus read) bandwidths of up to 70 GiB/s in the configuration with 10 NEXTGenIO nodes employed as DAOS servers. This was an outstanding result. For reference, the ECMWF's operational Lustre file system at that time, composed of 300 OST nodes, provided approximately 50 GiB/s of sustained aggregate throughput under operational workloads.

\begin{figure*}[htbp]
    \centering
    \begin{subfigure}[b]{212pt}
        \includegraphics[width=212pt]{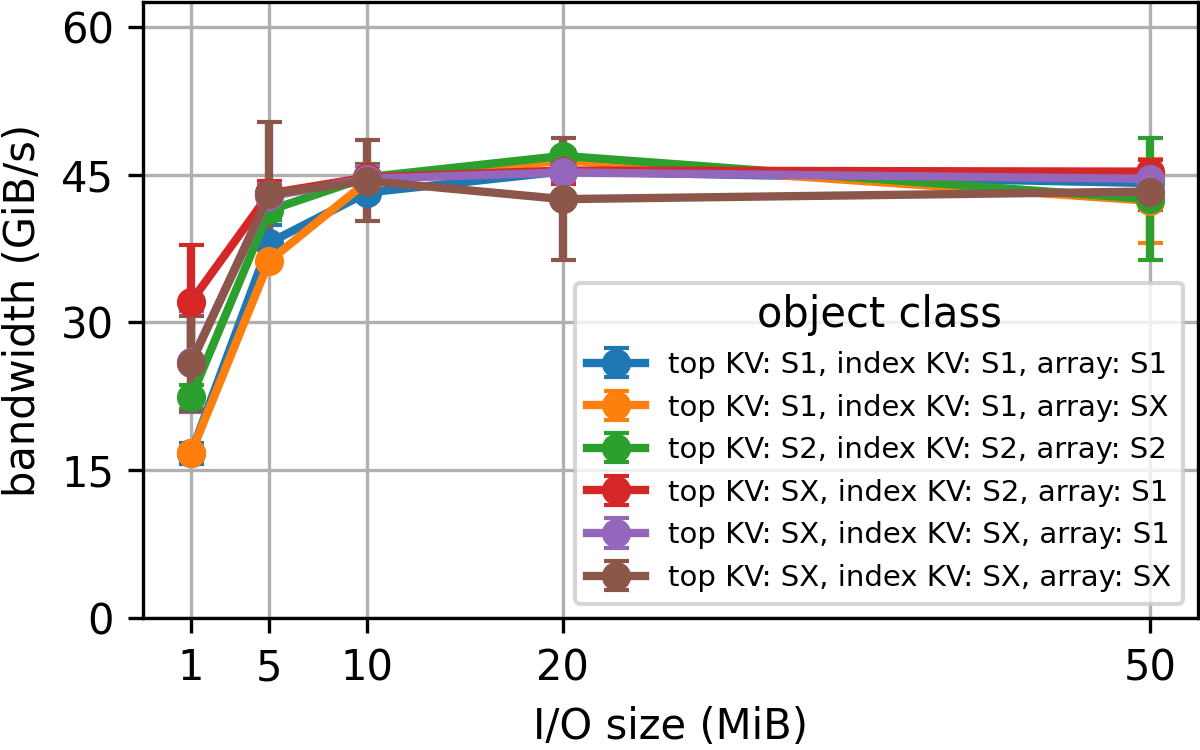}
        \caption{Write}
    \end{subfigure}
    \begin{subfigure}[b]{190pt}
        \includegraphics[width=190pt,trim={30pt 0 0 0},clip]{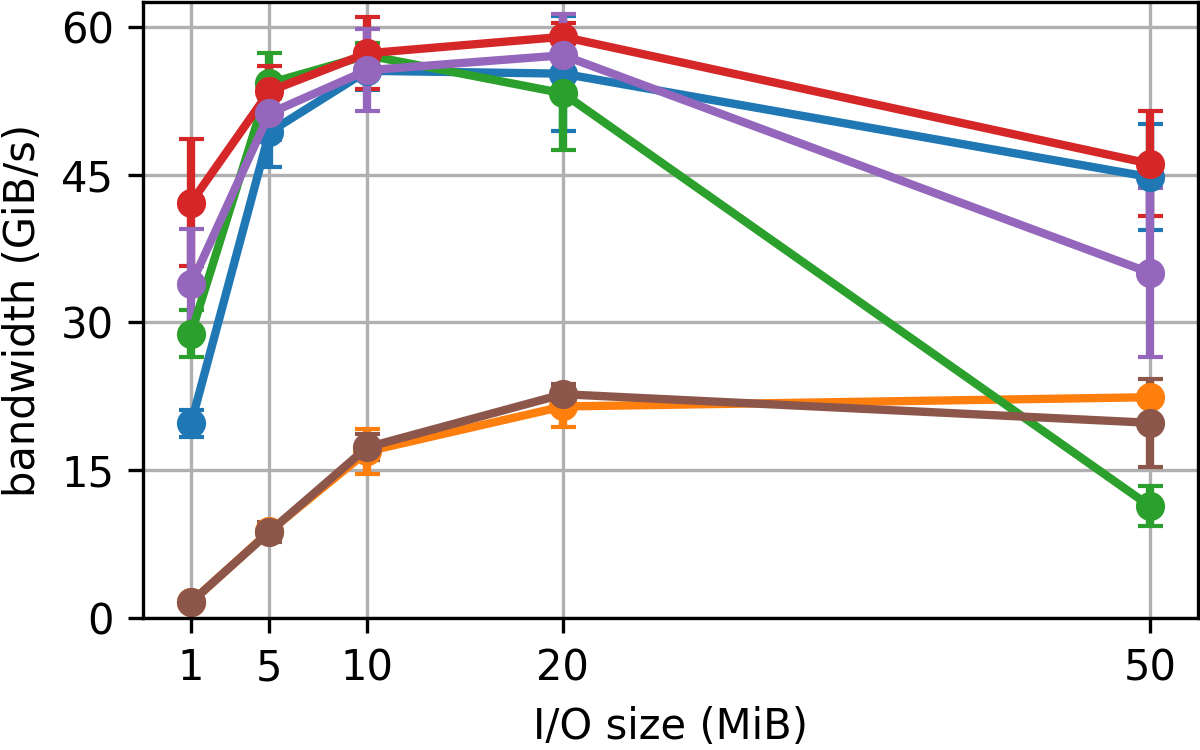}
        \caption{Read}
    \end{subfigure}
    \caption{Bandwidths of Field I/O runs on 16 client nodes, with no write+read contention, against a 8-node DAOS deployment in NEXTGenIO, varying field size and object sharding configuration. 36 processes were run in each client node. Every process wrote and indexed (or de-referenced and read) 100 x 1MiB weather fields. Tests were repeated 5 times.}
    \label{fig:ngio_fieldio_daos_oc_os}
\end{figure*}

Another relevant set of results from the benchmarking reported in Appendix B (Section VI D 4) were the bandwidths measured for Field I/O runs configured to write and read weather fields of larger sizes, beyond 1 MiB, and use different sharding configurations for the key-value and array objects. These runs were performed on 4 client nodes, against a 2-node DAOS deployment. Similar runs are shown here, in Fig. \ref{fig:ngio_fieldio_daos_oc_os}, but performed at larger scale on 16 client nodes, against a 8-node DAOS deployment.

These results show that using an object class of \verb!S1! for DAOS arrays (i.e. no sharding across DAOS engines) results in best performance, and increasing field size beyond 1 MiB ---as would be required for future resolution increases at the ECMWF--- improves rather than deteriorates performance.

Overall, the results and findings from Field I/O runs against DAOS deployments in NEXTGenIO were very promising, and justified and informed the later development of a DAOS backend for the ECMWF's FDB.

\subsubsection{Field I/O on Lustre}

One shortcoming of Field I/O was that, because it is designed to operate natively only on DAOS via \verb!libdaos!, there was no direct way of running this benchmark against Lustre deployments for comparison.

However, a dummy \verb!libdaos! library was developed for Continuous Integration and Continuous Development (CI/CD) purposes at the ECMWF. Dummy \verb!libdaos! implements the \verb!libdaos! API on top of POSIX file systems by mapping DAOS concepts such as pools, containers, key-values, and arrays, to POSIX directories and files. This enabled running Field I/O against Lustre deployments in NEXTGenIO by linking Field I/O to dummy \verb!libdaos! instead of proper \verb!libdaos!.

\begin{figure*}[htbp]
    \centering
    \begin{subfigure}[b]{212pt}
        \includegraphics[width=212pt]{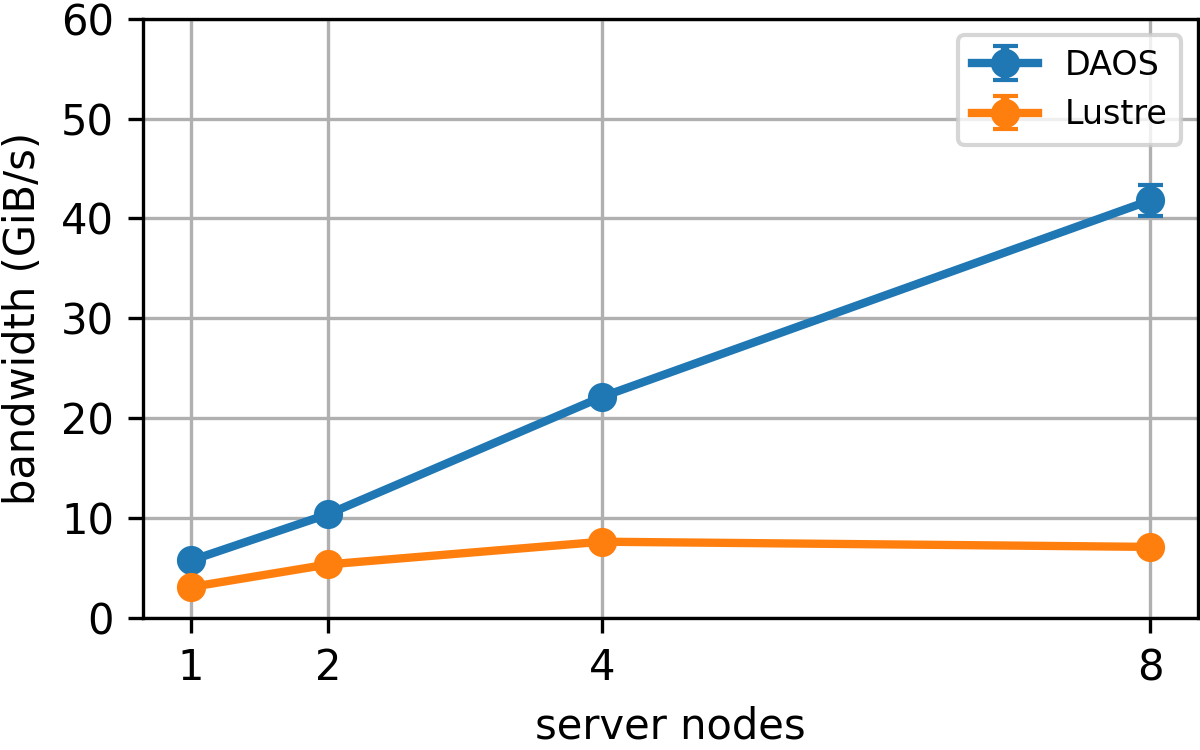}
        \caption{Write}
    \end{subfigure}
    \begin{subfigure}[b]{190pt}
        \includegraphics[width=190pt,trim={30pt 0 0 0},clip]{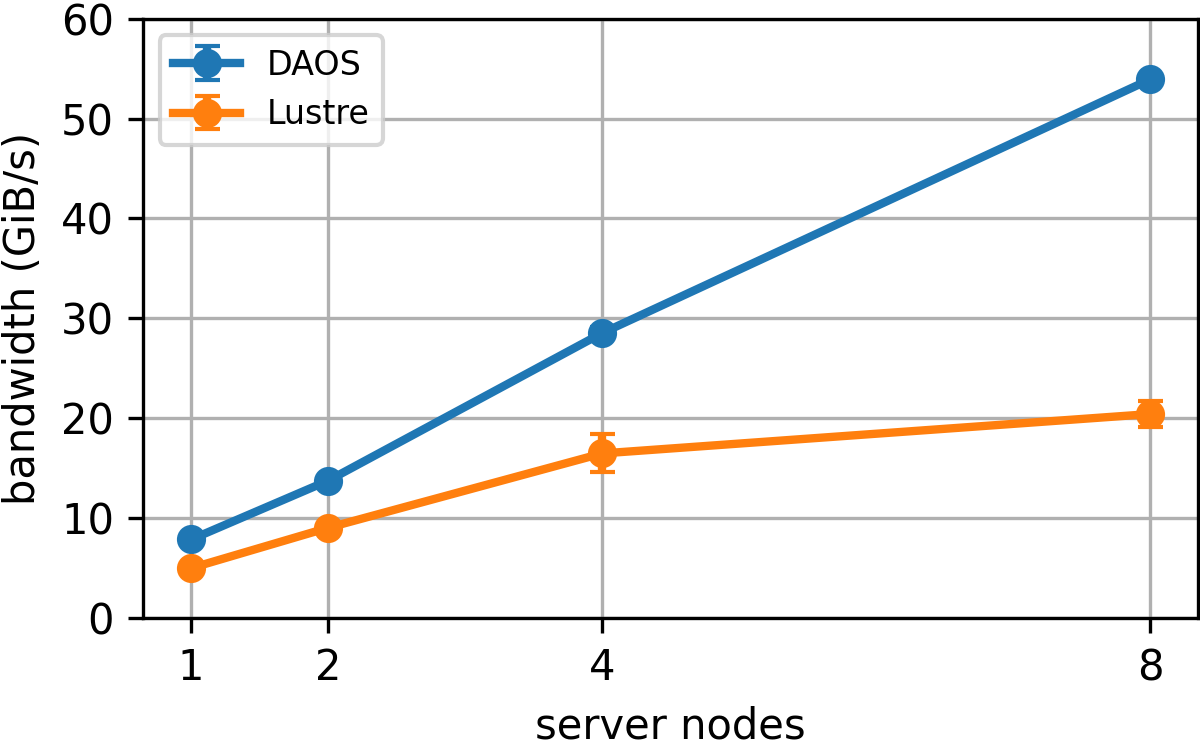}
        \caption{Read}
    \end{subfigure}
    \caption{Field I/O bandwidth scalability against increasingly large Lustre and DAOS deployments in NEXTGenIO. A ratio of 2-to-1 client-to-server nodes was used, and 24 to 48 processes were run in each client node. Every process wrote and indexed (or de-referenced and read) 2000 x 1MiB weather fields. Tests were repeated 5 times. \copyright 2022 IEEE.}
    \label{fig:ngio_daos_lustre_fieldio_scalability}
\end{figure*}

Appendix A describes this approach in more detail, including bandwidth measurements of Field I/O runs against Lustre deployments of various sizes. The results showed Field I/O on DAOS performed and scaled much better than on Lustre, and this was not surprising because dummy \verb!libdaos! made a rather abusive use of the POSIX file and directory APIs, e.g. by mapping every DAOS array to a separate file, or by mapping every key in a DAOS key-value to a separate file. This did not comply with some of the most important programming best practices for high-performance on POSIX file systems. A selection of the results comparing Field I/O performance and scalability on DAOS and Lustre is shown here in Fig. \ref{fig:ngio_daos_lustre_fieldio_scalability}. Lustre deployments used one more node than the corresponding DAOS deployments, for the MDTs.

Despite this comparison being unfair due to the POSIX I/O API abuse in dummy \verb!libdaos!, this provided evidence of the limitations of POSIX I/O and, more importantly, established that Field I/O in combination with dummy \verb!libdaos! were a representative example of non-well-behaved POSIX I/O. This benchmark was used again later in this work.

\subsection{FDB backend performance}

With the insight from Field I/O experiments, a pair of DAOS backends for the FDB ---described in Section \ref{sec:daos_backends}--- were developed. The \verb!fdb-hammer! benchmark, which performs I/O via the FDB library, was then run in NEXTGenIO against DAOS and Lustre deployments using the respective FDB backends. Because these backends were both carefully designed to make efficient use of DAOS and Lustre systems, \verb!fdb-hammer! allowed for direct and fair comparison of the performance potential of both systems for the ECMWF's NWP as well as for any other HPC applications performing similar I/O patterns.

\verb!fdb-hammer! runs as a set of parallel processes which write or read a sequence of weather fields using the FDB library. Just as in Field I/O, for every field written or read, a few key-value operations are performed and one array object is written or read.

The first set of \verb!fdb-hammer! tests aimed to address the parameter optimisation part of the methodology. The optimisation procedure was followed separately for \verb!fdb-hammer! on DAOS and for \verb!fdb-hammer! on Lustre. The results were covered thoroughly in Appendix C - Section 5.1. One result worth noting is that, in contrast to the IOR runs reported earlier, the optimal client-to-server node ratio was found to be 2 both for \verb!fdb-hammer! on Lustre and on DAOS. This meant there were sufficient nodes in the system to run scalability curves for Lustre deployments on up to 10 server nodes, unlike in earlier IOR scalability tests on Lustre where a client-to-server node ratio of 4 was required to fully saturate the system.

\begin{figure*}[htbp]
    \centering
    \begin{subfigure}[b]{212pt}
        \includegraphics[width=212pt]{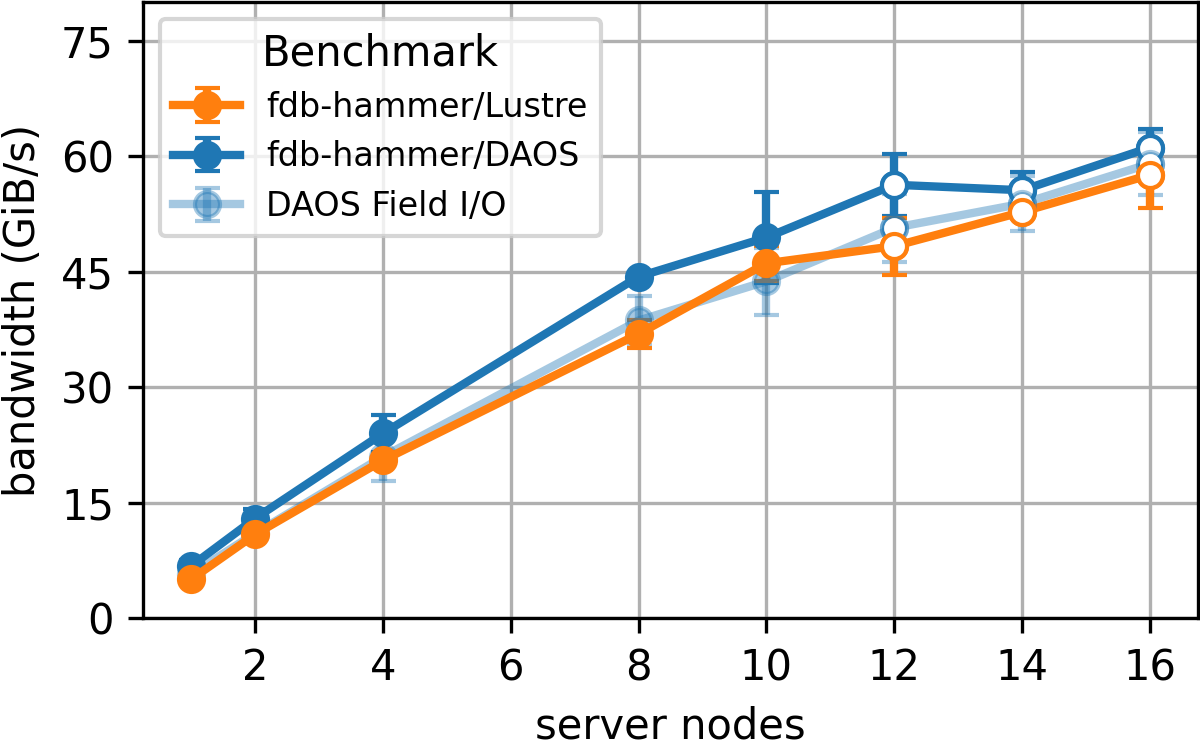}
        \caption{Write, no w+r contention}
    \end{subfigure}
    \begin{subfigure}[b]{190pt}
        \includegraphics[width=190pt,trim={30pt 0 0 0},clip]{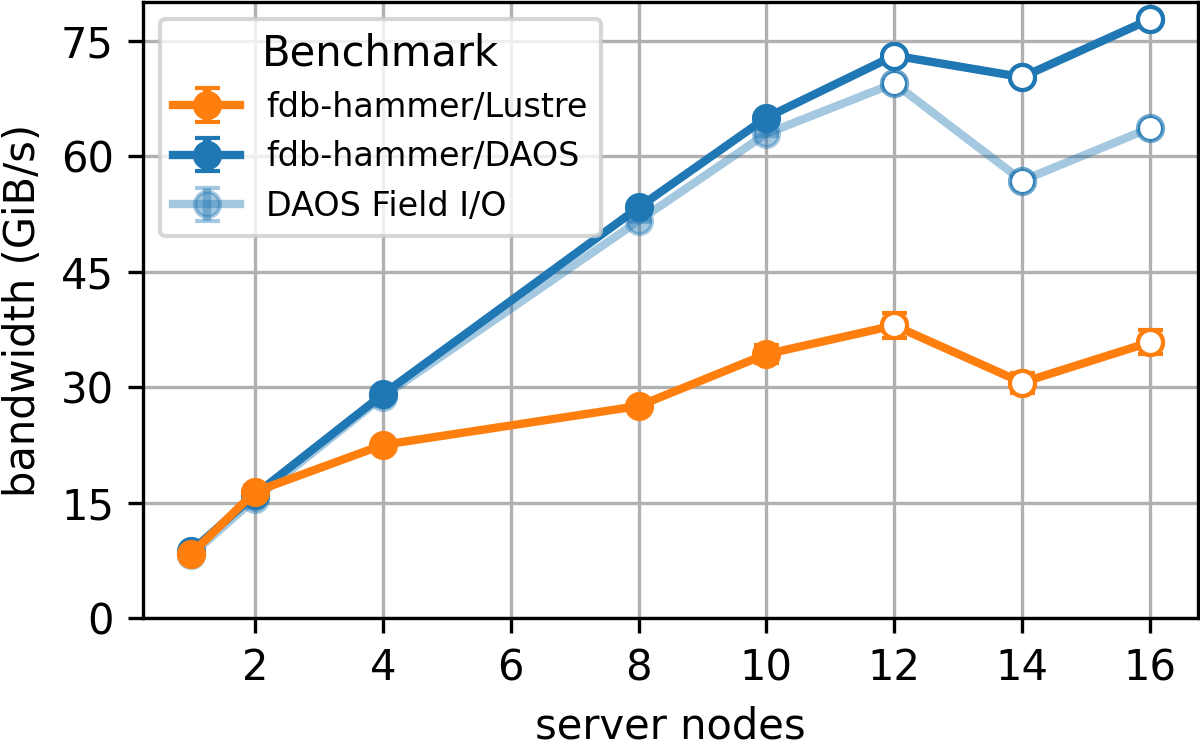}
        \caption{Read, no w+r contention}
    \end{subfigure}
    \caption{fdb-hammer and Field I/O bandwidth scalability, with no write+read contention, against increasingly large Lustre and DAOS deployments in NEXTGenIO. A ratio of 2-to-1 client-to-server nodes was used except for hollow dots, and 16 to 48 processes were run in each client node. Every process wrote and indexed (or de-referenced and read) 10000 x 1MiB weather fields. Tests were repeated 3 times.}
    \label{fig:ngio_daos_lustre_fdbh_fieldio_scalability_no_contention}
\end{figure*}

After several rounds of adjusting the benchmark configuration, optimising the FDB DAOS backends, and testing performance at scale ---which was covered in detail in Appendix C - Sections 5.1 to 5.3---, scalability curves were produced for \verb!fdb-hammer! runs with \textit{no write+read contention}, shown in Fig. \ref{fig:ngio_daos_lustre_fdbh_fieldio_scalability_no_contention} here. Field I/O scalability curves produced earlier were included in the graphs as well for direct comparison. Hollow dots indicate where the optimal client-to-server node ratio of 2 was not fulfilled, therefore not fully saturating the system.

The write performance results in (a) showed both DAOS and Lustre performed similarly and scaled linearly for write up to 10 server nodes, reaching IOR bandwidths. As in Fig. \ref{fig:ngio_ior_lustre_daos_scalability} (a), this suggested both storage systems were properly configured to exploit the underlying hardware, and both backends were properly configured to efficiently access the storage systems and optimised for best write performance. The write performance of DAOS at 10 server nodes showed variance and cast some doubt on the ability of DAOS to hold the linear scaling pattern at larger node counts. However, tests in the upper part of the range of variation did fall on the projected straight line, suggesting additional fine-tuning of DAOS and the benchmark might allow consistently reaching these bandwidths, although this was not pursued.
Tests beyond 10 server nodes resulted in good performance despite not using enough client nodes to fully saturate the system, also suggesting that the linear scaling would likely hold at these scales if sufficient nodes were available.

For read, in (b), DAOS performed remarkably well and scaled linearly, in line with previous Field I/O results, whereas Lustre showed a marked decline in scalability beyond 2 server nodes and reached only a half of the read performance of DAOS at larger scales.
The sub-optimal performance of \verb!fdb-hammer! on Lustre was first thought to be due to the design of the POSIX I/O backends, which favours write performance at the expense of read performance. However, as discussed earlier, \verb!fdb-hammer! does not initiate reader processes in a staggered way, as operational runs do, nor reproduces the transposed data access of operational readers, resulting in I/O workloads easier than the operational ones if using the POSIX I/O backends. 
The significant gap in read performance for \verb!fdb-hammer! runs on Lustre relative to IOR was therefore a slightly surprising result.

For both DAOS and Luste, read performance continued scaling with 12 server nodes despite not fulfilling the optimal 2-to-1 client-to-server node ratio. This was due to the fact that 20 client nodes were used for that configuration and this was not far from the optimal ratio. Performance dropped significantly at 14 and 16 nodes because a ratio of 1-to-1 was used for these configurations. Conversely, write performance did not drop as much at these scales because write is less sensitive to sub-optimal client-to-server node ratios, as shown in Appendix C - Section 5.1.

It is worth noting that \verb!fdb-hammer! on DAOS slightly exceeded IOR and Field I/O bandwidths, and this was explained by the fact that a few optimisations were implemented in the FDB DAOS backends which were not implemented in the IOR \verb!libdaos! backend nor in Field I/O. Specifically, these optimisations consisted in using DAOS arrays without attributes, which allow ditching array creation calls on write (see the documentation of \texttt{daos\_array\_open\_with\_attrs}), and avoiding unnecessary \texttt{daos\_array\_get\_size} calls on read.

\verb!fdb-hammer! was also run in its \verb!list()! mode for some of the configurations after the read phase completed, to compare the listing performance of both backends. The benchmark was configured to list all fields \verb!archive()!d by all writer processes during the write phase for the first simulation step ---1 out of 100 steps---, and executed from a single node.
Listing with the POSIX I/O backends was consistently twice as fast as with the DAOS backends, for all server node counts.
This was in line with expectations as, as explained in Section \ref{sec:daos_backends}, the POSIX I/O backends load all required indexing information in only a few \verb!read! calls, whereas the DAOS backends issue a \verb!daos_kv_get! operation for every matching entry, making the former backends more efficient for listing large sets of fields as done in this test.
The performance of FDB \verb!list()! with the DAOS backends was not further investigated or improved, as \verb!list()! would not be involved in operational runs on DAOS, as discussed in Section \ref{subsubsec:operational_nwp_pattern}.

\begin{figure*}[htbp]
    \centering
    \begin{subfigure}[b]{212pt}
        \includegraphics[width=212pt]{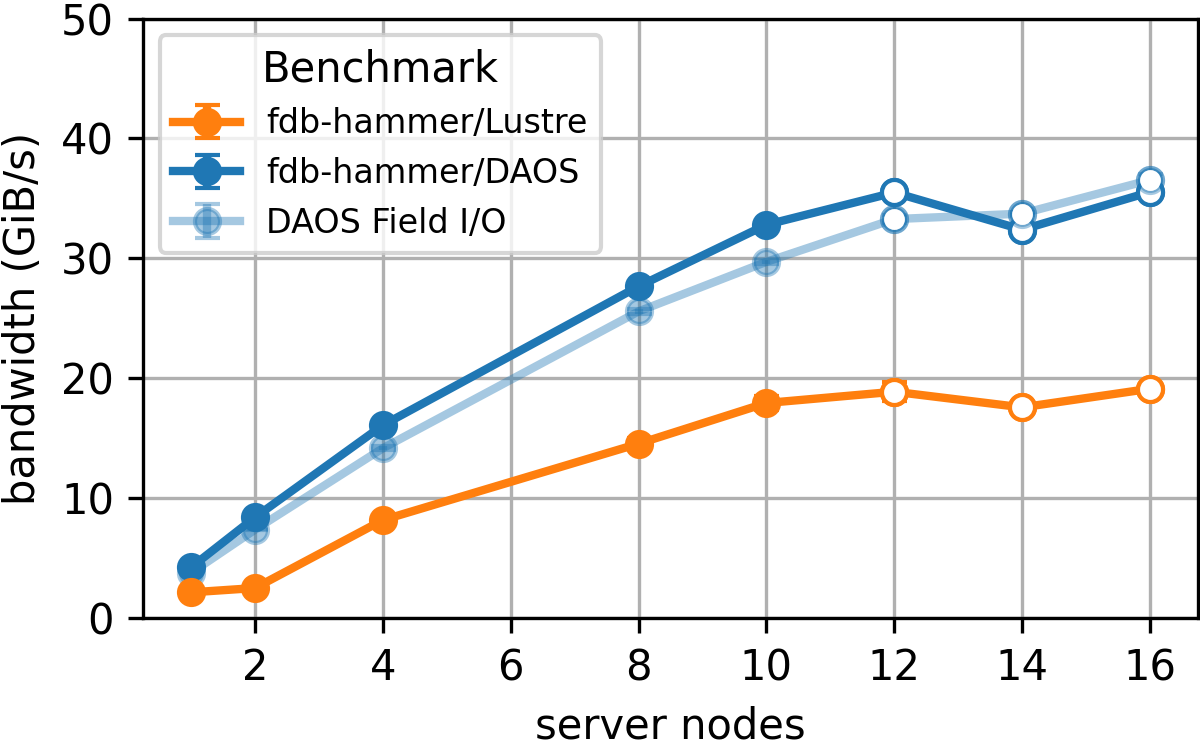}
        \caption{Write, w+r contention}
    \end{subfigure}
    \begin{subfigure}[b]{190pt}
        \includegraphics[width=190pt,trim={30pt 0 0 0},clip]{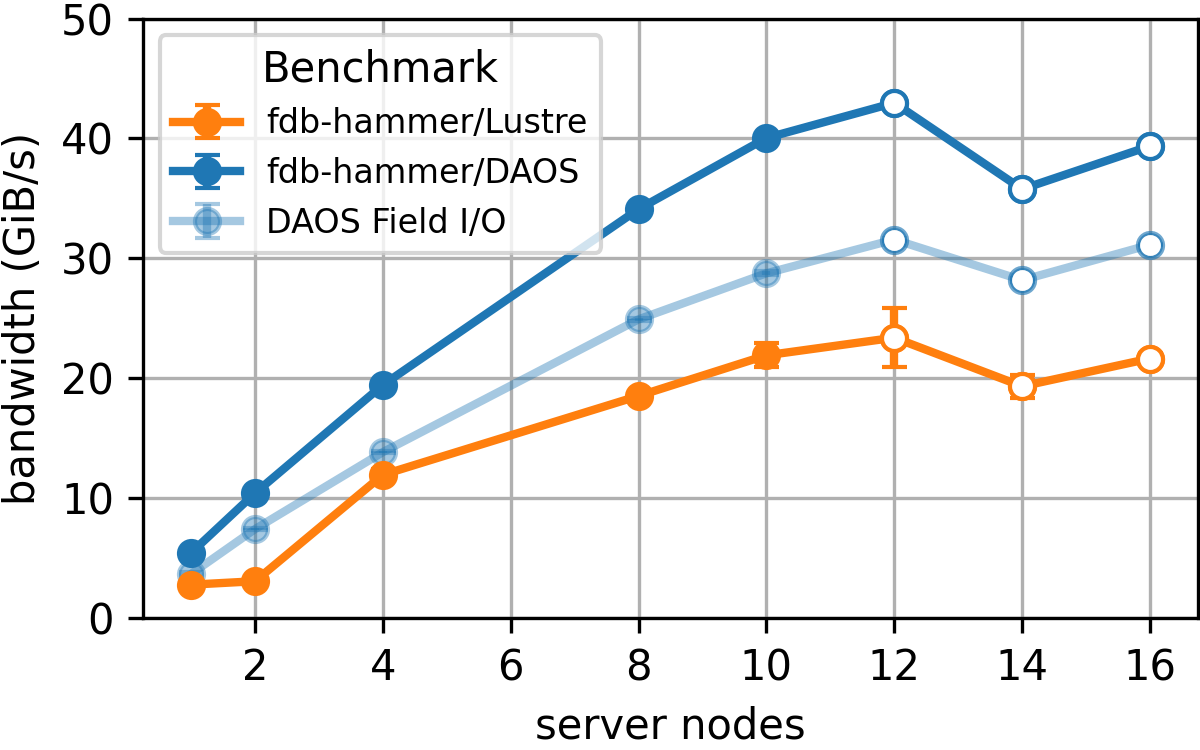}
        \caption{Read, w+r contention}
    \end{subfigure}
    \caption{fdb-hammer and Field I/O bandwidth scalability, with write+read contention, against increasingly large Lustre and DAOS deployments in NEXTGenIO. A ratio of 2-to-1 client-to-server nodes was used except for hollow dots, and 16 to 48 processes were run in each client node. Every process wrote and indexed (or de-referenced and read) 10000 x 1MiB weather fields. Tests were repeated 3 times.}
    \label{fig:ngio_daos_lustre_fdbh_fieldio_scalability_contention}
\end{figure*}

Scalability curves were also produced for \verb!fdb-hammer! runs with \textit{write+read contention}, shown in Fig. \ref{fig:ngio_daos_lustre_fdbh_fieldio_scalability_contention}. This access pattern more closely mimicked the I/O patterns of an operational NWP run at the ECMWF. As in Field I/O runs with contention, half of the allocated client nodes executed writer processes and the other half executed readers, simultaneously. Bandwidths for a given test in graphs (a) and (b) can be added to determine the total aggregate bandwidth the system delivered.

In this case, DAOS and Lustre showed different write performance behavior from each other. DAOS scaled nearly linearly, close to the expectations set by Field I/O under \textit{write+read contention}, although with a gentle decline as more server nodes were added. Remarkably, although Lustre scaled linearly for write ---neglecting the lack of performance increase at 2 server nodes, which was not investigated---, it reached significantly lower write bandwidths relative to DAOS. This was despite the POSIX I/O FDB backends being optimised for write, and despite the Lustre's write performance being very similar to DAOS's in the benchmark runs with \textit{no write+read contention}. This clearly indicated that the concurrent read workload was negatively impacting write performance on Lustre, effectively as though the writes and reads were mutually exclusive and never overlapped, while they fully overlapped on DAOS.

This low write performance on Lustre under contention was at first thought to be due to the Lustre's locking mechanisms engaging to ensure strong consistency of data and metadata across client-side caches and servers.
However, as described in Section \ref{subsubsec:fdbh_pattern}, \verb!fdb-hammer! with \textit{write+read contention} produces a type of write-read contention that should not result in intensive involvement of such locking mechanisms.
On one hand, this meant write bandwidth might be even lower if operational contention were fully reproduced. On the other hand, this also meant the low write performance was an unexpected result.

For read, in (b), the general performance behavior of both systems was similar to the write behavior, that is, with a gentle decline in DAOS's scalability and significantly lower but linearly scaling bandwidths for Lustre. The inferior read performance on Lustre was a slightly surprising result, for the same reasons as in the \textit{no write+read contention} case.
Also in (b), \verb!fdb-hammer! on DAOS performed significantly better than Field I/O, again likely due to the optimisation put in place in \verb!fdb-hammer! to avoid unnecessary array size checks.

\begin{figure*}[htbp]
    \centering
    \begin{subfigure}[b]{214pt}
        \includegraphics[width=214pt,trim={20pt 215pt 26pt 25pt},clip]{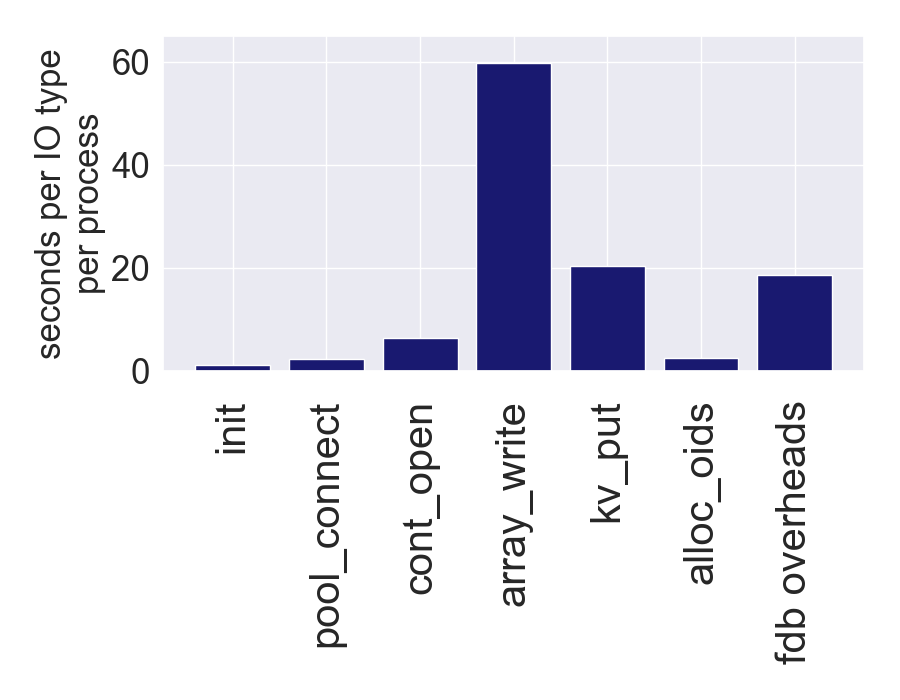}
        \caption{Writers, no w+r contention. Average per-\phantom{ } pro\-cess wall-clock time: 110.5s.}
    \end{subfigure}
    \begin{subfigure}[b]{188pt}
        \centering
        \includegraphics[width=177pt,trim={125pt 215pt 26pt 25pt},clip]{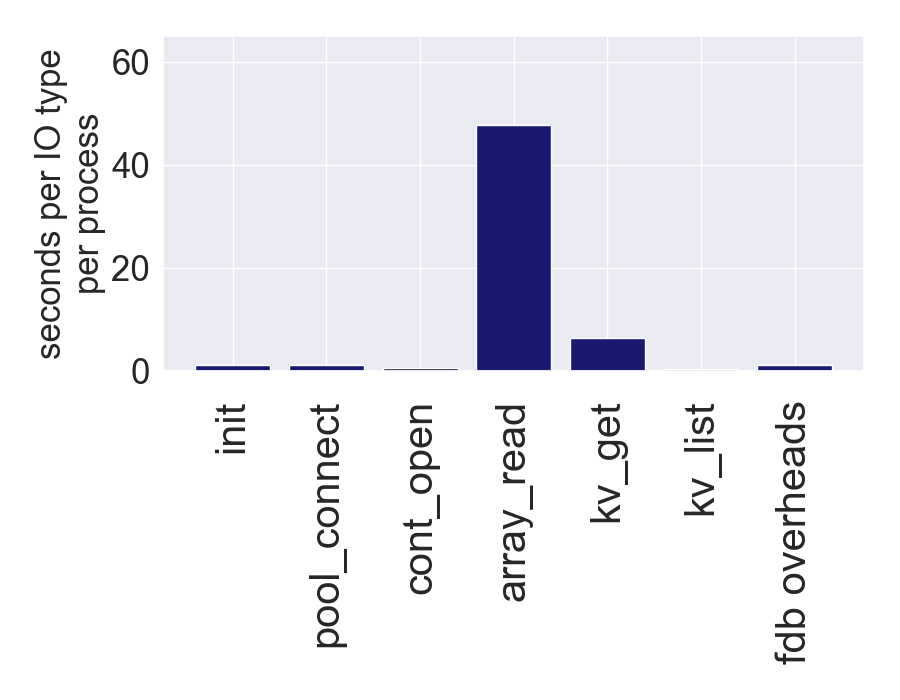}
        \caption{Readers, no w+r contention. Average per-process wall-clock time: 58.8s.}
    \end{subfigure}
    \vskip\baselineskip
    \begin{subfigure}[b]{214pt}
        \includegraphics[width=214pt,trim={20pt 25pt 26pt 25pt},clip]{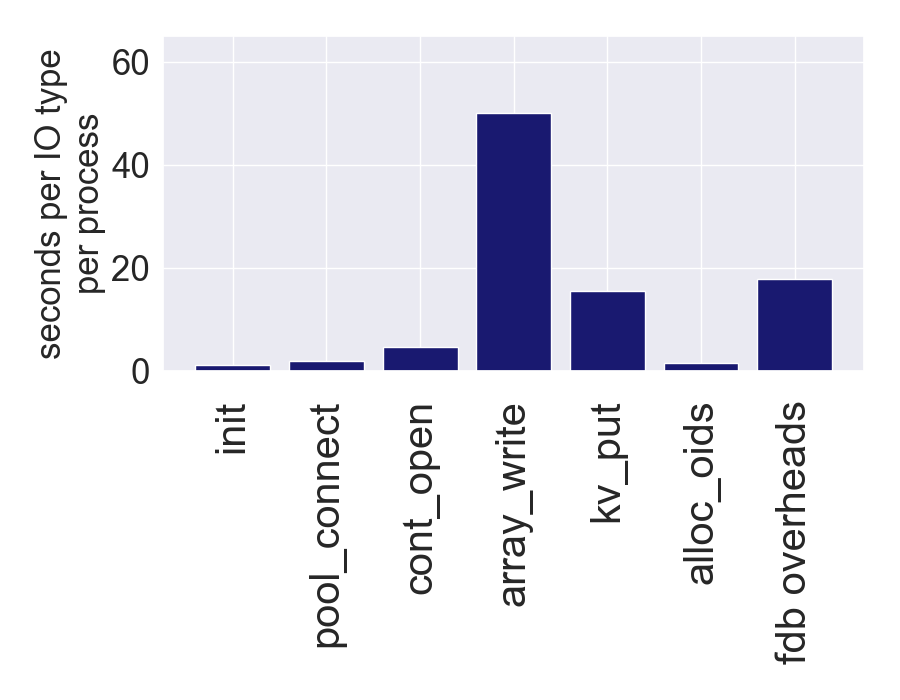}
        \caption{Writers, w+r contention. Average per-\phantom{ } process wall-clock time: 92.4s.}
    \end{subfigure}
    \begin{subfigure}[b]{188pt}
        \centering
        \includegraphics[width=177pt,trim={125pt 25pt 26pt 25pt},clip]{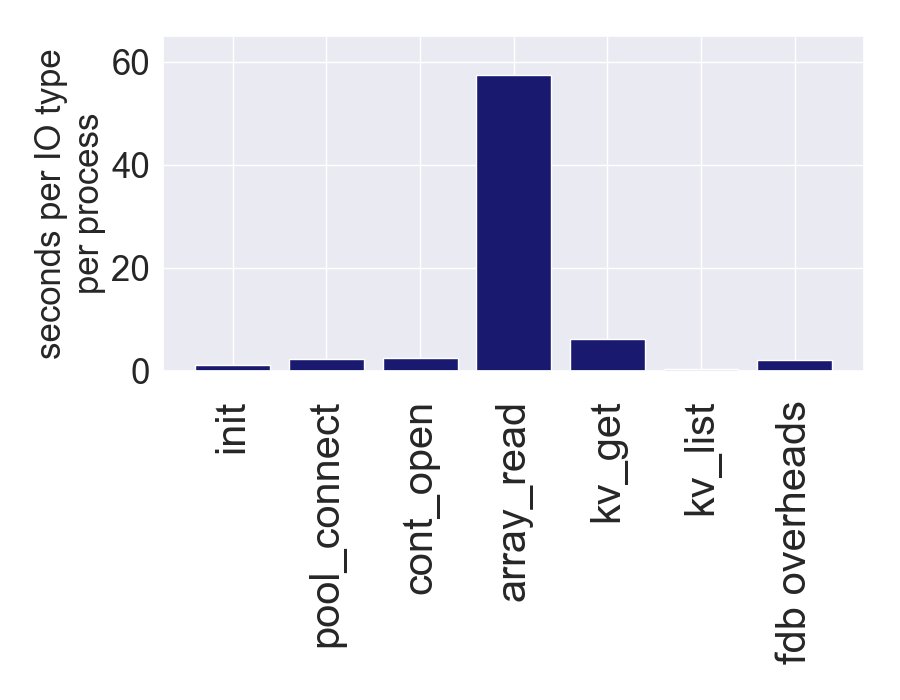}
        \caption{Readers, w+r contention. Average per-process wall-clock time: 71.5s.}
    \end{subfigure}
    \caption{Profiling results for fdb-hammer/DAOS runs without (top row) and with (bottom row) write+read contention, using 10 server and 20 client nodes; 32 processes per client node. Every process wrote and indexed (or de-referenced and read) 10000 x 1MiB weather fields.}
    \label{fig:ngio_fdbh_daos_profiling}
\end{figure*}

\begin{figure*}[htbp]
    \centering
    \begin{subfigure}[b]{224pt}
        \includegraphics[width=224pt,trim={26pt 265pt 26pt 25pt},clip]{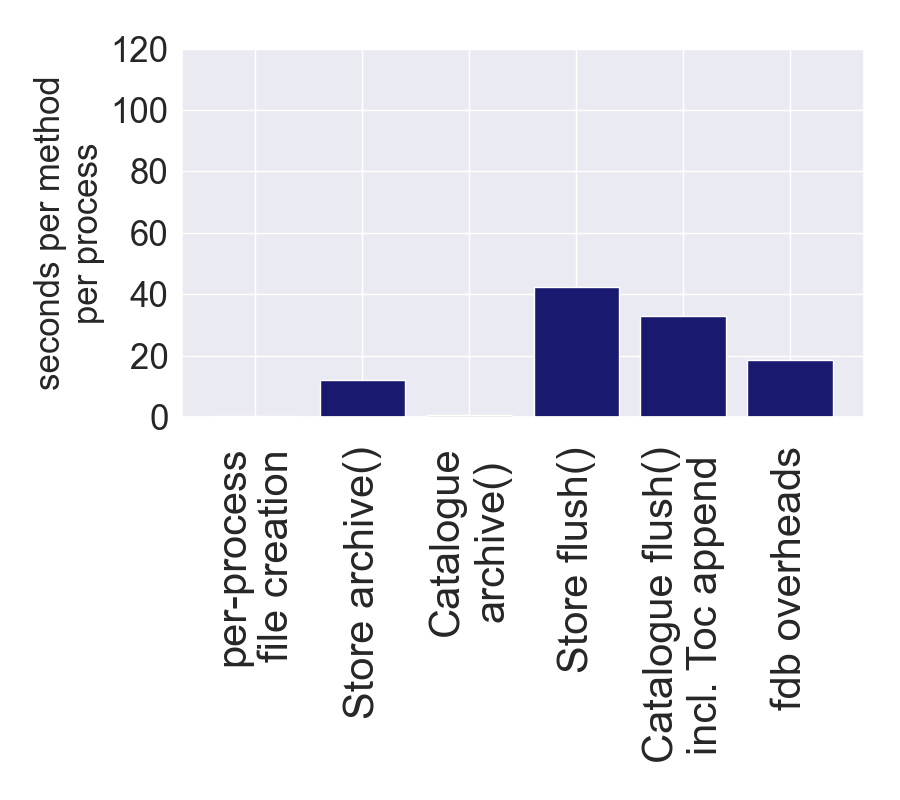}
        \caption{Writers, no w+r contention. Average per-\phantom{ } pro\-cess wall-clock time: 107.3s.}
    \end{subfigure}
    \begin{subfigure}[b]{178pt}
        \centering
        \includegraphics[width=128pt,trim={125pt 258pt 0pt 25pt},clip]{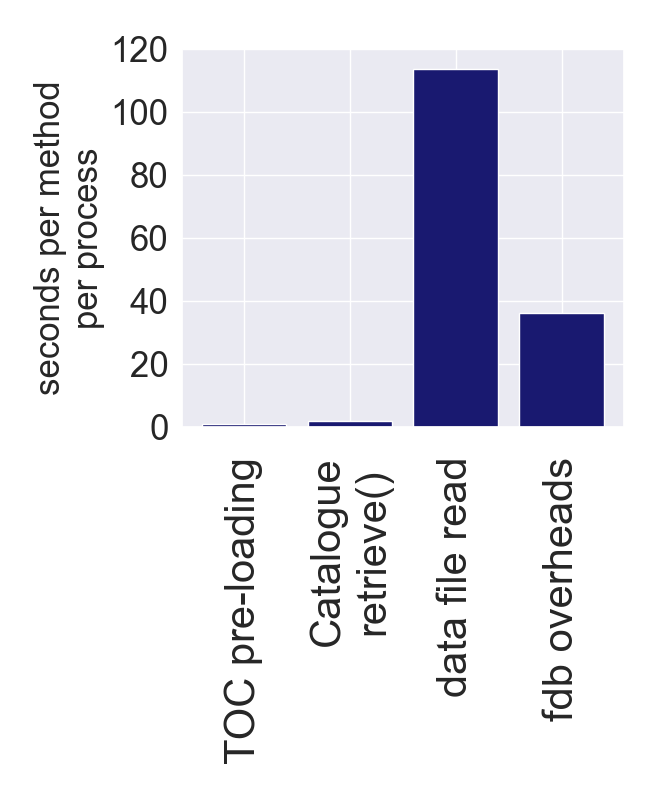}
        \caption{Readers, no w+r contention. Average per-process wall-clock time: 153s.}
    \end{subfigure}
    \vskip\baselineskip
    \begin{subfigure}[b]{224pt}
        \includegraphics[width=224pt,trim={26pt 25pt 26pt 25pt},clip]{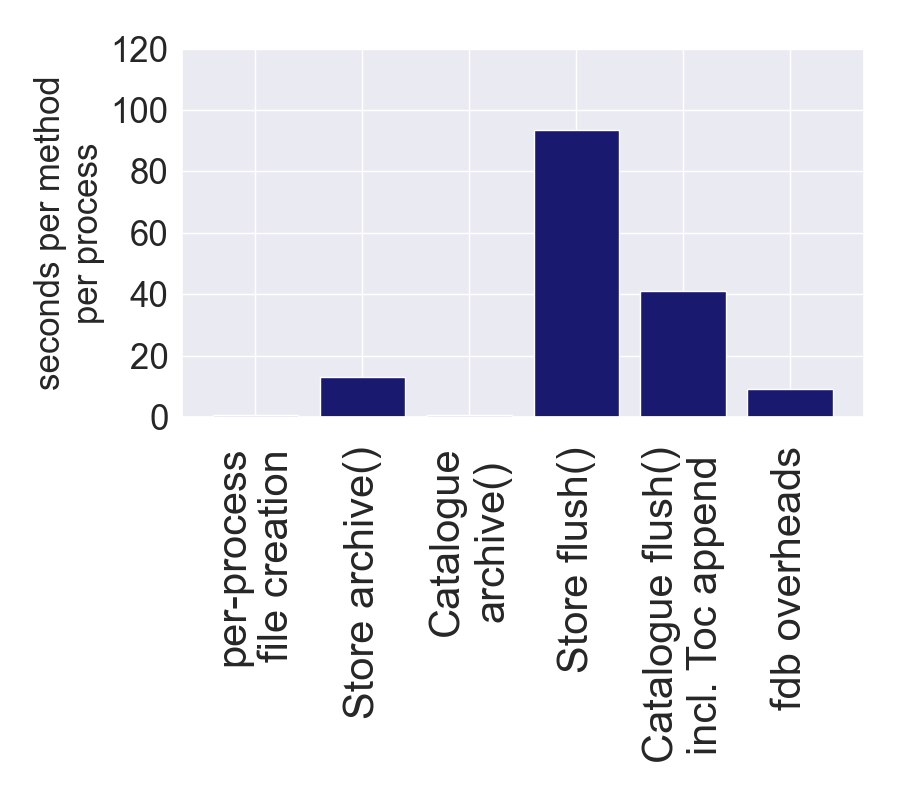}
        \caption{Writers, w+r contention. Average per-pro-\phantom{ } cess wall-clock time: 163s.}
    \end{subfigure}
    \begin{subfigure}[b]{178pt}
        \centering
        \includegraphics[width=128pt,trim={125pt 10pt 0pt 25pt},clip]{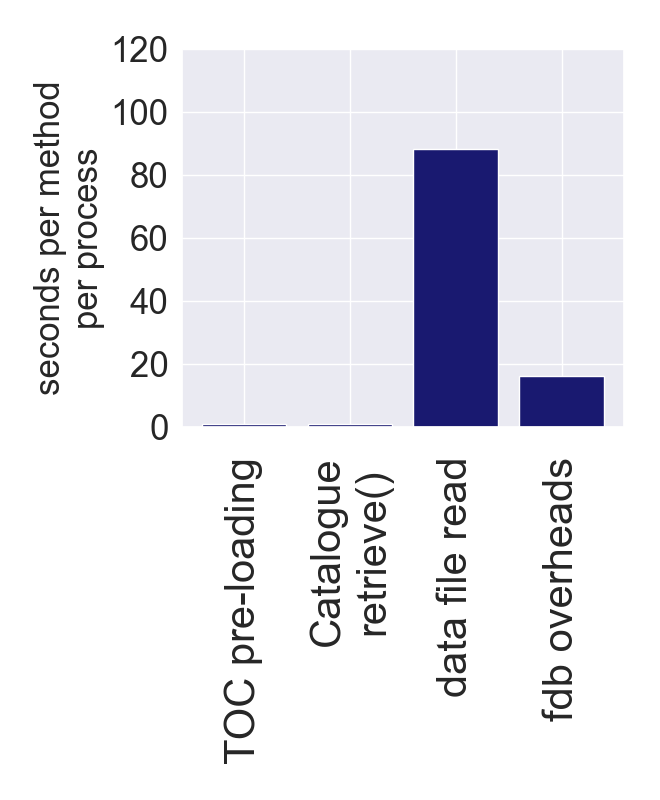}
        \caption{Readers, w+r contention. Average per-process wall-clock time: 112.4s.}
    \end{subfigure}
    \caption{Profiling results for fdb-hammer/Lustre runs without (top row) and with (bottom row) write+read contention, using 10 server and 20 client nodes; 32 processes per client node. Every process wrote and indexed (or de-referenced and read) 10000 x 1MiB weather fields.}
    \label{fig:ngio_fdbh_lustre_profiling}
\end{figure*}

To gain further insight on the behavior of both backends and the reasons why Lustre struggled under contention, \verb!fdb-hammer! and the DAOS and POSIX I/O backends were instrumented to report the amount of time spent on some of the most relevant operations performed. Fig. \ref{fig:ngio_fdbh_daos_profiling} and Fig. \ref{fig:ngio_fdbh_lustre_profiling} summarise the profiling information obtained for \verb!fdb-hammer! runs against 10-node DAOS and Lustre deployments, respectively, both with and without write+read contention.

The profiling results for DAOS, in Fig. \ref{fig:ngio_fdbh_daos_profiling}, showed that most of the time was spent in array write and read operations, which was a good sign as it meant DAOS and the FDB backends were able to prioritise data movement and persistence without being interfered by the surrounding metadata management, thus providing close-to-hardware bandwidths. The fact that the profiling results with write+read contention ---at the bottom row--- were very similar to the results without contention was an excellent result, meaning that neither writing nor reading on DAOS suffered a noticeable performance impact despite the benchmark fully reproducing the contention of a hypothetical operational run using the FDB DAOS backends.

The profiling results for Lustre, in Fig. \ref{fig:ngio_fdbh_lustre_profiling}, showed most of the time of the write phase ---both with and without contention--- was spent in the Store \verb!flush()! and Catalogue \verb!flush()! methods, and this was in line with expectations as these methods persist data or wait for data to be persisted into storage.

The fact that a similar potion of time was spent in Store \verb!archive()! and Catalogue \verb!flush()! for both patterns demonstrated that the contentious pattern was in fact not suffering from the Lustre's locking mechanisms. If the contentious pattern had captured the full operational contention, the Store \verb!archive()! and Catalogue \verb!flush()! methods would have issued \verb!write! calls ---on the data, sub-TOC, and index files--- which would have contended with \verb!read! calls on the same files issued as part of TOC pre-loading and the bulk data reading in the read phase, thus engaging the distributed locking mechanisms and resulting in higher wall-clock times for Store \verb!archive()! and Catalogue \verb!flush()!, but this was not the case.
Instead, surprisingly, Store \verb!flush()! ---which only waits for the bulk data to be transferred to the Lustre servers and persisted into storage media--- took more than double the time in the contentious pattern. This suggested that, possibly, the network or the storage media were not properly handling concurrent writes and reads, resulting in longer than expected persist wall-clock times. Given this issue did not occur in equivalent runs on DAOS, a possible explanation could be that Lustre might not be optimised to perform concurrent writes and reads on SCM, or alternatively that the PSM2 protocol used by Lustre did not properly handle concurrent transfers in both directions.

Regarding the read phase on Lustre, in Fig. \ref{fig:ngio_fdbh_lustre_profiling} (b) and (d), most of the time was spent on bulk reading from a single data file per process, in line with expectations.
However, the bulk reading time for runs with \textit{no write+read contention} seemed excessive, particularly if compared to the time spent on array reads in equivalent runs on DAOS, which overall read the exact same amount of data from storage.
This slowness in bulk reading was the main contributor to the inferior Lustre read performance observed in Fig. \ref{fig:ngio_daos_lustre_fdbh_fieldio_scalability_no_contention} and \ref{fig:ngio_daos_lustre_fdbh_fieldio_scalability_contention}, and was potentially due to Lustre not handling well the relatively large number ---albeit not as large as in operational runs--- of \verb!read! operations issued per process.
Another possible cause could be an undiscovered issue or inefficiency in the logic for merging \verb!DataHandles! returned by the sequence of FDB \verb!retrieve()! calls issued by a reader process, such that the same data file is re-opened and closed or its size queried several times unnecessarily.

A substantial amount of time was spent on FDB overheads --- that is, time spent by the FDB library on non-I/O operations such as calculations or in-memory operations. Although this was not investigated, as it was shadowed by the bulk read inefficiencies, this is discussed in more detail later on.

Neither the lower-than-expected read performance nor the write performance deterioration under contention for runs on Lustre were further investigated.

\subsection{Summary}

The I/O benchmarking so far has shown both Lustre and DAOS were capable of reaching similarly high and linearly scaling bandwidths for easy I/O workloads on a system with SCM.

The system has an Omni-Path fabric which DAOS could not exploit natively ---i.e. not bypassing the operating system for communications via the fabric---, but DAOS nevertheless achieved performance levels as high as Lustre ---which did natively exploit the fabric--- provided large parallel process counts were used.

Both storage systems showed an equally substantial gap in performance relative to hardware limits, and this was possibly due to the high sensitivity of the SCM storage hardware to small variations in I/O sizes and persistence horizon.

Both Lustre and DAOS, and the respective FDB backends, performed and scaled very well for NWP I/O workloads without write+read contention, although DAOS stood out reaching twice as high read bandwidths.
For contentious NWP workloads, DAOS reached twice as high write and read bandwidths.
DAOS's performance advantage should have been even more marked if the benchmark runs on Lustre had fully captured the operational access pattern and contention.

These results were encouraging as they showed DAOS can provide similar performance to Lustre for simple I/O workloads, and superior performance for complex applications requiring storage and indexing of relatively small data objects. This suggested DAOS has potential both for HPC storage in general as well as for the ECMWF's operational NWP.
The results, however, had to be taken with caution as, on one hand, the root cause of DAOS's advantage in read performance was not well identified and, on the other hand, the write performance advantage under contention seemed to be due to Lustre not being able to optimally exploit the specific hardware at hand, and might therefore not be entirely applicable to other systems with different hardware.

\clearpage
\section{DAOS, Ceph, and Lustre on NVMe SSDs}

All benchmarking so far was conducted in a system with Optane SCM and an Omni-Path network, but these hardware options became a less likely choice for new production systems after Omni-Path was temporarily discontinued in 2019 and 2020, and Optane was permanently discontinued in mid 2022.
This section presents I/O benchmarking experiments conducted against DAOS, Ceph, and Lustre systems deployed on machines with NVMe SSDs --- a commonplace storage hardware option nowadays.

\subsection{The Google Cloud Platform}

\begin{figure}[htbp]
\centerline{\includegraphics[width=300pt,trim={0pt 0 0 0},clip]{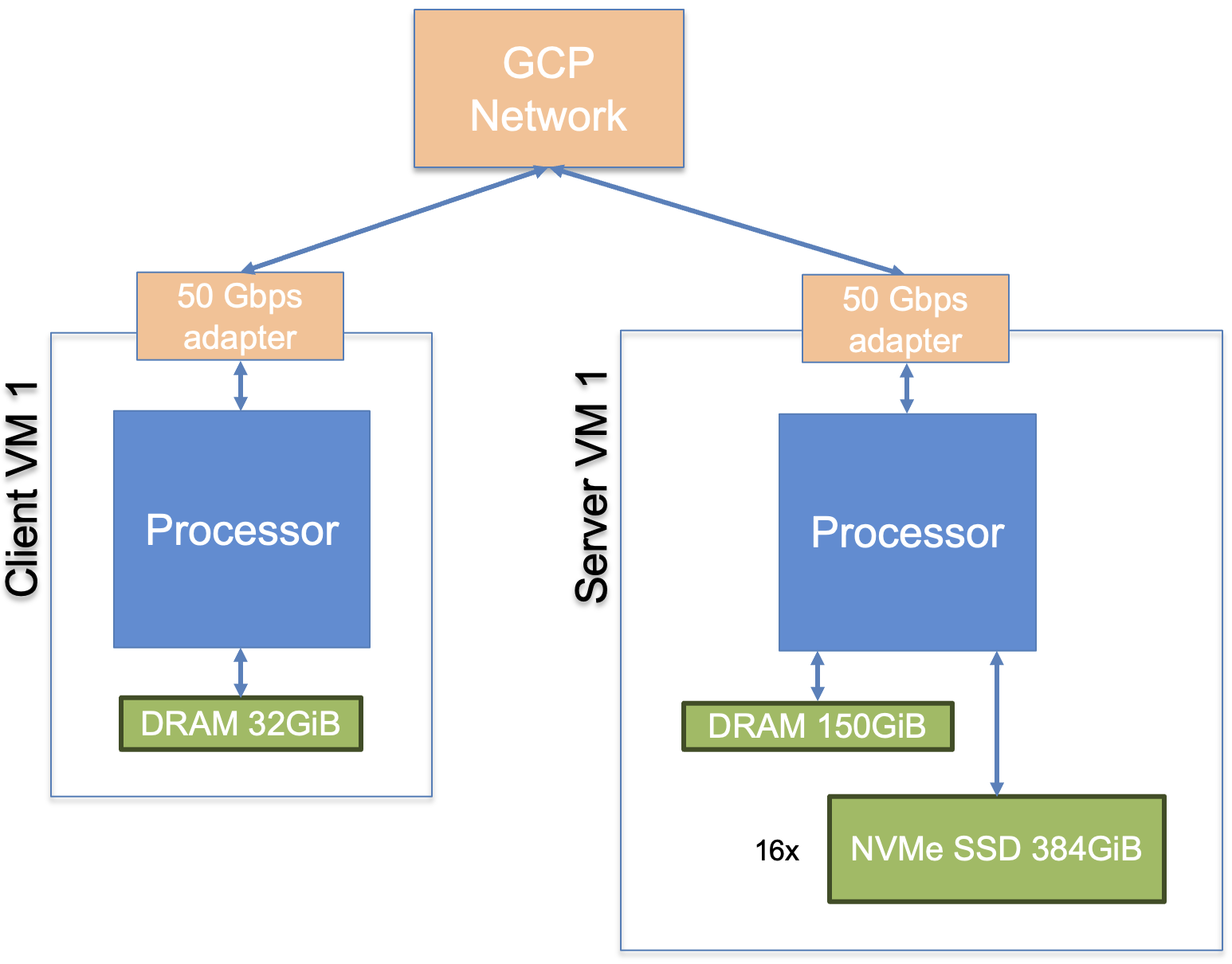}}
\caption{VM types used for benchmarking on GCP.}
\label{fig:gcp_architecture}
\end{figure}

The experiments were conducted on Google Cloud Platform (GCP)\cite{GooglePage} infrastructure. DAOS, Ceph, and Lustre systems were deployed on VMs ---also referred to as \textit{instances}--- of a custom type\cite{General-purposeEngine}, \verb!n2-custom-36-153600!, each with 36 logical cores, 150 GiB of DRAM, 6 TiB of local NVMe SSDs distributed in 16 logical devices, and a 50 Gbps network adapter. On the client side, the benchmarks were run in VMs of type \verb!n2-highcpu-32!, each with 32 logical cores, 32 GiB of DRAM, and a 50 Gbps network adapter. These VM types are illustrated in Fig. \ref{fig:gcp_architecture}.

Because the network adapters did not provide RDMA capability, the TCP protocol was used for all communication over the network.
To exploit all available network bandwidth, the VMs had to be configured with Simultaneous Multi-Threading (SMT)\cite{SetCore} enabled, and the benchmark processes pinned evenly across all available cores.

To minimise running costs, all VMs were provisioned with a \textit{spot} provisioning model\cite{SpotVMs}, implying these might not be immediately available during periods of high demand, and could be evicted at any time. This required putting mechanisms in place both client and server-side to wait for availability of spot resources and gracefully handle eviction.

For the server side, a few scripts were developed which programmatically provision VMs with SSDs and deploy one of the storage systems evaluated on these VMs. Upon eviction of one of the VMs, the scripts properly tear down the storage system and the rest of server VMs, wait for spot resources to be re-provisioned, and redeploy the storage system.

For the client side, an auto-scaling Slurm cluster was created with the help of \verb!cluster-toolkit!\cite{GoogleToolkit} --- a tool to declare and provision HPC and AI infrastructure-as-code on Google Cloud. The auto-scaling Slurm cluster automatically provisioned new spot client VMs as required upon submission of new I/O benchmarking jobs, thus ensuring the jobs did not run until enough spot resources were available, and automatically tore down the VMs if no new jobs were submitted. These client VMs were properly configured for high-performance access to the previously deployed storage systems.

Whenever any of the server or client VMs were evicted during a benchmark run, the described mechanisms engaged to re-provision as needed, and the test was re-run.

The infrastructure-as-code and scripts used to deploy the different storage systems ---except for DAOS, as these are property of Google Cloud--- and the \verb!cluster-toolkit! blueprints of the auto-scaling Slurm cluster are available in \cite{nicolau-manubens2024Ecmwf-projects/daos-tests:0.3.2}.

All provisioned client and server VMs ran images of the Rocky Linux 8 operating system.

\begin{figure}[htbp]
\centerline{\includegraphics[width=320pt,trim={0 0 0 0},clip]{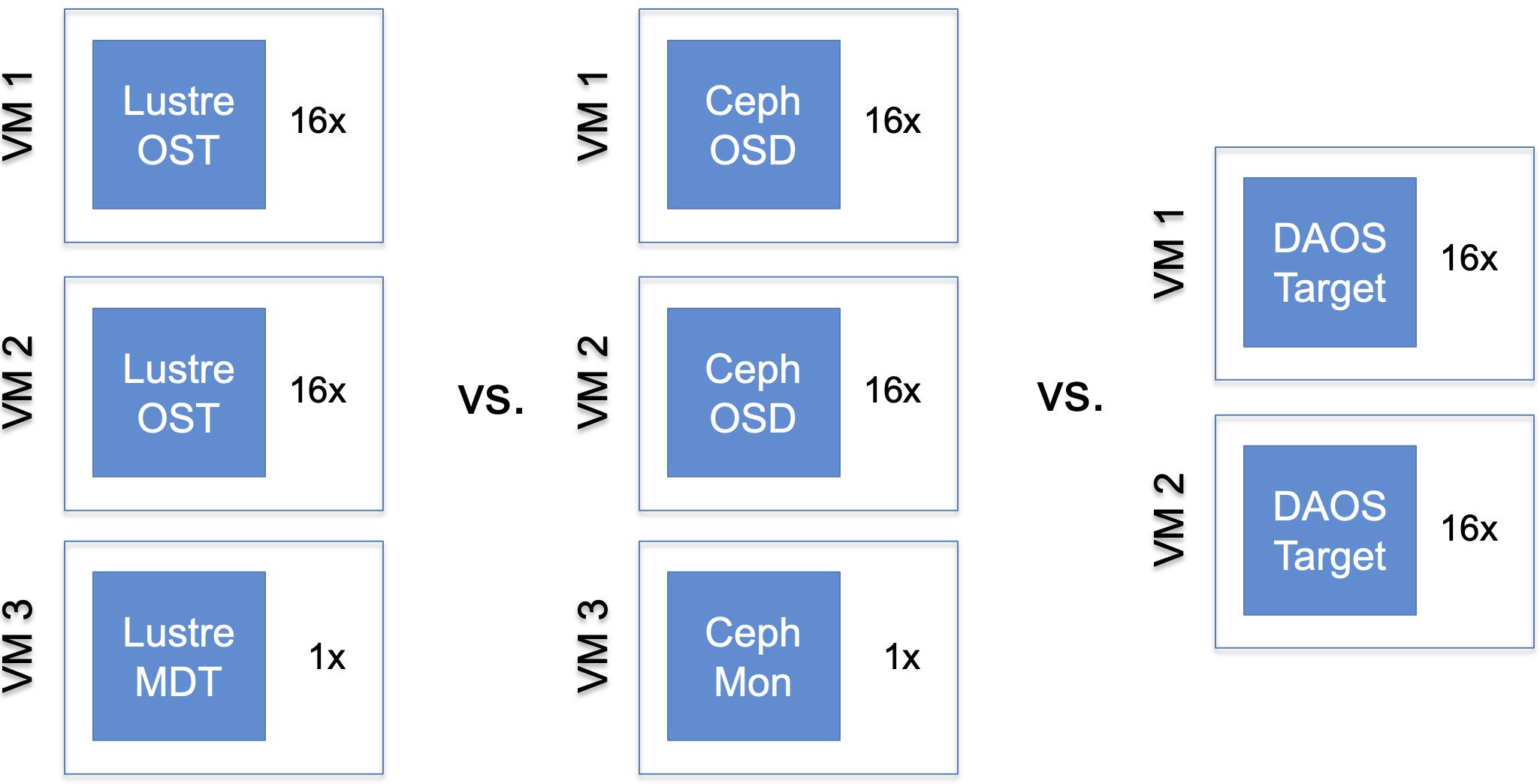}}
\caption{Example Lustre, Ceph, and DAOS configurations for comparison. An additional node was employed for Lustre MDTs and Ceph Monitors.}
\label{fig:lustre_vs_ceph_vs_daos}
\end{figure}

Ceph and Lustre deployments used one more VM than corresponding DAOS deployments. For instance, to compare to a DAOS deployment on two VMs ---each hosting one DAOS engine and 16 targets---, a Ceph deployment on two OSD VMs ---with 32 OSDs overall--- plus one Monitor VM was used, and a Lustre deployment on two OST VMs ---with 32 OSTs overall--- plus one MDT VM was used. This example is illustrated in Fig \ref{fig:lustre_vs_ceph_vs_daos}

This meant Ceph and Lustre had an advantage in terms of allocated resources, although DAOS had the advantage that it was configured to use the DRAM in the server VMs for metadata storage ---including any I/O operations smaller than 4 KiB---, whereas the other two persisted all metadata and small I/O into NVMe SDDs. DAOS currently provides the option to persist metadata into SSDs in the absence of SCM, but this feature was not yet available at the time the experiments were carried out.

Data protection mechanisms ---i.e. erasure-coding and replication--- were disabled for the three storage systems except for a few particular tests where the use of data protection is explicitly noted.
Furthermore, the storage devices in the server VMs were not configured as RAID arrays for any of the storage systems.

As a consequence of not using RAID setups, it was necessary to deploy a daemon ---be it a Lustre OST, a Ceph OSD, or a DAOS target--- to manage each storage device in every server VM, and this was done equally for the three storage systems.
Deploying several daemons per server VM can cause DRAM or CPU overheads with respect to creating RAID arrays and deploying only one or a few daemons per VM, although in this case the number of daemons deployed was small enough that the overheads were not expected to significantly impact I/O performance. Also, high OST/OSD/target counts usually improve rather than deteriorate performance with respect to RAID configurations in highly parallel I/O workloads like the ones produced by the benchmarks in this analysis.
For reference, the OST-to-OSS ratio used for the Lustre deployments in this analysis is within Lustre's practical range and close to that of a few high-performance production instances\cite{GeorgeUnderstandingInternals}.

Although Google Cloud provides VMs deployed on generic hardware, not necessarily tailored purely for HPC, it makes an excellent standard and neutral testbed enabling the analysis of software-level performance and scalability of the various storage systems being evaluated. While the analysis may not map directly to specialised systems, the same overall behavior should be expected on systems using similar storage and network technology.
Regarding contention with other users and performance stability, Google Cloud is committed to ensure the CPU, storage, and network specifications requested upon VM creation are fulfilled. Google Cloud should therefore not be perceived as a contended and fluctuating system. However, some fluctuation in performance can occur if running on spot instances, as done in this analysis, and for this reason the benchmarks were run multiple times to capture any performance variance.

\subsection{Hardware performance measurements}

The raw bulk I/O bandwidth of the NVMe SSDs on server instances was measured by mounting each of the 16 drives in one of the instances as an ext4 file system and then running the \texttt{dd} command in parallel for all of them, first writing and then reading 1000 blocks of 100 MiB. The measurements showed 3.86 GiB/s of aggregate write bandwidth and 7 GiB/s of aggregate read bandwidth.

\texttt{iperf} was used to measure raw network bandwidth between client and server instances, which was found to match the expected 50 Gbps (6.25 GiB/s) in both directions. The \verb!--parallel! option was necessary for \verb!iperf! to run multiple parallel streams and reach optimal bandwidths.

\begin{figure}[htbp]
\centerline{\includegraphics[width=270pt,trim={0 0 0 0},clip]{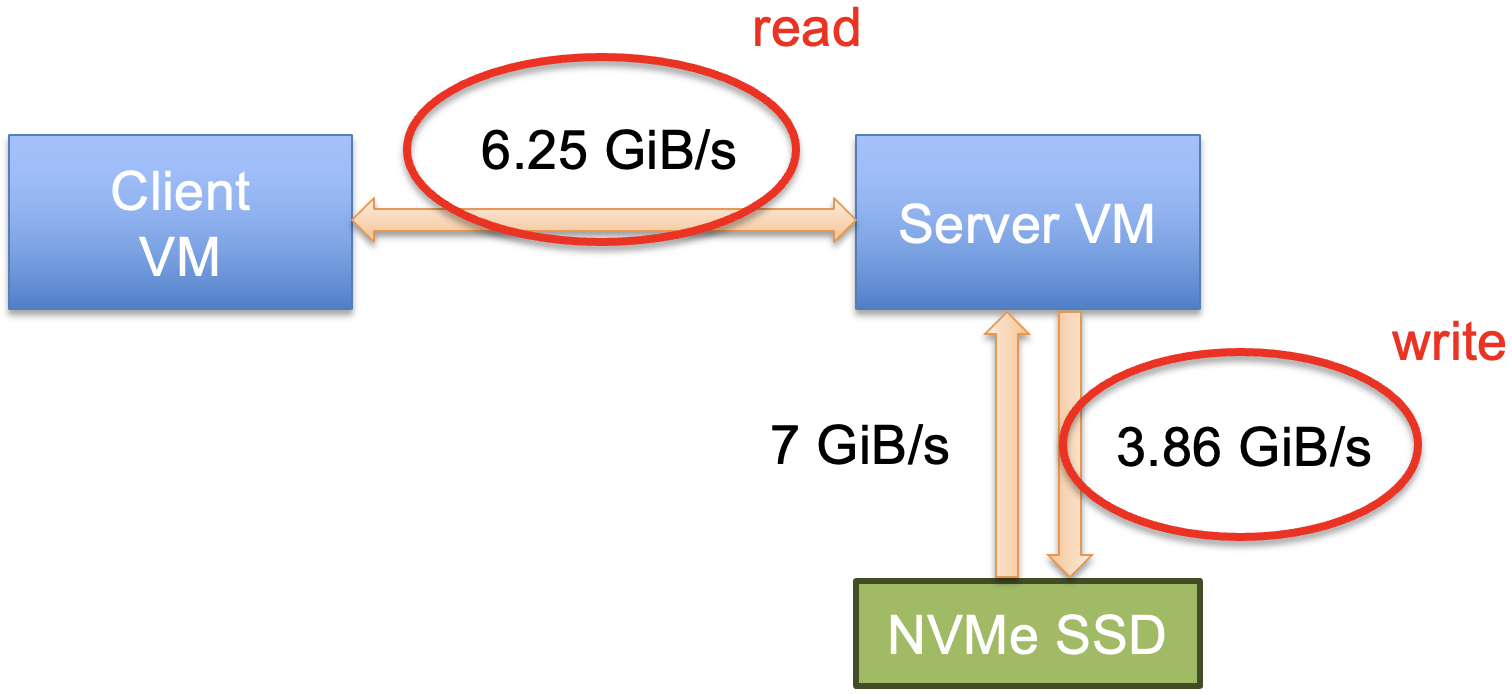}}
\caption{Ideal write and read bandwidths of a GCP VM of type n2-custom-36-153600 used as networked storage server.}
\label{fig:gcp_bottlenecks}
\end{figure}

These measurements indicated that every additional server instance employed for a DAOS, Ceph, or Lustre deployment could at best provide an additional 3.86 GiB/s for write, limited by the SSD bandwidth, and 6.25 GiB/s for read, limited by the network. This is represented in Fig. \ref{fig:gcp_bottlenecks}.

\subsection{IOR performance}

\begin{figure*}[htbp]
    \centering
    \begin{subfigure}[b]{214pt}
        \includegraphics[width=214pt,trim={0pt 16pt 0pt 0pt},clip]{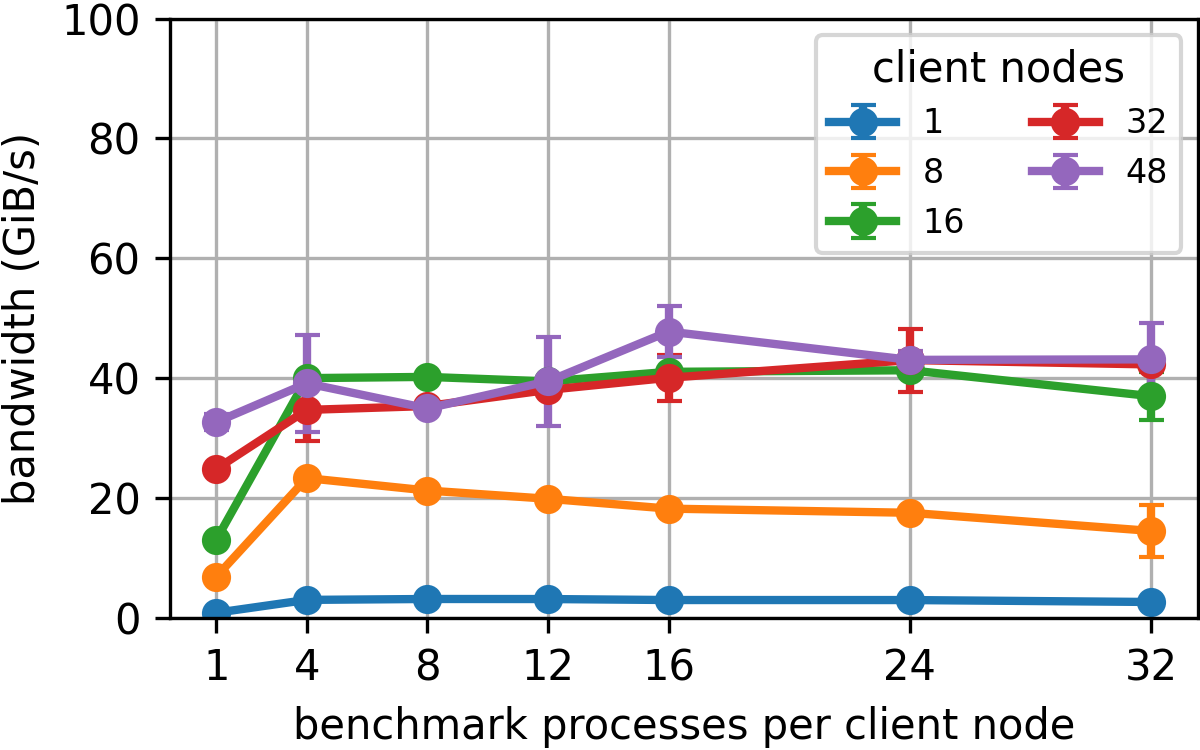}
        \caption{IOR/Lustre Write}
    \end{subfigure}
    \begin{subfigure}[b]{188pt}
        \includegraphics[width=188pt,trim={35pt 16pt 0 0pt},clip]{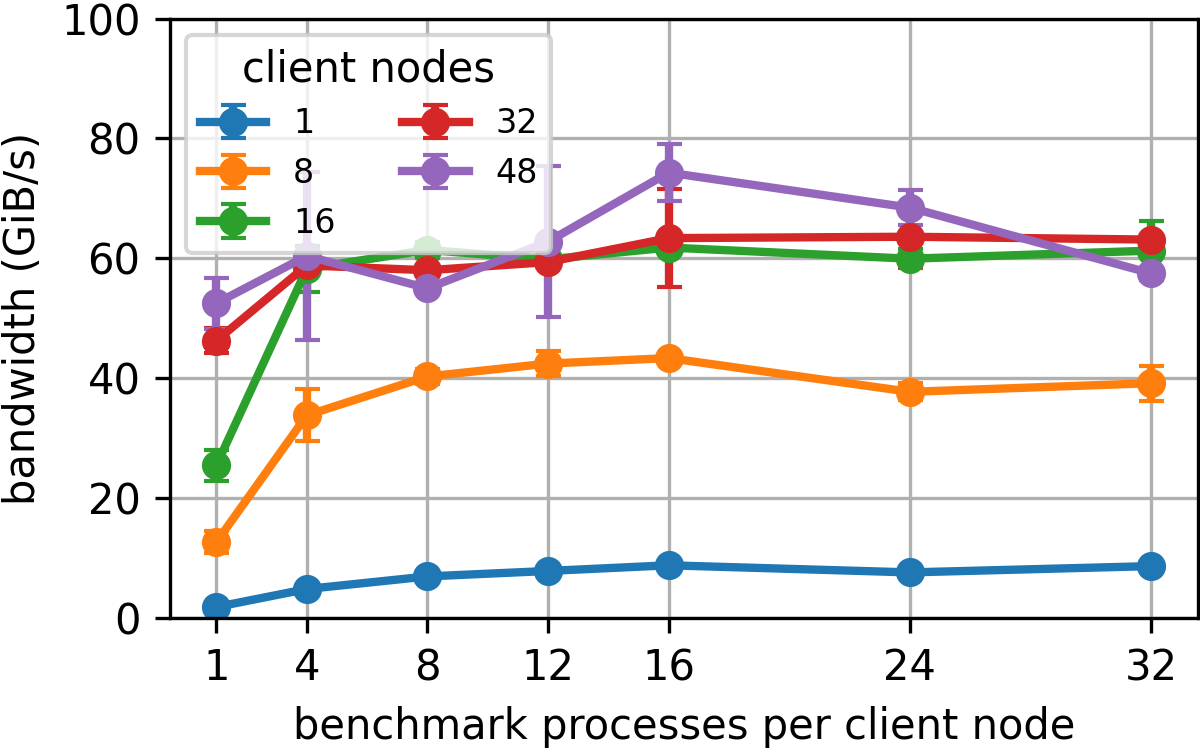}
        \caption{IOR/Lustre Read}
    \end{subfigure}
    \vskip\baselineskip
    \begin{subfigure}[b]{214pt}
        \includegraphics[width=214pt,trim={0pt 16pt 0pt 0pt},clip]{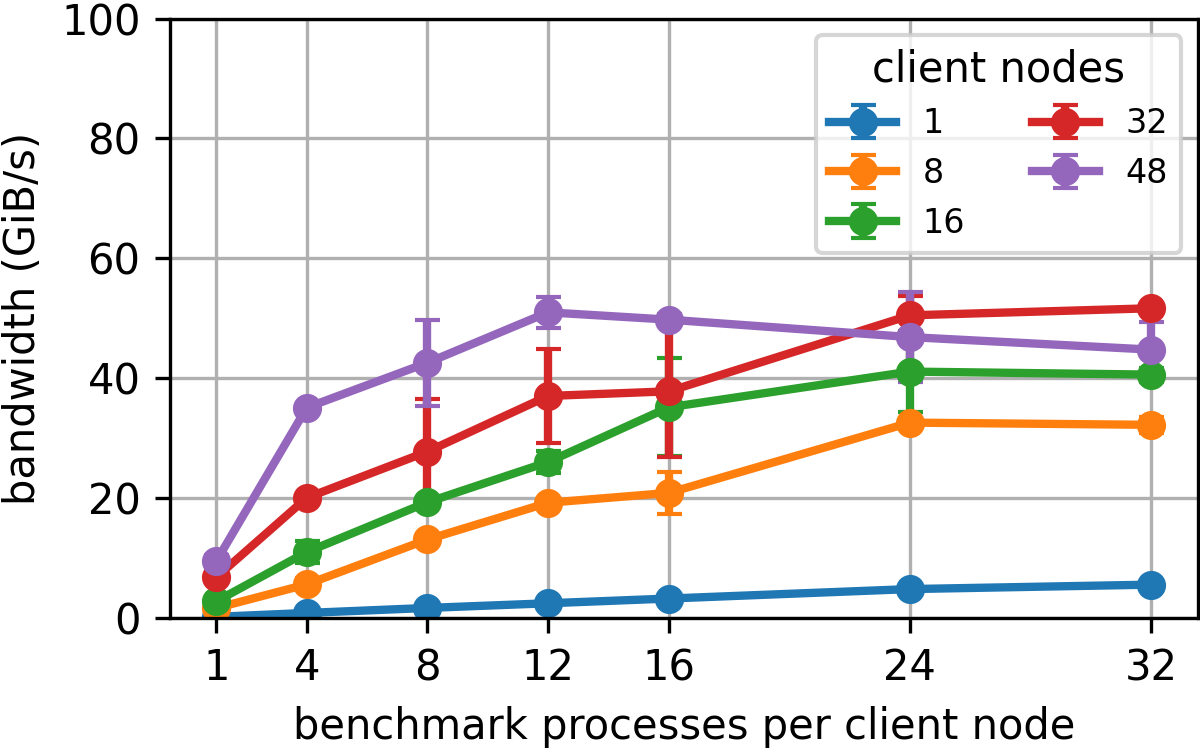}
        \caption{IOR/DAOS Write}
    \end{subfigure}
    \begin{subfigure}[b]{188pt}
        \includegraphics[width=188pt,trim={35pt 16pt 0 0pt},clip]{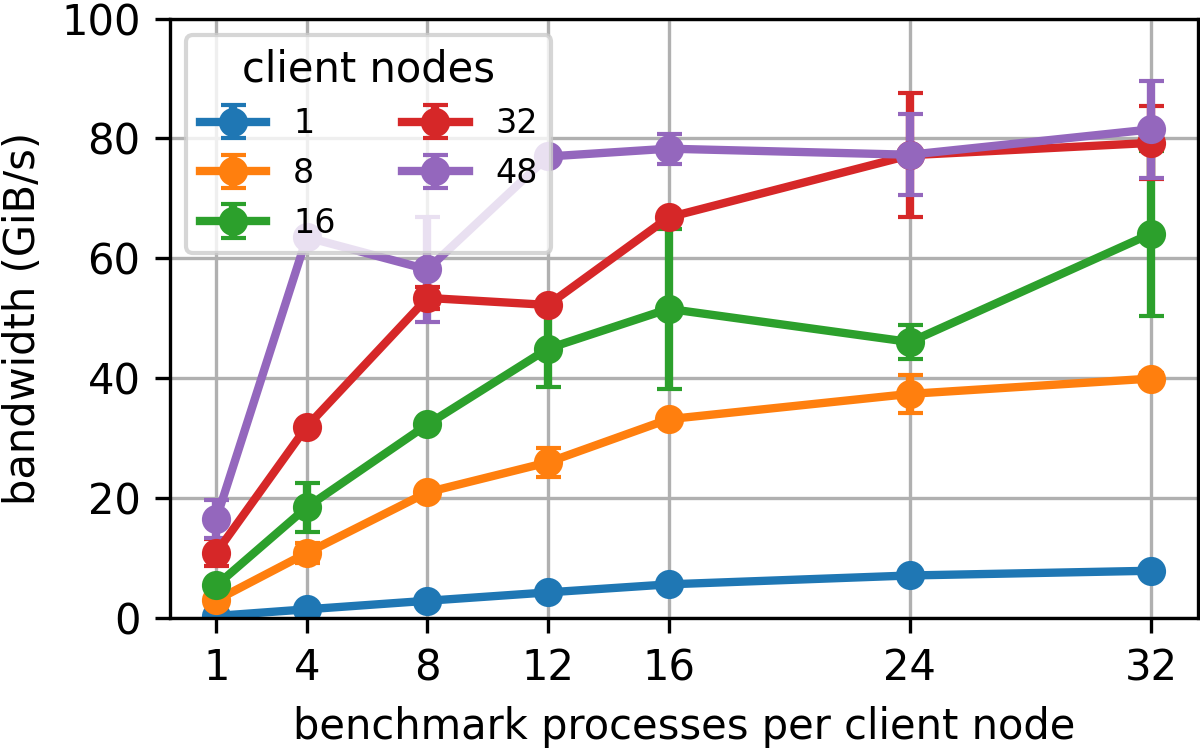}
        \caption{IOR/DAOS Read}
    \end{subfigure}
    \vskip\baselineskip
    \begin{subfigure}[b]{214pt}
        \includegraphics[width=214pt,trim={0pt 0pt 0pt 0pt},clip]{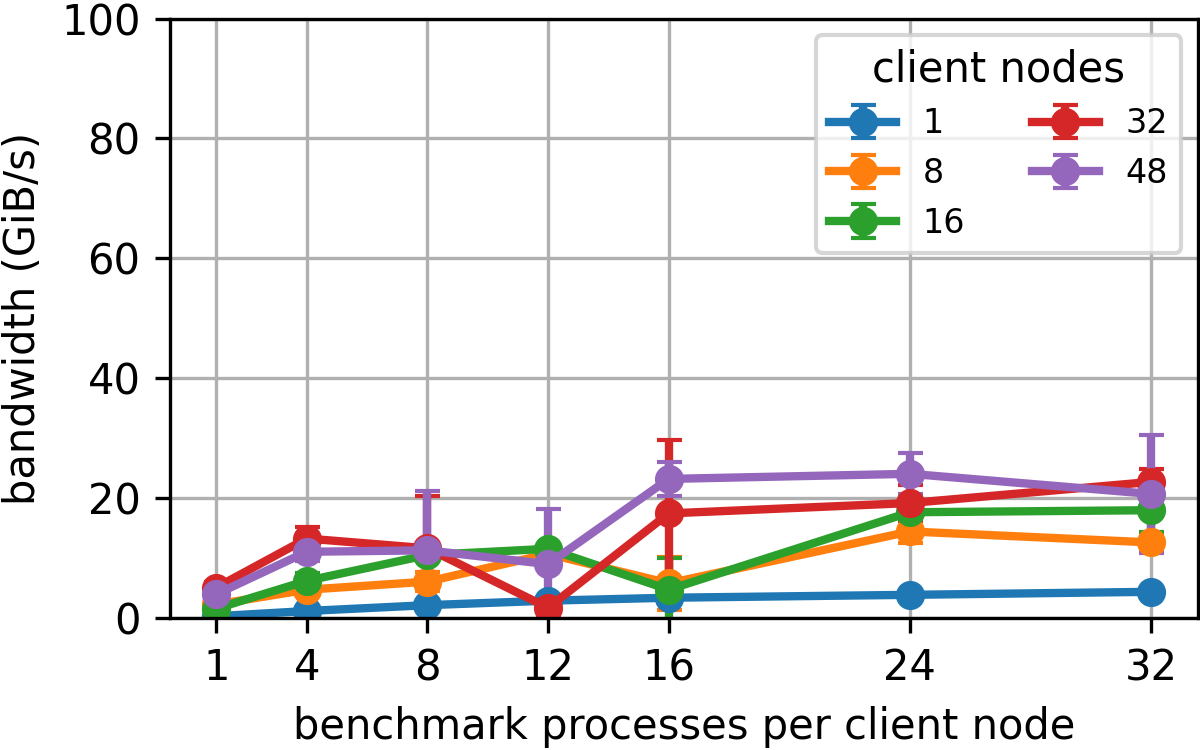}
        \caption{IOR/Ceph Write}
    \end{subfigure}
    \begin{subfigure}[b]{188pt}
        \includegraphics[width=188pt,trim={35pt 0pt 0 0pt},clip]{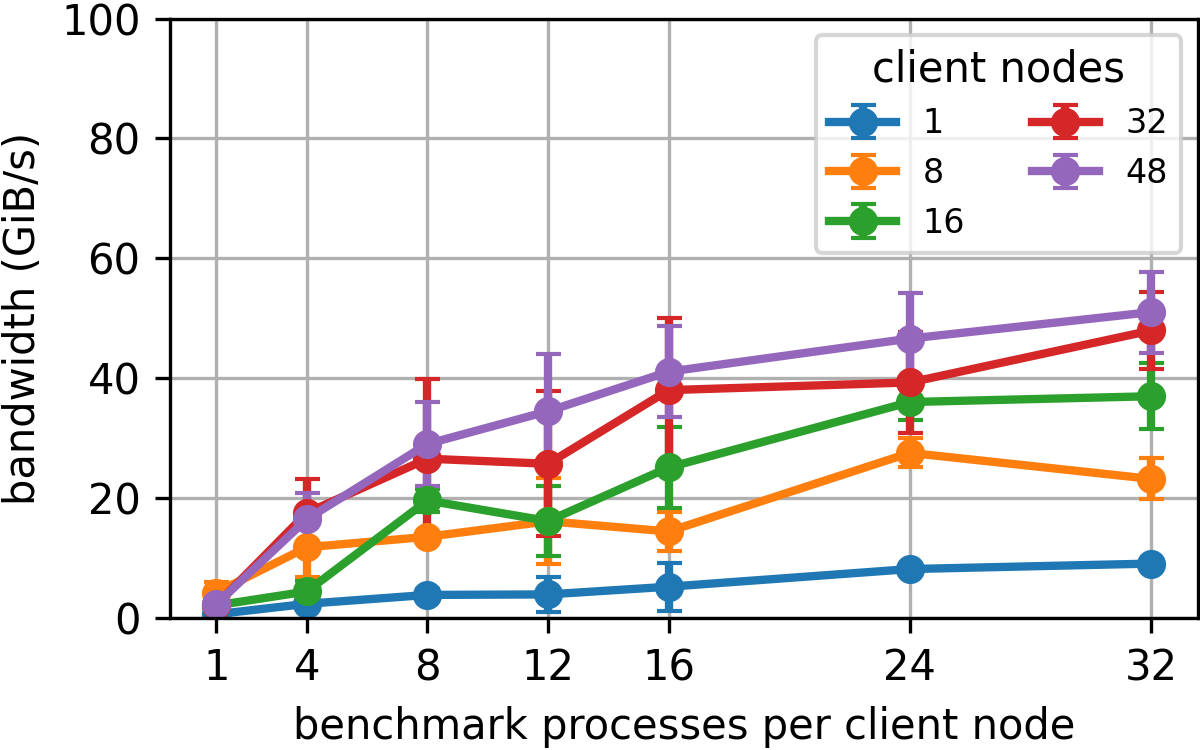}
        \caption{IOR/Ceph Read}
    \end{subfigure}
    \caption{Bandwidths for IOR runs against Lustre, DAOS, and Ceph deployments on 16 VMs (+1 for Lustre and Ceph). Every process performed 10000 x 1MiB I/O operations (100 x 1MiB for Ceph). Tests were repeated 3 times.}
    \label{fig:gcp_ior_lustre_daos_ceph_16sn_cn_cpcn}
\end{figure*}

DAOS, Ceph, and Lustre were deployed on 16 VMs (plus one for Ceph and Lustre), and the IOR benchmark was run using various amounts of client VMs and processes with the aim of determining the optimal client-to-server node ratio and process count, as per the parameter optimisation strategy, for every storage system.
In this case, in contrast to IOR runs on NEXTGenIO, every process was configured to perform 10000 I/O operations of 1 MiB each, also within a dedicated per-process file or object. Ceph objects, however, are by default limited to 128 MiB, and IOR runs on Ceph were consequently limited to 100 I/O operations per process.
Moreover, DAOS objects and Lustre files were sharded across all storage servers, whereas Ceph objects could not be sharded. The results are shown in Fig. \ref{fig:gcp_ior_lustre_daos_ceph_16sn_cn_cpcn}.

Based on the hardware bandwidth measurements earlier, 16 server VMs should ideally provide approximately 62 GiB/s for write and 100 GiB/s for read.
As shown in the first two rows of Fig. \ref{fig:gcp_ior_lustre_daos_ceph_16sn_cn_cpcn}, Lustre and DAOS reached close to these values, providing up to 50 GiB/s for write and 80 GiB/s for read.

Ceph, however, reached only a half of that write bandwidth, and up to 55 GiB/s for read.
One relevant aspect worth of scrutiny in regard to Ceph's performance, which one might think could be behind these far-from-optimal results, is the Placement Group (PG) configuration. However, in this case ---and in all following benchmark runs on Ceph--- PG auto-scaling was disabled, and several different PG counts were manually set and tested to determine the optimal value, thus leading to the PG configuration being discarded as a potential cause. For this particular test set, the optimal PG count was found to be 8192.

Instead, the observed sub-optimal performance was likely due to Ceph objects not being sharded, making it impossible for a single client VM to target all 256 OSDs, and less likely for the whole set of client processes to target all OSDs in a balanced way. Also, the small number of I/O operations issued per process prevented IOR processes from sustaining I/O to their corresponding objects for a long time period, thus likely not saturating the network nor the servers.

Some of the IOR runs on Ceph were repeated having Ceph configured with a large maximum object size ---which is discouraged--- and IOR configured to issue 10000 I/O operations per process. This option resulted in even lower write performance and was thereby discarded.
This could seem contradictory with the result in the 5th column of Fig. \ref{fig:ceph_backend_options_and_performance}, where the performance of the FDB Ceph backends was not impacted despite having Ceph configured with a large maximum object size. However, in that case, the objects were kept small, never actually approaching the maximum permitted size.

In fact, the results in Fig. \ref{fig:ceph_backend_options_and_performance} demonstrated that using multiple objects per process or even an object per I/O would result in better write performance ---presumably due to client processes targeting OSDs in a more balanced way---, but IOR did not support these arrangements.

Regarding client count optimisation, the three storage systems generally performed best in tests using 16 or 32 client VMs ---i.e. a client-to-server node ratio of 1 or 2---, and 16 to 32 benchmark processes per VM. 

The results for DAOS and Lustre demonstrated that despite using the TCP protocol, which involves the operating system in all network communications both client and server-side, high performance levels could be achieved if using large process counts in combination with a large enough transfer size, just as observed in NEXTGenIO.

\begin{figure*}[htbp]
    \centering
    \begin{subfigure}[b]{214pt}
        \includegraphics[width=214pt]{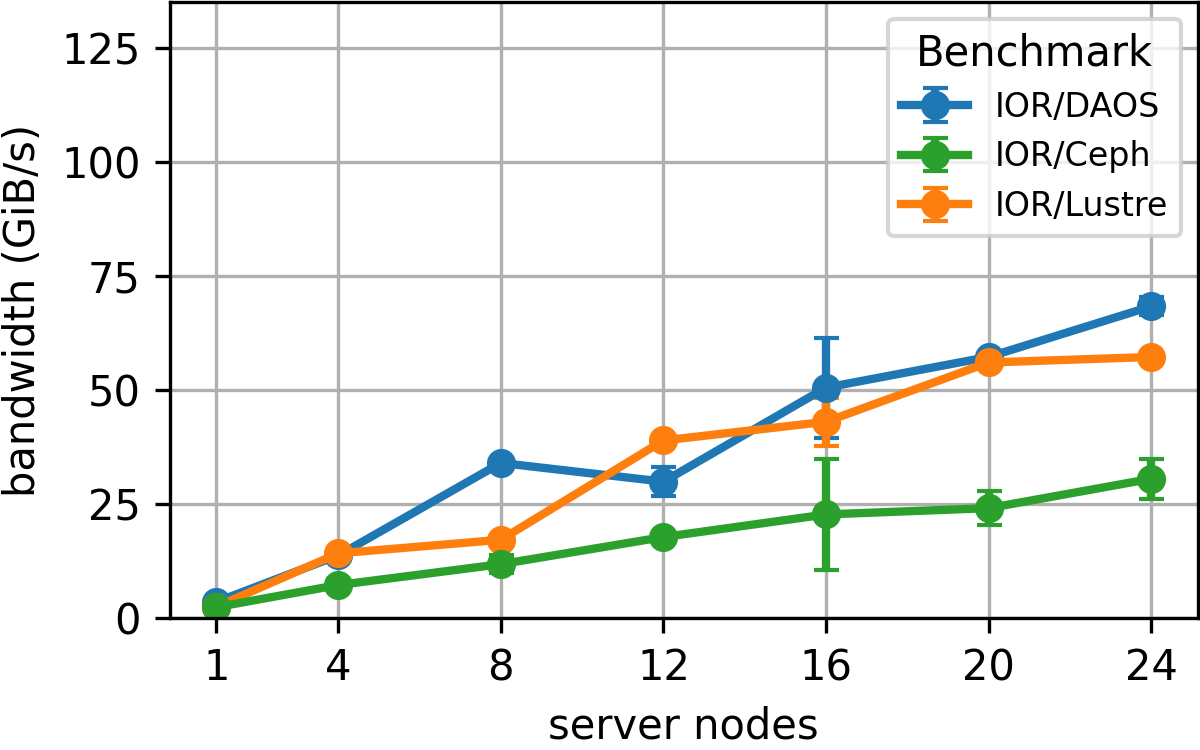}
        \caption{Write}
    \end{subfigure}
    \begin{subfigure}[b]{188pt}
        \includegraphics[width=188pt,trim={35pt 0 0 0},clip]{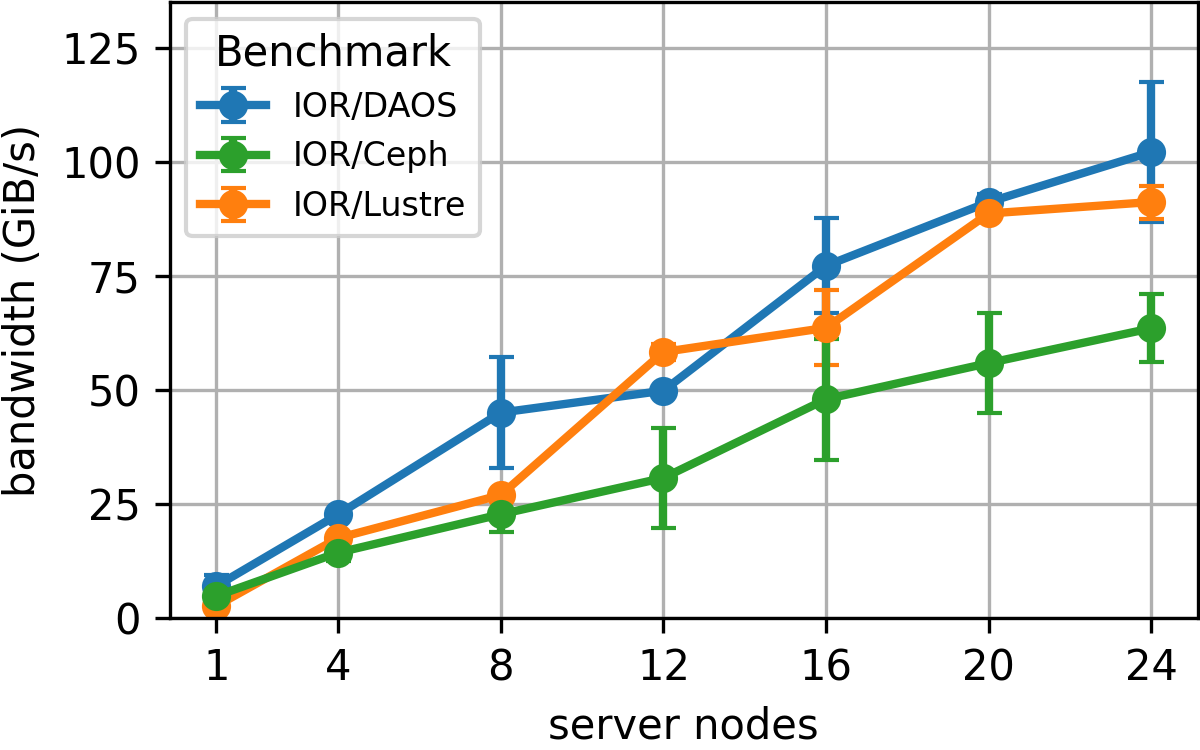}
        \caption{Read}
    \end{subfigure}
    \caption{IOR bandwidth scalability against increasingly large Lustre, Ceph, and DAOS deployments on GCP. A ratio of 2-to-1 client-to-server nodes was used for all tests, and 16 to 32 processes were run in each client node. Every process performed 10000 x 1MiB I/O operations (100 x 1 MiB for Ceph). Tests were repeated 3 times.}
    \label{fig:gcp_ior_scalability}
\end{figure*}

The client count optimisation procedure was repeated for deployments of the three storage systems on 4 server VMs (plus one for Ceph and Lustre), resulting in similar optimal client-to-server node ratios and process counts.
IOR was then run against deployments on increasing amounts of server VMs, using these optimal client counts.
The results are shown in Fig. \ref{fig:gcp_ior_scalability}.

These encouraging results demonstrated all storage systems were able to scale roughly linearly on NVMe SSDs, at least for IOR workloads.
At the largest scales tested, both DAOS and Lustre reached 80\% of the write bandwidth and 66\% of the read bandwidth the hardware could provide, whereas Ceph reached approximately a third of the write bandwidth and two thirds of the read bandwidth achieved by the other two storage systems.

The gap between the hardware bandwidths and the bandwidths for runs against DAOS or Lustre was significant, particularly for read, but this was not investigated further.
The difference in performance between Ceph and the other two was due to the same reasons discussed in the client count optimisation tests.

Despite the good scalability, all storage systems showed a slight decline beyond 16 server VMs, and this was more noticeable for Lustre than for the rest.

\subsection{FDB backend performance}

The methodology was followed again to produce scalability curves for \verb!fdb-hammer! runs against the three storage systems, using the respective FDB backends.

\begin{figure*}[htbp]
    \centering
    \begin{subfigure}[b]{214pt}
        \includegraphics[width=214pt]{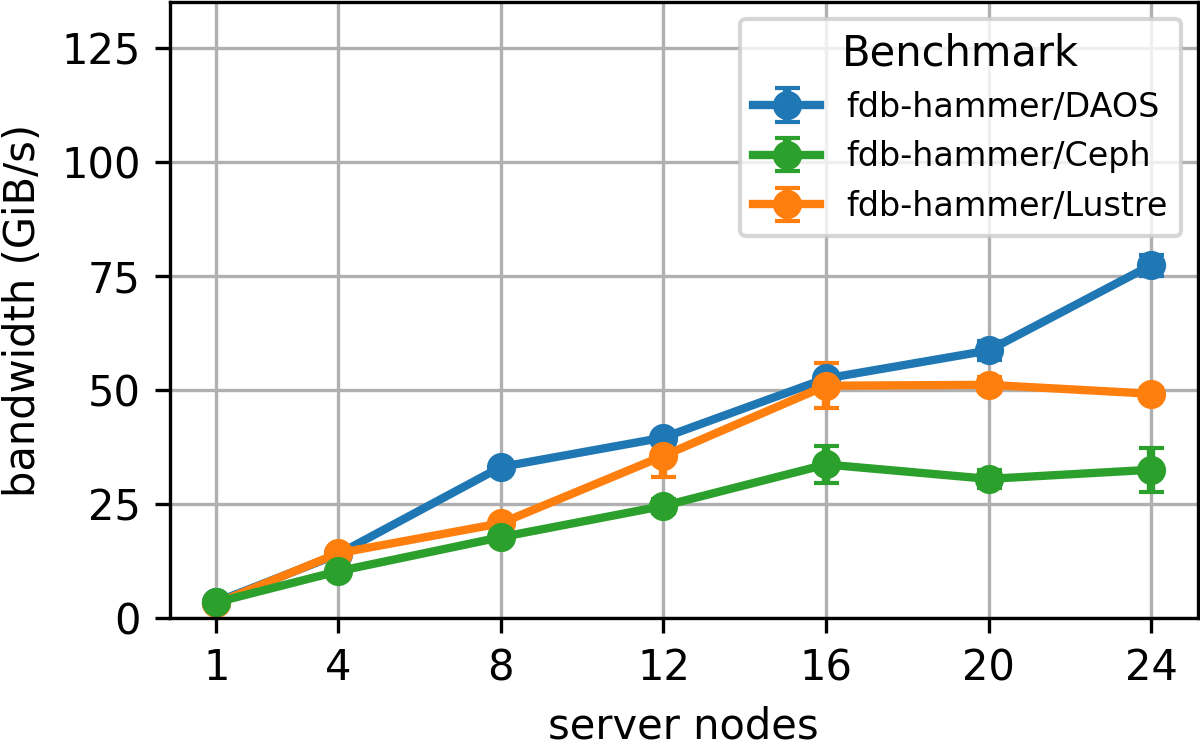}
        \caption{Write, no w+r contention}
    \end{subfigure}
    \begin{subfigure}[b]{188pt}
        \includegraphics[width=188pt,trim={35pt 0 0 0},clip]{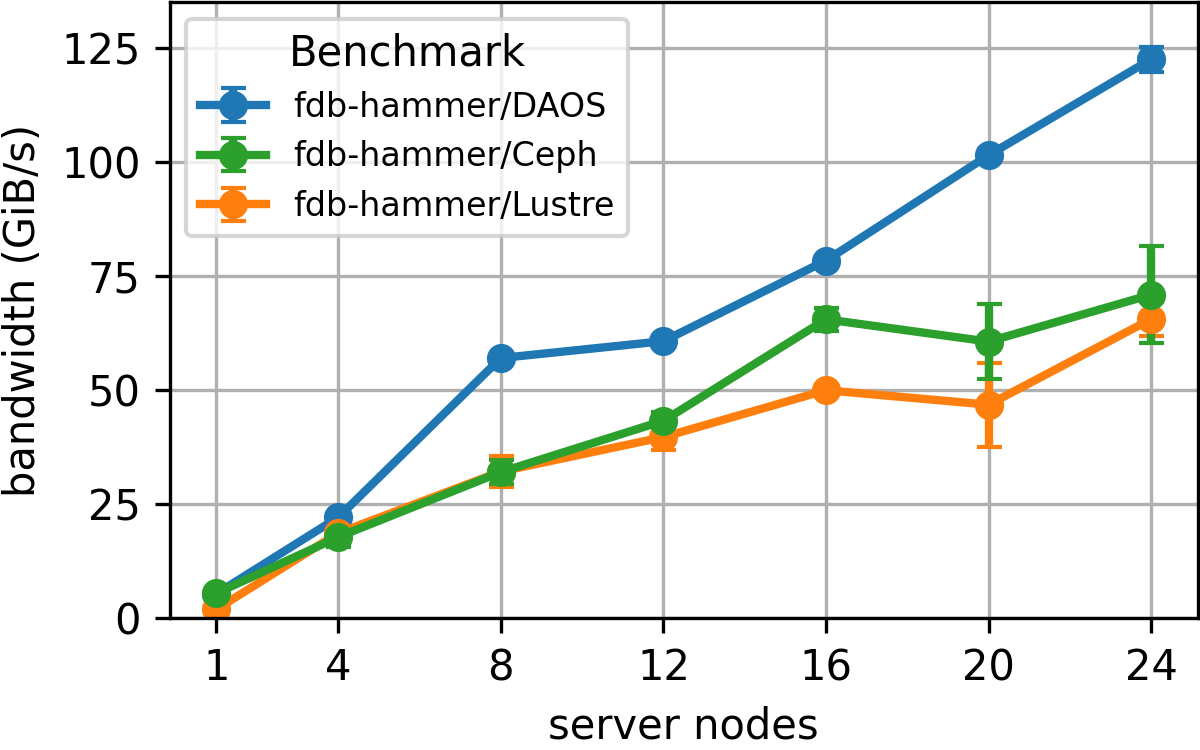}
        \caption{Read, no w+r contention}
    \end{subfigure}
    \caption{fdb-hammer bandwidth scalability, with no write+read contention, against increasingly large Lustre, Ceph, and DAOS deployments on GCP. A ratio of 2-to-1 client-to-server nodes was used for all tests, and 16 to 32 processes were run in each client node. Every process wrote and indexed (or de-referenced and read) 10000 x 1MiB weather fields. Tests were repeated 3 times.}
    \label{fig:gcp_fdbh_scalability_a}
\end{figure*}

The client count optimisation process was documented in Appendix D - Section III - B, E, and F, and the results shown in Figures 3 (g) and (h), 7, and 8 in that paper.
Similar to IOR, a client-to-server node ratio of 2 and 16 to 32 processes per node resulted in best performance.

\verb!fdb-hammer! was then run using the optimal client counts against deployments on increasing amounts of VMs. The resulting scalability curves are shown in Fig. \ref{fig:gcp_fdbh_scalability_a}.

DAOS and the respective backends scaled roughly linearly both for write and read, with bandwidths spiking above the trend for tests against 8 server VMs. The read bandwidths slightly exceeded those of previous IOR tests, just as observed in NEXTGenIO, likely due to the additional optimisations put in place in the DAOS backends.
These excellent results demonstrated that DAOS can achieve close to hardware performance at scale for complex I/O workloads such as the ECMWF's NWP I/O workloads, even if a large number of array and key-value operations are issued.
Also, these demonstrated high performance on DAOS is not an exclusive result of SCM, but can be achieved as well in systems with common NVMe SSDs.

\verb!fdb-hammer! runs on Ceph scaled linearly both for write and read up to 16 server VMs, reaching two thirds of the hardware write bandwidth and close to hardware read bandwidths, but plateaued at larger server counts.
The runs against up to 16 server VMs performed notably better than corresponding IOR runs, presumably due to the object per \verb!archive()! approach of the FDB Ceph backends ---as opposed to the object per process approach in IOR---, which resulted in a more balanced distribution of objects across OSDs and in turn allowed for longer benchmark runs with 10000 I/O operations per process.
Beyond 16 server VMs, write bandwidths plateaued, and read bandwidths seemed to continue scaling in line with the trend excluding a spike at 16 VMs.

For Lustre, the \verb!fdb-hammer! write bandwidths scaled linearly up to 16 server VMs, reaching close to DAOS and hardware bandwidths, but markedly plateaued at 50 GiB/s for larger server counts.
This plateauing was an unexpected result, on one hand because the POSIX I/O FDB backends are optimised for write performance, and on the other hand because equivalent tests in NEXTGenIO showed better write scalability.

Read bandwidths on Lustre scaled almost linearly throughout, with a slight decline beyond 12 server VMs and a spike under the trend for 20 server VMs, and were notably lower than those observed in corresponding IOR tests.
This gap relative to IOR was slightly surprising as \verb!fdb-hammer! produces a relatively easy read workload, involving only a few files per process.
A similar unexpected gap had been observed for runs in NEXTGenIO, and this strongly suggested that the lower \verb!fdb-hammer! performance was likely unrelated to the hardware of either system and was rather caused by a potential software or configuration issue at the storage system level or in the POSIX I/O FDB backends, such as Lustre not handling well the relatively large number of \verb!read! operations, or an undiscovered issue in the logic for merging \verb!DataHandle!s.

\begin{figure*}[htbp]
    \centering
    \begin{subfigure}[b]{212pt}
        \includegraphics[width=212pt]{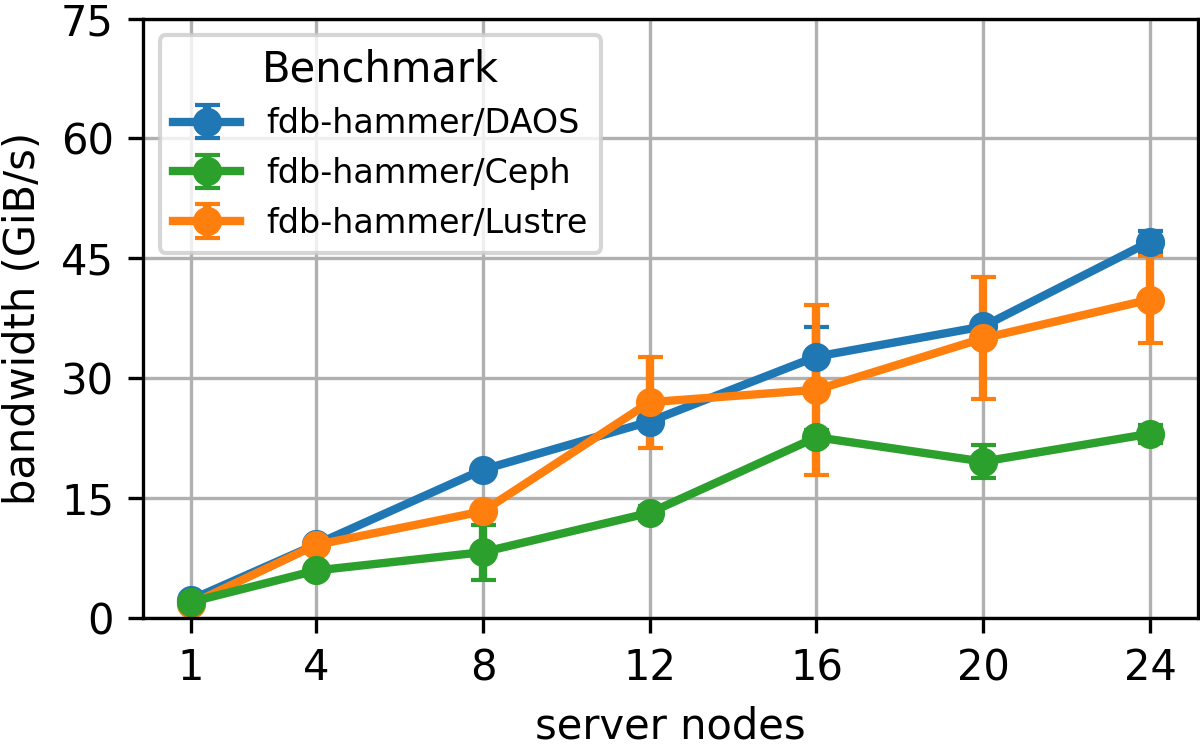}
        \caption{Write, w+r contention}
    \end{subfigure}
    \begin{subfigure}[b]{190pt}
        \includegraphics[width=190pt,trim={30pt 0 0 0},clip]{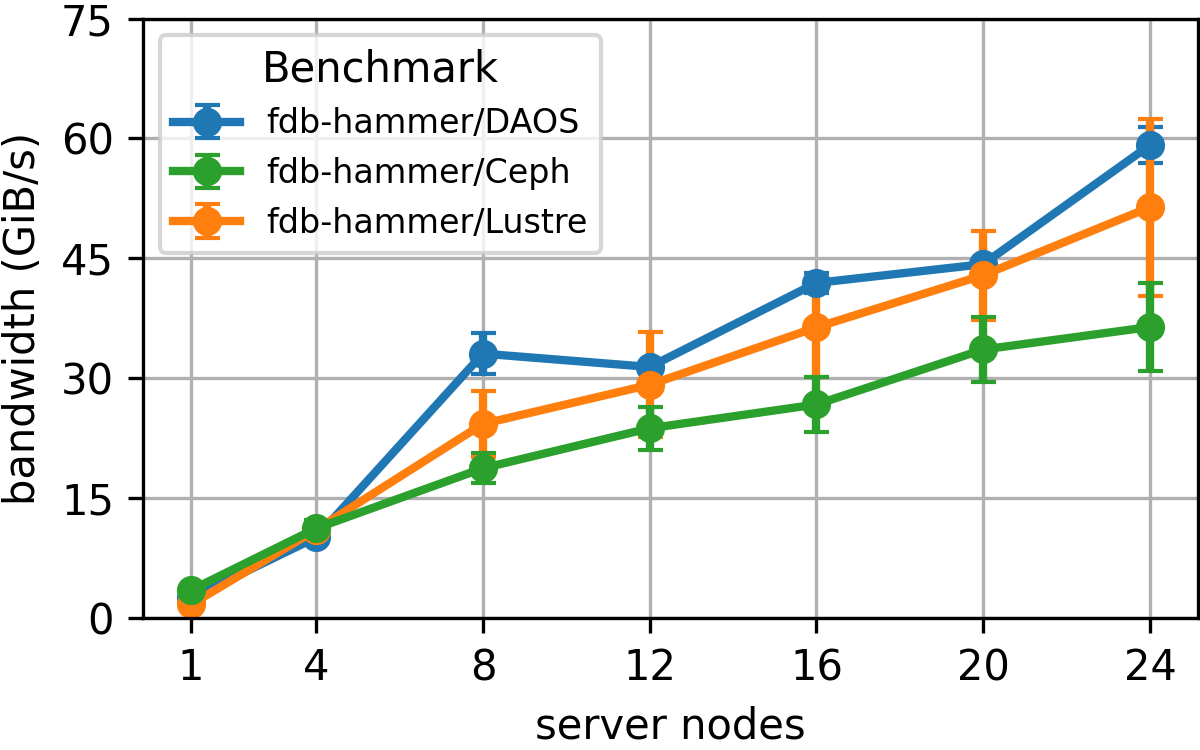}
        \caption{Read, w+r contention}
    \end{subfigure}
    \caption{fdb-hammer bandwidth scalability, with write+read contention, against increasingly large Lustre, Ceph, and DAOS deployments on GCP. A ratio of 2-to-1 client-to-server nodes was used for all tests, and 16 to 32 processes were run in each client node. Every process wrote and indexed (or de-referenced and read) 10000 x 1MiB weather fields. Tests were repeated 3 times.}
    \label{fig:gcp_fdb_scalability_c}
\end{figure*}

Scalability curves for \verb!fdb-hammer! runs with \textit{write+read contention} were also produced. One half of the client nodes ran writer processes, and the other half ran reader processes. These are shown in Fig. \ref{fig:gcp_fdb_scalability_c}.

Just as observed previously in NEXTGenIO, DAOS and the respective backends scaled roughly linearly both for write and read ---omitting a few spikes---, thus were not significantly impacted in terms of scalability by operational contention, which was fully captured by \verb!fdb-hammer! in this case.
In terms of performance, writing performed better than initially expected --- runs with \textit{no write+read contention}, in Fig. \ref{fig:gcp_fdbh_scalability_a}, reached up to 75 GiB/s using 24 server and 48 client VMs, and therefore the runs here, which had writers running on only a half of the client VMs presumably being served by a half of the server resources, were expected to reach up to a half of these bandwidths.
Instead, writers exceeded these bandwidths, reaching up to 45 GiB/s, and this was due to readers ending sooner and releasing server resources which immediately proceeded to serve writers.

Ceph also scaled roughly linearly, although with a progressive decline in read performance.
The bandwidths achieved were significantly lower than with DAOS, but this was within expectations based on previous tests with \textit{no write+read contention}.
The results here demonstrated that, just as DAOS, Ceph's performance was not significantly impacted by operational contention levels, although the performance would very likely plateau at larger server VM counts based on tests with \textit{no write+read contention}.
These results were moderately encouraging. Although Ceph does not provide optimal performance, it ought not to be discarded as an option for high-performance storage for small to medium-scale applications.

\verb!fdb-hammer! runs on Lustre scaled nearly linearly, reaching bandwidths similar to DAOS, although the performance would have likely plateaued or declined at larger server VM counts based on tests with \textit{no write+read contention}, and would have further deteriorated if \verb!fdb-hammer! on Lustre had more precisely captured the operational contention and read transposition. 
Nevertheless, the positive results here contrasted with equivalent tests on NEXTGenIO, where both the write and read bandwidths on Lustre halved under contention.
This reinforced the idea that Lustre did not properly exploit the specialised hardware in NEXTGenIO for contentious workloads, whereas it did for the more commonplace hardware in GCP.

\begin{figure*}[htbp]
    \centering
    \begin{subfigure}[b]{214pt}
        \includegraphics[width=214pt,trim={20pt 215pt 26pt 25pt},clip]{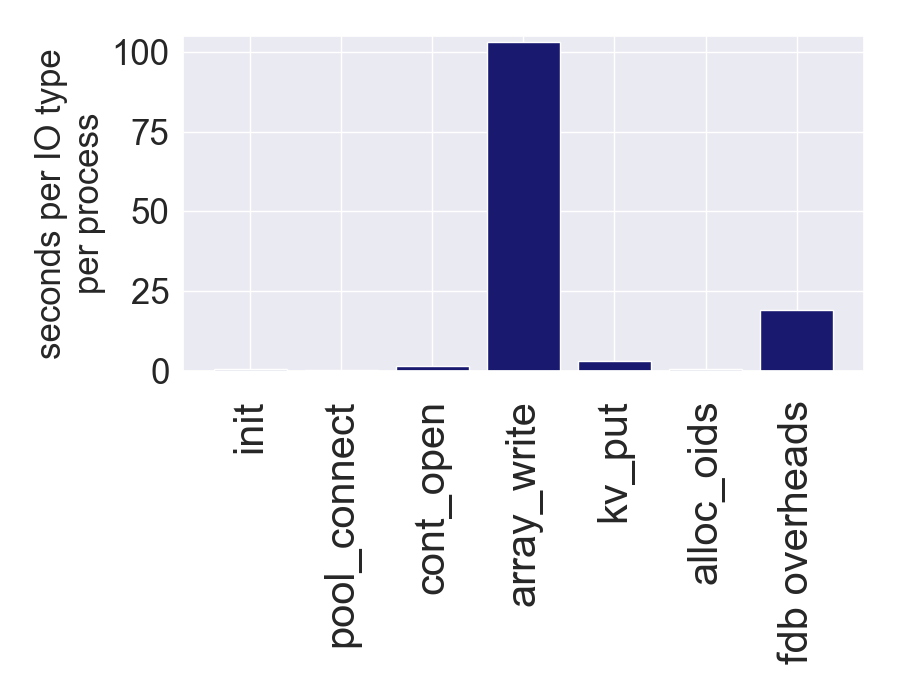}
        \caption{Writers, no w+r contention. Average per-\phantom{ } pro\-cess wall-clock time: 108.5s.}
    \end{subfigure}
    \begin{subfigure}[b]{188pt}
        \centering
        \includegraphics[width=177pt,trim={125pt 215pt 26pt 25pt},clip]{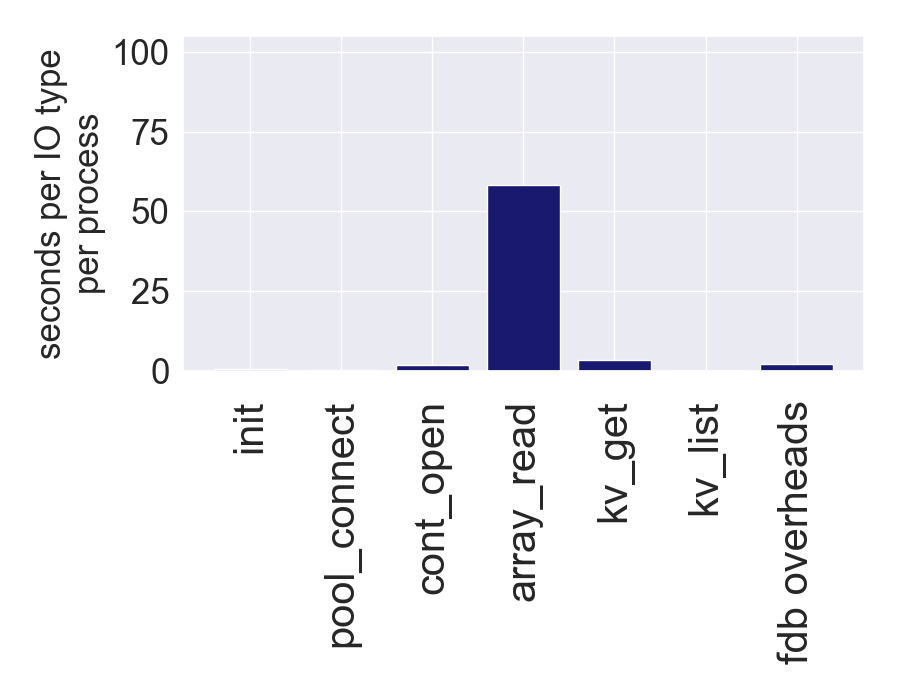}
        \caption{Readers, no w+r contention. Average per-process wall-clock time: 65.8s.}
    \end{subfigure}
    \vskip\baselineskip
    \begin{subfigure}[b]{214pt}
        \includegraphics[width=214pt,trim={20pt 25pt 26pt 25pt},clip]{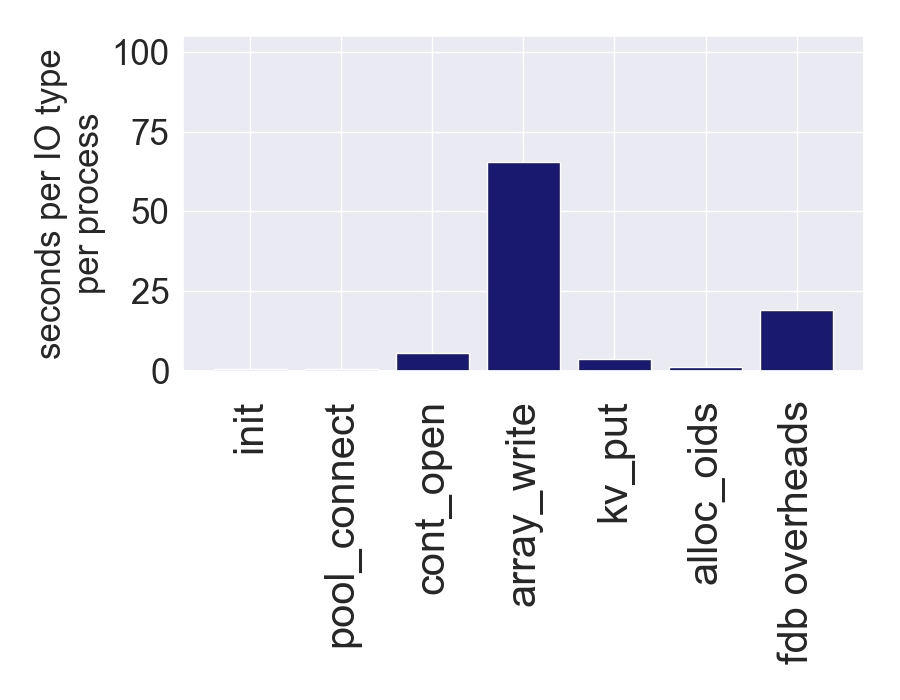}
        \caption{Writers, w+r contention. Average per-\phantom{ } process wall-clock time: 95.4s.}
    \end{subfigure}
    \begin{subfigure}[b]{188pt}
        \centering
        \includegraphics[width=177pt,trim={125pt 25pt 26pt 25pt},clip]{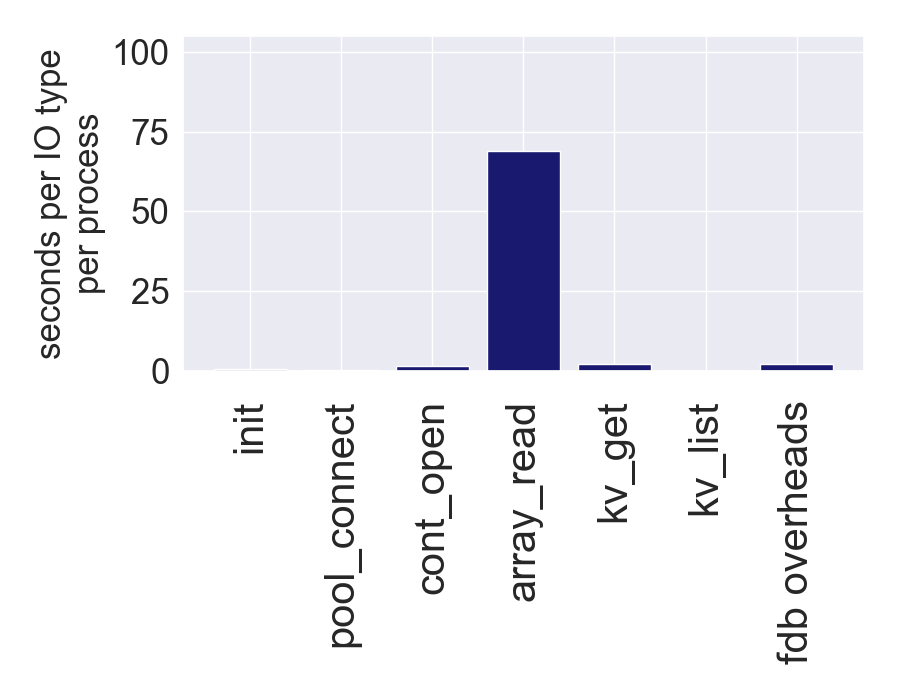}
        \caption{Readers, w+r contention. Average per-process wall-clock time: 75.2s.}
    \end{subfigure}
    \caption{Profiling results for fdb-hammer/DAOS runs without (top row) and with (bottom row) write+read contention, using 20 server and 40 client nodes; 24 processes per client node. Every process wrote and indexed (or de-referenced and read) 10000 x 1MiB weather fields.}
    \label{fig:gcp_fdbh_daos_profiling}
\end{figure*}

\begin{figure*}[htbp]
    \centering
    \begin{subfigure}[b]{178pt}
        \centering
        \includegraphics[width=152pt,trim={14pt 235pt 26pt 25pt},clip]{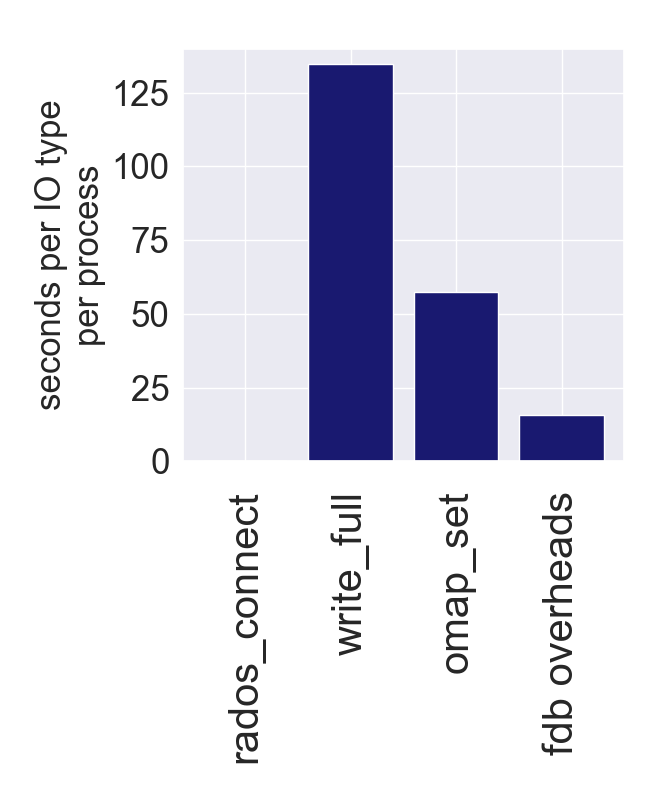}
        \caption{Writers, no w+r contention. Avg. per-pro\-cess wall-clock time: 208.6s.}
    \end{subfigure}
    \begin{subfigure}[b]{224pt}
        \centering
        \hspace{-10pt}\includegraphics[width=183pt,trim={125pt 235pt 0pt 25pt},clip]{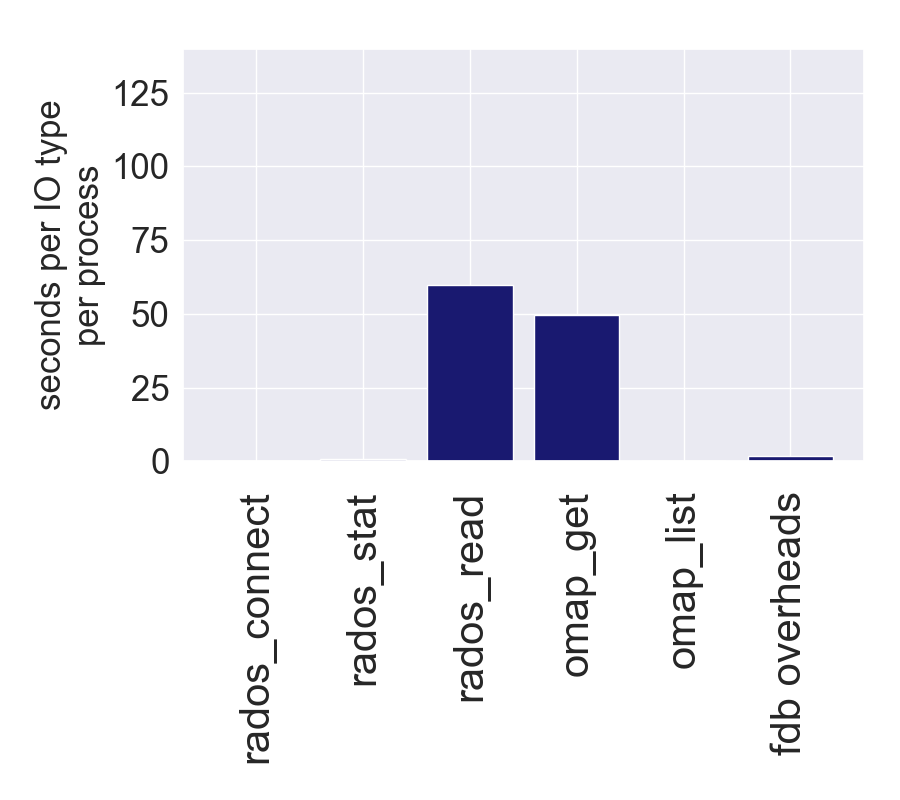}
        \caption{Readers, no w+r contention. Average per-process wall-clock time: 112.3s.}
    \end{subfigure}
    \vskip\baselineskip
    \begin{subfigure}[b]{178pt}
        \centering
        \includegraphics[width=152pt,trim={14pt 25pt 26pt 25pt},clip]{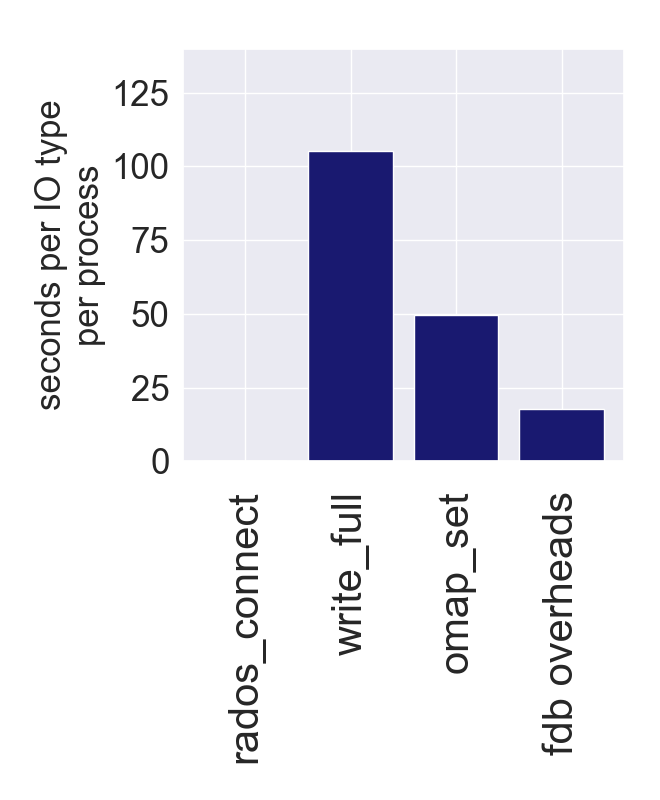}
        \caption{Writers, w+r contention. Average per-process wall-clock time: 172.7s.}
    \end{subfigure}
    \begin{subfigure}[b]{224pt}
        \centering
        \hspace{-10pt}\includegraphics[width=183pt,trim={125pt 25pt 0pt 25pt},clip]{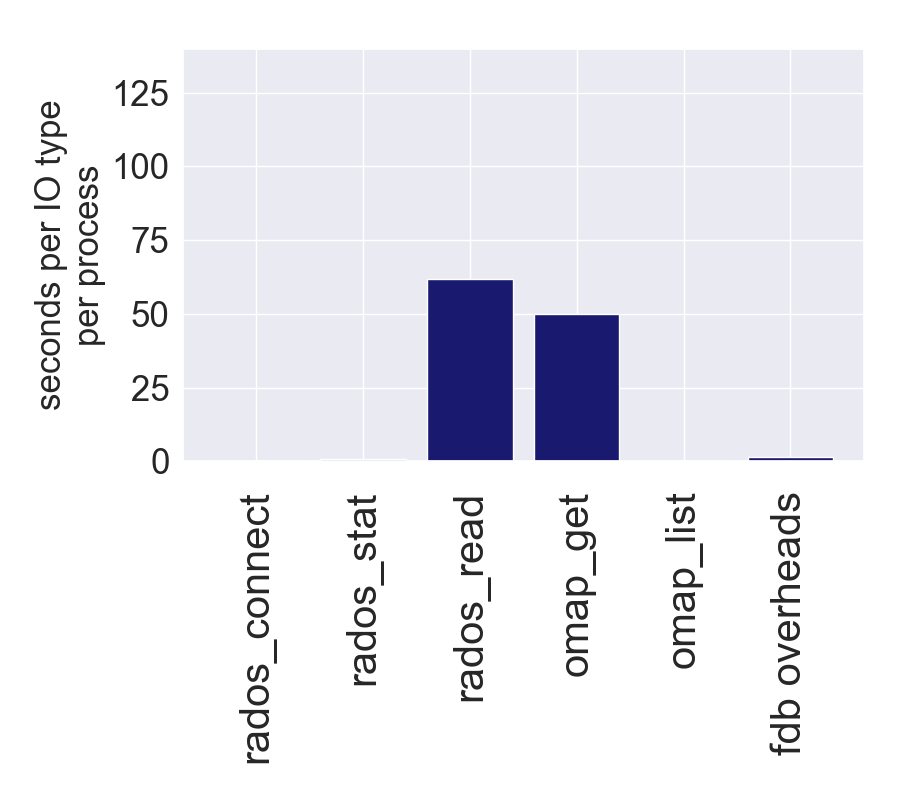}
        \caption{Readers, w+r contention. Average per-process wall-clock time: 114.5s.}
    \end{subfigure}
    \caption{Profiling results for fdb-hammer/Ceph runs without (top row) and with (bottom row) write+read contention, using 20 server and 40 client nodes; 24 processes per client node. Every process wrote and indexed (or de-referenced and read) 10000 x 1MiB weather fields.}
    \label{fig:gcp_fdbh_ceph_profiling}
\end{figure*}

\begin{figure*}[htbp]
    \centering
    \begin{subfigure}[b]{224pt}
        \includegraphics[width=224pt,trim={26pt 265pt 26pt 25pt},clip]{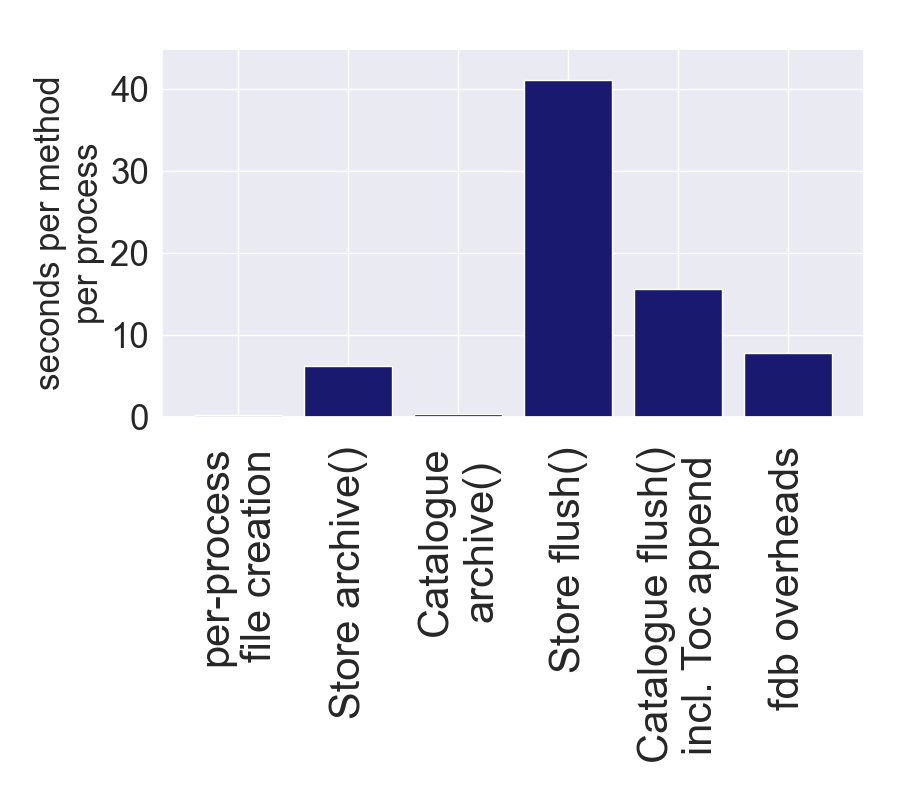}
        \caption{Writers, no w+r contention. Average per-\phantom{ } pro\-cess wall-clock time: 71.8s.}
    \end{subfigure}
    \begin{subfigure}[b]{178pt}
        \centering
        \includegraphics[width=128pt,trim={125pt 258pt 0pt 25pt},clip]{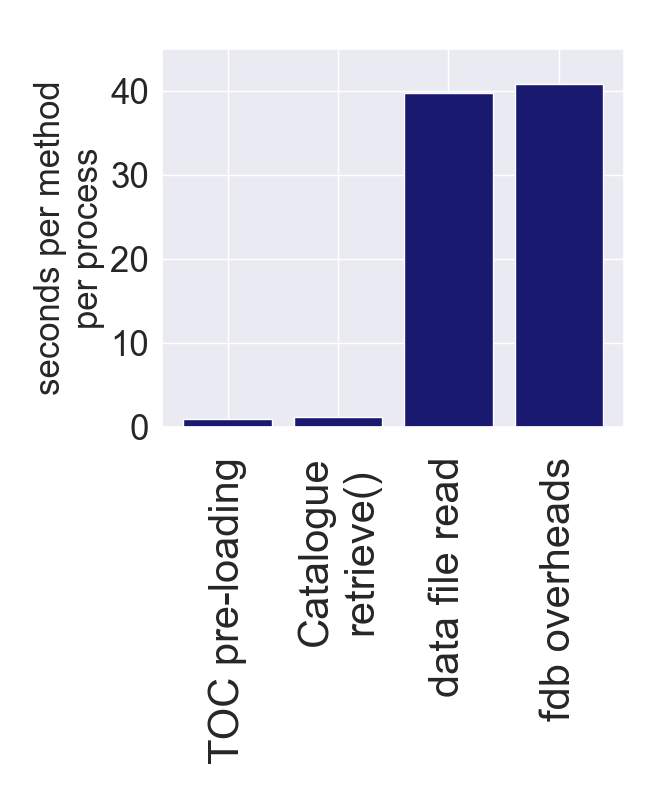}
        \caption{Readers, no w+r contention. Average per-process wall-clock time: 82.8s.}
    \end{subfigure}
    \vskip\baselineskip
    \begin{subfigure}[b]{224pt}
        \includegraphics[width=224pt,trim={26pt 25pt 26pt 25pt},clip]{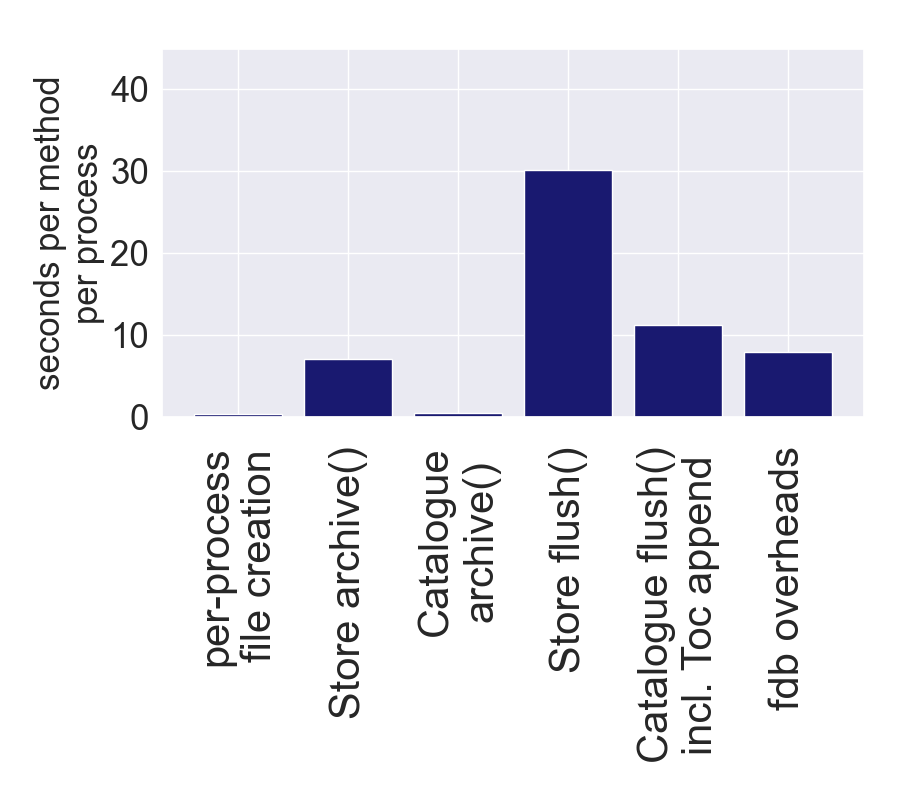}
        \caption{Writers, w+r contention. Average per-pro-\phantom{ } cess wall-clock time: 57.4s.}
    \end{subfigure}
    \begin{subfigure}[b]{178pt}
        \centering
        \includegraphics[width=128pt,trim={125pt 10pt 0pt 25pt},clip]{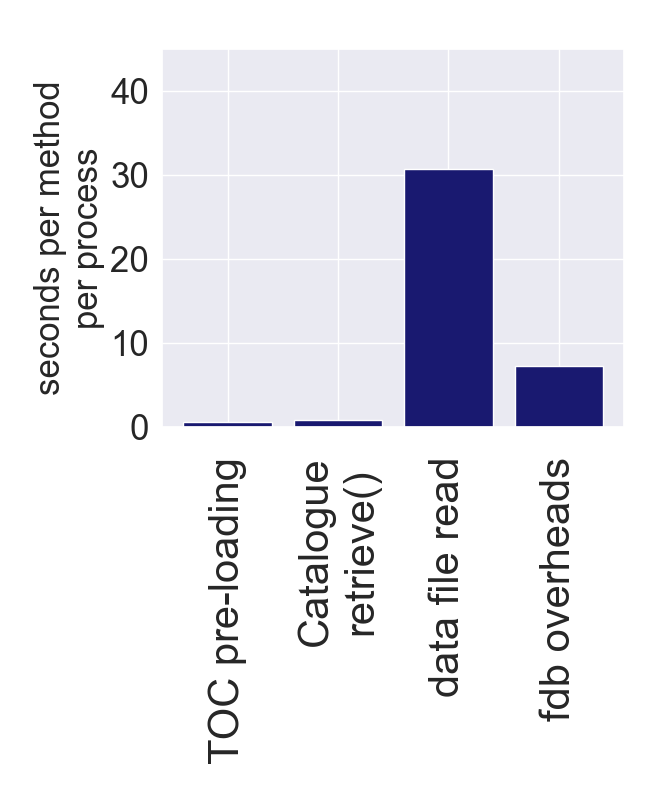}
        \caption{Readers, w+r contention. Average per-process wall-clock time: 39.6s.}
    \end{subfigure}
    \caption{Profiling results for fdb-hammer/Lustre runs without (top row) and with (bottom row) write+read contention, using 20 server and 40 client nodes; 12 processes per client node. Every process wrote and indexed (or de-referenced and read) 10000 x 1MiB weather fields.}
    \label{fig:gcp_fdbh_lustre_profiling}
\end{figure*}

To gain further insight on the behaviour of the storage systems tested, and particularly on why Ceph and Lustre performed and scaled worse than DAOS, some of the \verb!fdb-hammer! runs at scale ---with and without write-read contention--- were profiled and the results summarised in Fig. \ref{fig:gcp_fdbh_daos_profiling}, \ref{fig:gcp_fdbh_ceph_profiling}, and \ref{fig:gcp_fdbh_lustre_profiling}.

For runs against DAOS, in Fig. \ref{fig:gcp_fdbh_daos_profiling}, most of the time was spent on array write and read operations, which permitted bandwidths to reach close to hardware limits.
In contrast to equivalent runs on NEXTGenIO, a nearly negligible amount of time was spent on key-value \verb!get!s and \verb!put!s, likely due to the fact that DAOS deployments in GCP exploited DRAM for storage of metadata and small I/O operations such as index key-value \verb!get!s and \verb!put!s, whereas deployments in NEXTGenIO used SCM.
DAOS's performance might be lower or plateau at larger scales if enabling the feature to persist metadata and small I/O into NVMe SSDs, although previous analysis has shown that, due to its write-ahead-log (WAL) approach, the impact of this feature should only be noticeable for extremely metadata-operation-intensive workloads\cite{Hennecke2023DAOSResults}.

For runs with \textit{write+read contention} on DAOS, the portion of time spent on array write operations ---in Fig. \ref{fig:gcp_fdbh_daos_profiling} (c)--- was significantly smaller than for runs without contention ---in Fig. \ref{fig:gcp_fdbh_daos_profiling} (a)---, and this was again due to the read phase ending sooner and releasing server resources which immediately proceeded to handle pending writes.
This occurred to some extent as well for runs against Ceph and Lustre, but occurred to a much lesser extent for equivalent runs in NEXTGenIO as the write and read performance levels were closer to each other in that system, resulting in closer end times for the write and read phases.

For runs on Ceph, summarised in Fig. \ref{fig:gcp_fdbh_ceph_profiling}, most of the time was similarly spent on object writes and reads, although a more significant portion was spent on Omap \verb!set!s and \verb!get!s, likely due to Ceph being configured to persist all metadata and Omaps into NVMe SSDs instead of holding these in DRAM.

Smaller-scale runs on Ceph ---the profiling results of which have been omitted in favour of conciseness--- spent a less significant portion of time on object \verb!write! operations, which indicated that the plateauing observed in Fig. \ref{fig:gcp_fdbh_scalability_a} was caused by such \verb!write! operations rather than by Omap \verb!set!s and \verb!get!s.
The Omap overheads, instead, were equally significant in smaller-scale runs, and hence were the main contributor to the consistently inferior performance levels reached by Ceph in the experiments in Fig. \ref{fig:gcp_fdbh_scalability_a} and \ref{fig:gcp_fdb_scalability_c}.

This suggested that Ceph could potentially have reached performance levels as high DAOS at small scales if Ceph had been configured to hold Omaps in DRAM,
although even in that case, and even in the best-case scenario that the time spent on Omap operations on DRAM were negligible,
Ceph would have likely suffered a scalability decline beyond 16 server VMs due to object writes ---which are persisted on NVMe SSDs both by Ceph and DAOS--- being slow at scale.

The profiling for runs with \textit{write+read contention}, in Fig. \ref{fig:gcp_fdbh_ceph_profiling} (c) and (d), did not significantly differ, except for the slightly lower portion of time spent on object write operations --- again due to the server resources that served readers being released earlier.
This demonstrated that, like DAOS, Ceph is not sensitive to contention, at least for the type of I/O workload tested here.

Profiling results for runs on Lustre are shown in Fig. \ref{fig:gcp_fdbh_lustre_profiling}. 
Lower client counts had to be used on Lustre for optimal performance, which resulted in smaller I/O workloads being issued than in previous tests, hence the wall-clock times shown for Lustre runs cannot be directly compared to previous results.

The results in (a) and (b) correspond to runs with \textit{no write+read} contention against 20 server VMs, for which both write and read bandwidths plateaued or diverged from DAOS, as shown in Fig. \ref{fig:gcp_fdbh_scalability_a}.
These results should therefore reflect any overheads causing the performance deterioration on Lustre.

For write, in (a), the proportions of time spent on the different methods were reasonable and within expectations, with most of the time being spent on writing ---in Store \verb!archive()! and the first part of Catalogue \verb!flush()!--- and waiting for persistence --- in Store \verb!flush()! and the last part of Catalogue \verb!flush()!.
However, profiling results for runs at smaller scale, not included here, showed a less prominent portion of time being spent on Catalogue \verb!flush()!, and the write scalability decline was therefore attributed to Lustre not handling well the large number of small per-process file \verb!write!s issued by that method.
If Lustre had been configured to hold metadata and small I/O operations in DRAM, as done for the DAOS system, this decline might have been alleviated.

Compared to equivalent runs in NEXTGenIO, the distribution of time over the different methods was similar, except that a larger portion of time was spent on Store \verb!flush()! relative to the other methods. This was likely due to the different characteristics of the storage devices, with NVMe SSDs being slower to persist the data than SCM.

For read, in (b), two aspects stood out. The first one was the large amount of time spent on data file reads.
In the write phase, approximately 45 seconds were spent on bulk data writes ---in Store \verb!archive()! and \verb!flush()!---, and in the read phase 40 seconds were spent on the bulk data reads.
However, the bulk reading should have been ideally twice as fast as the writing, based on the profiling of \verb!array_write!s and \verb!array_reads! for runs against DAOS.
A similar slowness was observed for equivalent runs in NEXTGenIO, although the slowness in that system was more marked --- double the time was spent on bulk reading relative to bulk writing. This difference was again likely due to the different characteristics of the hardware at hand --- the hardware read bandwidths in GCP were higher than in NEXTGenIO, relative to hardware write bandwidths.
This slowness was potentially due to inefficiencies in \verb!DataHandle! merging, or due to Lustre not handling well the relatively large number ---albeit not as large as in operational runs--- of \verb!read! operations issued per process.

The other aspect that stood out were the very substantial FDB overheads. That time was spent mostly in the \verb!axis()! method of the POSIX I/O backends described in \ref{sec:posix_backends}, essentially manipulating in-memory metadata.
These overheads were less prominent for runs at smaller scale, and grew exponentially as a function of the number of reader processes used for benchmarking.
This pointed out that there was room for further optimisation in the POSIX I/O backends, and the \verb!fdb-hammer! read performance at scale on Lustre, in Fig. \ref{fig:gcp_fdbh_scalability_a}, would have been closer to DAOS if these inefficiencies had been addressed.
Runs with \textit{no write+read contention} were not noticeably affected by this issue because in these cases only a half of the client VMs were used to run reader processes, preventing the excessive \verb!axis()! overheads from manifesting. Similarly, these overheads were not as marked in \verb!fdb-hammer! runs against Lustre on NEXTGenIO as these used only up to 20 client nodes.

Profiling results for runs with \textit{write+read contention}, in Fig. \ref{fig:gcp_fdbh_lustre_profiling} (c) and (d), showed smaller portions of time being spent on Store and Catalogue \verb!flush()!. Just like for runs against DAOS and Ceph, this was due to part of the server resources being released as soon as the read phase ended.
This was a positive sign indicating that Lustre was able to efficiently use the available resources under contention, and reaffirmed the hypothesis that it did not properly exploit the specialised hardware in NEXTGenIO, where contentious runs spent twice as much time in Store \verb!flush()! as runs with \textit{no write+read contention}.

For read, in (d), the portion of time spent on bulk data reads was smaller than in (b), likely due to the smaller number of \verb!read! operations issued which alleviated the RPC pressure on Lustre.
Also, the \verb!axis()! overheads reduced very significantly due to the smaller scale of the read workload, allowing \verb!fdb-hammer! to scale better than in runs with \textit{no write+read contention}.

\subsection{Small object size performance}

Tests so far issued medium size (O(MiB)) I/O operations, and showed all system performed relatively well, with DAOS standing out reaching higher bandwidths.
Many HPC applications, however, operate with object and I/O sizes orders of magnitude smaller or larger.
While workloads with large I/O size are generally easier ---due to the reduced metadata or RPC pressure--- and should not negatively impact the performance observed so far, workloads with small I/O size might significantly affect the performance behaviour of the storage systems, and the analysis so far did not provide insight on whether such applications would benefit more from the POSIX I/O approach or the object storage one.

\begin{figure*}[htbp]
    \centering
    \begin{subfigure}[b]{214pt}
        \includegraphics[width=214pt]{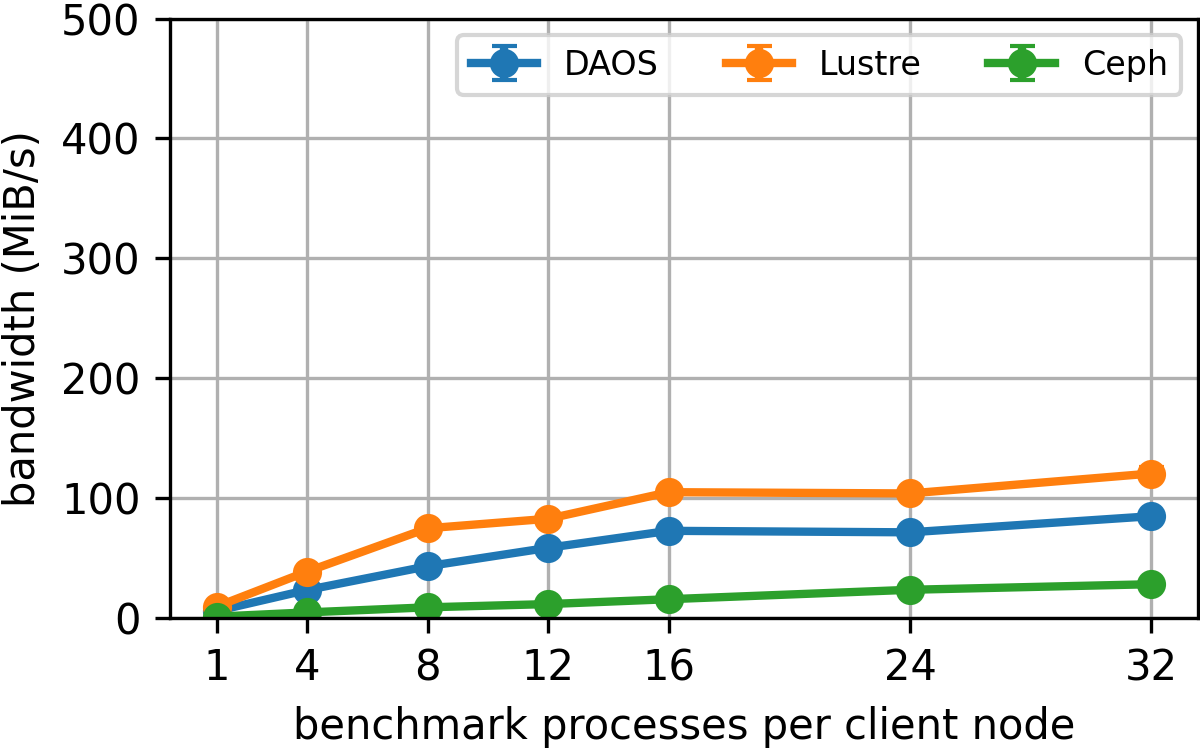}
        \caption{Write, no w+r contention}
    \end{subfigure}
    \begin{subfigure}[b]{188pt}
        \includegraphics[width=188pt,trim={35pt 0 0 0},clip]{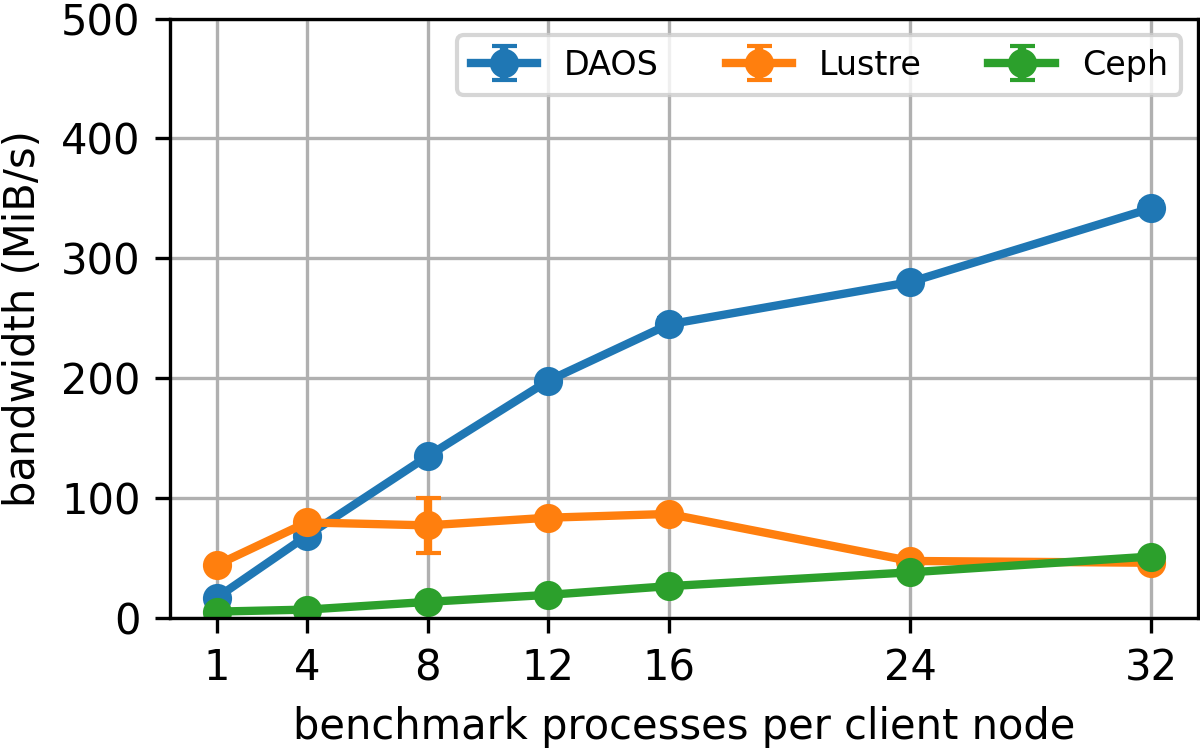}
        \caption{Read, no w+r contention}
    \end{subfigure}
    \caption{Bandwidths for fdb-hammer runs with small object size on 8 client VMs against Lustre, DAOS, and Ceph deployments on 4 VMs (+1 for Lustre and Ceph). Every process performed 10000 x 1KiB I/O operations. Tests were repeated 3 times.}
    \label{fig:gcp_fdbh_small_size}
\end{figure*}

An additional set of \verb!fdb-hammer! tests was run using small I/O size ---1 KiB instead of 1 MiB--- against deployments of the three storage systems on 4 VMs. The results are shown in Fig. \ref{fig:gcp_fdbh_small_size}. 

For write, \verb!fdb-hammer! on Lustre performed better than on the other storage systems, very likely due to the fact that the POSIX I/O backends use large files to contain all the small I/Os and this results in a much smaller metadata workload compared to the object storage backends which create a separate object per I/O operation.
DAOS performed only slightly worse than Lustre, and although this was a very good result given the object per I/O approach, it was likely attributable to DAOS holding many ---if not all--- I/O operations on DRAM server-side given their small size.

For read, \verb!fdb-hammer! on DAOS performed clearly better than the rest. Even if DAOS had been configured to persist metadata and small I/O operations into NVMe devices, it would have cached a copy of such operations on DRAM server-side for fast access, resulting in the same read performance as observed here.
Lustre can similarly cache data server-side, but in this experiment the bandwidths on Lustre sharply plateaued again likely due to inefficiencies in FDB \verb!DataHandle! merging or due to Lustre not handling well the large number of I/O operations issued.
Ceph performed an order of magnitude worse than DAOS likely due to accessing NVMe devices on most Omap and regular object operations. 

These results highlighted the importance of understanding whether the storage systems persist or cache metadata operations server-side or not, and hinted at DAOS's superior capability for metadata-intensive read workloads.

\subsection{DAOS and Ceph data redundancy}

Data safety mechanisms such as replication and erasure-coding are paramount for storage systems to prevent permanent data loss as well as to minimise the impact on running applications in the event of various types of hardware failure or unexpected storage node reboots.

For storage systems intended as long-term permanent storage ---for example backing home directories---, enabling such mechanisms is a must to prevent permanent data loss.
If solely intended as high-performance storage for short-lived applications ---also referred to as scratch storage---, however, enabling such mechanisms is not strictly necessary.
In fact, not enabling data safety mechanisms for scratch storage can be a sensible choice to avoid the associated overheads, although if a failure occurs applications must detect or be made aware of it and start from scratch to regenerate any temporarily unavailable or permanently lost data.
For short-lived and time-critical applications such as the ECMWF's operational NWP, however, a rerun from scratch in the event of a hardware failure generally cannot be afforded, as the execution would likely overrun the operational deadlines.
Data safety mechanisms are therefore a necessary choice for the ECMWF's production setup and their performance impact should be assessed.

Data safety mechanisms can be implemented at a hardware or software level.
Production Lustre setups usually rely on specialised storage hardware implementing such mechanisms, although Lustre can also leverage software-level disk arrays created with \verb!mdadm!, with the disadvantage that these arrays cannot be configured to span multiple devices distributed over the network, and hence are a less likely configuration for production setups.
DAOS and Ceph, conversely, do currently provide integrated production-ready software-level data safety options.

To gain insight on the performance impact of the data safety options provided by DAOS and Ceph, the \verb!fdb-hammer! benchmark was run against deployments of the two storage systems on increasing amounts of VMs in GCP, as done in Fig. \ref{fig:gcp_fdbh_scalability_a}, this time having the systems configured to either replicate or erasure-code all objects.

\begin{figure*}[htbp]
    \centering
    \begin{subfigure}[b]{214pt}
        \includegraphics[width=214pt]{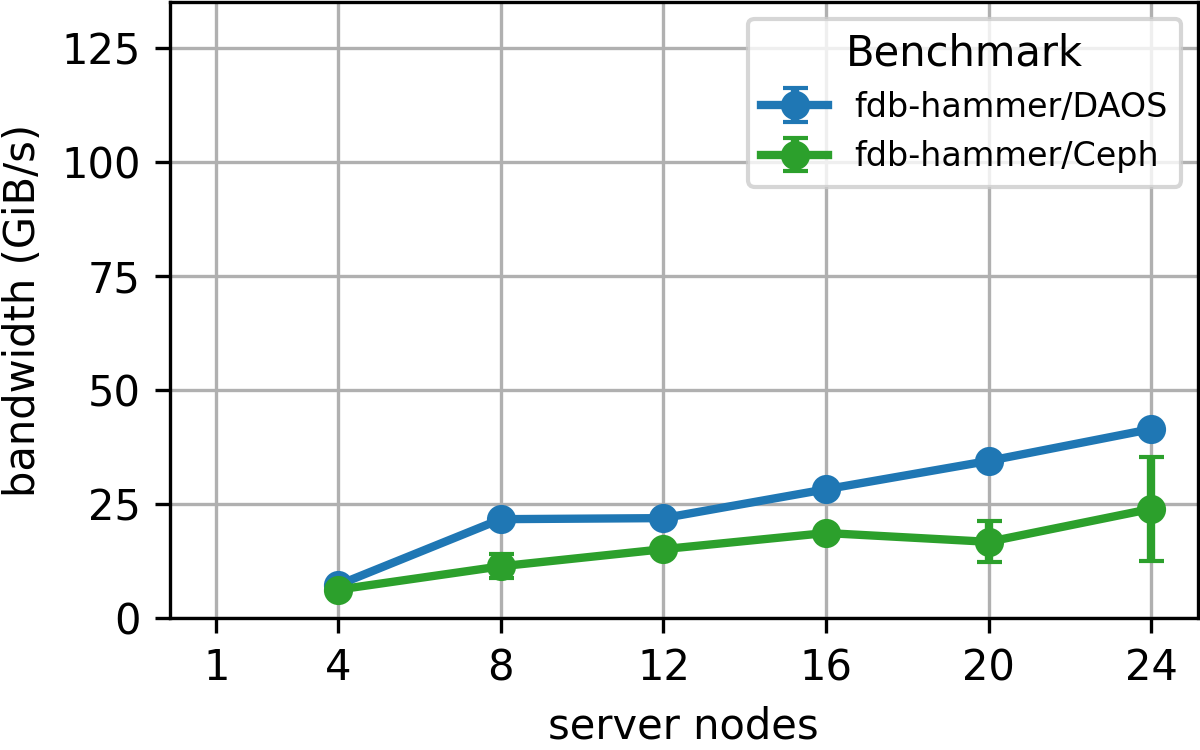}
        \caption{Write, no w+r contention}
    \end{subfigure}
    \begin{subfigure}[b]{188pt}
        \includegraphics[width=188pt,trim={35pt 0 0 0},clip]{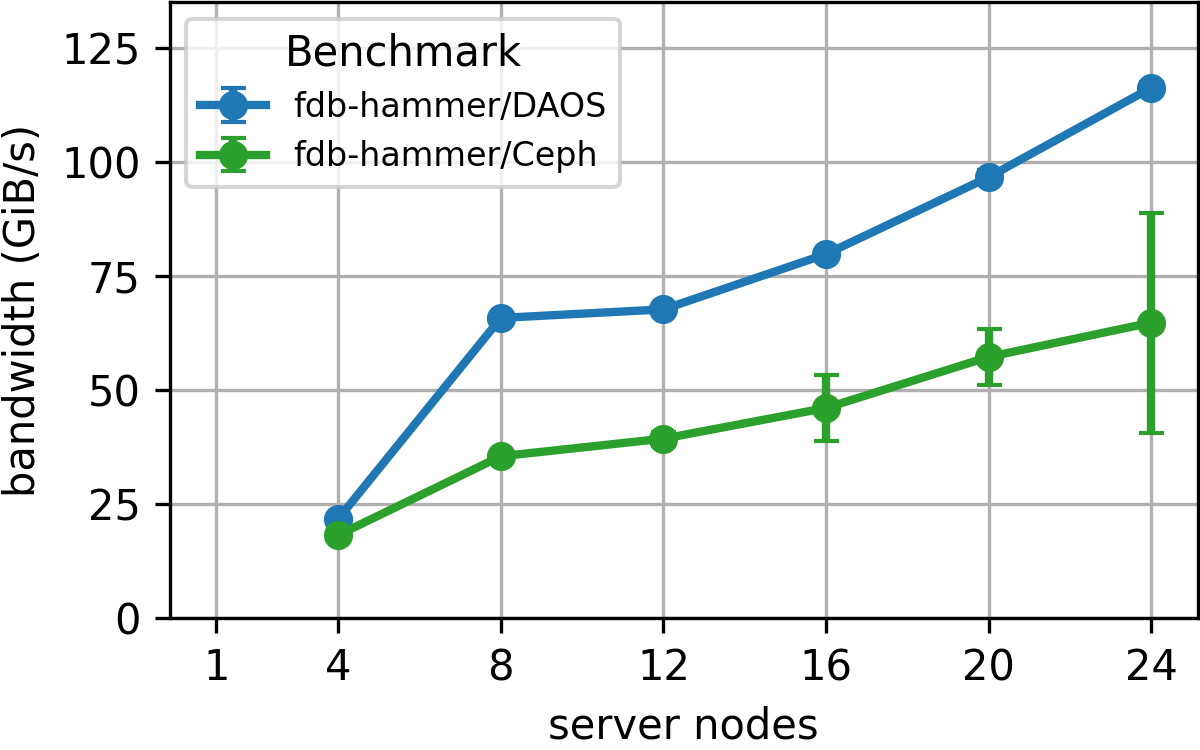}
        \caption{Read, no w+r contention}
    \end{subfigure}
    \caption{fdb-hammer bandwidth scalability, with no write+read contention, against increasingly large Ceph, and DAOS deployments on GCP, using a replication factor of 2. A ratio of 2-to-1 client-to-server nodes was used for all tests, and 16 to 32 processes were run in each client node. Every process wrote and indexed (or de-referenced and read) 10000 x 1MiB weather fields. Tests were repeated 3 times.}
    \label{fig:gcp_fdbh_scalability_a_rp}
\end{figure*}

Fig. \ref{fig:gcp_fdbh_scalability_a_rp} shows results for runs having the storage systems maintain one redundant replica for every object, be it a key-value or a bulk data object.

Relative to Fig. \ref{fig:gcp_fdbh_scalability_a}, the write performance of both DAOS and Ceph reduced by approximately 40\%, and the scalability behavior did not change significantly.
This decrease in performance was an expected result. Either storage system, if mostly idle, should be able to absorb write operations as fast as without replication since the replicas for a given write operation should be written concurrently on multiple storage devices. Nevertheless, when the I/O workload saturates the system ---as done in the experiments here---, half of the storage devices are busy handling writes of redundant copies, therefore reducing the effective write bandwidth to a half if one replica is maintained for every object.

For read, performance was not significantly affected on either system, and this was in line with expectations --- having redundant copies should improve rather than deteriorate performance, as reader processes can target more server VMs to retrieve data in parallel for a given object.

\begin{figure*}[htbp]
    \centering
    \begin{subfigure}[b]{214pt}
        \includegraphics[width=214pt]{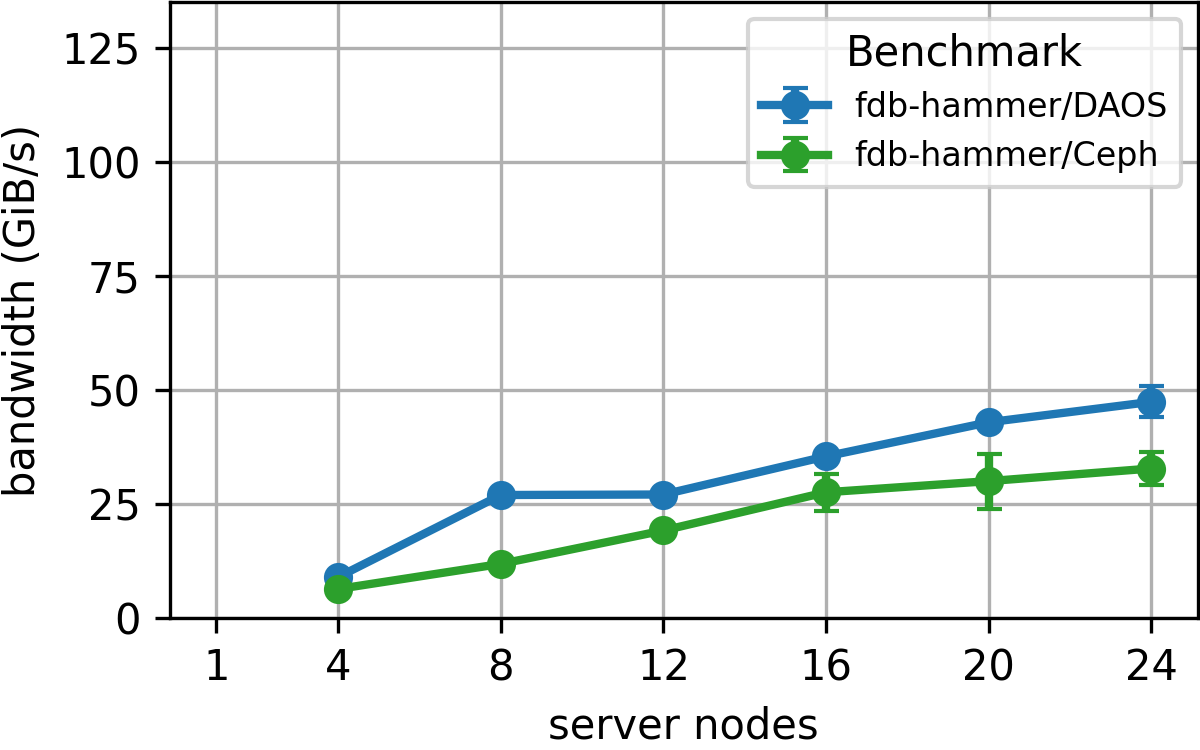}
        \caption{Write, no w+r contention}
    \end{subfigure}
    \begin{subfigure}[b]{188pt}
        \includegraphics[width=188pt,trim={35pt 0 0 0},clip]{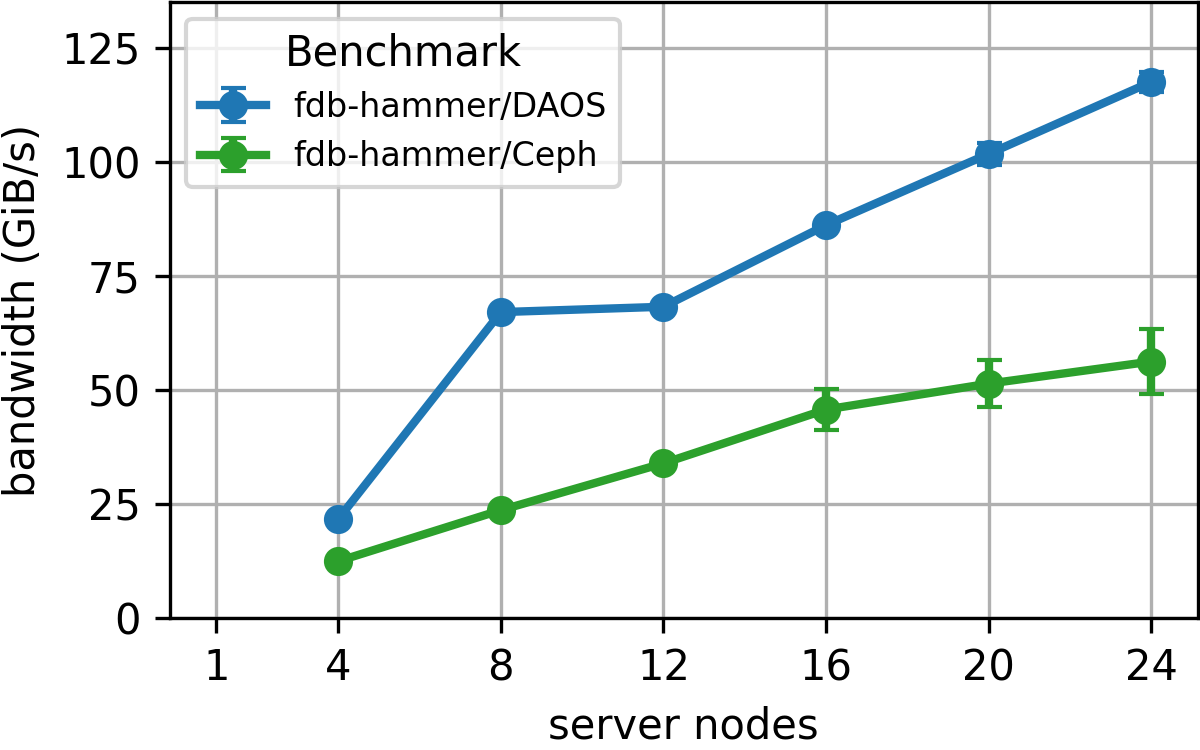}
        \caption{Read, no w+r contention}
    \end{subfigure}
    \caption{fdb-hammer bandwidth scalability, with no write+read contention, against increasingly large Ceph, and DAOS deployments on GCP, using 2+1 erasure-coding. A ratio of 2-to-1 client-to-server nodes was used for all tests, and 16 to 32 processes were run in each client node. Every process wrote and indexed (or de-referenced and read) 10000 x 1MiB weather fields. Tests were repeated 3 times.}
    \label{fig:gcp_fdbh_scalability_a_ec}
\end{figure*}

Results for runs with 2+1 erasure-coding ---that is, with objects being split in two parts, an additional part with redundant data being created, and the three parts being stored in different devices--- are shown in Fig. \ref{fig:gcp_fdbh_scalability_a_ec}. Since Ceph does not support erasure-coded Omaps, the Omaps were kept in a separate pool configured for replication just as in the previous experiment. 

Write bandwidths on both DAOS and Ceph were slightly higher than in the replication experiment.
Runs on DAOS reached up to a 66\% of the bandwidths measured for experiments without redundancy, in Fig. \ref{fig:gcp_fdbh_scalability_a}. This was again in line with expectations, as an erasure-coding factor of 2+1 implies 50\% more data is written into storage devices, and therefore one third of the storage devices are busy handling these additional data.
Runs on Ceph, however, reached similar bandwidths as without redundancy, and this was an indicator that Ceph was likely not being limited by the storage device bandwidth.

For read, bandwidths on DAOS were unaffected with respect to runs without erasure-coding, although Ceph suffered a noticeable decline in performance at larger scales.

These results showed both storage systems could perform well despite enabling their built-in data safety mechanisms, and this was particularly true for DAOS, which stood out reaching bandwidths as high as the hardware permitted.

\subsection{DAOS interfaces}

So far, DAOS demonstrated outstanding performance when used natively via libdaos. Porting existing POSIX I/O applications to libdaos, however, can be a daunting task ---the API and semantics are radically different, and there are challenges associated to porting as showcased in this work--- and is not always a viable option.

DAOS provides a FUSE module for transparent access via the standard POSIX I/O infrastructure, namely DFUSE, with potential to mitigate the challenges of porting to DAOS.
DFUSE, however, is not a silver bullet as it is currently not fully POSIX-compliant.
Among other limitations, it currently does not provide advisory locks, and does not support features such as the atomicity of \verb!mkdir! or the \verb!O_APPEND! flag of the \verb!open! system call.
For instance, due to these limitations, running \verb!fdb-hammer! or any other application using the POSIX I/O FDB backends on DFUSE is not possible.
Nevertheless, a large portion of existing POSIX I/O applications can run directly on DFUSE and, in fact, DAOS is often presented as a new type of file system, and is by default exposed as a file system in early-adopting production HPC systems such as Aurora.

This rises the question of what is the performance impact of accessing DAOS via DFUSE relative to native libdaos access, and how POSIX I/O applications perform on DAOS via DFUSE compared to a high-performance distributed file system deployed on the same hardware.

Appendix D - Section III - B and C looked into the performance and scalability of DFUSE compared to libdaos, and demonstrated that using DFUSE in conjunction with \verb!libioil! ---a POSIX I/O interception library distributed alongside DAOS--- resulted in as good performance and scalability as accessing DAOS natively via libdaos, at least for IOR workloads.

\begin{figure*}[htbp]
    \centering
    \begin{subfigure}[b]{214pt}
        \includegraphics[width=214pt,trim={0pt 16pt 0pt 0pt},clip]{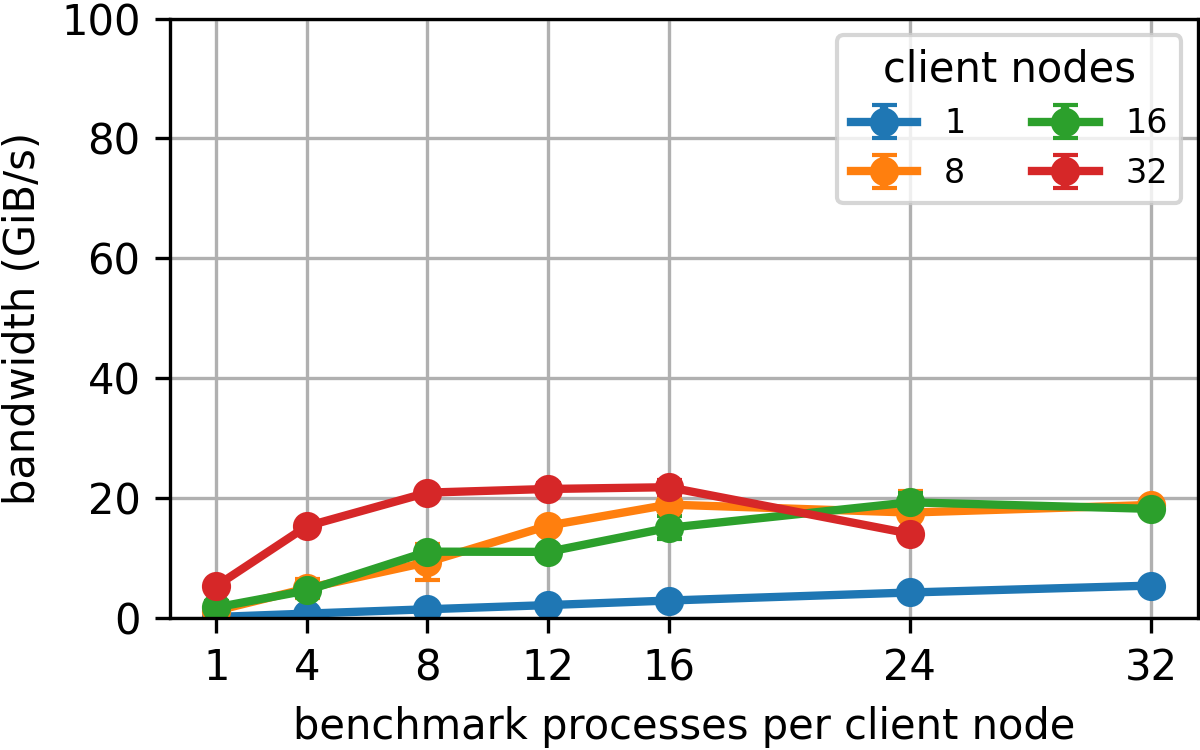}
        \caption{HDF5 on DAOS, Write}
    \end{subfigure}
    \begin{subfigure}[b]{188pt}
        \includegraphics[width=188pt,trim={35pt 16pt 0 0pt},clip]{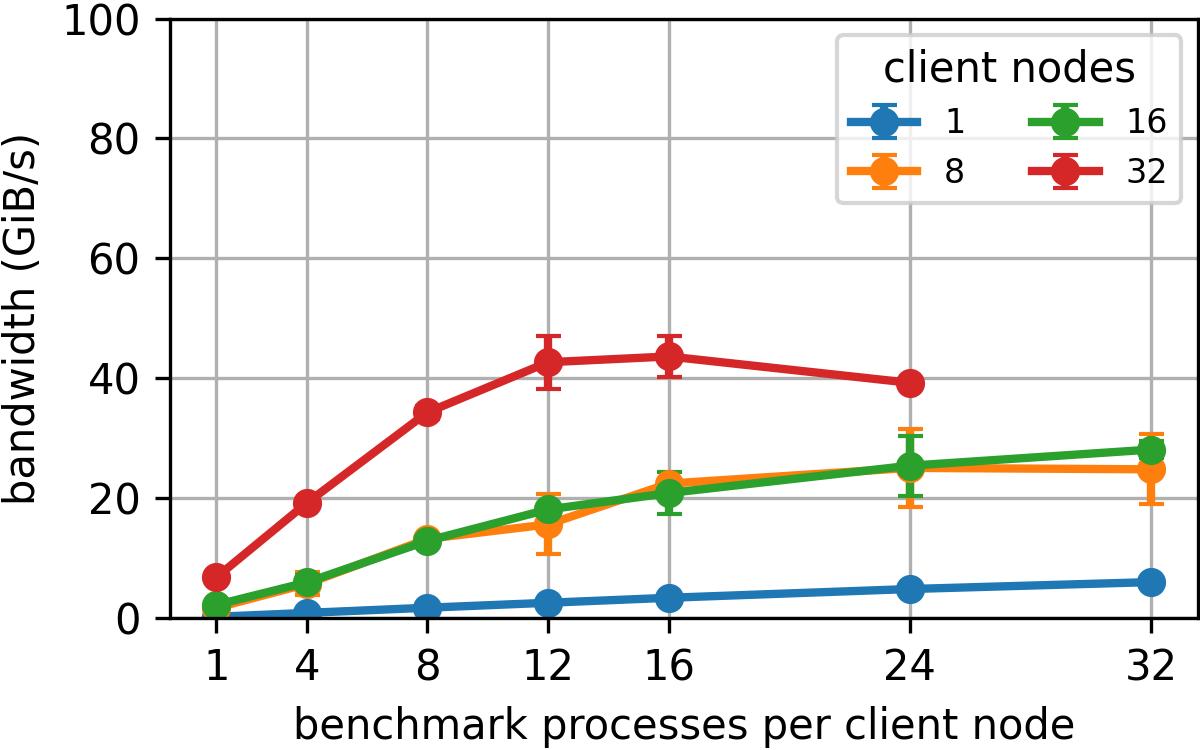}
        \caption{HDF5 on DAOS, Read}
    \end{subfigure}
    \vskip\baselineskip
    \begin{subfigure}[b]{214pt}
        \includegraphics[width=214pt,trim={0pt 16pt 0pt 0pt},clip]{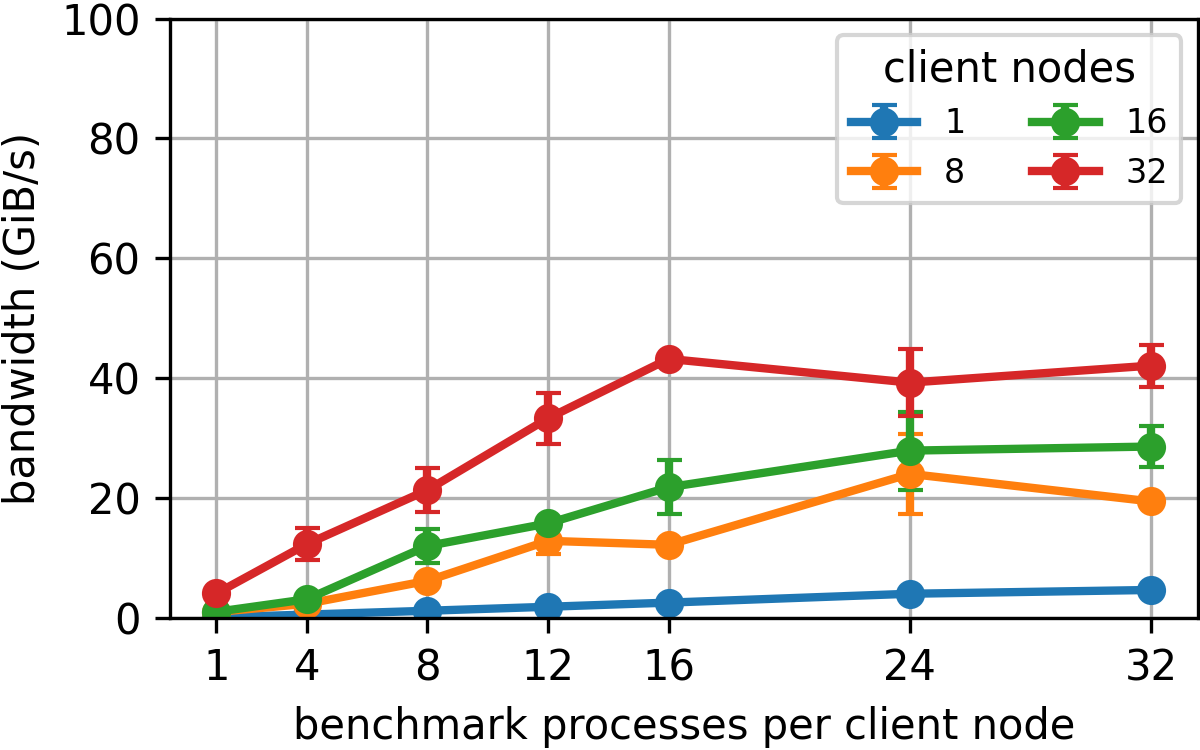}
        \caption{HDF5 POSIX on DFUSE+IL, Write}
    \end{subfigure}
    \begin{subfigure}[b]{188pt}
        \includegraphics[width=188pt,trim={35pt 16pt 0 0pt},clip]{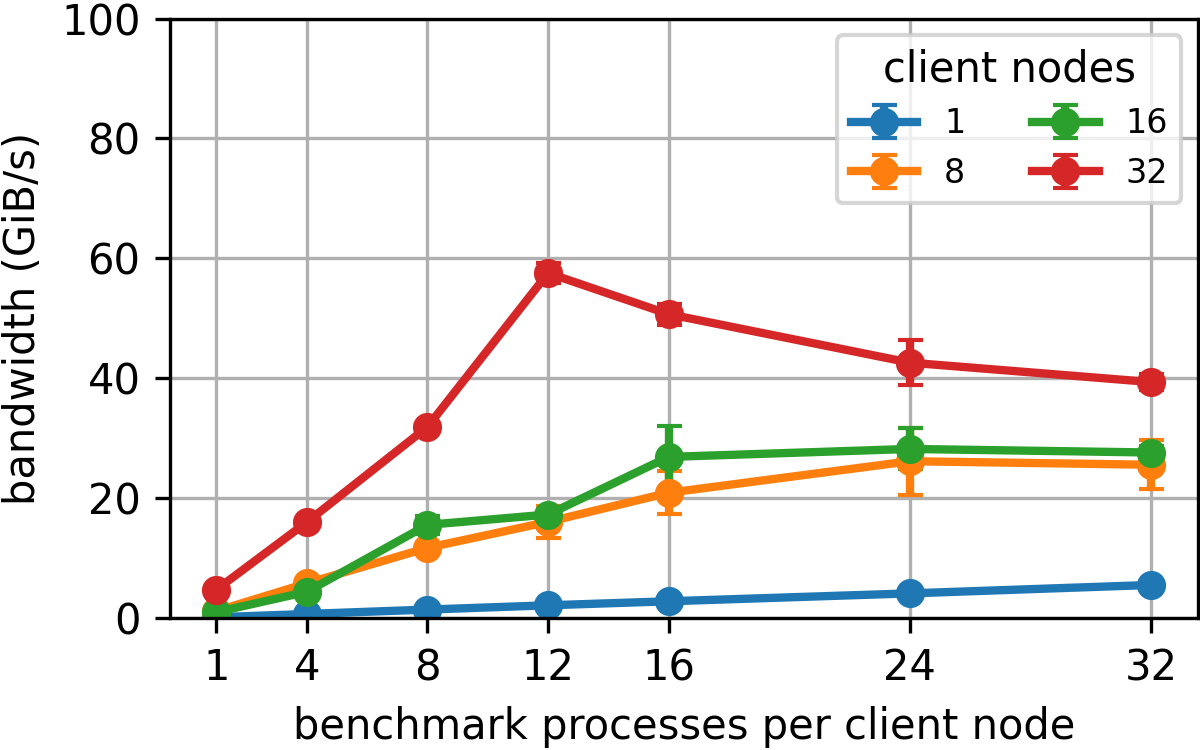}
        \caption{HDF5 POSIX on DFUSE+IL, Read}
    \end{subfigure}
    \vskip\baselineskip
    \begin{subfigure}[b]{214pt}
        \includegraphics[width=214pt,trim={0pt 0pt 0pt 0pt},clip]{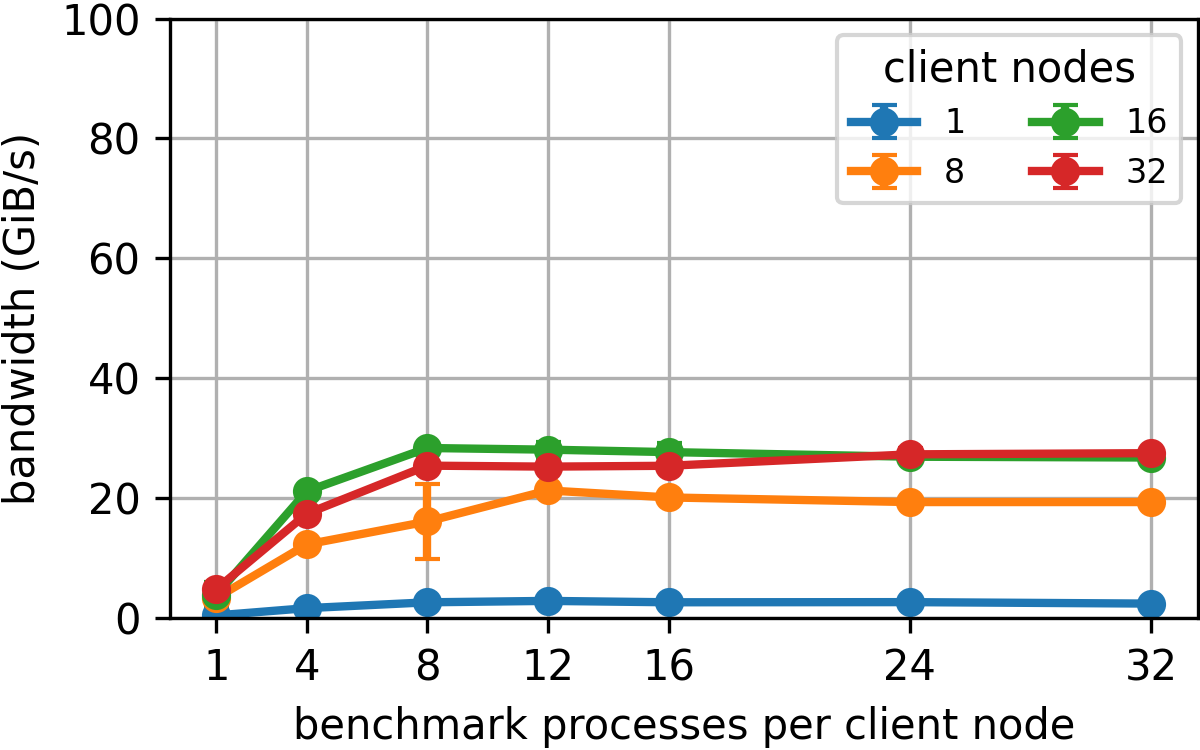}
        \caption{HDF5 POSIX on Lustre, Write}
    \end{subfigure}
    \begin{subfigure}[b]{188pt}
        \includegraphics[width=188pt,trim={35pt 0pt 0 0pt},clip]{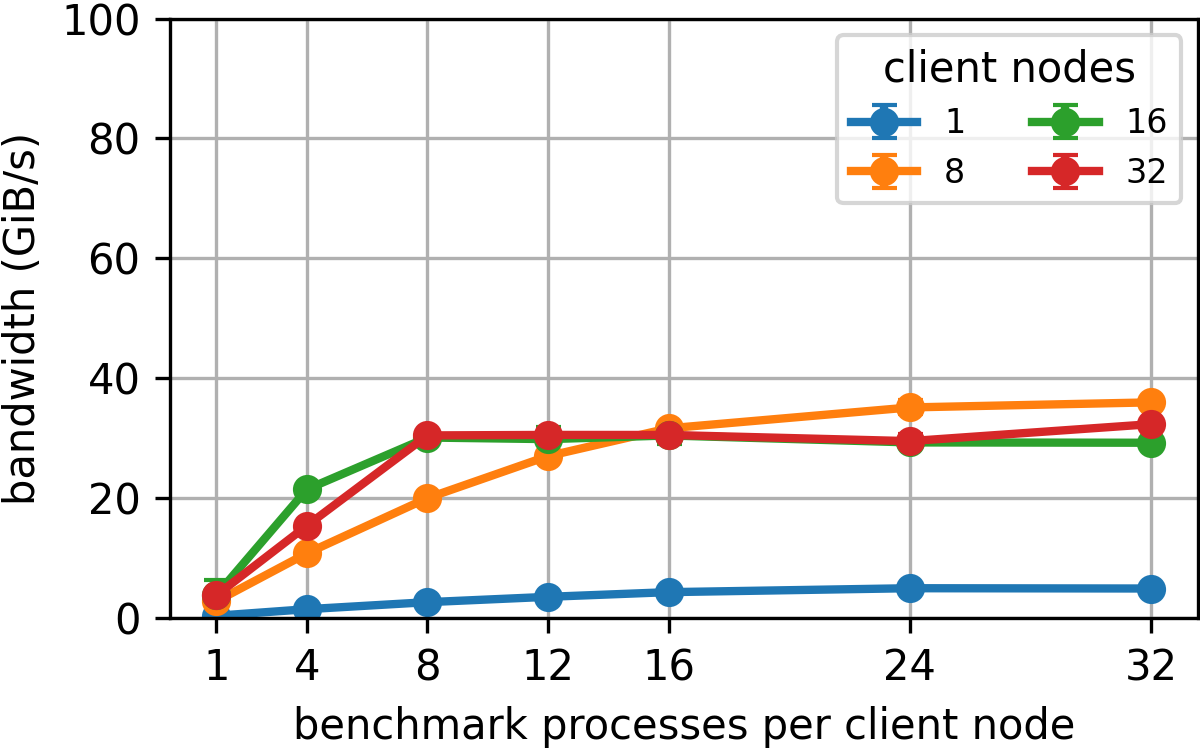}
        \caption{HDF5 POSIX on Lustre, Read}
    \end{subfigure}
    \caption{Bandwidths for IOR/HDF5 runs against DAOS and Lustre deployments on 16 VMs (+1 for Lustre). Every process performed 10000 x 1MiB I/O operations. Tests were repeated 3 times.}
    \label{fig:gcp_ior_hdf5_lustre_daos_16sn_cn_cpcn}
\end{figure*}

The paper also tested a more involved HDF5 benchmark which provides backends for native operation on DAOS via libdaos as well as for efficient operation on POSIX file systems. The benchmark was run in GCP against DAOS deployments on 16 server VMs, both natively and via DFUSE using the POSIX backend. The results have been included here in Fig. \ref{fig:gcp_ior_hdf5_lustre_daos_16sn_cn_cpcn} (a), (b), (c), and (d).

Surprisingly, runs on DFUSE resulted in higher performance than via libdaos, but this was due to the HDF5 native backend making use of a large number of containers, which was shown to deteriorate performance earlier. In spite of this, these results suggested that DFUSE has potential to provide high performance also for complex POSIX I/O applications with zero porting effort.

To gain further insight on whether runs on DFUSE performed well or not, additional HDF5 tests were run against Lustre deployed on the same hardware, using HDF5's POSIX I/O backend.
The results, in Fig. \ref{fig:gcp_ior_hdf5_lustre_daos_16sn_cn_cpcn} (e) and (f), showed that HDF5 on Lustre reached notably lower bandwidths than runs via DFUSE, both for write and read.
Although write runs on DFUSE might have taken advantage to an extent of the fact that DAOS held metadata and small I/O operations in DRAM server-side instead of persisting these into NVMe devices, read runs both on DFUSE and Lustre benefited from the same amount of server-side DRAM available for caching data and optimising read operations.
This was yet another indicator that DFUSE can provide high performance levels for complex POSIX I/O applications without porting, in this case higher than a high-performance file system could provide.

HDF5 and its POSIX I/O backend were designed to make efficient use of the underlying distributed file system.
However, often the reality at HPC centres is that large part of the applications do not perform easy or well-behaved POSIX I/O, either because these do not adhere to programming best practices for high performance on distributed file systems, or because these, by nature, cannot be made to conform to well-behaved POSIX I/O patterns.
One example of this are some of the emerging AI applications, which require accessing many small and sparse subsets of the data domain.

\begin{figure*}[htbp]
    \centering
    \begin{subfigure}[b]{211pt}
        \includegraphics[width=211pt,trim={0pt 0pt 0pt 0pt},clip]{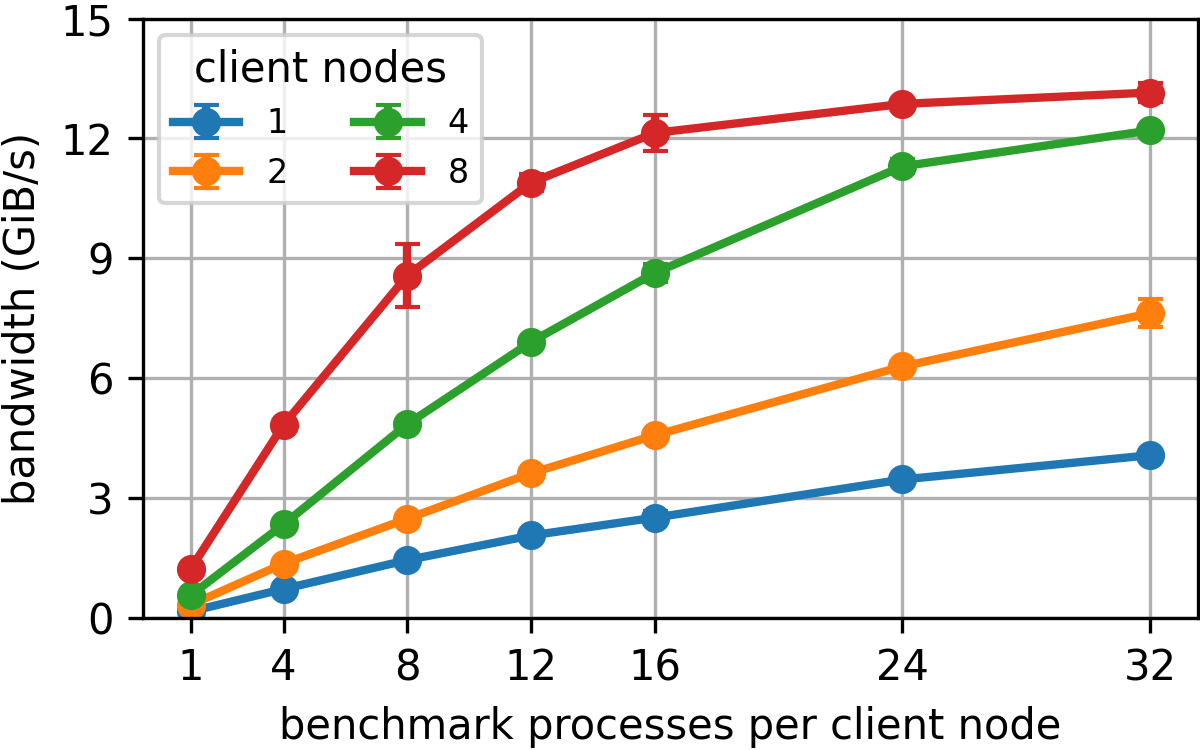}
        \caption{Field I/O on DFUSE+IL, Write}
    \end{subfigure}
    \begin{subfigure}[b]{190pt}
        \includegraphics[width=190pt,trim={28pt 0pt 0 0pt},clip]{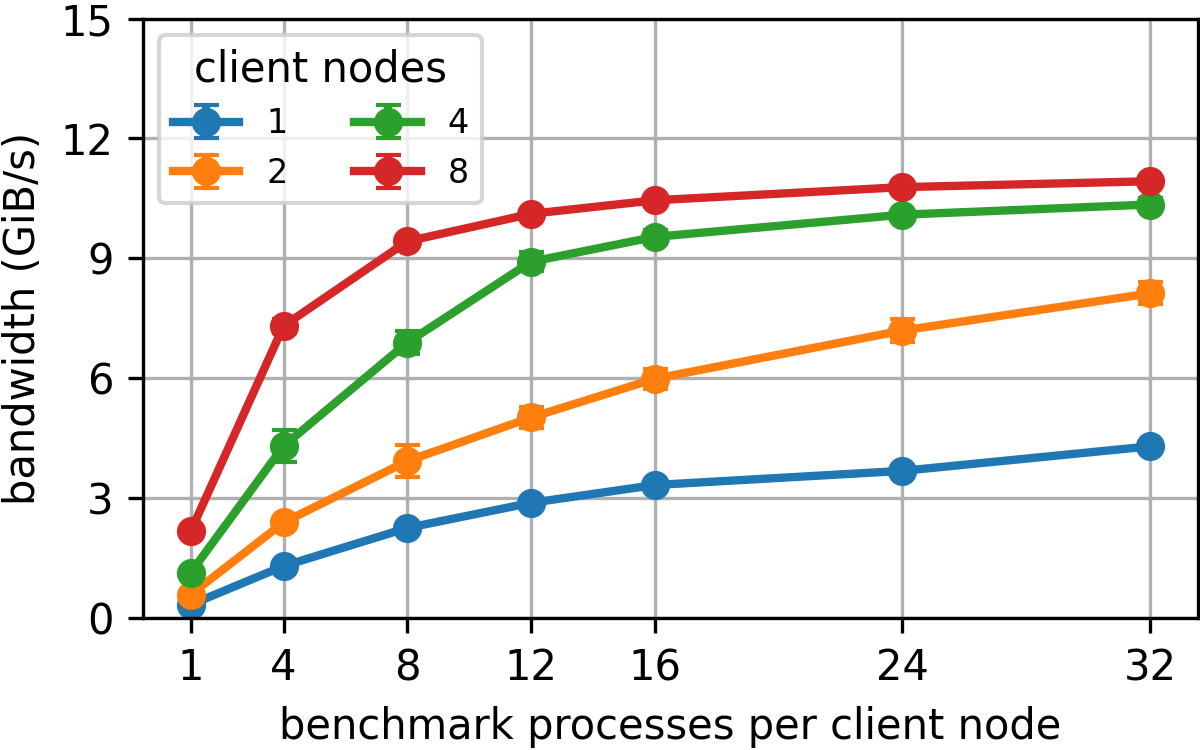}
        \caption{Field I/O on DFUSE+IL, Read}
    \end{subfigure}
    \vskip\baselineskip
    \begin{subfigure}[b]{211pt}
        \includegraphics[width=211pt,trim={0pt 0pt 0pt 0pt},clip]{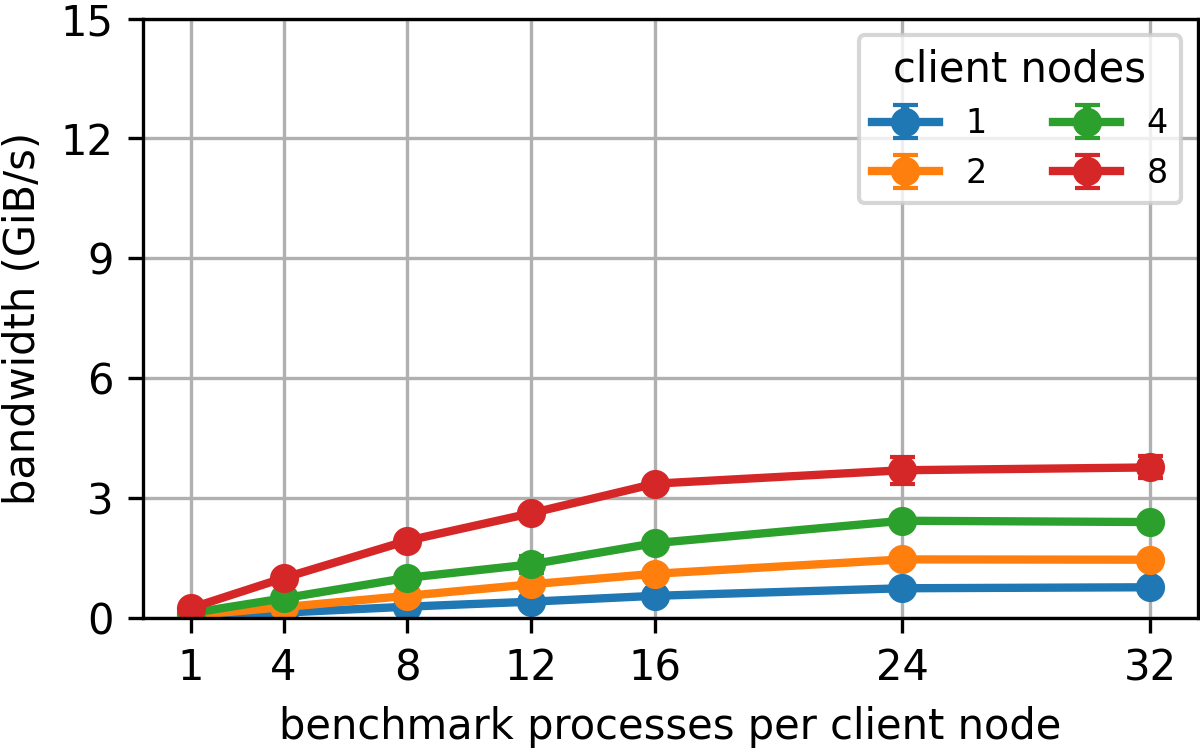}
        \caption{Field I/O on Lustre, Write}
    \end{subfigure}
    \begin{subfigure}[b]{190pt}
        \includegraphics[width=190pt,trim={28pt 0pt 0 0pt},clip]{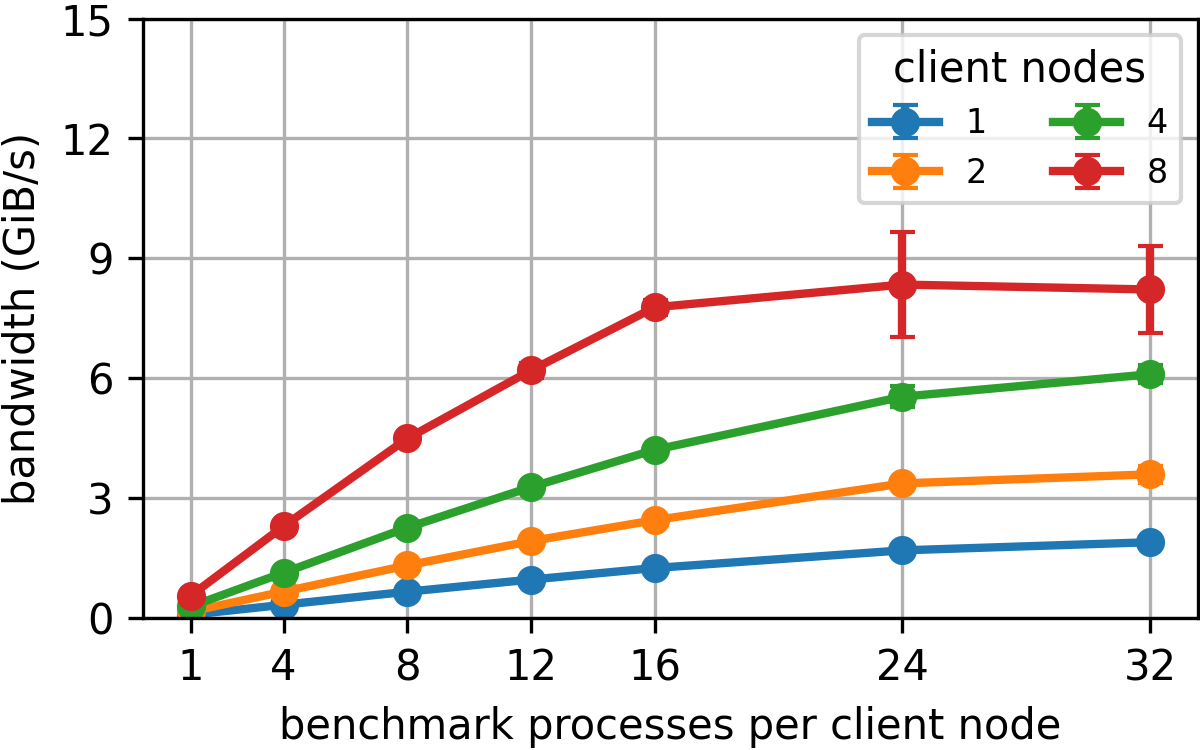}
        \caption{Field I/O on Lustre, Read}
    \end{subfigure}
    \caption{Bandwidths for Field I/O runs using dummy libdaos against DAOS and Lustre deployments on 4 VMs (+1 for Lustre). Every process performed 1000 x 1MiB I/O operations. Tests were repeated 3 times.}
    \label{fig:gcp_fieldio_dummy_daos_lustre_daos_4sn_cn_cpcn}
\end{figure*}

To analyse the performance of DFUSE when exposed to this type of non-well-behaved applications, the Field I/O benchmark and the dummy DAOS library ---which were shown to make an abusive use of the POSIX I/O APIs in Section \ref{sec:ngio_fieldio}--- were rescued and run against DAOS (via DFUSE) and Lustre deployments on 4 server VMs in GCP.
The resulting bandwidths for Field I/O runs on DFUSE are shown at the top row of Fig. \ref{fig:gcp_fieldio_dummy_daos_lustre_daos_4sn_cn_cpcn}, and bandwidths for equivalent runs on Lustre are shown at the bottom row.

The write bandwidths for runs on DFUSE were significantly higher than for runs on Lustre, this time likely mostly due to DAOS holding metadata and small I/O operations on DRAM server-side. For read, both storage systems reached similar bandwidths, although DAOS was slightly ahead.

These results suggested that DAOS and DFUSE do not harm performance or can even result in slight performance improvements for non-well-behaved POSIX I/O applications without porting effort.
In turn, this meant that the POSIX I/O APIs and semantics might not be a causing factor of the known performance limitations of POSIX file systems at scale, and these limitations might instead be a result of the design and implementation of the internals of the file systems so far backing such APIs and semantics.

\subsection{Summary}

The I/O benchmarking in GCP revealed that the three storage systems evaluated ---Lustre, DAOS, and Ceph--- were capable of scaling nearly linearly in performance for IOR workloads if deployed on NVMe SSDs, albeit with a slight decline at the largest scales tested.
Ceph reached significantly lower bandwidths than the other two, but this was partly due to the lack of a feature for IOR to use multiple objects per process.

In contrast to previous experiments in NEXTGenIO, the three storage systems used the TCP protocol this time. Using large client process counts unlocked high performance, although tests at larger scales showed a significant gap relative to hardware limits, particularly for read, and the reasons of this were not investigated.

For NWP I/O workloads, DAOS and the respective FDB backends performed remarkably well, scaling nearly linearly both for write and read despite the benchmark closely capturing the access pattern and contention of an operational run at the ECMWF.

Ceph and the respective backends scaled linearly up to a certain point, reaching higher bandwidths than equivalent IOR tests due to I/O being performed over several objects per process.
Nevertheless, the bandwidths were significantly lower than for DAOS, and this was likely partly due to Ceph immediately persisting metadata and Omaps into NVMe SDDs, whereas DAOS held these in DRAM.
Also, more importantly, the write bandwidths plateaued in the larger tests due to object \verb!write! operations not scaling well on Ceph, which conversely did scale well on DAOS despite being equally persisted into SSDs.

Lustre and the respective backends performed and scaled very well for write up to a certain point, but bandwidths plateaued at the largest scales tested due to large amounts of small \verb!write! operations being issued periodically by the backends and \verb!sync!ed into NVMe SSDs.
For read, Lustre scaled relatively well but reached significantly lower performance than DAOS, although this was largely due to inefficiencies in the POSIX I/O backends, and if these had been addressed, read bandwidths on Lustre would have been close to DAOS's.
One aspect worth noting, however, is that, unfortunately, the \verb!fdb-hammer! benchmark did not capture well the operational contention nor the read access transposition when run on Lustre with the POSIX I/O backends. If these had been captured, both the write and read performance ---but particularly the read performance--- would have likely been lower on Lustre.

All things considered, the performance curves in Fig. \ref{fig:gcp_fdbh_scalability_a} fairly represented the performance potential of the different storage systems on NVMe, with a few exceptions:
\begin{itemize}[leftmargin=*]

    \item DAOS's write performance might be lower at larger scales if enabling DAOS's feature to persist metadata and small I/O operations into NVMe SSDs.
    \item Lustre's write and particularly read performance would likely be lower, with a potential scalability deterioration, if the benchmark fully captured the operational contention and read transposition.
    \item Lustre's read performance would likely be higher if the inefficiencies identified in the POSIX I/O backends were addressed. Combined with the previous exception, the resulting read performance on Lustre would likely be slightly higher than in the graphs.

\end{itemize}

These results were moderately positive for Ceph, which performed and scaled relatively well both for contentious and non-contentious workloads, and demonstrated it ought not be discarded for high-performance at small to medium scales.

The results were also moderately positive for Lustre, as these demonstrated Lustre on NVMe can perform better than it did in NEXTGenIO, and showed it can provide competitive bandwidths and scalability relative to DAOS, with the exception that it can suffer scalability issues for writing at scale.

The results were very positive for DAOS, as these demonstrated that the outstanding performance and scalability was not exclusive of SCM or other specialised hardware, and further consolidated DAOS as the evaluated storage system with the most potential to handle the ECMWF's operational NWP or other similar HPC I/O workloads at scale.

This benchmarking in GCP also looked into the data safety mechanisms provided by DAOS and Ceph, and showed that both systems can provide replication and erasure-coding without significantly harming performance or scalability, although DAOS stood out reaching bandwidths close to hardware limits despite the data safety mechanisms being enabled.

Finally, the performance of DFUSE ---an interface provided by DAOS for transparent POSIX-like access, although not fully POSIX compliant--- was shown to reach close to hardware performance for simple POSIX I/O workloads, and slightly higher performance than Lustre for POSIX I/O applications performing well-behaved and non-well-behaved I/O patterns.

    \chapter{Conclusion}

Driven by scientific and industry ambition, HPC and AI applications such as operational NWP require processing and storing ever-increasing data volumes as fast as possible.
Whilst POSIX distributed file systems and NVMe SSDs are currently a common HPC storage configuration providing I/O to applications,~new storage solutions have proliferated or gained traction over the last decade with potential to address performance limitations POSIX file systems manifest at scale for certain I/O workloads.

This work has primarily aimed to assess the suitability and performance of two object storage systems ---namely DAOS and Ceph--- for ECMWF's operational~NWP as well as for HPC and AI applications in general.
New software-level adapters have been developed which enable ECMWF's NWP to leverage these systems, and extensive I/O benchmarking has been conducted on a few computer systems, comparing the performance delivered by the object stores to that of equivalent Lustre file system deployments on the same hardware.
Challenges of porting to object storage and its benefits with respect to the traditional POSIX I/O approach have been discussed and, where possible, domain-agnostic performance analysis has been conducted, leading to insight also of relevance to I/O practitioners and the broader HPC community.
This work has also investigated the options and challenges the S3 storage protocol provides by porting the ECMWF's NWP to it. This allowed direct comparison to the object storage approaches that have been the main focus of this research, and in turn broadened the range of storage solutions supported by the ECMWF's NWP.

The main findings and conclusions drawn from this work are summarised in the following.

Both DAOS and Ceph were found to provide enough features for a full porting of the ECMWF's NWP, which was ultimately demonstrated with the development of the mentioned adapters.
Also, both were shown to provide strong consistency, free of data corruption or inconsistencies when exercised with the ECMWF's highly contentious I/O workloads.

Porting to these object stores was not trivial. Whilst using their native programming APIs was initially not particularly complicated, unlocking the highest performance the stores could provide required making non-obvious configuration and development choices, and several rounds of adjustment and performance evaluation at scale. This was more marked for DAOS than for Ceph. Some examples included choosing the right granularity of objects, properly pinning DAOS server processes to available cores, finding the optimal number of Ceph placement groups, distributing the data across the right number of DAOS containers or Ceph pools, and selecting the optimal DAOS object classes. This experience led to the production of recommendations, encapsulated in documentation and training material made openly available to the community.

Nevertheless, atomic key-values stood out as a particularly useful feature available in both object stores to implement well-performing indexing of NWP data elements, which is currently difficult to achieve in POSIX file systems. The programming APIs and features offered by the DAOS and Ceph object stores enabled a more flexible and natural expression of I/O patterns, in turn resulting in simpler code for the newly developed object storage adapters than for the previously-existing POSIX ones.

Both object stores and the respective adapters performed well or very well under NWP I/O workloads, but DAOS stood out reaching higher performance than Lustre and Ceph and scaling linearly up to the largest scale tested, whereas Lustre and Ceph showed signs of a write scalability decline at the largest scales.
Specifically, in a system with up to 24 server nodes equipped with NVMe SSDs, DAOS performed slightly better than Lustre for write workloads and notably better (40\%) for read, whereas Ceph performed notably worse (40\%) than Lustre for write, and slightly better for read. The write scalability of Lustre and Ceph declined starting at 16 server nodes, with write bandwidths plateauing beyond that scale. 
In another system with Intel's Optane SCM devices ---Optane is unfortunately discontinued now---, both Lustre and DAOS scaled linearly throughout, although DAOS reached significantly higher bandwidths particularly under contentions I/O workloads. Nevertheless, this was in part due to Lustre not efficiently exploiting the specialised hardware in that system.

Remarkably, both object stores achieved such performance despite being exercised with operational-like NWP workloads, and despite very frequently persisting a large number of small (1~MiB) objects and performing several persistent key-value operations to index and de-index those objects. This type of storage access pattern would have been too expensive to perform at scale with POSIX file systems, and this demonstrated object stores, and DAOS in particular, enable new access patterns at scale involving many I/O operations and small-sized data elements.

Furthermore, DAOS was found to also perform well for both large and very small objects, making it the only evaluated storage system able to provide high performance at KiB, MiB, and GiB object sizes. This emphasized DAOS's unique flexibility, unlocking a wider range of access patterns whilst preserving high performance for traditional ones.

Regarding software-level data redundancy, of relevance for production setups, neither DAOS nor Ceph showed significant deterioration of their baseline performance or scalability when enabling replication of objects, although DAOS stood out maintaining its superior performance levels despite the redundancy mechanisms being engaged.

Further performance tests demonstrated that DFUSE ---DAOS's POSIX inter\-face--- can provide native performance for simple, large-file-per-process POSIX I/O applications without porting, showing no significant performance differences relative to Lustre.
Other more complex applications, performing more involved POSIX I/O patterns, showed slightly better performance when run on DFUSE without porting compared to equivalent runs on Lustre.
Although for some of the tests DAOS benefited from holding small I/O operations on DRAM server-side, these encouraging results strongly suggested that DAOS, via DFUSE, can provide performance levels as high or slightly higher than Lustre both for easy or well-behaved POSIX I/O workloads as well as for workloads not adhering to well-behaved I/O patterns. Nevertheless, this can only be taken advantage of if the application does not use certain POSIX I/O features that are currently not supported by DFUSE.

Overall, both evaluated object stores were found to be suitable and beneficial for the ECMWF's NWP as they both provided sufficient functionality, consistency guarantees, and performance. Due to their different performance behavior, however, they were deemed suitable for different purposes. Where a Ceph system is available ---not uncommon in Cloud environments--- it can provide high-performance read access to applications such as NWP data services at small to medium scale. DAOS, instead, has potential for operational NWP due to its superior performance at scale.

More broadly, this work characterized and shed some light on the performance differences between DAOS, Ceph, and Lustre, which had not been previously documented in such a rigorous apples-to-apples comparison. On one hand, it demonstrated the advantage and potential of DAOS as HPC storage, enabling a wide range of access patterns and providing better performance and scalability, and, on the other hand, it showed Ceph is~also an option worth of consideration, enabling new patterns as well, although providing lower write performance and suffering from scalability limitations similar to Lustre.

These positive results do not mean object stores can easily replace POSIX file systems in HPC centres as typically, in these contexts, several existing applications using POSIX I/O must continue to be supported whilst minimizing performance impact if the storage system is replaced.
Ceph can provide fully POSIX-compliant access via CephFS, and therefore is able to support existing applications without disruption, but may provide significantly lower performance than Lustre, as seen in this work.
In the DAOS case, 
DFUSE is not currently fully POSIX compliant, and porting all unsupported POSIX I/O applications to object storage is generally not an option. Nevertheless, where a complementary DAOS system is provisioned, any applications ported natively may see an increase in performance and scalability, as observed with the ECMWF's NWP I/O benchmark. Furthermore, any POSIX I/O applications supported by DFUSE may see a notable performance increase if they perform complex I/O workloads that were not optimized for POSIX file systems or do not fit well-behaved POSIX I/O patterns --- as is usually the case in AI applications.

All evaluated storage systems, however, are continually evolving, and the view may change in the years to come. For instance, there are ongoing efforts in Ceph and Lustre to address some of their performance limitations, as there are in DAOS's DFUSE to extend POSIX compliance.

The feasibility of porting the ECMWF's NWP to the storage systems and protocols evaluated in this work was a great result. This demonstrated that these offer enough features to implement consistent indexing and bulk-storage functionality on top of them, and was an indicator of the good quality of the design and abstractions put in place in the FDB ---the ECMWF's I/O middleware for NWP applications--- allowing easy implementation of new adapters. This work has contributed enabling the FDB and the ECMWF's NWP to leverage additional storage solutions and opening doors for the ECMWF to consider these in future HPC system procurements.

\section{Future work}

One relevant result of this work has been the demonstration of the advantage and potential of DAOS as HPC storage, and this has been based on its observed superior performance and scalability, compared to Ceph and Lustre, in systems with NVMe SSDs or SCM. There are a few factors that could potentially modify or even disprove this result, and should therefore be examined thoroughly with additional performance tests.

In the first place, only up to 24 server nodes were employed for the scalability analysis. Additional tests should be conducted at scales more representative of production HPC setups, ideally using up to hundreds of server nodes, to determine whether the scalability behaviors observed so far hold at such larger scales. This should include running the ECMWF's Kronos benchmark, or even running full operational NWP workloads, potentially at very high resolution, against all storage systems deployed at scale, and preferably enabling software-level data redundancy.

Secondly, both Lustre and Ceph persisted all metadata immediately on SSDs, whereas DAOS did not yet have the capability of doing so and, instead, it stored metadata and small I/O operations under 4 KiB on DRAM server-side --- at that time, DAOS was still transitioning after Optane was discontinued. A newer version of DAOS should be employed and configured to immediately persist all metadata and I/O on SSDs for a fairer comparison. Nevertheless, prior work has shown that this new feature of DAOS should only have a noticeable performance impact for corner-case metadata-intensive workloads, and should in any case mostly impact only write workloads as DAOS caches metadata and small I/O operations, even if the new feature is enabled, to faster serve read operations.

In the third place, all Lustre deployments used for this analysis were set up with only one MDS (metadata server) regardless of the number of OSTs (object storage targets) deployed. However, ideally, the MDSs should be grown alongside the OSTs, for example using a ratio of 1 MDS to 8 OSTs, to ensure metadata servers do not bottleneck I/O performance.  

Lastly, more advanced Ceph setups and configuration could be explored to ensure the maximum potential of Ceph is unlocked. Some ideas include deploying Ceph OSDs in a dedicated storage network; adjusting the depth of the Write-Ahead Logs in the OSDs; and adjusting the amount of DRAM used by OSDs.

Regarding the assessment of the performance benefits of object storage for the ECMWF's operations, the NWP I/O benchmark used did not fully capture the operational contention nor access pattern, particularly when run on Lustre. The benchmark should be enhanced to better do so, or alternatively other more complex benchmarks or full operational workloads should be used, and this would likely result in a more marked performance advantage on object stores, even after the read performance issue in the POSIX I/O backends is addressed.

One limitation of this work is that it has followed a markedly empirical approach for the performance analysis, relying mostly on analysis of bandwidths measured in benchmark runs and analysis of application-level profiling. Further in-depth analysis should be conducted, collecting abundant metrics on the server side such as queue depths and RPC counts, with the aim of providing more detailed explanations of what mechanisms may be being triggered and causing bottlenecks internally in the servers.

Also, the performance analysis has focused on rather short-lived benchmark runs of up to tens of minutes each. However, storage systems can manifest oscillation or changes in performance at longer time scales as internal buffers fill up and data structures are further populated, and this should be addressed by running benchmarks of longer duration. Likewise, the impact of storage device or node failures on storage performance should be assessed, as well as the impact of intense and short-lived I/O workloads triggered while other longer-lived workloads are active. These scenarios are common in production contexts.

As previously discussed, the new object storage adapters for the ECMWF's NWP, as currently implemented, have the writer clients persist every single data element and key-value operation immediately on non-volatile storage on the server side, as opposed to the approach followed by the POSIX adapters used currently operationally, where clients build local caches of data and indexing information which are only ensured to be persisted from time to time, when strictly necessary. Although the immediately persistent approach of the object storage adapters has some advantages, it may be unnecessarily conservative. New object storage adapters could be implemented which follow a caching approach based on that of the POSIX adapters, and this might further improve the performance and scalability of object stores for the ECMWF's operational NWP.

Finally, given the potential of object stores to provide high performance for I/O workloads involving many I/O operations and small-sized data elements, and for less well-behaved POSIX I/O workloads, further benchmarking should be conducted to measure the likely positive impact of object storage and the new adapters on machine-learning-based NWP at the ECMWF.

    \backmatter
    \cleardoublepage
    \renewcommand{\baselinestretch}{1.0}
    \renewcommand*{\bibfont}{\small}
    \begin{spacing}{1}  
        \printbibliography[heading=bibintoc]

@inproceedings{Smart2019AClimate,
    title = {{A High-Performance Distributed Object-Store for Exascale Numerical Weather Prediction and Climate}},
    year = {2019},
    booktitle = {Proceedings of the Platform for Advanced Scientific Computing Conference},
    author = {Smart, Simon D. and Quintino, Tiago and Raoult, Baudouin},
    month = {6},
    pages = {1--11},
    publisher = {ACM},
    address = {New York, NY, USA},
    isbn = {9781450367707},
    doi = {10.1145/3324989.3325726}
}

@inproceedings{Smart2017AData,
    title = {{A Scalable Object Store for Meteorological and Climate Data}},
    year = {2017},
    booktitle = {Proceedings of the Platform for Advanced Scientific Computing Conference},
    author = {Smart, Simon D. and Quintino, Tiago and Raoult, Baudouin},
    month = {6},
    pages = {1--8},
    publisher = {ACM},
    address = {New York, NY, USA},
    isbn = {9781450350624},
    doi = {10.1145/3093172.3093238}
}

@misc{AboutForecasts,
    title = {{About Our Forecasts}},
    booktitle = {(Accessed: 28 January 2025)},
    url = {https://www.ecmwf.int/en/forecasts/documentation-and-support}
}

@article{Soumagne2022AcceleratingDAOS,
    title = {{Accelerating HDF5 I/O for Exascale Using DAOS}},
    year = {2022},
    journal = {IEEE Transactions on Parallel and Distributed Systems},
    author = {Soumagne, Jerome and Henderson, Jordan and Chaarawi, Mohamad and Fortner, Neil and Breitenfeld, Scot and Lu, Songyu and Robinson, Dana and Pourmal, Elena and Lombardi, Johann},
    number = {4},
    pages = {903--914},
    volume = {33},
    doi = {10.1109/TPDS.2021.3097884}
}

@phdthesis{Neuwirth2019AcceleratingEnvironments,
    title = {{Accelerating Network Communication and I/O in Scientific High Performance Computing Environments}},
    year = {2019},
    author = {Neuwirth, Sarah},
    school = {Universit{\"{a}}t Heidelberg}
}

@misc{Lombardi2021AcceleratingDAOS,
    title = {{Accelerating Storage with Optane {\&} DAOS}},
    year = {2021},
    booktitle = {19th Workshop on High-Performance Computing in Meteorology},
    author = {Lombardi, Johann and Manubens, Nicolau},
    month = {9},
    url = {https://events.ecmwf.int/event/169/timetable/}
}

@book{Stevens2013AdvancedEnvironment,
    title = {{Advanced Programming in the UNIX Environment}},
    year = {2013},
    author = {Stevens, W Richard and Rago, Stephen A},
    edition = {3rd},
    publisher = {Addison-Wesley Professional},
    isbn = {0321637739}
}

@misc{AmazonReference,
    title = {{Amazon S3 API Reference}},
    booktitle = {(Accessed: 29 July 2025)},
    url = {https://docs.aws.amazon.com/AmazonS3/latest/API}
}

@inproceedings{Weiland2019AnApplications,
    title = {{An early evaluation of Intel's optane DC persistent memory module and its impact on high-performance scientific applications}},
    year = {2019},
    booktitle = {Proceedings of the International Conference for High Performance Computing, Networking, Storage and Analysis},
    author = {Weiland, Michèle and Brunst, Holger and Quintino, Tiago and Johnson, Nick and Iffrig, Olivier and Smart, Simon and Herold, Christian and Bonanni, Antonino and Jackson, Adrian and Parsons, Mark},
    month = {11},
    pages = {1--19},
    publisher = {ACM},
    address = {New York, NY, USA},
    isbn = {9781450362290},
    doi = {10.1145/3295500.3356159}
}

@inproceedings{RajaChandrasekar2017AnSupercomputers,
    title = {{An Exploration into Object Storage for Exascale Supercomputers}},
    year = {2017},
    booktitle = {Cray User Group},
    author = {Raja Chandrasekar, Raghunath and Evans, Lance and Wespetal, Robert}
}

@article{Gadban2021AnalyzingWorkloads,
    title = {{Analyzing the Performance of the S3 Object Storage API for HPC Workloads}},
    year = {2021},
    author = {Gadban, Frank and Kunkel, Julian},
    doi = {10.3390/app11188540}
}

@misc{Manubens2021AssessmentFDB,
    title = {{Assessment of DAOS as a Backend for ECMWF's FDB}},
    year = {2021},
    booktitle = {DAOS User Group Meeting},
    author = {Manubens, Nicolau},
    month = {11},
    url = {https://daosio.atlassian.net/wiki/spaces/DC/pages/11015454821/DUG21}
}

@misc{CephReference,
    title = {{Ceph Network Configuration Reference}},
    booktitle = {(Accessed: 29 July 2025)},
    url = {https://docs.ceph.com/en/latest/rados/configuration/network-config-ref/#general-settings}
}

@misc{Weil2019CephCeph,
    title = {{Ceph Tech Talk - Introduction to Ceph}},
    year = {2019},
    booktitle = {(Accessed: 29 July 2025)},
    author = {Weil, Sage},
    month = {6},
    url = {https://www.youtube.com/watch?v=PmLPbrf-x9g}
}

@misc{Nelson2024Ceph:TiB/s,
    title = {{Ceph: A Journeyto 1 TiB/s}},
    year = {2024},
    booktitle = {(Accessed: 29 July 2025)},
    author = {Nelson, Mark},
    month = {1},
    url = {https://ceph.io/en/news/blog/2024/ceph-a-journey-to-1tibps/}
}

@inproceedings{Weil2006Ceph:System,
    title = {{Ceph: a scalable, high-performance distributed file system}},
    year = {2006},
    booktitle = {Proceedings of the 7th Symposium on Operating Systems Design and Implementation},
    author = {Weil, Sage A and Brandt, Scott A and Miller, Ethan L and Long, Darrell D E and Maltzahn, Carlos},
    pages = {307--320},
    series = {OSDI '06},
    publisher = {USENIX Association},
    address = {USA},
    isbn = {1931971471}
}

@inproceedings{Lofstead2016DAOSSystem,
    title = {{DAOS and Friends: A Proposal for an Exascale Storage System}},
    year = {2016},
    booktitle = {SC '16: Proceedings of the International Conference for High Performance Computing, Networking, Storage and Analysis},
    author = {Lofstead, Jay and Jimenez, Ivo and Maltzahn, Carlos and Koziol, Quincey and Bent, John and Barton, Eric},
    number = {},
    pages = {585--596},
    volume = {},
    doi = {10.1109/SC.2016.49}
}

@misc{DAOSArchitecture,
    title = {{DAOS Architecture}},
    booktitle = {(Accessed: 29 July 2025)},
    url = {https://docs.daos.io/latest/overview/architecture}
}

@inproceedings{Manubens2023DAOSPrediction,
    title = {{DAOS as HPC Storage: a View From Numerical Weather Prediction}},
    year = {2023},
    booktitle = {2023 IEEE International Parallel and Distributed Processing Symposium (IPDPS)},
    author = {Manubens, Nicolau and Quintino, Tiago and Smart, Simon D. and Danovaro, Emanuele and Jackson, Adrian},
    month = {5},
    pages = {1029--1040},
    publisher = {IEEE},
    isbn = {979-8-3503-3766-2},
    doi = {10.1109/IPDPS54959.2023.00106}
}

@inproceedings{Jackson2023DAOSInterfaces,
    title = {{DAOS as HPC Storage: Exploring Interfaces}},
    year = {2023},
    booktitle = {2023 IEEE International Conference on Cluster Computing Workshops (CLUSTER Workshops)},
    author = {Jackson, Adrian and Manubens, Nicolau},
    number = {},
    pages = {8--10},
    volume = {},
    doi = {10.1109/CLUSTERWorkshops61457.2023.00011}
}

@inproceedings{Hennecke2023DAOSResults,
    title = {{DAOS Beyond Persistent Memory: Architecture and Initial Performance Results}},
    year = {2023},
    booktitle = {High Performance Computing: ISC High Performance 2023 International Workshops, Hamburg, Germany, May 21–25, 2023, Revised Selected Papers},
    author = {Hennecke, Michael and Olivier, Jeff and Nabarro, Tom and Zhen, Liang and Niu, Yawei and Wang, Shilong and Liu, Xuezhao},
    pages = {353--365},
    publisher = {Springer-Verlag},
    url = {https://doi.org/10.1007/978-3-031-40843-4_26},
    address = {Berlin, Heidelberg},
    isbn = {978-3-031-40842-7},
    doi = {10.1007/978-3-031-40843-4{\_}26}
}

@misc{DAOSSystem,
    title = {{DAOS File System}},
    booktitle = {(Accessed: 29 July 2025)},
    url = {https://docs.daos.io/latest/user/filesystem/}
}

@inproceedings{Liang2020DAOS:Memory,
    title = {{DAOS: A Scale-Out High Performance Storage Stack for Storage Class Memory}},
    year = {2020},
    booktitle = {Supercomputing Frontiers: 6th Asian Conference, SCFA 2020, Singapore, February 24–27, 2020, Proceedings},
    author = {Liang, Zhen and Lombardi, Johann and Chaarawi, Mohamad and Hennecke, Michael},
    pages = {40--54},
    publisher = {Springer-Verlag},
    url = {https://doi.org/10.1007/978-3-030-48842-0_3},
    address = {Berlin, Heidelberg},
    isbn = {978-3-030-48841-3},
    doi = {10.1007/978-3-030-48842-0{\_}3}
}

@misc{nicolau-manubens2024Ecmwf-projects/daos-tests:0.3.2,
    title = {{ecmwf-projects/daos-tests: 0.3.2}},
    year = {2024},
    author = {{nicolau-manubens}},
    month = {9},
    publisher = {Zenodo},
    url = {https://doi.org/10.5281/zenodo.13757427},
    doi = {10.5281/zenodo.13757427}
}

@misc{Manubens2024ECMWFsData,
    title = {{ECMWF’s FDB: a Versatile High-Performance Store for Earth System Data}},
    year = {2024},
    booktitle = {3rd Destination Earth User Exchange Poster},
    author = {Manubens, Nicolau and Kremer, Tobias and {\c{C}}akırcalı, Metin and Bradley, Chris and Danovaro, Emanuele and Smart, Simon and Quintino, Tiago},
    month = {10},
    url = {https://destination-earth.eu/event/3rd-destination-earth-user-exchange/}
}

@inproceedings{SarpangalaVenkatesh2023EnhancingContext,
    title = {{Enhancing Metadata Transfer Efficiency: Unlocking the Potential of DAOS in the ADIOS context}},
    year = {2023},
    booktitle = {Proceedings of the SC '23 Workshops of the International Conference on High Performance Computing, Network, Storage, and Analysis},
    author = {Sarpangala Venkatesh, Ranjan and Eisenhauer, Greg and Klasky, Scott and Gavrilovska, Ada},
    pages = {1223--1228},
    series = {SC-W '23},
    publisher = {Association for Computing Machinery},
    url = {https://doi.org/10.1145/3624062.3624193},
    address = {New York, NY, USA},
    isbn = {9798400707858},
    doi = {10.1145/3624062.3624193}
}

@book{Acquaviva2025ETP4HPCStorage,
    title = {{ETP4HPC SRA 6 White Paper-I/O and Storage}},
    year = {2025},
    author = {Acquaviva, Jean-Thomas and Golasowski, Martin and Hennecke, Michael and Jackson, William Adrian and Leibovici, Thomas and Luettgau, Jakob and Nou, Ramon},
    editor = {Neuwirth, Sarah and Deniel, Philippe},
    series = {ETP4HPC SRA},
    doi = {10.5281/zenodo.14605692},
    language = {English}
}

@inproceedings{Jackson2023EvaluatingSwansong,
    title = {{Evaluating the latest Optane memory: A glorious swansong?}},
    year = {2023},
    booktitle = {4th Workshop on Heterogeneous Memory Systems, SC23},
    author = {Jackson, Adrian}
}

@inproceedings{Liu2018EvaluationSystems,
    title = {{Evaluation of HPC Application I/O on Object Storage Systems}},
    year = {2018},
    booktitle = {2018 IEEE/ACM 3rd International Workshop on Parallel Data Storage {\&} Data Intensive Scalable Computing Systems (PDSW-DISCS)},
    author = {Liu, Jialin and Koziol, Quincey and Butler, Gregory F and Fortner, Neil and Chaarawi, Mohamad and Tang, Houjun and Byna, Suren and Lockwood, Glenn K and Cheema, Ravi and Kallback-Rose, Kristy A and Hazen, Damian and Prabhat, Mr},
    number = {},
    pages = {24--34},
    volume = {},
    doi = {10.1109/PDSW-DISCS.2018.00005}
}

@inproceedings{Manubens2024ExploringPerformance,
    title = {{Exploring DAOS Interfaces and Performance}},
    year = {2024},
    booktitle = {SC24-W: Workshops of the International Conference for High Performance Computing, Networking, Storage and Analysis},
    author = {Manubens, Nicolau and Lombardi, Johann and Smart, Simon D. and Danovaro, Emanuele and Quintino, Tiago and Hildebrand, Dean and Jackson, Adrian},
    month = {11},
    pages = {1340--1348},
    publisher = {IEEE},
    isbn = {979-8-3503-5554-3},
    doi = {10.1109/SCW63240.2024.00175}
}

@misc{FDB,
    title = {{FDB}},
    booktitle = {GitHub repository (Accessed: 28 January 2025)},
    url = {https://github.com/ecmwf/fdb}
}

@inproceedings{Aghayev2019FileEvolution,
    title = {{File systems unfit as distributed storage backends: lessons from 10 years of Ceph evolution}},
    year = {2019},
    booktitle = {Proceedings of the 27th ACM Symposium on Operating Systems Principles},
    author = {Aghayev, Abutalib and Weil, Sage and Kuchnik, Michael and Nelson, Mark and Ganger, Gregory R and Amvrosiadis, George},
    pages = {353--369},
    series = {SOSP '19},
    publisher = {Association for Computing Machinery},
    url = {https://doi.org/10.1145/3341301.3359656},
    address = {New York, NY, USA},
    isbn = {9781450368735},
    doi = {10.1145/3341301.3359656}
}

@misc{General-purposeEngine,
    title = {{General-purpose machine family for Compute Engine}},
    booktitle = {(Accessed: 29 July 2025)},
    url = {https://cloud.google.com/compute/docs/general-purpose-machines}
}

@misc{GooglePage,
    title = {{Google Cloud Landing Page}},
    booktitle = {(Accessed: 29 July 2025)},
    url = {https://cloud.google.com}
}

@misc{GoogleToolkit,
    title = {{Google Cluster Toolkit}},
    booktitle = {(Accessed: 29 July 2025)},
    url = {https://github.com/GoogleCloudPlatform/cluster-toolkit}
}

@inproceedings{Schmuck2002GPFS:Clusters,
    title = {{GPFS: A Shared-Disk File System for Large Computing Clusters}},
    year = {2002},
    booktitle = {Conference on File and Storage Technologies},
    author = {Schmuck, Frank and Haskin, Roger}
}

@misc{TheHDFGroupHierarchicalSoftware,
    title = {{Hierarchical Data Format, version 5 [Computer Software]}},
    booktitle = {(Accessed: 29 July 2025)},
    author = {{The HDF Group}},
    url = {https://github.com/HDFGroup/hdf}
}

@misc{Jackson2024High-PerformanceEra,
    title = {{High-Performance Object Storage: I/O for the Exascale Era}},
    year = {2024},
    booktitle = {SC'24 Conference Tutorial},
    author = {Jackson, Adrian and Chaarwi, Mohamad and Lombardi, Johann and Manubens, Nicolau and Hildebrand, Dean},
    month = {11},
    url = {https://sc24.conference-program.com/presentation/?id=tut143&sess=sess417}
}

@misc{HPCRepository,
    title = {{HPC IO Benchmark Repository}},
    booktitle = {(Accessed: 29 July 2025)},
    url = {https://github.com/hpc/ior}
}

@inproceedings{Latham2025InitialAurora,
    title = {{Initial Experiences with DAOS Object Storage on Aurora}},
    year = {2025},
    booktitle = {Proceedings of the SC '24 Workshops of the International Conference on High Performance Computing, Network, Storage, and Analysis},
    author = {Latham, Rob and Ross, Robert B and Carns, Philip and Snyder, Shane and Harms, Kevin and Velusamy, Kaushik and Coffman, Paul and McPheeters, Gordon},
    pages = {1304--1310},
    series = {SC-W '24},
    publisher = {IEEE Press},
    url = {https://doi.org/10.1109/SCW63240.2024.00171},
    isbn = {9798350355543},
    doi = {10.1109/SCW63240.2024.00171}
}

@misc{IntroductionLibrados,
    title = {{Introduction to librados}},
    booktitle = {(Accessed: 29 July 2025)},
    url = {https://docs.ceph.com/en/latest/rados/api/librados-intro/}
}

@misc{IO500,
    title = {{IO500}},
    booktitle = {(Accessed: 13 December 2025)},
    url = {https://io500.org/about}
}

@misc{Bez2024IO500Lists,
    title = {{IO500 Lists}},
    year = {2024},
    booktitle = {(Accessed: 28 January 2024)},
    author = {Bez, Jean Luca and Dilger, Andreas and Hildebrand, Dean and Kunkel, Julian and Lofstead, Jay and Markomanolis, George},
    month = {1},
    url = {https://io500.org/releases}
}

@misc{LustrePractices,
    title = {{Lustre Best Practices}},
    booktitle = {(Accessed: 28 January 2025)},
    url = {https://www.nas.nasa.gov/hecc/support/kb/lustre-best-practices_226.html}
}

@misc{NEXTGenIOApplications,
    title = {{NEXTGenIO User Guide and Applications}},
    booktitle = {(Accessed: 29 July 2025)},
    url = {https://ngioproject.github.io/nextgenio-docs/html/index.html}
}

@inproceedings{Manubens2022PerformanceApproaches,
    title = {{Performance Comparison of DAOS and Lustre for Object Data Storage Approaches}},
    year = {2022},
    booktitle = {2022 IEEE/ACM International Parallel Data Systems Workshop (PDSW)},
    author = {Manubens, Nicolau and Smart, Simon D. and Quintino, Tiago and Jackson, Adrian},
    month = {11},
    pages = {7--12},
    publisher = {IEEE},
    isbn = {978-1-6654-7562-4},
    doi = {10.1109/PDSW56643.2022.00007}
}

@misc{Jackson2024PerformanceSharding,
    title = {{Performance Impacts of Replication and Sharding}},
    year = {2024},
    booktitle = {DAOS User Group Meeting},
    author = {Jackson, Adrian and Manubens, Nicolau},
    month = {11},
    url = {https://daos.io/event/daos-user-group-at-sc24-dug24}
}

@article{Haerder1983PrinciplesRecovery,
    title = {{Principles of transaction-oriented database recovery}},
    year = {1983},
    journal = {ACM Computing Surveys},
    author = {Haerder, Theo and Reuter, Andreas},
    number = {4},
    month = {12},
    pages = {287--317},
    volume = {15},
    doi = {10.1145/289.291},
    issn = {0360-0300}
}

@misc{Jackson2023ProfilingDAOS,
    title = {{Profiling and Identifying Bottlenecks in DAOS}},
    year = {2023},
    booktitle = {DAOS User Group Meeting},
    author = {Jackson, Adrian and Manubens, Nicolau},
    month = {11},
    url = {https://www.research.ed.ac.uk/en/activities/profiling-and-identifying-bottlenecks-in-daos}
}

@inproceedings{Manubens2024ReducingDAOS,
    title = {{Reducing the Impact of I/O Contention in Numerical Weather Prediction Workflows at Scale Using DAOS}},
    year = {2024},
    booktitle = {Proceedings of the Platform for Advanced Scientific Computing Conference},
    author = {Manubens, Nicolau and Smart, Simon D. and Danovaro, Emanuele and Quintino, Tiago and Jackson, Adrian},
    month = {6},
    pages = {1--12},
    publisher = {ACM},
    address = {New York, NY, USA},
    isbn = {9798400706394},
    doi = {10.1145/3659914.3659926}
}

@misc{SetCore,
    title = {{Set the number of threads per core}},
    booktitle = {(Accessed: 29 July 2025)},
    url = {https://cloud.google.com/compute/docs/instances/set-threads-per-core}
}

@misc{SpotVMs,
    title = {{Spot VMs}},
    booktitle = {(Accessed: 29 July 2025)},
    url = {https://cloud.google.com/solutions/spot-vms}
}

@misc{ECMWF2025SupercomputerFacility,
    title = {{Supercomputer facility}},
    year = {2025},
    booktitle = {(Accessed: 13 December 2025)},
    author = {{ECMWF}},
    month = {12},
    url = {https://www.ecmwf.int/en/computing/our-facilities/supercomputer-facility}
}

@article{Luttgau2018SurveyComputing,
    title = {{Survey of Storage Systems for High-Performance Computing}},
    year = {2018},
    journal = {Supercomputing Frontiers and Innovations},
    author = {L{\"{u}}ttgau, Jakob and Kuhn, Michael and Duwe, Kira and Alforov, Yevhen and Betke, Eugen and Kunkel, Julian and Ludwig, Thomas},
    number = {1},
    month = {3},
    volume = {5},
    doi = {10.14529/jsfi180103},
    issn = {23138734}
}

@misc{Braam2019TheArchitecture,
    title = {{The Lustre Storage Architecture}},
    year = {2019},
    author = {Braam, Peter},
    url = {httpsarxiv.orgabs1903.01955},
    arxivId = {1903.01955}
}

@article{Bauer2015ThePrediction,
    title = {{The quiet revolution of numerical weather prediction}},
    year = {2015},
    journal = {Nature},
    author = {Bauer, Peter and Thorpe, Alan and Brunet, Gilbert},
    number = {7567},
    month = {9},
    pages = {47--55},
    volume = {525},
    doi = {10.1038/nature14956},
    issn = {0028-0836}
}

@misc{TimelineComputers,
    title = {{Timeline Of Computer History - Computers}},
    booktitle = {(Accessed: 19 January 2024)},
    url = {https://www.computerhistory.org/timeline/computers/}
}

@misc{TimelineStorage,
    title = {{Timeline Of Computer History - Memory {\&} Storage}},
    booktitle = {(Accessed: 19 January 2024)},
    url = {https://www.computerhistory.org/timeline/memory-storage/}
}

@misc{WernerMeuerTOP500Lists,
    title = {{TOP500 Lists}},
    booktitle = {(Accessed: 19 January 2024)},
    author = {Werner Meuer, Hans and Strohmaier, Erich and Dongarra, Jack and Simon, Horst and Meuer, Martin},
    url = {https://top500.org/lists/top500/}
}

@inproceedings{Hennecke2023UnderstandingScalability,
    title = {{Understanding DAOS Storage Performance Scalability}},
    year = {2023},
    booktitle = {Proceedings of the HPC Asia 2023 Workshops},
    author = {Hennecke, Michael},
    pages = {1--14},
    series = {HPCAsia '23 Workshops},
    publisher = {Association for Computing Machinery},
    url = {https://doi.org/10.1145/3581576.3581577},
    address = {New York, NY, USA},
    isbn = {9781450399890},
    doi = {10.1145/3581576.3581577}
}

@inproceedings{Paul2020UnderstandingStatistics,
    title = {{Understanding HPC Application I/O Behavior Using System Level Statistics}},
    year = {2020},
    booktitle = {2020 IEEE 27th International Conference on High Performance Computing, Data, and Analytics (HiPC)},
    author = {Paul, Arnab K and Faaland, Olaf and Moody, Adam and Gonsiorowski, Elsa and Mohror, Kathryn and Butt, Ali R},
    number = {},
    pages = {202--211},
    volume = {},
    doi = {10.1109/HiPC50609.2020.00034}
}

@misc{GeorgeUnderstandingInternals,
    title = {{Understanding Lustre Internals}},
    booktitle = {(Accessed: 29 July 2025)},
    author = {George, Anjus and Mohr, Rick},
    url = {https://wiki.lustre.org/Understanding_Lustre_Internals}
}

@misc{WelcomeCeph,
    title = {{Welcome to Ceph}},
    booktitle = {(Accessed: 29 July 2025)},
    url = {https://docs.ceph.com/en/latest/}
}
    \end{spacing}

\chapter{Appendix A -- Performance Comparison of DAOS and Lustre for Object Data Storage Approaches}

\includepdf[pages=-]{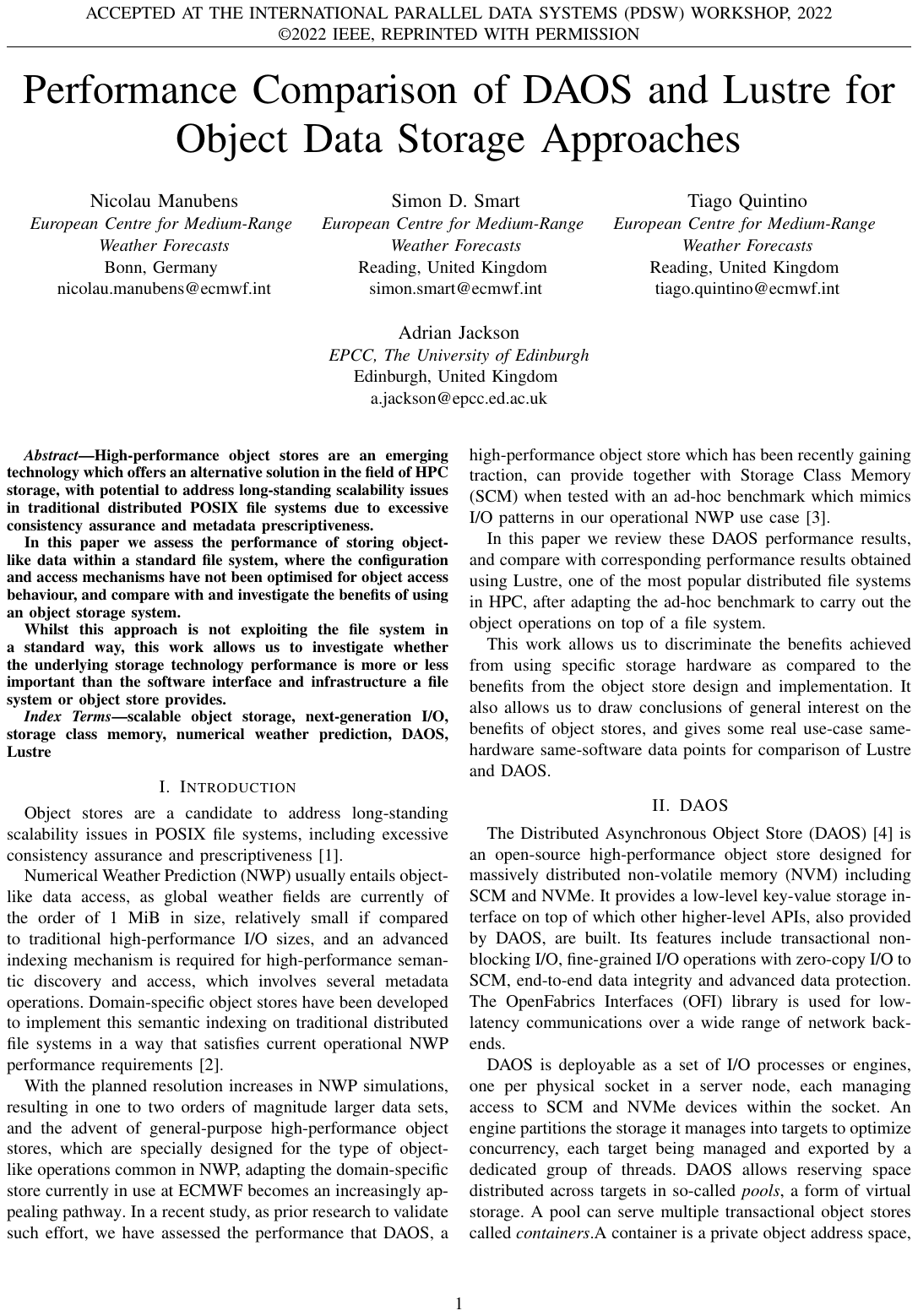}
\newpage\null\thispagestyle{plain}\newpage

    \chapter{Appendix B -- DAOS as HPC Storage: a View from Numerical Weather Prediction}

\includepdf[pages=-]{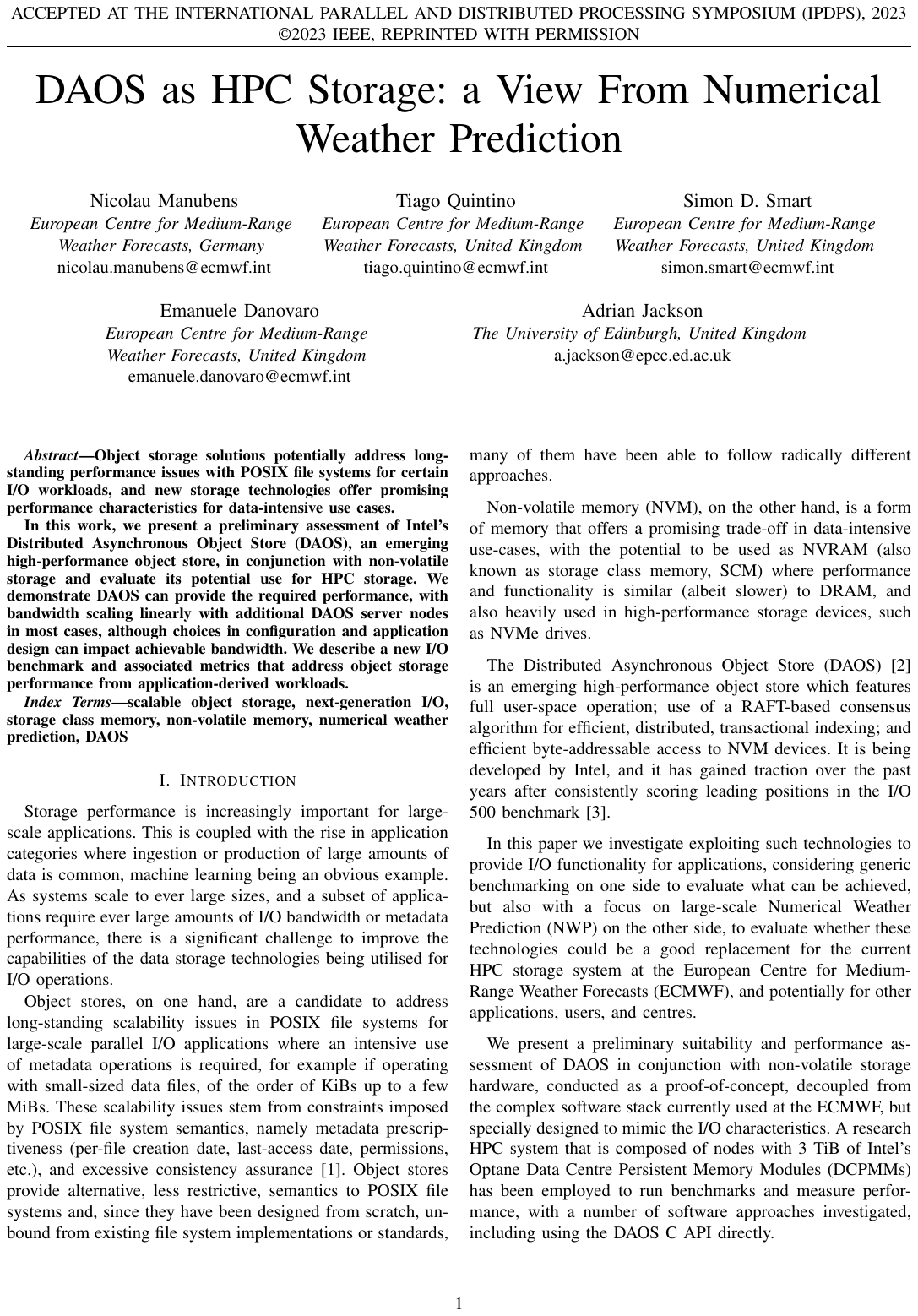}
\newpage\null\thispagestyle{plain}\newpage
    \chapter{Appendix C -- Reducing the Impact of I/O Contention in Numerical Weather Prediction Workflows at Scale Using DAOS}

\includepdf[pages=-]{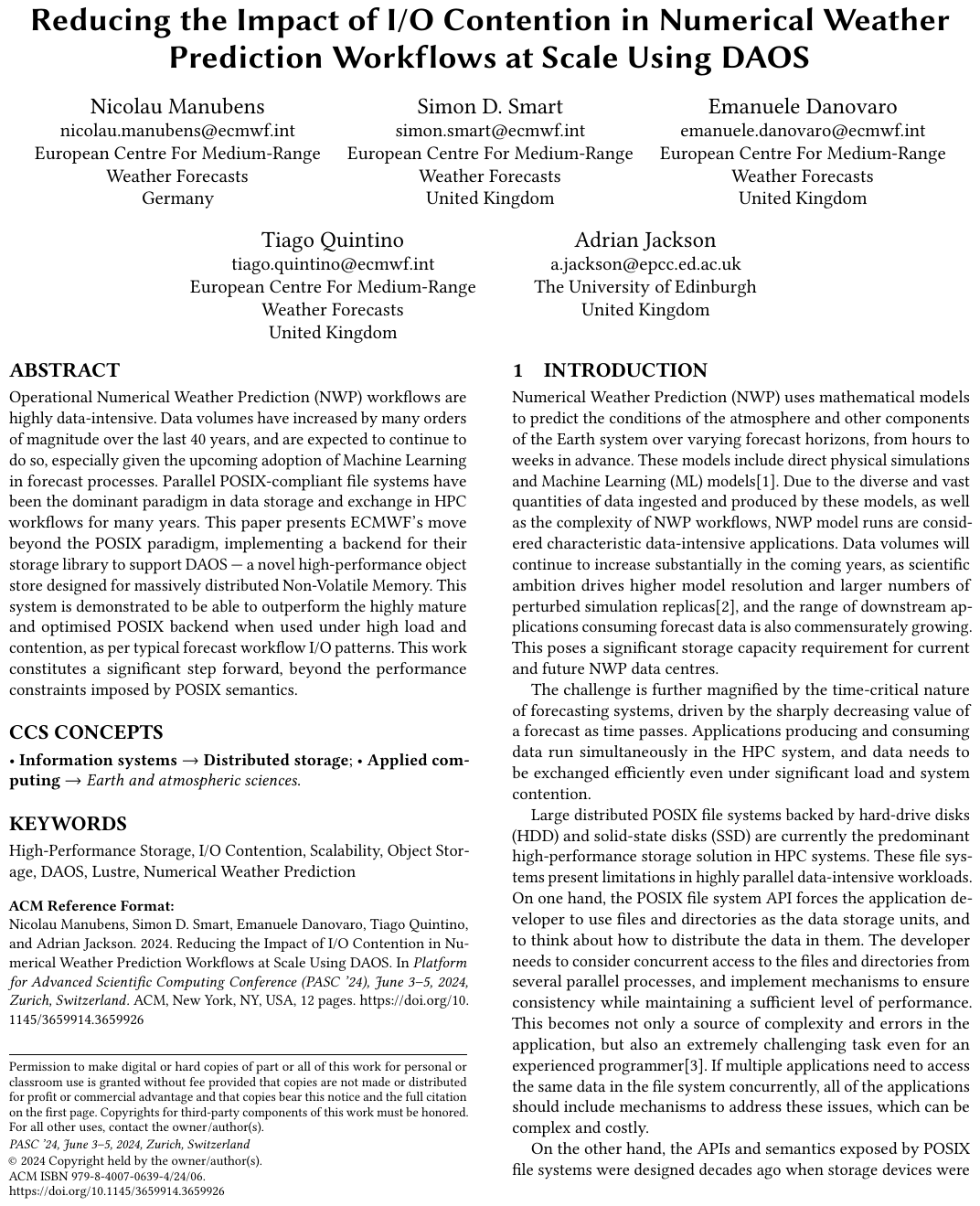}
\newpage\null\thispagestyle{plain}\newpage
    \chapter{Appendix D -- Exploring DAOS Interfaces and Performance}

\includepdf[pages=-]{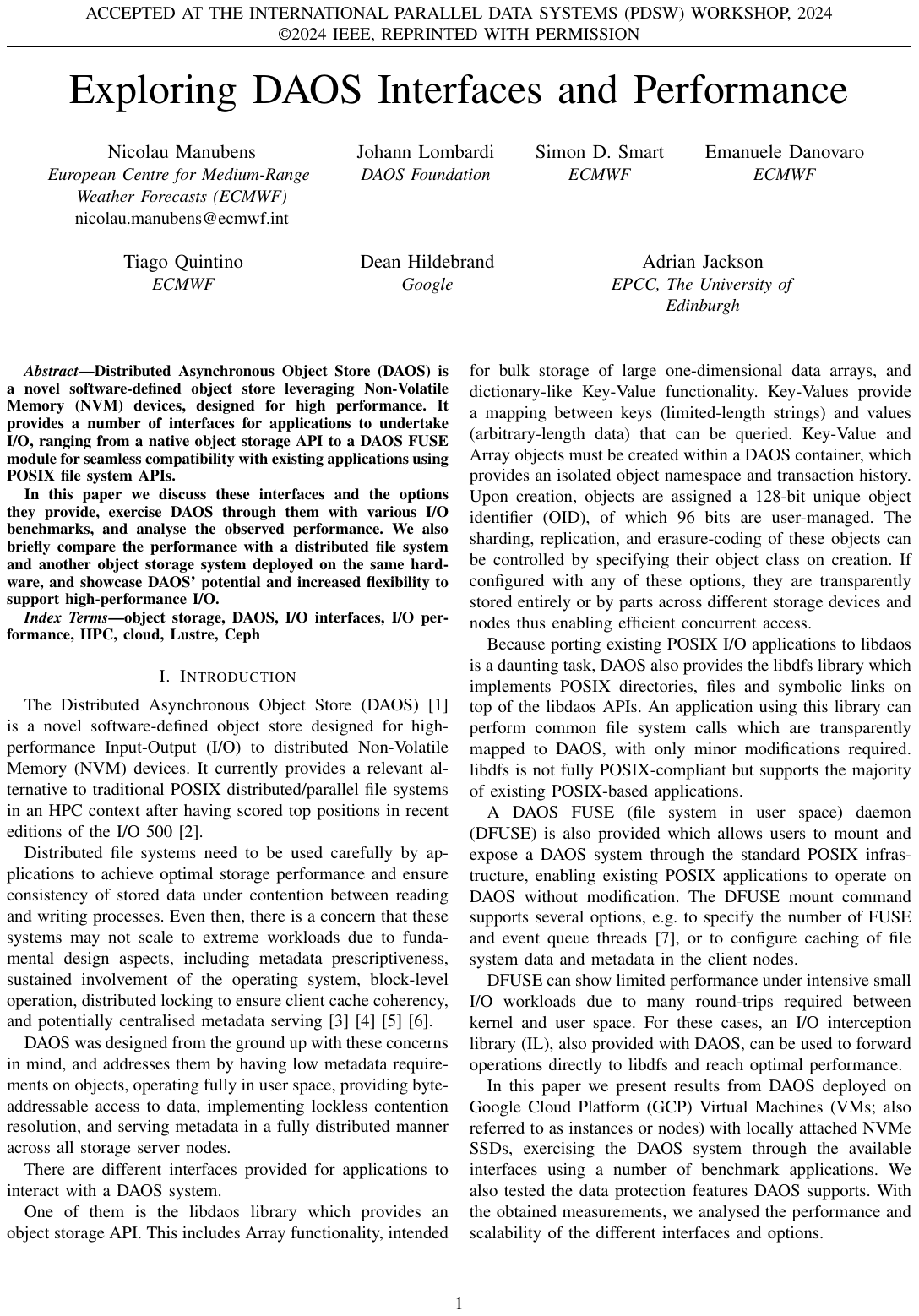}

\end{document}